\documentclass[12pt,a4paper]{article}    
\usepackage{jheppub}
\makeatletter
\renewcommand\@fpheader{\color{white}'}
\usepackage[T1]{fontenc}
\usepackage[utf8]{inputenc}
\usepackage[british]{babel}
\usepackage{amssymb,amstext,amsthm,
bbm,mathrsfs,amscd,bbold,pifont, tensor}
\usepackage{mathtools}
\usepackage{array}
\usepackage[svgnames,table]{xcolor}
\usepackage[T1]{fontenc}
\usepackage[utf8]{inputenc}
\usepackage{enumitem}
\usepackage{subcaption}
\usepackage{cleveref}
\crefformat{footnote}{#2\footnotemark[#1]#3}

\usepackage{booktabs,caption}
\usepackage{verbatim}
\usepackage{braket}
\usepackage{graphicx}
\graphicspath{{figures/}}
\usepackage{mdframed}
\usepackage{tikz,pgfplots}
\usetikzlibrary{calc,arrows,cd}

\theoremstyle{plain}

\theoremstyle{definition}

\ifdefined\C
\renewcommand{\C}{\boldsymbol{C}}
\else
\newcommand{\C}{\boldsymbol{C}}
\fi
\newcommand{\J}{\boldsymbol{J}}

\newcommand{\be}{\begin{equation}}
\newcommand{\ee}{\end{equation}}

\definecolor{gris}{rgb}{0.5,0.5,0.5}
\definecolor{darkgreen}{rgb}{0.0,0.5,0.0}

\allowdisplaybreaks[0]
\numberwithin{equation}{section}
\usetikzlibrary{shapes,decorations}
\pgfdeclarelayer{nodelayer}
\pgfdeclarelayer{edgelayer}
\pgfsetlayers{edgelayer,nodelayer,main}
\tikzstyle{ghost}=[fill=none, draw=none, shape=circle]
\tikzstyle{cross}=[fill=none, draw=black, shape=circle]
\tikzstyle{grey line}=[-, draw={rgb,255: red,128; green,128; blue,128}]
\tikzstyle{blue line}=[-, draw=blue, fill={rgb,255: red,162; green,246; blue,255}]
\tikzstyle{dash blue}=[-, draw=blue, dashed]
\tikzstyle{dash grey}=[-, draw={rgb,255: red,128; green,128; blue,128}, dashed]
\tikzstyle{thick black line}=[-, thick]
\tikzstyle{blue fill}=[-, draw=none, fill={rgb,255: red,126; green,214; blue,255}]
\tikzstyle{black line}=[-]
\tikzstyle{red fill}=[-, fill={rgb,255: red,255; green,162; blue,164}, draw={rgb,255: red,255; green,0; blue,4}]
\tikzstyle{new edge style 0}=[-, draw=none, fill={rgb,255: red,255; green,162; blue,164}]
\pgfplotsset{compat=1.16}

\begin{document}

\title{
\bf{Black holes and 2-group symmetry from blackfolds in minimal supergravity
}}

\author[1,2,3,4]{Jay Armas,}
\author[5,6]{Giorgos Batzios,}
\author[7]{Pedro Vicente Marto,}
\author[8,9]{Chiara Toldo}

\affiliation[1]{Institute for Theoretical Physics, University of Amsterdam, \\ 1090 GL Amsterdam, The Netherlands}
\affiliation[2]{Dutch Institute for Emergent Phenomena,\\ 
1090 GL Amsterdam, The Netherlands}
\affiliation[3]{Institute for Advanced Study, University of Amsterdam,\\ 
Oude Turfmarkt 147, 1012 GC Amsterdam, The Netherlands}
\affiliation[4]{Niels Bohr International Academy, The Niels Bohr Institute,\\
University of Copenhagen, Blegdamsvej 17, DK-2100 Copenhagen \O, Denmark}
\affiliation[5]{Dipartimento di Fisica, Università di Genova, via Dodecaneso 33, I-16146, Genova, Italy}
\affiliation[6]{I.N.F.N. - Sezione di Genova, via Dodecaneso 33, I-16146, Genova, Italy}

\affiliation[7]{Institute for Theoretical Physics, Utrecht University, Princetonplein 5, 3584 CC Utrecht, The Netherlands}
\affiliation[8]{Physique Théorique et Mathématique and International Solvay Institutes,\\
Université Libre de Bruxelles (ULB), C.P. 231, 1050 Brussels, Belgium.}
\affiliation[9]{Departament de Matem\`atiques \& SYMCREA, Universitat Polit\`ecnica de Catalunya, Av. Diagonal, 647, 08028 Barcelona,  Spain}


\emailAdd{j.armas@uva.nl}
\emailAdd{g.batzios@uva.nl}
\emailAdd{p.vicentemarto@uu.nl}
\emailAdd{chiara.toldo@upc.edu}

\abstract{We investigate the black hole phase diagram of five-dimensional minimal supergravity using the blackfold approach and the framework of generalized symmetries. Starting from supergravity coupled to probe sources, we derive the 2-group (non-)conservation laws, construct the equilibrium action for a two-dimensional 2-group fluid, and compute the associated global conserved charges. We then match the thermodynamics of the general non-extremal black string in the ungauged theory to that of this fluid, organizing the blackfold effective theory and its solutions in terms of symmetry invariants. In the non-extremal ungauged sector, we perturbatively construct new black rings, Saturns, di-rings, and bi-rings, including a black Saturn with seven independent parameters. Taking appropriate extremal limits, we identify several extremal branches, some previously unexplored and others anticipated, as well as BPS branches that approximate known supersymmetric families. Finally, we apply the same framework to construct extremal black rings in AdS$_5$, identifying three regular branches that are promising candidates for an exact completion.}

\maketitle

\section{Introduction}
\textcolor{black}{Higher-dimensional gravity reveals physical possibilities absent in four dimensions.} Certain geometric identities only hold in low dimensions, and their absence or modification in higher dimensions permit new interactions and physically admissible states. Primary examples of this observation are a number of theorems, commonly referred to as uniqueness theorems, that severely constrain black hole solutions in four spacetime dimensions \cite{PhysRevLett.26.331, PhysRevLett.34.905, Hawking:1971vc}, while many of these constraints relax in higher dimensions. For instance, Hawking's restriction to spherical horizon topology for four-dimensional asymptotically flat black holes relies on applying the Gauss-Bonnet theorem to their two-dimensional horizon cross-sections \cite{Hawking:1971vc}. This argument does not extend directly to higher dimensions. Under the appropriate energy and horizon hypotheses, the higher-dimensional topology theorem instead requires horizon cross-sections to admit a metric of positive scalar curvature, a condition compatible with a broader range of horizon topologies \cite{Galloway:2005mf}. In parallel, increasing the dimensionality of spacetime allows new consistent interactions. For instance, in five dimensions a cubic Chern-Simons interaction can appear.

The historical example of a black hole with non-spherical horizon topology is the one of \cite{Emparan:2001wn,Pomeransky:2006bd} with horizon topology $\mathbb{S}^1\times \mathbb{S}^2$, hence the name "black ring", constructed in five-dimensional vacuum gravity. A black ring can surround a central black hole with spherical horizon topology, forming a black Saturn with two disconnected horizons \cite{Elvang:2007rd, armas2026doublyspinningblacksaturns}. More exotic horizon topology classes manifest themselves via concentric \cite{Iguchi:2007is, Evslin_2009} and orthogonal \cite{Izumi:2007qx}, as well as multiple rings \cite{Emparan_2010}. Adding a Maxwell field provides new ways for balancing a black hole, enlarges the phase space, and enriches the thermodynamics \cite{PhysRevD.35.455,Emparan_2004, Yazadjiev_2007, Copsey:2005dipole,Yazadjiev_2008}\footnote{For example, electrostatic repulsion can balance two extremal black holes with spherical horizon \cite{PhysRevD.35.455}, yielding the disconnected topology $\mathbb{S}^3\sqcup\mathbb{S}^3$.}. Adding a cubic Chern-Simons term introduces a self-interaction of the gauge field that couples electric and magnetic charges, with interesting implications from a symmetry perspective, as we explain in the core of this paper. The resulting Einstein-Maxwell-Chern-Simons theory contains the minimal bosonic field content and interactions admitting a supersymmetric completion\footnote{Supersymmetry fixes, in particular, the coefficient of the Chern-Simons term.}, leading to five-dimensional minimal supergravity. Black holes of this theory is the main subject of the present paper.

Regular, stationary, asymptotically flat black holes in five-dimensional minimal supergravity can carry four independent asymptotic conserved charges: the mass, two angular momenta, and an electric charge. Constructing a regular black ring carrying all four charges independently has proved challenging and only recently an explicit solution was obtained in \cite{Suzuki:2024generalring} (see also \cite{Feldman_2014}). Fixing the horizon topology and asymptotic charges does not, however, generally determine a unique solution, even in vacuum \cite{Emparan:2001wn, Elvang:2007rd}. Local magnetic dipole charges enter the thermodynamics, and can distinguish solutions with identical asymptotic charges, providing further instances of continuous or discrete non-uniqueness \cite{Emparan_2004,Yazadjiev_2007}. This rich phase structure, particularly in the non-extremal sector, has motivated efforts to classify black holes in terms of appropriate symmetry invariants \cite{Gibbons_2002,Hollands_2008, Figueras_2010,Harmark_2009,Tomizawa_2009,Tomizawa_2010, Armas_2010,Hollands:2012xy,Kunduri_2014,Armas_2015, khuri2019stationaryvacuumblackholes, alaee2023existenceuniquenessstationarysolutions}.
For example, in the non-extremal, biaxisymmetric sector, the results of \cite{alaee2023existenceuniquenessstationarysolutions} suggest that a balanced, asymptotically flat black Saturn with standard exterior topology could form a nine-parameter family, with four asymptotic conserved charges.

In this paper, we address the existence of, and perturbatively construct, new black hole solutions in five-dimensional minimal supergravity, by combining the blackfold approach \cite{Emparan:2009at, Emparan:2011hg, Camps:2012hw, Armas:2016mes} with the framework of generalized symmetries \cite{Baez_2010, Gaiotto_2015}. Our results extend our knowledge of the black hole phase diagram within various horizon topology classes, and towards various directions. As an illustrative example, we present a construction of an asymptotically flat, biaxial black Saturn with standard exterior topology, obtained through a perturbative treatment of the field equations. The solution possesses four non-vanishing independent conserved charges and one magnetic dipole charge, and it is characterized by seven independent parameters. In that sense, to date, it represents the most general biaxial black Saturn solution of five-dimensional minimal supergravity recorded in the literature.


In the BPS sector of five-dimensional minimal supergravity substantial progress has been made with respect to classifying and constructing asymptotically flat supersymmetric black objects. Following the classification of all possible near-horizon geometries of bosonic supersymmetric solutions of \cite{Reall:2002bh}, and the more general classification of \cite{Gauntlett:2002nw}, the first asymptotically flat supersymmetric black ring was found in \cite{Elvang:2004rt}. This solution is specified by its electric charge and two independent angular momenta, and also carries a magnetic dipole charge. It admits a higher-dimensional brane description in terms of three-charge supertubes \cite{Elvang:2004ds,Elvang:2004xi,Bena:2004de}. In an appropriate infinite-radius limit, it reduces to the supersymmetric black string previously obtained in \cite{Bena:2004wv}, whose construction built on earlier work on supertubes \cite{Bena:2004wt,Mateos:2001qs}. The black ring construction was subsequently extended to supersymmetric configurations involving concentric rings and black Saturns \cite{Gauntlett:2004wh,Gauntlett:2004qy}. Regular supersymmetric black lenses with horizon topology $L(n,1)=\mathbb{S}^3/\mathbb{Z}_n$ were later constructed \cite{PhysRevLett.113.211101, Tomizawa_2016}, realizing further topologies compatible with the general restrictions of \cite{Galloway:2005mf}. In this work, we obtain supersymmetric black rings, Saturns, di-rings, and bi-rings at leading order in the blackfold expansion by taking suitable extremal limits of more general finite-temperature branches. Comparisons with the corresponding exact solutions show excellent agreement within the regime of validity of our approximation.

\begin{figure}[h]
 \begin{center}
 \includegraphics[width=\linewidth]{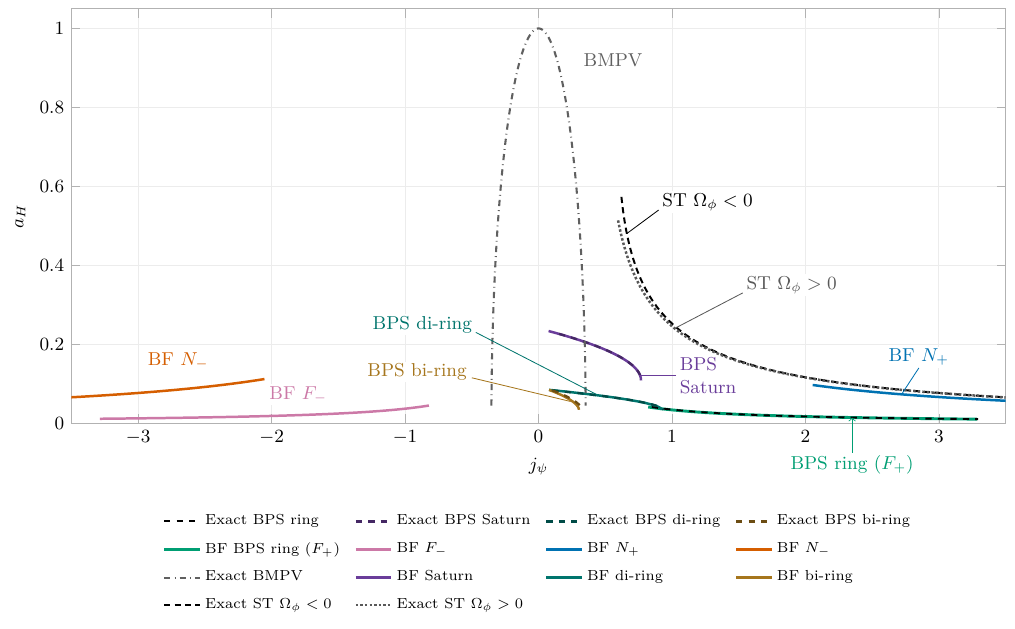}
 \end{center}
 \caption{Overview of extremal asymptotically flat black rings and Saturns considered in this paper. The reduced total horizon area $a_H$ is plotted against the signed reduced angular momentum $j_\psi$, defined in
\eqref{eq:phase_overview_variables}. Solid coloured curves
show blackfold constructions, while dashed, dotted and
dash-dotted curves denote exact solutions, including the
BMPV black hole, BPS rings and composite
configurations, and the extremal Suzuki--Tomizawa (ST) rings.
The four isolated blackfold ring branches are the BPS
finite-velocity branch $F_+$ (green), the non-BPS
finite-velocity branch $F_-$ (pink), the charged null-wave
branch $N_+$ (blue), and the purely magnetic null-wave
branch $N_-$ (orange). They arise from extremal limits of
the string of \cite{Compere:2010fm}, as described in
Sec.~\ref{sec:flat_ring_extremal}.
Blackfold constructions of BPS Saturns, di-rings
and bi-rings are also shown.
The selected curves do not all have the same conserved
charges, so their relative areas do not establish
thermodynamic preference.
Parameter choices are given in
Sec.~\ref{sec:phase_extremal} and, for the BPS composites,
in the caption of Figure~\ref{fig:phase_bps_comparison}.}
 \label{fig:extremal_ph_sp}
\end{figure}

Beyond the supersymmetric sector, classification efforts have focused primarily on near-horizon geometries \cite{Kunduri_2009, Hollands_2010, Kunduri_2011, Kunduri_2013, Alaee_2019}. Our (extremal) solutions include thin-ring horizons whose local near-horizon geometry is inherited, at least in our leading order analysis\footnote{\textcolor{black}{The point is that}, for the non-BPS solutions constructed here, the leading approximation does not establish whether this local form persists once finite-radius corrections are included. Such corrections can affect the near-horizon geometry even at extremality, as illustrated by the dipole rings discussed in \cite{Kunduri_2011}. For our BPS solutions, on the other hand, this form is protected by supersymmetry \cite{Reall:2002bh}.}, from the near-horizon geometry of the black string of \cite{Compere:2010fm}. Their near-horizon geometry is therefore locally $\mathrm{AdS}_3\times\mathbb{S}^2$, placing them within the class of near-horizon solutions studied in \cite{Kunduri_2011, Kunduri_2013, Alaee_2019}.  A representative picture of our main asymptotically flat extremal results, including the BPS sector, is provided in Figure \ref{fig:extremal_ph_sp}.

Five-dimensional minimal $\mathcal{N}=2$ gauged supergravity is obtained by gauging a $U(1)$ subgroup of the $SU(2)$ R-symmetry of the parent minimal theory. The resulting theory has a negative cosmological constant term and admits an AdS$_5$ vacuum, but its black hole phase diagram remains less well understood than that of its ungauged counterpart. In the supersymmetric sector, regular asymptotically AdS$_5$ black rings are excluded by near-horizon classifications, first under the assumption of two rotational symmetries \cite{Kunduri:2006uh}, and subsequently without this assumption \cite{Grover:2013hja}\footnote{Paper \cite{Kunduri:2007qy} constructed near-horizon solutions whose spatial cross-sections have topology \(\mathbb{S}^1 \times \mathbb{S}^2\). These solutions involve constant scalar fields and are admissible only for restricted choices of the parameters defining the moduli space; the minimal gauged supergravity theory, in particular, does not admit such configurations.}. 
\textcolor{black}{Within the blackfold approach, an obstruction was found by bending a static supersymmetric
string into a balanced ring \cite{Caldarelli:2008pz}. Here we show in a more complete leading order analysis that
using a boosted string as the seed yields new balanced black ring configurations,
which do not preserve supersymmetry, in agreement with \cite{Kunduri:2006uh, Grover:2013hja}.} Furthermore, a separate non-existence theorem excludes biaxisymmetric extremal black rings in five-dimensional pure Einstein gravity with a nonzero cosmological constant \cite{Lucietti_2018}. The same work conjectures the absence of biaxisymmetric charged extremal AdS rings, including in minimal gauged supergravity, further motivating the investigation of the extremal sector. In particular, in this paper, we find four balanced extremal ring branches, three out of which provide promising candidates for regular, non-supersymmetric AdS$_5$ completions.

The richness of the higher-dimensional black hole phase diagram, together with the difficulty of constructing exact solutions, motivates the use of perturbative methods.
From the perspective of solving the gravitational equations, the blackfold approach \cite{Emparan:2009at, Emparan:2011hg, Camps:2012hw, Armas:2016mes} provides a method for exploring the black hole phase diagram by exploiting a hierarchy of scales. The latter represents a generic feature of black hole physics in higher dimensions. Black rings and Saturns in AdS illustrate the value of this method: constructing exact solutions of black rings and Saturns in AdS space proved to be generally hard, and while numerical solutions include \cite{Figueras:2014dta}, the blackfold approach has enabled a range of analytical constructions \cite{Caldarelli_2011, Armas:2010hz, Armas:2015qsv, Armas:2024dtq}. More broadly, the blackfold approach has been proven to be a powerful method for constructing previously inaccessible D/M-brane configurations in supergravity that address a variety of problems in string theory and holography \cite{Armas_2012, Armas_2019, PhysRevLett.122.181601, Nguyen_2022, Armas:2022bkh}. For many applications, the underlying idea is to model the black hole as a \textit{black} brane wrapped around a compact submani\textit{fold} of spacetime. Near the horizon, locally, the black hole is approximated by a uniform black brane solution and at the same time, globally, the brane is bent into a submanifold of the background spacetime via its embedding maps. The field equations, organized in a derivative expansion, provide an effective theory on the brane's worldvolume. Solving the equations of this effective theory on a given background is at least necessary (and, in all cases considered in the literature so far, sufficient) for the existence of a regular deformed brane solution at first derivative order, which asymptotes to this given background \cite{Emparan:2007wm, Camps:2012hw, Armas:2016mes, Nguyen_2022, PhysRevD.94.026009}. This perturbative method is natural when the intrinsic scales of the brane, such as its energy and charge density, can be parametrically smaller than its geometric scales, such as its extrinsic curvature, and the scales of the background.

Symmetries are the primary organizing principle of effective field theories. We show, following \cite{Armas:2024caa}, that five-dimensional minimal supergravity coupled to probe point particles and strings furnishes the conservation laws of a 2-group symmetry\footnote{For recent discussions of higher-form and higher-group symmetries in supergravity, see \cite{barbagallo2025higherformsymmetriessupergravityscalar, Intriligator:2026xul}. In Section~\ref{sec: Ward}, we briefly discuss how the local symmetry structure used here and in \cite{Armas:2024caa} relates to the global issues raised in \cite{Intriligator:2026xul}.}. The "mixing" between electric and magnetic charge induced by the Chern-Simons term is essential here. We find that the 2-group symmetry is inherited to the effective worldsheet theory of a string embedded in a minimal supergravity background. We thus formulate the equilibrium action of the non-extremal string of \cite{Compere:2010fm} as a 2-group invariant fluid action. This, in turn, allows a clear organization of the effective theory and its solutions in terms of symmetry invariants. Matching this effective fluid description to the thermodynamics of the black string considered here, allows us to use this framework to provide general formulae for the asymptotic conserved charges of stationary black holes in minimal supergravity. In this way, the generalized symmetry structure connects the local dynamics of charged strings to the global properties of the black hole families explored in this paper.

\vspace{3mm}

This paper is organized as follows: in Section \ref{sec:highergroup} we derive the 2-group conservation laws directly from the coupled minimal supergravity theory. We then discuss aspects of the local structure of such symmetries, paying attention to subtleties related to gauge invariance, and explain their implementation in classical supergravity. The effective theory on the worldsheet is formulated as a 2-group invariant theory for an effective fluid. We finish this section with the computation of the global conserved charges that arise from the conservation laws of a higher-group. Section \ref{ch: THERMOGENERAL} is devoted to the thermodynamic description of the non-extremal black string of \cite{Compere:2010fm}, including its supersymmetric and extremal limits, as a 2-group fluid. This string will be the probe object that will be bent into a black ring. A comprehensive analysis of the validity of our method is also included. In Section \ref{sec:constructed_horizons} we show how to apply the formalism developed in the previous sections to balance black rings and Saturns in asymptotically flat space, and in appropriate limits recover known solutions whenever possible. We analyze the thermodynamic properties of our solutions and provide the relevant phase diagrams. Section \ref{sec:ads_rings} deals instead with new solutions of extremal black rings in AdS$_5$, pointing out certain limitations in finding solutions preserving supersymmetry. We conclude the main body of our paper with an overview of our results, mentioning a few future directions. We also provide appendices that detail the computations needed for important results in the main text. In particular, appendix \ref{app:nonext_multirings} presents the computation and the thermodynamic analysis of our (non-)extremal bi- and di- ring solutions.

\section{The continuous 2-group symmetry of five-dimensional minimal supergravity} \label{sec:highergroup}
This section develops the 2-group description of strings coupled to five-dimensional minimal supergravity. Section~\ref{sec: sugra_ward_identities} derives the 2-group Ward identities from the supergravity field equations. Section~\ref{sec: Ward} treats the variational formulation, and explains the role of the dual two-form potential in supergravity. Section~\ref{sec:equilibrium_action} constructs the equilibrium action and derives the finite-temperature constitutive relations. Finally, Section~\ref{sec: conserved_charges} establishes the global conserved charges, first law and Smarr relations in the presence of background fluxes.

\subsection{The 2-group (non-)conservation laws from supergravity}\label{sec: sugra_ward_identities}
As explained in \cite{Armas:2024caa}, the presence of Chern-Simons terms in a given supergravity theory leads to a continuous higher-group symmetry structure, when the theory is coupled to spatially extended probe sources. Here, we will witness an example where this mechanism takes place, namely five-dimensional minimal supergravity, following the reasoning of \cite{Armas:2016mes, Armas:2024caa}. Another primary example of this mechanism is eleven-dimensional supergravity \cite{Armas:2024caa}.\\
\\
 The bosonic truncation of minimal five-dimensional ungauged supergravity has the action 
\begin{equation}\label{eq:L}
    S=\frac{1}{16\pi G_5}\int\left(R\star 1-\frac{1}{2}\star F\wedge F+\frac{\kappa}{3}F\wedge F\wedge A\right)\,,
\end{equation}
where, following the conventions of \cite{Compere:2010fm}, $\kappa=\frac{1}{\sqrt{3}}$ is the structure constant\footnote{Note that here we are interested in the local structure of the 2-group and do not treat global issues. For an ordinary, compact $U(1)^{(0)} \times_{\kappa}U(1)^{(1)}$ for example, the (appropriately normalized) structure constant is further constrained, for instance for unit flux of the 2-form field strength it must be an integer \cite{C_rdova_2019}. See also our discussion in the next section.} that will enter the 2-group conservation laws and constitutive relations. $A$ is a $U(1)$ gauge field, $F=dA$ its field strength satisfying the Bianchi identity $dF=0$, and $\star$ is the Hodge dual with respect to the five-dimensional metric. Since the bosonic truncation of the gauged theory differs from the one of the ungauged theory just by a cosmological constant term, meaning at the level of equations of motion just by a constant, an identical derivation with the one below trivially shows that the gauged theory also possesses the same 2-group local symmetry structure.\\
\\
Let us now introduce arbitrary sources that modify the equations of motion and Bianchi identities of the theory. We use mostly-plus signature and the Hodge convention $\alpha_p\wedge\star\beta_p=(\alpha_p\cdot\beta_p)\star1$, where $\alpha_p\cdot\beta_p=\alpha_{\mu_1\ldots\mu_p}\beta^{\mu_1\ldots\mu_p}/p!$. The signs of the electric and magnetic currents are fixed by the source variations in Section~\ref{sec: Ward}. For the equation of motion and Bianchi identity of the gauge field, we write
\begin{equation}\label{eq:EOM}
    \text{d}\star F-\kappa\, F\wedge F=16\pi G_5\star \boldsymbol{J}\,.
\end{equation}
\begin{equation}\label{eq:Bianchi}
    \text{d}F=-16\pi G_5\star  \boldsymbol{\mathcal{J}}\,.
\end{equation}
Acting with the exterior derivative on equations \eqref{eq:EOM} and \eqref{eq:Bianchi} and using the latter into the former, we obtain the current (non-)conservation equations of this theory:
\begin{equation}\label{eq: 2-grouplaws}
        \text{d}\star \boldsymbol{J}=2\kappa\, \star\boldsymbol{\mathcal{J}}\wedge F \quad, \quad
        \text{d}\star\boldsymbol{\mathcal{J}}=0~.
\end{equation}
Derived in this way, equations \eqref{eq: 2-grouplaws} represent general identities that the electric source $\boldsymbol{\J}$ and the magnetic source $\boldsymbol{\mathcal{J}}$ must obey. We now associate these sources to probes. For the case of five-dimensional supergravity, these sources can be associated to probe point particles and strings. Now taking the probe limit of \eqref{eq: 2-grouplaws}, we obtain the equations of motion for the currents $\boldsymbol{J}$ and $\boldsymbol{\mathcal{J}}$ in an asymptotic background of the five-dimensional metric $g_{\mu \nu}$ and gauge field $A$. Equations \eqref{eq: 2-grouplaws} are then precisely the (non)-conservation laws defining the local structure of the Abelian $U(1)^{(0)} \times_{\kappa}U(1)^{(1)}$ 2-group, see \cite{Armas:2024caa}. Point particles are charged under the $U(1)^{(0)}$ symmetry, while strings are charged under the $U(1)^{(1)}$ symmetry. Equations \eqref{eq: 2-grouplaws} say that point-particle charge is sourced by the conserved string charge in the presence of a non-vanishing background field strength $F$.\\
\\
Let us now look at the equation that a stress tensor $\boldsymbol{T}^{\mu \nu}$ sourcing the Einstein equations obeys. The sourced equation of motion for the five-dimensional metric reads
\begin{equation}\label{eq: sourcedEINSTEIN}
    G_{\mu \nu}-8\pi G_{5}\boldsymbol{\Theta}_{\mu \nu}=8\pi G_{5}\boldsymbol{T}_{\mu \nu}~,
\end{equation}
where $\Theta_{\mu \nu}$ denotes the "matter" contribution, in our case just the standard Maxwell contribution of the gauge field, since the Chern-Simons term is independent of the metric. $G_{\mu \nu}$ is the Einstein tensor. One now acts with the covariant derivative on \eqref{eq: sourcedEINSTEIN}, and upon using the Bianchi identity $\nabla_{\mu}G^{\mu\nu}=0$ as well as the sourced equation of motion and Bianchi identity for the gauge field, one finds
\begin{equation}\label{eq: 2-groupstresslaw}
    \nabla_{\mu}\boldsymbol{T}^{\mu \nu}=F^{\nu \mu}\boldsymbol{J}_{\mu}+\frac{1}{2}\mathcal{F}^{\nu \mu \sigma}\boldsymbol{\mathcal{J}}_{\mu \sigma}~,
\end{equation}
where $\mathcal{F}=-\star F$. As previously, now taking the probe limit of \eqref{eq: 2-groupstresslaw} the five-dimensional metric and $F, \mathcal{F}$ are understood as background fields. Equation \eqref{eq: 2-groupstresslaw} has precisely the form of the stress tensor non-conservation equation of the $U(1)^{(0)} \times_{\kappa}U(1)^{(1)}$ 2-group \cite{Armas:2024caa}.  We have thus derived the 2-group Ward identities, namely \eqref{eq: 2-grouplaws} and \eqref{eq: 2-groupstresslaw}, from the coupled supergravity theory. Note, however, that on-shell supergravity restricts the permissible backgrounds via the relation $\mathcal{F}=-\star F$.\\
\\
Throughout this paper we are interested in sources which are localized on the worldsheet of a probe string.  We emphasize however that \eqref{eq: 2-grouplaws}, \eqref{eq: 2-groupstresslaw} can accommodate far more general choices. Here, we thus specialize the general sources in equations \eqref{eq: 2-grouplaws} and \eqref{eq: 2-groupstresslaw} to a single worldsheet geometry carrying both electric and magnetic charge. To this end, we introduce the spacetime scalar and worldsheet reparametrization invariant delta function expressing localization on the worldsheet
\begin{equation}
\tilde{\delta}_{\Sigma_2}(x)
=
\int_{\Sigma_2} d^2\sigma\,\sqrt{-\gamma(\sigma)}\,
\frac{\delta^{(D)}\!\left(x-X(\sigma)\right)}
     {\sqrt{-g(x)}} ~,
\end{equation}
where $x^{\mu}$ are co-ordinates in the background spacetime $\mathcal{M}$ with metric $g_{\mu \nu}$ whose determinant is denoted by $g$. The embedding maps $X^{\mu}:\Sigma_{2} \rightarrow \mathcal{M}$ specify the position of the worldsheet $\Sigma_{2}$ parametrized by $\sigma^{a}$, in the background spacetime $\mathcal{M}$. We have also defined the induced metric on the worldsheet $\gamma_{ab}(\sigma)=g_{\mu\nu}\!\left(X(\sigma)\right)
\partial_a X^\mu(\sigma)\partial_b X^\nu(\sigma)$ whose determinant is denoted by $\gamma$. With $\delta ^{(D)}(x-X)$ we denote the ordinary $D$-dimensional Dirac distribution. The dimension of the background spacetime for us here is set to $D=5$. The specialization of the stress tensor $\boldsymbol{T}^{\mu \nu}$ then reads in a compact notation
\begin{equation}\label{eq: multipole}
    \boldsymbol{T}^{\mu \nu}=T^{\mu \nu}\tilde{\delta}_{\Sigma_2}+\mathcal{O}(\partial)~,
\end{equation}
where worldsheet integration is understood. The symbol $\mathcal{O}(\partial)$ here denotes corrections involving derivatives of the localized delta function $\tilde{\delta}_{\Sigma_2}$. It is there to emphasize that, within the choice of localized sources made here, one can formally write a multipole expansion for each such source \cite{Armas_2013ele}. In this paper we work at leading derivative order and truncate the expansion so that it includes only the leading (monopole) contribution, where no derivatives of the delta function appear.  Similarly, for the p-form sources we write 
\begin{equation}\label{eq: multipole_currents}
    \boldsymbol{J}^{\mu}=J^{\mu}\tilde{\delta}_{\Sigma_2}+\mathcal{O}(\partial)\quad,\quad \boldsymbol{\mathcal{J}}^{\mu \nu}= J^{\mu \nu}\tilde{\delta}_{\Sigma_2}+\mathcal{O}(\partial)~,
\end{equation}
where we truncate again at monopole order. Furthermore, the continuity equations \eqref{eq: 2-grouplaws}, \eqref{eq: 2-groupstresslaw} imply that at leading order the currents $T^{\mu \nu}, J, \mathcal{J}$ are purely tangential to the worldsheet, for instance $T^{\mu \nu}=T^{a b}\partial_{a}X^{\mu} \partial_{b} X^{\nu}$ \cite{Vasili__2006, Armas_2013}.\\
\\
For our choice of sources and at the probe level, equations \eqref{eq: 2-grouplaws}, \eqref{eq: 2-groupstresslaw} constitute the equations of motion of a probe string in an asymptotic background. A probe brane can accommodate degrees of freedom on its worldvolume, and at the same time move and bend in the background spacetime. To study these two physical sectors we project \eqref{eq: 2-grouplaws} and \eqref{eq: 2-groupstresslaw} to directions parallel and orthogonal to the worldsheet. We refer to \cite{carter2001essentialsclassicalbranedynamics, Emparan:2009at} for a related treatment of calculus of submanifolds. Introducing the transverse projector $n^{i}_{\ \mu}$, by definition orthogonal to the tangential projector $n^{i}_{\ \mu} \partial_{a}X^{\mu}=0$, the extrinsic curvature tensor of the submanifold along a transverse direction $i$ is given by $K^{\ \ i}_{a b}=n^{i}_{\ \mu} \nabla_{a} \partial_{b}X^{\mu}$. Since we truncate at monopole order, noting that $\nabla_\mu T^{\mu\nu}=\left(\partial_{b} X^\nu \nabla_{a}T^{ab}+n^{\ \nu}_{i}T^{ab}K_{ab}{}^i \right)$, we find
\begin{equation}\label{eq: blackfoldstress}
     T^{ab}K_{ab}^{\,\,\,\,i} =F^{i c}J_{c}+ \frac{1}{2}\mathcal{F}^{i c d}\mathcal{J}_{c d}\quad, \quad  \nabla_{a} T^{ab}= F_{}^{bc}{J}_{c}~.
\end{equation}
The first equation above is the extrinsic equation which expresses the mechanical balance of forces that an allowed embedding of the probe should satisfy. The second equation above is the intrinsic equation and expresses the local balance of energy and longitudinal momentum for the tangential degrees of freedom carried by the string. In the same leading order approximation, the equations for the p-form currents reduce to
\begin{equation}\label{eq: blackfoldcurrents}
    d\ast J=2\kappa \star\mathcal{J}\wedge F\quad,\quad d\ast \mathcal{J}=0~,
\end{equation}
where $\ast$ is the Hodge dual with respect to the induced metric. In particular, only the tangential projection survives. Thus, the 2-group symmetry is inherited to the worldsheet theory from the (single) 2-group symmetry \eqref{eq: 2-grouplaws} acting across the whole spacetime \cite{Armas:2024caa}. Together \eqref{eq: blackfoldstress} and \eqref{eq: blackfoldcurrents} are the leading order blackfold equations of five-dimensional minimal supergravity. These equations will be central to our analysis of equilibrium actions for two-dimensional 2-group fluids and to our search for new perturbative black hole solutions.


\subsection{Aspects of the 2-group local structure}\label{sec: Ward}
In this section we discuss general 2-group Ward identities as obtained from the variation of an effective action. We carefully clarify subtleties related to the gauge invariance of the currents. Finally, we explain how the local structure of the 2-group symmetry discussed in this paper is implemented in supergravity, making contact with the derivation in the previous section.\\
\subsubsection{Ward identities and gauge invariance \label{sec:sugra_ward_identities}}
At the infinitesimal level, we describe the local higher-group structure via the transformation rules of the corresponding collection of background fields. Let $A$ be a 1-form potential and $B$ a 2-form potential probing the $U(1)^{(0)}$ and $U(1)^{(1)}$ sectors of the $U(1)^{(0)} \times_{\kappa}U(1)^{(1)}$ 2-group, respectively. Their transformation rules read
\begin{equation}\label{eq: ward_gauge_transformations}
    \delta_{g} A=d\lambda\quad,\quad
    \delta_{g} B=d\Lambda-\kappa\,d\lambda\wedge A~,
\end{equation}
where $\lambda$ is a scalar and $\Lambda$ is a one-form gauge parameter. The symbol $\delta_{g}$ denotes a gauge variation. Observe that the gauge parameter of the lower rank symmetry enters the transformation law of the background field corresponding to the higher-rank symmetry. This results in a mixing of the two higher-form symmetries participating in the higher-group, as discussed in \cite{Armas:2024caa}. This mixing is reflected on the definition of the gauge invariant field strengths
\begin{equation}\label{eq: ward_field_strengths}
    F=dA\quad,\quad \mathcal{F}=dB-\kappa A\wedge F~,
\end{equation}
as well as on their Bianchi identities
\begin{equation}\label{eq:BianchiF}
     dF=0 \quad, \quad d\mathcal{F}=-\kappa F\wedge F~.
\end{equation}
In particular, observe the term proportional to $\kappa$ in the definition of the field strength of the higher-rank symmetry $\mathcal{F}$, which is responsible for the non-closure of $\mathcal{F}$. From a more formal standpoint the local structure of the continuous 2-group we consider in this paper corresponds to an adjusted differential crossed module \cite{Rist_2026}\footnote{For related discussions of global realizations see for instance \cite{baez2004higherdimensionalalgebrav2groups,Bunk_2021}.}.\\
\\
We now comment on the relation between the local higher-group structure described in \cite{Armas:2024caa} and in the present work, and the global issues examined in \cite{Intriligator:2026xul}. At the two-derivative level, the relevant bosonic sectors of five- and eleven-dimensional supergravity share the same schematic Maxwell-Chern-Simons form. Let $Q_e$ and $Q_m$ denote the elementary electric and magnetic charges, respectively, and define
\[
a=Q_eA\,,\qquad b=Q_mB\,,\qquad
x=\frac{Q_eF}{2\pi}\,,\qquad
g=\frac{Q_m\mathcal{F}}{2\pi}\,.
\]
Here $x=da/(2\pi)$. In regions away from the sources, the modified Bianchi identity takes the form
\begin{equation}\label{eq: Bianchinorma}
    dg=-n\,x\wedge x\,,
    \qquad
    n=2\pi\kappa\,\frac{Q_m}{Q_e^2}\,.
\end{equation}
This fixes the characteristic polynomial of the lower-rank background. A global realization of the higher-group symmetry specified by the conservation laws \eqref{eq: blackfoldcurrents} and modified Bianchi identity \eqref{eq: Bianchinorma} requires to impose the quantization conditions for the charges. A question one can ask is whether there exists a universal integral cohomology class whose de Rham representative is given by the right-hand side of \eqref{eq: Bianchinorma} upon imposing the quantization conditions. Consider units for which $16\pi G_{D}=1$ where $G_{D}$ is Newton's constant in $D=5,11$ dimensions. For the case of the unit M-theory charges, for example, setting  $Q_{m}=T_{5}$, $Q_{e}=T_{2}$ where $T_{5}$ is the tension of the M5 brane and $T_{2}$ is the tension of the M2 brane, as well as $\kappa=-1/2$ \cite{Armas:2024caa}, we find
\[
    n=2\pi\kappa\,\frac{T_5}{T_2^2}
     =-\frac12\,.
\]
 The Dirac condition then reads $Q_{m}Q_{e}=T_{2}T_{5}=2\pi$. At the same time, the Postnikov class of the ordinary compact $U(1)^{(2)} \times_{n}U(1)^{(5)}$ requires $n \in \mathbb{Z}$ \cite{Intriligator:2026xul}. Already at this stage we can conclude that a global realization of the 6-group symmetry discussed in \cite{Armas:2024caa} is not universally provided by the usual compact $U(1)^{(2)} \times_{n}U(1)^{(5)}$. Related issues arise for electric and magnetic charges in five-dimensional Maxwell-Chern-Simons theory, as discussed in \cite{Intriligator:2026xul}. As mentioned in the work of \cite{Armas:2024caa}, and further explained in the present paper, our focus lies on a treatment of the local structure of the higher-group symmetry, leaving certain global issues unspecified. Nevertheless, this is enough to study Ward identities and provide a symmetry-based effective theory for classical black holes and M-brane configurations. In addition, in particular for the case of M-theory, a compact 3-form $A$ is not a usual Abelian gauge field, but rather a "shifted" compact field \cite{Witten_1997}. Locally, however, this shift can be invisible and the 3-form acts as an Abelian gauge field, namely it transforms according to \eqref{eq: ward_gauge_transformations} (where the appropriate form degrees are understood). In ref. \cite{Intriligator:2026xul} the higher-derivative gravitational correction $A\wedge I_{8}(R)$ was also considered, which lies beyond the two-derivative supergravity approximation used in \cite{Armas:2024caa}. We also mention that our local treatment of the symmetry structure allows for certain global statements. For instance, in section \ref{sec: conserved_charges} we provide the global conserved charges stemming from the higher-group conservation laws. These compute correctly the charges of the black holes considered in this paper, as our results in the next sections show. \\
 \\
Let $S_{\mathrm{eff}}=S[A, B, g_{\mu \nu}]$ be an effective action labeled by the background fields $A, B, g_{\mu \nu}$. In the absence of anomalies, this action must be invariant under (spacetime) diffeomorphisms and the 2-group transformations \eqref{eq: ward_gauge_transformations}. In what follows, the variational principle defining the sources is defined on-shell so that variations of the dynamical fields do not contribute. We shall omit boundary terms assuming that the variations have compact support or that suitable boundary conditions are imposed. Following the common lore for independent higher-form symmetries, a direct definition of the source currents would give
\begin{equation}\label{eq: ward_ordinary_variation}
    \delta S_{eff}
    =\frac{1}{2}\int_{\mathcal{M}}d^Dx\sqrt{-g}\,
      \boldsymbol{T}^{\mu\nu}\delta g_{\mu\nu}
    +\int_{\mathcal{M}}\left(
      \delta A\wedge\star\boldsymbol{j}
      +\delta B\wedge\star\boldsymbol{\mathcal{J}}\right)~.
\end{equation}
We have denoted the current obtained by varying $A$ at fixed $B$ by $\boldsymbol{j}$, to distinguish it from the gauge-invariant current $\boldsymbol{J}$ used in the previous subsection. The distinction is necessary because an arbitrary variation $\delta B$ is itself not gauge invariant. To wit,
\begin{equation}\label{eq: ward_second_variation_sources}
    \delta_{g}(\delta A)=\delta (\delta_{g}A)=0\quad,\quad
    \delta_{g}(\delta B)=\delta(\delta_{g}B)
    =-\kappa\,d\lambda\wedge\delta A~,
\end{equation}
since the gauge parameters are held fixed under $\delta$. We now require that the variational principle is itself gauge-invariant, that is
\begin{equation}\label{eq: secondvariationvanish}
    \delta_{g}(\delta S_{eff})=0~.
\end{equation}
For the variational principle \eqref{eq: ward_ordinary_variation}, and using \eqref{eq: ward_second_variation_sources}, we get
\begin{equation}\label{eq:ward_second_variation_action}
\begin{aligned}
   \delta_{g}(\delta S_{eff})
    ={}&\frac{1}{2}\int_{\mathcal{M}}d^Dx\sqrt{-g}\,
       \delta g_{\mu\nu}\,\delta_{g}\boldsymbol{T}^{\mu\nu}+\int_{\mathcal{M}}\delta B\wedge
       \delta_{g}(\star\boldsymbol{\mathcal{J}})\\
    &+\int_{\mathcal{M}}\delta A\wedge
       \left[\delta_{g}(\star\boldsymbol{j})
       +\kappa\,d\lambda\wedge\star\boldsymbol{\mathcal{J}}\right]~.
\end{aligned}
\end{equation}
The source variations are arbitrary, so imposing \eqref{eq: secondvariationvanish} implies
\begin{equation}\label{eq: ward_ordinary_current_transformation}
    \delta_{g}(\star\boldsymbol{\mathcal{J}})=0\quad,\quad
    \delta_{g}\boldsymbol{T}^{\mu\nu}=0 \quad,\quad \delta_{g}(\star\boldsymbol{j})
       =-\kappa\,d\lambda\wedge\star\boldsymbol{\mathcal{J}}~~.
\end{equation}
Thus the ordinary functional derivative with respect to $A$ defines a current which transforms under the background gauge symmetry, that is a current which is not gauge invariant. This reflects the mixing of the sources. Note however that this feature is nonetheless compatible with gauge invariance of the action. \\
\\
Nevertheless, a well-defined observable is required to be gauge invariant. We then search for a variational principle which, upon demanding \eqref{eq: secondvariationvanish}, produces gauge invariant currents. It is easy to see that we need to define
\begin{equation}\label{eq: ward_invariant_variation}
    \delta S_{eff}=\frac{1}{2}\int_{\mathcal{M}}d^Dx\sqrt{-g}\,
      \boldsymbol{T}^{\mu\nu}\delta g_{\mu\nu}
    +\int_{\mathcal{M}}\left(
      \delta A\wedge\star\boldsymbol{J}
      +(\delta B+\kappa A\wedge\delta A)\wedge\star\boldsymbol{\mathcal{J}}\right)~.
\end{equation}
This results from \eqref{eq: ward_ordinary_variation} by replacing $\delta B \rightarrow \Delta B=\delta B+\kappa A\wedge\delta A$ so that $\delta_{g}\Delta B=0$.
All currents appearing in \eqref{eq: ward_invariant_variation} are now gauge invariant. This is precisely the variational principle of \cite{Armas:2024caa} (specialized to the 2-group) used to define gauge invariant currents in phases characterized by a given field content and pattern of spontaneous symmetry breaking. Our simple computation here using the requirement \eqref{eq: secondvariationvanish} shows that the variational principle \eqref{eq: ward_invariant_variation} is a general feature of effective field theories invariant under higher-group symmetries, independently of symmetry breaking and dynamical field content. \\
\\
We may now derive the Ward identities obeyed by the currents $\boldsymbol{J}, \boldsymbol{\mathcal{J}}$ by setting $\delta=\delta_{g}$ in \eqref{eq: ward_invariant_variation}. Invariance of the effective action under \eqref{eq: ward_gauge_transformations} implies
\begin{equation}\label{eq: ward_gauge_action_variation}
    \int_{\mathcal{M}}\left(
      (-\lambda)\left(d\star\boldsymbol{J}
           -2\kappa F\wedge\star\boldsymbol{\mathcal{J}}\right)
      +(\Lambda-2\kappa\lambda A)
          \wedge d\star\boldsymbol{\mathcal{J}}\right)=0~.
\end{equation}
where we integrated by parts. The gauge parameters are arbitrary and independent among themselves, yielding
\begin{equation}\label{eq:ward_current_identities}
    d\star\boldsymbol{J}
       =2\kappa F\wedge\star\boldsymbol{\mathcal{J}}
       \quad,\quad d\star\boldsymbol{\mathcal{J}}=0~,
\end{equation}
in agreement with \eqref{eq: 2-grouplaws}. If one instead used the current $\boldsymbol{j}$, its Ward identity would read $d\star\boldsymbol{j}=\kappa F\wedge\star\boldsymbol{\mathcal{J}}$. The factor of 2 in \eqref{eq:ward_current_identities} is therefore a consequence of the gauge invariant current definition, and is not merely a convenient redefinition of the structure constant $\kappa$.\\
\\
Finally, let us consider an infinitesimal diffeomorphism generated by a vector field $\zeta^{\mu}$. The metric varies as $\delta_{\zeta}g_{\mu\nu}=2\nabla_{(\mu}\zeta_{\nu)}$, while the gauge fields vary by their Lie derivatives. Using Cartan's formula, the gauge field variations entering \eqref{eq: ward_invariant_variation} are given by
\begin{equation}\label{eq:ward_diffeomorphism_sources}
\begin{aligned}
    \delta_{\zeta}A&=\iota_{\zeta}F+d\chi~,\\
    \Delta_{\zeta}B&=\iota_{\zeta}\mathcal{F}
                       +d\Psi+2\kappa A\wedge d\chi~,
\end{aligned}
\end{equation}
where $\iota_{\zeta}$ denotes contraction with $\zeta$, $\chi=\iota_{\zeta}A$ and $\Psi=\iota_{\zeta}B+\kappa\chi A$. The terms involving $\chi$ and $\Psi$ disappear from the variation upon using \eqref{eq:ward_current_identities}. The remaining variation can be written as
\begin{equation}\label{eq:ward_diffeomorphism_action}
   \delta_{\zeta}S_{{eff}}
      =\int_{\mathcal{M}}d^Dx\sqrt{-g}\,\zeta_{\nu}
       \left[-\nabla_{\mu}\boldsymbol{T}^{\mu\nu}
       +F^{\nu\mu}\boldsymbol{J}_{\mu}
       +\frac{1}{2}\mathcal{F}^{\nu\mu\rho}
                           \boldsymbol{\mathcal{J}}_{\mu\rho}\right]~.
\end{equation}
We thus obtain the stress tensor Ward identity
\begin{equation}\label{eq:ward_stress_identity}
    \nabla_{\mu}\boldsymbol{T}^{\mu\nu}
      =F^{\nu\mu}\boldsymbol{J}_{\mu}
       +\frac{1}{2}\mathcal{F}^{\nu\mu\rho}
                           \boldsymbol{\mathcal{J}}_{\mu\rho}~.
\end{equation}
 Equation \eqref{eq:ward_stress_identity} has precisely the form of \eqref{eq: 2-groupstresslaw}, now expressed entirely in terms of the invariant background field strengths and currents.\\

\subsubsection{Supergravity implementation}\label{sec: sugra_implementation}
In section \ref{sec: sugra_ward_identities} we derived the form of the 2-group Ward identities from the coupled supergravity theory, upon the on-shell identification $\mathcal{F}=-\star F$. We will now see why this identification is natural from the point of view of the local structure of the 2-group discussed in this section and how it is implemented on the worldsheet theory. Given this, we proceed to formulate the variational principle on the embedded worldsheet. \\
\\
Let us start by explaining the origin of the two-form potential needed to couple to the string current. The minimal supergravity action \eqref{eq:L} is written in terms of the electric potential $A$, while the string is a magnetic source for this field, as expressed by \eqref{eq:Bianchi}. In five dimensions its dual description uses a two-form potential $B$, to which a string can couple electrically through its two-dimensional worldsheet.\\
\\
To construct $B$, take the probe limit and work on a contractible patch of the background away from electric and magnetic sources. The source-free equations following from \eqref{eq:EOM} and \eqref{eq:Bianchi} imply
\begin{equation}\label{eq:ward_closed_dual_form}
    dF=0\quad,\quad d\star F=\kappa F\wedge F
    \quad\Longrightarrow\quad
    d\left(\star F-\kappa A\wedge F\right)=0~.
\end{equation}
The Chern-Simons term thus prevents $\star F$ itself from being closed. It is the combination $\star F-\kappa A\wedge F$ that admits a local potential by the Poincare lemma, namely
\begin{equation}\label{eq:ward_dual_potential}
    dB=-\star F+\kappa A\wedge F\quad,\quad
    \mathcal{F}\equiv dB-\kappa A\wedge F=-\star F~.
\end{equation}
 Choosing the sign of that potential to agree with the conventions of \cite{Armas:2024caa}, we define
\begin{equation}
      \mathcal{F}= dB-\kappa A\wedge F=-\star F~.
\end{equation}
This introduces no additional local propagating degrees of freedom into the supergravity theory: on a supergravity solution the field strength of $B$ is fixed by $F$ and the metric. The construction is local. In the presence of magnetic flux, the potentials must in general be defined on overlapping patches.\\
\\
The transformation of $B$ now follows from its definition. Under $\delta A=d\lambda$, the field strength $F$ and its Hodge dual are invariant, so \eqref{eq:ward_dual_potential} requires $d(\delta B)=\kappa\,d\lambda\wedge F=d(-\kappa\,d\lambda\wedge A)$. Up to the ordinary two-form gauge freedom $d\Lambda$, we therefore obtain
\eqref{eq: ward_gauge_transformations}. The Chern-Simons coupling has therefore induced a mixing between the gauge transformations of the electric potential and its magnetic dual. This is precisely the $q=0$ case of the continuous Abelian higher-group $U(1)^{(q)}\times_{\kappa}U(1)^{(2q+1)}$ discussed in \cite{Armas:2024caa}.\\
\\
For the variational description of the probe, we now regard $g_{\mu\nu}$, $A$ and $B$ as independent background fields. In particular, $B$ acts as a source for the one-form symmetry current $\boldsymbol{\mathcal{J}}$ carried by the string. This independent-source description allows us to define the electric and string currents separately. It does not make $B$ an additional independent bulk field of minimal supergravity. The duality relation in \eqref{eq:ward_dual_potential} is imposed only after taking the source variations and deriving the Ward identities, when restricting the backgrounds to supergravity solutions.\\

\subsection*{Worldsheet variational principle}
For an embedded string geometry in a background spacetime, the effective action also includes the embedding fields $X^{\mu}$. Equation \eqref{eq: ward_invariant_variation} can now be understood as a variational principle at fixed embedding fields, $\delta X^{\mu}=0$.  It is convenient to compactly denote the pullback of a field $F$ onto the worldvolume as $X^*(F)$. Inserting the localized monopole sources \eqref{eq: multipole}, \eqref{eq: multipole_currents} into \eqref{eq: ward_invariant_variation}, the result is
\begin{equation}\label{eq: localized_variation}
    \delta S_{eff}=\frac{1}{2}\int_{\Sigma_{2}}d^2\sigma \sqrt{-\gamma}T^{a b} [X^*(\delta g)]_{a b}+\int_{\Sigma_{2}} X^*(\delta A)\wedge \ast J+X^*(\Delta B)\wedge \ast \mathcal{J}~,
\end{equation}
where we performed the integration over $\mathcal{M}$. Here $\ast$ denotes the Hodge dual with respect to the induced metric. Since we consider variations for which the embedding is kept fixed, the pullback commutes with the variation of the background fields, and we can write \eqref{eq: localized_variation} also as
\begin{equation}\label{eq: ward_worldsheet_variation}
\begin{aligned}
   \delta S_{eff}
    =&\frac{1}{2}\int_{\Sigma_2}d^2\sigma\sqrt{-\gamma}\,
               T^{ab}\delta\gamma_{ab}
   +\int_{\Sigma_2}
         \delta X^*(A)\wedge\ast J\\
        &+\int_{\Sigma_{2}}(\delta X^*(B)+\kappa X^*(A)\wedge\delta X^*(A))
                  \wedge\ast\mathcal{J}~.
\end{aligned}
\end{equation}
Let us derive explicitly the equations that follow from the worldsheet variational principle. Note that an (internal) gauge transformation automatically falls into the class of variations we consider, because $\delta_{g} X^{\mu}=0$ is already part of how such symmetries acts. Furthermore, the pullback commutes with the exterior derivative, so the gauge transformations of the pulled back background fields retain the form \eqref{eq: ward_gauge_transformations}, with $\lambda$ and $\Lambda$ also pulled back on the worldsheet. Substituting the pulled back version of \eqref{eq: ward_gauge_transformations} into \eqref{eq: ward_worldsheet_variation} and demanding invariance, upon partial integration we get
\begin{equation}\label{eq: ward_worldsheet_current_equations}
    d\ast J=2\kappa X^*(F)\wedge\ast\mathcal{J}\quad,\quad
    d\ast\mathcal{J}=0~.
\end{equation}
 Since $\mathcal{J}$ is a top-form on the worldsheet, its equation admits the unique solution
\begin{equation}\label{eq: ward_worldsheet_top_form}
    \mathcal{J}=Q_{(1)}\ast1\quad,\quad dQ_{(1)}=0~.
\end{equation}
Thus the string charge is constant on each connected worldsheet, while the electric current can be sourced by the conserved string charge when there exists a non-vanishing pullback of the background 2-form flux.\\
\\
On the geometric side, here we perform Lagrangian-type variations \cite{Battye_1995, Armas_2018_act}. These are obtained from the variations under spacetime diffeomorphisms with the embedding maps held fixed. We thus vary a spacetime point $x^{\mu}$ by an arbitrary ambient vector $\xi^{\mu}$ with $\delta_{\xi} X^{\mu}=0=\delta_{\xi} \partial_{a}X^{\mu}$. The background potentials transform by their Lie derivatives
\begin{equation}
    \delta_{\xi}g_{\mu \nu}=\mathcal{L}_{\xi}g_{\mu \nu}\quad,\quad \delta A=\mathcal{L}_{\xi}A\quad,\quad \delta B=\mathcal{L}_{\xi}B~.
\end{equation}
To study the two physical sectors, namely the intrinsic and the extrinsic, we decompose the vector $\xi^{\mu}$ with respect to the worldsheet $\xi^{\mu}=\partial_{a}X^{\mu}\xi^{a}+n_{i}^{\ \mu}\xi^{i}$. We then calculate the variations induced to the pullback of the background fields appearing in \eqref{eq: ward_worldsheet_variation}. For the induced metric we get
\begin{equation}\label{eq: ward_worldsheet_metric_deformation}
    \delta_\xi\gamma_{ab}
      =2\nabla_{(a}\xi_{b)}-2K_{ab}{}^i\xi_i~.
\end{equation}
Applying Cartan's formula to the pullbacks of the background gauge potentials, we find 
\begin{equation}\label{eq: ward_worldsheet_source_deformation}
\begin{aligned}
    \delta_\xi X^*(A)&=X^*(\iota_{\xi}F)+d\chi_{\xi}~,\\
    \Delta_\xi X^*(B)&=X^*(\iota_{\xi}\mathcal{F})
                  +d\Psi_{\xi}+2\kappa X^*(A)\wedge d\chi_{\xi}~,
\end{aligned}
\end{equation}
where $\Delta_\xi X^*(B)=\delta_\xi X^*(B)+\kappa X^*(A)\wedge\delta_\xi X^*(A)$, $\chi_{\xi}=X^*(\iota_{\xi}A)$ and $\Psi_{\xi}=X^*(\iota_{\xi}B)+\kappa \chi_{\xi} X^*(A)$. The contractions on the right-hand side are performed in the ambient spacetime before taking the pullback. This order of operations is essential. Substituting \eqref{eq: ward_worldsheet_metric_deformation} and \eqref{eq: ward_worldsheet_source_deformation} into \eqref{eq: ward_worldsheet_variation}, the terms involving $\chi_{\xi}$ and $\Psi_{\xi}$ drop out by \eqref{eq: ward_worldsheet_current_equations}, just as in the spacetime derivation. Integrating the tangential metric variation by parts, gives
\begin{equation}\label{eq:ward_worldsheet_embedding_variation}
\begin{aligned}
    \delta_X S_{eff}
      ={}&\int_{\Sigma_2}d^2\sigma\sqrt{-\gamma}\,\xi_b
       \left[-\nabla_a T^{ab}+F^{bc}J_c
                    +\frac{1}{2}\mathcal{F}^{bcd}\mathcal{J}_{cd}\right]\\
       &+\int_{\Sigma_2}d^2\sigma\sqrt{-\gamma}\,\xi_i
       \left[-T^{ab}K_{ab}{}^i+F^{ic}J_c
                    +\frac{1}{2}\mathcal{F}^{icd}\mathcal{J}_{cd}\right]~.
\end{aligned}
\end{equation}
Tangential Lagrangian variations induce worldsheet diffeomorphisms of the induced metric and pulled-back background fields. Their contribution therefore vanishes by worldsheet reparametrization invariance, with the intrinsic fields on-shell. Transverse Lagrangian variations change the background relative to the fixed surface, and requiring stationarity under arbitrary such variations yields the extrinsic equation of motion. Moreover, $\mathcal{F}^{bcd}=0$ on a two-dimensional worldsheet. We consequently obtain
\begin{equation}\label{eq:ward_worldsheet_stress_equations}
\begin{aligned}
    \nabla_aT^{ab}&=F^{bc}J_c~,\\
    T^{ab}K_{ab}{}^i&=F^{ic}J_c
                   +\frac{1}{2}\mathcal{F}^{icd}\mathcal{J}_{cd}~.
\end{aligned}
\end{equation}
Together with \eqref{eq: ward_worldsheet_current_equations}, these reproduce the intrinsic and extrinsic equations \eqref{eq: blackfoldstress}--\eqref{eq: blackfoldcurrents}. Although the pullback of $\mathcal{F}$ itself vanishes, the pullback of $\iota_{\xi}\mathcal{F}$ need not vanish for a transverse displacement. Its mixed components $\mathcal{F}^{iab}$ therefore exert a transverse force, retaining the non-triviality of the 2-group symmetry in the worldsheet theory via the extrinsic equation, see \cite{Armas:2024caa}. We will use \eqref{eq: ward_worldsheet_variation} to extract the currents and stress tensor from the equilibrium action in the next subsection.
\\

In this section we performed Lagrangian variations where the spacetime background sources vary and the embedding is kept fixed, since these arise more natural from a spacetime point of view. It is instructive to note that one can keep the ambient sources fixed and vary the embedding. These two type of variations are equivalent \cite{Armas_2018_act}.


\subsection{Equilibrium action for a two-dimensional 2-group fluid}\label{sec:equilibrium_action}
We now construct the leading-order equilibrium action whose variations give the string currents and stress tensor. It turns out that this action can be obtained from the $q=0$ specialization of the T-U phase action of \cite{Armas:2024caa}. For a zero-form symmetry there is no separate temporal Goldstone, and this sector describes an ordinary charged fluid. In particular, spontaneous breaking in two dimensions is forbidden for such symmetries by the usual Mermin-Wagner theorem. The one-form symmetry is also unbroken \cite{lake2018higherformsymmetriesspontaneoussymmetry}, with its conserved top-form current fixing the string charge. The scalar Stueckelberg field introduced below merely implements gauge invariance, without adding a propagating worldsheet mode. Throughout this subsection the background fields are understood as pullbacks to the worldsheet, and transverse spin is set to zero for simplicity. Generalizing the action to include a non-vanishing transverse spin can be done using for instance \cite{armas2025chiralanomalyanomalousspin}.\\
\\
Let $k^a$ be a timelike worldsheet Killing vector and $T_0$ the constant global temperature. The thermal vector, local temperature and fluid velocity are
\begin{equation}\label{eq:equilibrium_thermal_vector}
    \beta^a=\frac{k^a}{T_0}\quad,\quad
    T=\frac{1}{\sqrt{-\gamma_{ab}\beta^a\beta^b}}
      =\frac{T_0}{|k|}\quad,\quad
    u^a=T\beta^a~.
\end{equation}
Equilibrium requires invariance under a thermal translation combined with a background gauge transformation. Keeping the corresponding scalar thermal potential $\Lambda_\beta$ and one-form thermal potential $\Lambda_{\beta a}$ explicit, this condition reads
\begin{equation}\label{eq:equilibrium_stationarity}
\begin{aligned}
    \mathcal{L}_\beta\gamma_{ab}&=0~,\\
    \mathcal{L}_\beta A_a+\nabla_a\Lambda_\beta&=0~,\\
    \mathcal{L}_\beta B_{ab}+2\nabla_{[a}\Lambda_{\beta b]}
           -2\kappa(\nabla_{[a}\Lambda_\beta)A_{b]}&=0~.
\end{aligned}
\end{equation}
These potentials specify the gauge transformation accompanying the thermal flow.\footnote{The $\Lambda_q$ and $\Lambda_{2q+1}$ in eq.~(3.4) of \cite{Armas:2024caa} are residual background gauge parameters in a stationary gauge. Here $\Lambda_\beta$ and $\Lambda_{\beta a}$ are their thermal counterparts, namely the compensating parameters in \eqref{eq:equilibrium_stationarity}. Keeping them explicit restores the chemical potentials before choosing a gauge with strictly time-independent sources.} Under a background transformation with parameters $(\lambda,\Lambda_a)$ as in \eqref{eq: ward_gauge_transformations}, we can choose their transformations to be
\begin{equation}\label{eq:equilibrium_thermal_gauge_transformations}
\begin{aligned}
    \delta\Lambda_\beta&=-\mathcal{L}_\beta\lambda~,\\
    \delta\Lambda_{\beta a}&=-\mathcal{L}_\beta\Lambda_a
          +\kappa\Lambda_\beta\nabla_a\lambda
          -\kappa\lambda\nabla_a\Lambda_\beta~.
\end{aligned}
\end{equation}
These preserve \eqref{eq:equilibrium_stationarity}. Introduce also a scalar $\varphi$ with gauge transformation and equilibrium condition
\begin{equation}\label{eq:equilibrium_stueckelberg}
    \delta\varphi=-\beta^a\Lambda_a-\kappa\Lambda_\beta\lambda
    \quad,\quad
    \mathcal{L}_\beta\varphi
       =\beta^a\Lambda_{\beta a}+\kappa\Lambda_\beta^2~.
\end{equation}
The gauge-invariant chemical potentials are then
\begin{equation}\label{eq:equilibrium_invariants}
\begin{aligned}
    \mu_0&=T\left(\beta^a A_a+\Lambda_\beta\right)~,\\
    \mu_a&=T\left[\beta^bB_{ba}+\Lambda_{\beta a}-\nabla_a\varphi
          -\kappa\left(\beta^b A_b+2\Lambda_\beta\right)A_a\right]
          -\kappa\mu_0^2 u_a~,\\
    \mu_1&=v^a\mu_a\quad,\quad
    v^a=\epsilon^{ab}u_b~.
\end{aligned}
\end{equation}
Here $\epsilon^{ab}$ is the worldsheet volume tensor, with orientation chosen so that $(u\wedge v)_{ab}=\epsilon_{ab}$. Thus $v^av_a=1$, $u^av_a=0$ and $\gamma^{ab}=-u^au^b+v^av^b$. The equilibrium condition for $\varphi$ ensures $u^a\mu_a=0$, so that $\mu_a=\mu_1v_a$. Gauge invariance of \eqref{eq:equilibrium_invariants} follows directly from \eqref{eq:equilibrium_thermal_gauge_transformations} and \eqref{eq:equilibrium_stueckelberg}. In particular, restoring the thermal potentials requires the term proportional to $2\Lambda_\beta A_a$ in $\mu_a$. In a local stationary gauge with $\Lambda_\beta=\Lambda_{\beta a}=0$, these expressions reduce to
\begin{equation}\label{eq:equilibrium_stationary_gauge}
    \mu_0=u^aA_a\quad,\quad
    \mu_a=u^bB_{ba}-\kappa\mu_0P_a{}^bA_b-T\nabla_a\varphi
    \quad,\quad P_{ab}=\gamma_{ab}+u_au_b~,
\end{equation}
with $\mathcal{L}_\beta\varphi=0$. This gauge choice can be made locally, while thermal holonomies are retained through the chemical potentials.\\
\\
At ideal order the action is determined by the pressure $P(T,\mu_0,\mu_1)$. We use its Lorentzian continuation when taking source variations, consistently with \eqref{eq: ward_worldsheet_variation}, and write
\begin{equation}\label{eq:equilibrium_action}
    S_{\mathrm{eq}}=\int_{\Sigma_2}d^2\sigma\sqrt{-\gamma}\,
                         P(T,\mu_0,\mu_1)~.
\end{equation}
The Euclidean equilibrium functional follows by continuation and integration over the thermal circle. Define the entropy and charge densities by $s=\partial P/\partial T$, $\rho_0=\partial P/\partial\mu_0$ and $\rho_1=\partial P/\partial\mu_1$, holding the other thermodynamic variables fixed. The Gibbs-Duhem relation, Euler relation and local first law are
\begin{equation}\label{eq:equilibrium_thermodynamics}
\begin{aligned}
    dP&=s\,dT+\rho_0\,d\mu_0+\rho_1\,d\mu_1~,\\
    \epsilon+P&=Ts+\mu_0\rho_0+\mu_1\rho_1~,\\
    d\epsilon&=T\,ds+\mu_0\,d\rho_0+\mu_1\,d\rho_1~.
\end{aligned}
\end{equation}
The magnetic density is constant on a connected equilibrium worldsheet, as we verify below. Thermodynamic variations of $\rho_1$ compare different charge sectors. Within a fixed sector $\rho_1=Q_{(1)}$, the T-U functional takes the form
\begin{equation}\label{eq:equilibrium_fixed_charge_pressure}
    P=\widehat P(T,\mu_0,Q_{(1)})+Q_{(1)}\mu_1\quad,\quad
    \widehat P=P-\mu_1\rho_1~.
\end{equation}
The dependence on $\mu_1$ is then strictly linear, as in \cite{Armas:2024caa}. A grand-canonical equation of state $P(T,\mu_0,\mu_1)$ provides a convenient description of a family of such sectors. Its magnetic Legendre transform satisfies $d\widehat P=s\,dT+\rho_0\,d\mu_0-\mu_1\,d\rho_1$. The two descriptions give the same constitutive relations upon fixing the magnetic density.\\
\\
To vary \eqref{eq:equilibrium_action}, hold $\beta^a$, $\Lambda_\beta$ and $\Lambda_{\beta a}$ fixed. The spacelike vector $v^a$ is determined by the metric and $u^a$, and must also be varied. In particular,
\begin{equation}\label{eq:equilibrium_vector_variations}
\begin{aligned}
    \delta T&=\frac{T}{2}u^au^b\delta\gamma_{ab}~,
    &\delta u^a&=\frac{1}{2}u^au^bu^c\delta\gamma_{bc}~,\\
    \delta v^a&=\left(u^au^{(b}v^{c)}
              -\frac{1}{2}v^av^bv^c\right)\delta\gamma_{bc}~.
\end{aligned}
\end{equation}
Using \eqref{eq:equilibrium_invariants} and the equilibrium condition for $\varphi$, the chemical potentials vary as
\begin{equation}\label{eq:equilibrium_potential_variations}
\begin{aligned}
    \delta\mu_0={}&u^a\delta A_a
                    +\frac{\mu_0}{2}u^au^b\delta\gamma_{ab}~,\\
    \delta\mu_a={}&u^b\Delta B_{ba}-2\kappa\mu_0P_a{}^b\delta A_b
                         -T\nabla_a\delta\varphi\\
                &+\left(\frac{\mu_a}{2}u^bu^c
                      -\kappa\mu_0^2P_a{}^{(b}u^{c)}\right)
                         \delta\gamma_{bc}~,\\
    \delta\mu_1={}&v^a\delta\mu_a
                    -\frac{\mu_1}{2}v^av^b\delta\gamma_{ab}~,
\end{aligned}
\end{equation}
where $\Delta B_{ab}=\delta B_{ab}+\kappa(A_a\delta A_b-A_b\delta A_a)$ and $\mathcal{L}_\beta\delta\varphi=0$. Comparing the resulting source variation with \eqref{eq: ward_worldsheet_variation} yields
\begin{equation}\label{eq:equilibrium_constitutive_relations}
\begin{aligned}
    T^{ab}={}&(\epsilon+P)u^au^b+P\gamma^{ab}
                  -\mu_1\rho_1v^av^b
                  -2\kappa\mu_0^2\rho_1u^{(a}v^{b)}~,\\
    J^a={}&\rho_0u^a-2\kappa\mu_0\rho_1v^a~,\\
    \mathcal{J}^{ab}={}&\rho_1(u^av^b-u^bv^a)~.
\end{aligned}
\end{equation}
There is no independent spatial direction beyond the one fixed by the thermal velocity and the orientation of the worldsheet. Nevertheless, the 2-group coupling produces a spatial electric current and an energy flux even in the thermal frame. Defining the longitudinal mechanical pressure $\Pi$ and the energy flux $h$ by
\begin{equation}\label{eq:equilibrium_mechanical_pressure}
    \Pi=P-\mu_1\rho_1\quad,\quad h=-\kappa\mu_0^2\rho_1~,
\end{equation}
the stress tensor can also be written as $T^{ab}=\epsilon u^au^b+\Pi v^av^b+2h u^{(a}v^{b)}$. Thus $T^{ab}u_b=-\epsilon u^a-hv^a$, and the thermal frame generally differs from the Landau frame. The thermodynamic pressure $P$ in \eqref{eq:equilibrium_thermodynamics} is distinct from the longitudinal stress $\Pi$.\\
\\
Let us check explicitly the intrinsic equations. The variation with respect to the Stueckelberg field gives
\begin{equation}\label{eq:equilibrium_stueckelberg_equation}
    \delta_\varphi S_{\mathrm{eq}}
       =\int_{\Sigma_2}d^2\sigma\sqrt{-\gamma}\,
                     \nabla_a(T\rho_1v^a)\delta\varphi
    \quad\Longrightarrow\quad \nabla_a(T\rho_1v^a)=0~.
\end{equation}
Only the thermal vector $\beta^a=u^a/T$ is assumed to be Killing. To see why $\nabla_a(Tv^a)=0$ follows without a spatial Killing vector, define $a_a=u^b\nabla_bu_a=a_vv_a$. The thermal Killing equation gives $\nabla_aT=-Ta_a$ and $\nabla_au^a=0$. On a two-dimensional worldsheet the latter condition, together with $u^au_a=-1$, implies $\nabla_au_b=-u_aa_b$. Differentiating $v^a=\epsilon^{ab}u_b$ therefore yields
\begin{equation*}
\begin{gathered}
    \nabla_av_b=-a_vu_au_b\quad,\quad
    \nabla_av^a=a_v~,\\
    \nabla_a(Tv^a)=v^a\nabla_aT+T\nabla_av^a
                  =-Ta_v+Ta_v=0~.
\end{gathered}
\end{equation*}
In particular, $\nabla_{(a}v_{b)}=-a_vu_au_b$ is generally nonzero, so $v^a$ need not be Killing. Equation~\eqref{eq:equilibrium_stueckelberg_equation} thus reduces to $v^a\nabla_a\rho_1=0$. Together with thermal stationarity, $u^a\nabla_a\rho_1=0$, this implies $\nabla_a\rho_1=0$, and consequently $\nabla_a\mathcal{J}^{ab}=0$. At fixed $Q_{(1)}$ the Stueckelberg equation is identically satisfied, and the $\varphi$ dependence of the action is a boundary term. This explains why $\varphi$ carries no propagating mode in this sector.\\
\\
Including the gauge-field stationarity condition in \eqref{eq:equilibrium_stationarity} and defining $E=v^aF_{ab}u^b$, the hydrostatic identities can be collected as
\begin{equation}\label{eq:equilibrium_hydrostatic_identities}
\begin{aligned}
    \nabla_aT&=-Ta_a~,
    &T\nabla_a\left(\frac{\mu_0}{T}\right)&=F_{ab}u^b~,\\
    \nabla_au_b&=-u_aa_b~,
    &\nabla_av^a&=a_v~,
    &v^a\nabla_a\mu_0+\mu_0a_v&=E~.
\end{aligned}
\end{equation}
All thermodynamic scalars are stationary along $u^a$. It follows immediately from \eqref{eq:equilibrium_constitutive_relations} that
\begin{equation}\label{eq:equilibrium_electric_check}
    \nabla_aJ^a=-2\kappa\rho_1
          \left(v^a\nabla_a\mu_0+\mu_0a_v\right)
       =-2\kappa\rho_1E
       =\kappa F_{ab}\mathcal{J}^{ab}~.
\end{equation}
For the stress tensor, the Gibbs-Duhem and Euler relations give
\begin{equation}\label{eq:equilibrium_stress_check}
\begin{aligned}
    v^a\nabla_a\Pi+(\epsilon+\Pi)a_v&=\rho_0E~,\\
    v^a\nabla_ah+2ha_v&=-2\kappa\mu_0\rho_1E~,
\end{aligned}
\end{equation}
where $d\Pi=s\,dT+\rho_0\,d\mu_0$ along a fixed-charge equilibrium configuration. Hence $\nabla_aT^{ab}=E(\rho_0v^b-2\kappa\mu_0\rho_1u^b)=F^{bc}J_c$. The intrinsic equations are therefore satisfied once the magnetic constraint is imposed. The remaining equation determines the equilibrium embedding,
\begin{equation}\label{eq:equilibrium_embedding_equation}
    \left(\epsilon u^au^b+\Pi v^av^b+2h u^{(a}v^{b)}\right)K_{ab}{}^i
       =F^{ic}\left(\rho_0u_c-2\kappa\mu_0\rho_1v_c\right)
          +\rho_1\mathcal{F}^{icd}u_cv_d~.
\end{equation}
The last term retains the ambient magnetic force although the fully tangential three-form vanishes on the worldsheet.

\subsection{The conserved charges of higher-group symmetry and Smarr relations}\label{sec: conserved_charges}
Here we construct the conserved charges associated 2-group Ward identities and derive the first law of thermodynamics.

\subsubsection{Conserved charges of higher-group symmetry}
We now construct the global charges in the presence of background fluxes. We use the gauge-invariant currents and their Ward identities established in Section~\ref{sec: Ward}, with the same structure constant $\kappa$ throughout. The construction applies at monopole order and does not depend on the equation of state. When discussing thermodynamics, we specialise to the equilibrium configurations of Section~\ref{sec:equilibrium_action}. We take the spatial worldsheet slice $B_1$ to be closed and connected. The corresponding expressions for a non-compact string require a regulator and boundary conditions which exclude charge flux through its ends.\\
\\
The electric Ward identity immediately gives a conserved current,
\begin{equation}\label{modified_J}
    \widetilde J^a=J^a-2\kappa A_b\mathcal J^{ab}
    \quad,\quad \nabla_a\widetilde J^a=0~.
\end{equation}
Indeed, $\nabla_a(A_b\mathcal J^{ab})=\frac12F_{ab}\mathcal J^{ab}$, which cancels the source in \eqref{eq: ward_worldsheet_current_equations}. This is a Page-type current. It differs both from the gauge-invariant current $J^a$ and from the ordinary variational current $j^a=J^a-\kappa A_b\mathcal J^{ab}$ of Section~\ref{sec: Ward}. Under $\delta A=d\lambda$ it changes by the identically conserved improvement $\delta\widetilde J^a=-2\kappa\nabla_b(\lambda\mathcal J^{ab})$. Its integrated charge is therefore invariant under single-valued gauge transformations on a closed slice, although it can change under large gauge transformations. We fix the gauge branch and the reference values of the potentials when comparing charges.\\
\\
Let $k^\mu$ be a Killing vector of the ambient metric which preserves the gauge sources up to a 2-group gauge transformation. Introduce a scalar $\lambda_k$ and a one-form $\Lambda_k$ such that
\begin{equation}\label{eq:global_background_symmetry}
    \mathcal L_k g_{\mu\nu}=0\quad,\quad
    \mathcal L_k A+d\lambda_k=0\quad,\quad
    \mathcal L_k B+d\Lambda_k-\kappa\,d\lambda_k\wedge A=0~.
\end{equation}
These compensators specify how the spacetime symmetry acts on the gauge potentials. Their reference values form part of the definition of the conserved generator. Define
\begin{equation}\label{eq:global_symmetry_potentials}
\begin{aligned}
    \alpha_k&=\iota_k A+\lambda_k~,\\
    \eta_k&=\iota_k B+\Lambda_k
                 -\kappa(\iota_k A+2\lambda_k)A~.
\end{aligned}
\end{equation}
Cartan's formula and \eqref{eq: ward_field_strengths} then imply
\begin{equation}\label{eq:global_symmetry_potential_identities}
    d\alpha_k=-\iota_k F\quad,\quad
    d\eta_k=-\iota_k\mathcal F-2\kappa\alpha_k F~.
\end{equation}
Contractions are taken in the ambient spacetime before pulling these expressions back to the worldsheet. In particular, $\iota_k\mathcal F$ can contribute when $k^\mu$ has a normal component, despite the vanishing of the fully tangential three-form.\\
\\
The conserved current associated with this symmetry is
\begin{equation}\label{Ptilde}
    \widetilde P_k^a=T^{ab}k_b+\alpha_kJ^a
                          +\eta_{kb}\mathcal J^{ab}
    \quad,\quad \nabla_a\widetilde P_k^a=0~,
\end{equation}
where $k_b=\partial_bX^\mu k_\mu$ and all gauge quantities are pulled back. To verify conservation, use the intrinsic and extrinsic equations in \eqref{eq:ward_worldsheet_stress_equations}, together with the ambient Killing equation, to obtain
\begin{equation}\label{eq:global_killing_current_check}
    \nabla_a(T^{ab}k_b)
       =-J^a\nabla_a\alpha_k-\mathcal J^{ab}\nabla_a\eta_{kb}
          -\kappa\alpha_kF_{ab}\mathcal J^{ab}~.
\end{equation}
The remaining terms cancel against the divergence of the gauge contributions in \eqref{Ptilde}. This extends the conserved-current construction for branes in flux backgrounds \cite{Armas_2012} to the 2-group Ward identities of \cite{Armas:2024caa}. In a stationary gauge with $\lambda_k=\Lambda_k=0$, the result becomes
\begin{equation}\label{eq:global_current_stationary_gauge}
    \widetilde P_k^a=T^{ab}k_b
       +(\iota_k A)\left(J^a-\kappa A_b\mathcal J^{ab}\right)
       +(\iota_k B)_b\mathcal J^{ab}~.
\end{equation}
Thus the electric improvement entering this expression has coefficient $\kappa$, while the separately conserved electric current \eqref{modified_J} has coefficient $2\kappa$. Both follow from the same Ward identities. No change of the 2-group coupling is involved.\\
\\
The compensators transform under background gauge transformations as in \eqref{eq:equilibrium_thermal_gauge_transformations}, with $\beta$ replaced by $k$. Consequently $\alpha_k$ is invariant and $\delta\eta_k=-d(\iota_k\Lambda+\kappa\lambda\lambda_k)$. The corresponding change in $\widetilde P_k^a$ is a divergence of an antisymmetric tensor, so its charge is unchanged on a closed slice for single-valued gauge parameters. These formulae assume a compatible choice of potentials and compensators along the worldsheet. If several gauge patches are needed, their transition data must also be included in the charge integrals.\\
\\
Let $n^a$ be the future-directed unit normal to $B_1$ within the worldsheet and $dV_1$ its proper length element. The normal need not be a Killing vector. For a time-translation generator $\xi^\mu$ and rotational generators $\chi_i^\mu$, with fixed normalisations and compensators, the mass and angular momenta are
\begin{equation}\label{eq:mass}
    M=\int_{B_1}dV_1\,n_a\widetilde P_\xi^a~,
\end{equation}
\begin{equation}\label{eq:momentum}
    J_i=-\int_{B_1}dV_1\,n_a\widetilde P_{\chi_i}^a~.
\end{equation}
These include the coupling energy and momentum in the external gauge fields. They describe the probe contribution relative to the chosen background. The electric and string charges are
\begin{equation}\label{eq:0-formcharge}
    Q_{(0)}=-\int_{B_1}dV_1\,n_a\widetilde J^a~,
\end{equation}
\begin{equation}\label{eq:string_charge}
    \mathcal J^{ab}=Q_{(1)}\epsilon^{ab}
    \quad,\quad \nabla_aQ_{(1)}=0
    \quad,\quad Q_{(1)}=\rho_1~.
\end{equation}
The equality $Q_{(1)}=\rho_1$ follows by comparing the top-form current in \eqref{eq: ward_worldsheet_top_form} with the constitutive relation in \eqref{eq:equilibrium_constitutive_relations}. The string charge is the constant coefficient of the top-form current and is not multiplied by the length of $B_1$. For a ring it measures a dipole charge, determined by magnetic flux through a transverse sphere. Choosing the oriented unit tangent $e^a=-n_b\epsilon^{ba}$ on $B_1$, the electric charge can equivalently be written as
\begin{equation}\label{eq:global_page_charge}
    Q_{(0)}=-\int_{B_1}dV_1\,n_aJ^a
                    -2\kappa Q_{(1)}\oint_{B_1}A~.
\end{equation}
In particular, a large gauge transformation shifts this Page charge by $-2\kappa Q_{(1)}\oint_{B_1}d\lambda$. This dependence is retained when choosing the thermodynamic ensemble.\\

\subsubsection{First law of thermodynamics and generalized Smarr relations}\label{sec: Smarr}
We next identify the global chemical potentials. Let the equilibrium generator and its thermal vector be
\begin{equation}\label{eq:global_thermal_generator}
    k^\mu=\xi^\mu+\sum_i\Omega_i\chi_i^\mu
    \quad,\quad \beta^a=\frac{k^a}{T_0}
    \quad,\quad u^a=\frac{k^a}{|k|}
    \quad,\quad T=\frac{T_0}{|k|}~,
\end{equation}
where $k^\mu$ is tangent to the equilibrium worldsheet and $T_0,\Omega_i$ are constant. The reference compensators for $k$ are the same linear combination of those used in \eqref{eq:mass}--\eqref{eq:momentum}. Comparing them with the thermal compensators of Section~\ref{sec:equilibrium_action} defines a scalar and a one-form,
\begin{equation}\label{eq:global_thermal_zero_modes}
\begin{aligned}
    \Phi_0&=T_0\Lambda_\beta-\lambda_k~,
       &d\Phi_0&=0~,\\
    \zeta_a&=T_0\Lambda_{\beta a}-\Lambda_{ka}
                                      -T_0\nabla_a\varphi~,
       &d\zeta&=0~.
\end{aligned}
\end{equation}
Their closure follows by subtracting the two stationarity conditions \eqref{eq:equilibrium_stationarity} and \eqref{eq:global_background_symmetry}, with the latter pulled back to the worldsheet. Using \eqref{eq:equilibrium_invariants}, they are equivalently
\begin{equation}\label{eq:global_local_potentials}
\begin{aligned}
    \Phi_0&=|k|\mu_0-\alpha_k~,\\
    \zeta_a&=|k|\left(\mu_1v_a+\kappa\mu_0^2u_a\right)
                  -\eta_{ka}+2\kappa\Phi_0 A_a~.
\end{aligned}
\end{equation}
The global electric potential is therefore the constant $\Phi_0$, while the potential conjugate to the string charge is the period
\begin{equation}\label{eq:global_magnetic_potential}
    \Phi_1=\oint_{B_1}\zeta
          =\int_{B_1}dV_1\,e^a
             \left[|k|\left(\mu_1v_a+\kappa\mu_0^2u_a\right)
                         -\eta_{ka}+2\kappa\Phi_0A_a\right]~.
\end{equation}
The two terms proportional to $\kappa$ are required by the mixed gauge transformations and the spatial electric current in \eqref{eq:equilibrium_constitutive_relations}. The local chemical potential $\mu_1$ alone does not give the full conjugate potential in a general frame and gauge background. Under single-valued gauge transformations, $\Phi_0$ is invariant and $\zeta$ changes by an exact form, so $\Phi_1$ is also invariant. The reference compensators fix the zero of these potentials.\\
\\
These identifications can be checked directly against the equilibrium action. Define the entropy and the positive lapse of the chosen slicing by
\begin{equation}\label{eq:global_entropy_lapse}
    S=-\int_{B_1}dV_1\,s\,n_au^a
    \quad,\quad N=-n_ak^a~.
\end{equation}
The constitutive relations \eqref{eq:equilibrium_constitutive_relations}, the Euler relation and \eqref{eq:global_local_potentials} give the local identity
\begin{equation}\label{eq:global_gibbs_current_identity}
    \widetilde P_k^a+\Phi_0\widetilde J^a
                +\zeta_b\mathcal J^{ab}
       =P k^a-T_0s u^a~.
\end{equation}
In particular, the terms proportional to $\kappa\mu_0^2\rho_1v^a$ cancel between the stress and the two gauge contributions. Contracting with $n_a$ and integrating yields
\begin{equation}\label{eq:global_gibbs_free_energy}
    G=M-T_0S-\sum_i\Omega_iJ_i-\Phi_0Q_{(0)}-\Phi_1Q_{(1)}
      =-\int_{B_1}dV_1\,N P~.
\end{equation}
The Euclidean action is $I_E=G/T_0$. The measure follows from $\sqrt{-\gamma}\,d^2\sigma=N\,dt\,dV_1$ in coordinates adapted to the thermal flow. This extends the usual relation between the stationary blackfold action and its thermodynamic potential \cite{Caldarelli_2011} to the present 2-group fluid. At fixed $Q_{(1)}$, the appropriate magnetic Legendre transform is $G+\Phi_1Q_{(1)}$, consistently with the fixed-sector description in \eqref{eq:equilibrium_fixed_charge_pressure}.\\
\\
For variations through equilibrium solutions in a fixed background and gauge branch, the on-shell action obeys
\begin{equation}\label{eq:global_gibbs_variation}
    dG=-S\,dT_0-\sum_iJ_i\,d\Omega_i
                    -Q_{(0)}\,d\Phi_0-Q_{(1)}\,d\Phi_1~.
\end{equation}
The embedding variation vanishes by the extrinsic equation, while the intrinsic field equations remove the variations of the worldsheet fields. Equations \eqref{eq:global_gibbs_free_energy} and \eqref{eq:global_gibbs_variation} then give the global first law,
\begin{equation}\label{eq:global_first_law}
    dM=T_0\,dS+\sum_i\Omega_i\,dJ_i
                      +\Phi_0\,dQ_{(0)}+\Phi_1\,dQ_{(1)}~.
\end{equation}
Here a variation of $Q_{(1)}$ compares neighbouring charge sectors, rather than changing the string charge along a single worldsheet.\\
\\
\noindent\textbf{Tension and tensional energy.}
The stress tensor also determines the mechanical tension of the string. In the thermal frame it is the negative longitudinal stress,
\begin{equation}\label{eq:global_rest_tension}
    \tau=-T^{ab}v_av_b=-\Pi=\mu_1\rho_1-P~.
\end{equation}
For global quantities, however, the spatial direction is the unit tangent $e^a$ to the chosen slice $B_1$. Define its spatial projector and the corresponding tension by
\begin{equation}\label{eq:global_slice_tension}
    q_{ab}=\gamma_{ab}+n_an_b=e_ae_b
    \quad,\quad \tau_{(n)}=-q_{ab}T^{ab}~.
\end{equation}
These definitions include the full stress tensor, including its equilibrium energy flux. To make the frame dependence explicit, introduce a signed rapidity $\eta$ through $u^a=\cosh\eta\,n^a+\sinh\eta\,e^a$ and $v^a=\sinh\eta\,n^a+\cosh\eta\,e^a$. Substitution of \eqref{eq:equilibrium_constitutive_relations} gives
\begin{equation}\label{eq:global_boosted_tension}
    \tau_{(n)}=(\mu_1\rho_1-P)\cosh^2\eta
          -\epsilon\sinh^2\eta
          +2\kappa\mu_0^2\rho_1\sinh\eta\cosh\eta~.
\end{equation}
In particular, $\tau_{(n)}=\tau$ when $n^a=u^a$. The last term is the contribution of the 2-group energy flux and cannot be dropped in a moving frame.\\
\\
The total mechanical tensional energy is
\begin{equation}\label{eq:global_tensional_energy}
    \mathcal T_{\mathrm{tot}}
       =\int_{B_1}dV_1\,N\tau_{(n)}
       =-\int_{B_1}dV_1\,N(\gamma_{ab}+n_an_b)T^{ab}~.
\end{equation}
This is the string version of eq.~(2.48) of~\cite{Emparan:2011hg}. In the slicing used there, $\xi^a=R_0n^a$ and the lapse is $N=R_0$. Here $N$ remains the lapse defined in \eqref{eq:global_entropy_lapse}. Unlike $\tau_{(n)}$, which has units of energy per length, $\mathcal T_{\mathrm{tot}}$ has units of energy. Its variational interpretation follows directly from \eqref{eq: ward_worldsheet_variation}. A uniform spatial strain at fixed lapse, shift, thermal data and pulled-back gauge sources obeys $\delta\gamma_{ab}=2\lambda q_{ab}$ and gives
\begin{equation}\label{eq:global_tensional_strain}
    \delta G=-\frac12\int_{B_1}dV_1\,N T^{ab}\delta\gamma_{ab}
             =\lambda\mathcal T_{\mathrm{tot}}~.
\end{equation}
Thus the tensional energy measures the response to a fractional change of spatial length.\\
\\
A physical variation of the surrounding geometry can also change the gauge sources. Denote the independent external data by $b^I$. These include compactification lengths, curvature scales and flux parameters. At fixed global potentials their responses are
\begin{equation}\label{eq:global_background_work}
    \mathcal X_I=\left(\frac{\partial G}{\partial b^I}\right)_{
               T_0,\Omega_i,\Phi_0,\Phi_1,b^{J\ne I}}~.
\end{equation}
On a fixed coordinate domain, with the reference compensators fixed, the source variation gives the explicit expression
\begin{equation}\label{eq:global_background_response}
\begin{aligned}
    \mathcal X_I&=-\int_{B_1}dV_1\,N
        \left[\frac12T^{ab}\partial_I\gamma_{ab}
              +J^a\partial_I A_a
              +\frac12\mathcal J^{ab}\Delta_I B_{ab}\right]~,\\
    \Delta_I B_{ab}&=\partial_I B_{ab}
                        +2\kappa A_{[a}\partial_I A_{b]}~.
\end{aligned}
\end{equation}
The variations of the equilibrium fields and embedding vanish on shell. The metric term contains the mechanical strain response, while the remaining terms give the work of the gauge sources in the same ensemble. The extended first law is therefore
\begin{equation}\label{eq:global_extended_first_law}
    dM=T_0\,dS+\sum_i\Omega_i\,dJ_i
          +\Phi_0\,dQ_{(0)}+\Phi_1\,dQ_{(1)}
          +\sum_I\mathcal X_I\,db^I~.
\end{equation}
For a length parameter $L$, its conjugate $\mathcal T=\mathcal X_L$ is the thermodynamic tension~\cite{Armas:2015qsv}. An equilibrium ring radius is an embedding variable and does not supply an independent work term.\\
\\
For a uniform compact string in its asymptotic rest frame, with vanishing longitudinal background potentials, $G=-LP$ and $\Phi_1=L\mu_1$. Varying $L$ at fixed $T_0,\Phi_0,\Phi_1$ therefore yields
\begin{equation}\label{eq:global_compact_string_tension}
    \mathcal T=\left(\frac{\partial(-LP)}{\partial L}\right)_{
                         T_0,\Phi_0,\Phi_1}
       =-P+\mu_1\rho_1=\tau
    \quad,\quad \mathcal T_{\mathrm{tot}}=L\mathcal T~.
\end{equation}
Holding the global potential $\Phi_1$ fixed requires $\partial_L\mu_1=-\mu_1/L$. This produces the magnetic contribution to the tension. Holding $\mu_1$ fixed instead would give $-P$, which is a different response.\\
\\
Suppose the family of backgrounds and probe solutions is closed under a rescaling of all lengths, with $G_5$ and $\kappa$ fixed. In five dimensions the scaling weights of $(M,S,J_i,Q_{(0)},Q_{(1)})$ are $(2,3,3,2,1)$. If $b^I$ has weight $w_I$, define the total scaling response by
\begin{equation}\label{eq:global_scaling_tension}
    \boldsymbol{\mathcal T}=\sum_I w_I b^I\mathcal X_I~.
\end{equation}
For length parameters this is a sum of length times tension. Flux parameters contribute with their own scaling weights. Euler's relation then gives
\begin{equation}\label{eq:global_smarr}
    2M-3T_0S-3\sum_i\Omega_iJ_i
                    -2\Phi_0Q_{(0)}-\Phi_1Q_{(1)}
       =\boldsymbol{\mathcal T}~.
\end{equation}
The string charge is a flux rather than a charge integrated along the string, which accounts for the different coefficients of the two electromagnetic work terms.\\
\\
It remains to relate the right-hand side to the mechanical tensional energy in \eqref{eq:global_tensional_energy}. Choose the rotational generators to be tangent to $B_1$, so that $n_a\chi_i^a=0$, and write the tangential time-translation generator as $\xi^a=Nn^a+\nu^a$, with $n_a\nu^a=0$. The vector $\nu^a$ is its spatial shift. Using the charge definitions, \eqref{eq:global_local_potentials} and the local Euler relation gives
\begin{equation}\label{eq:global_tension_smarr_identity}
    \boldsymbol{\mathcal T}
        =\mathcal T_{\mathrm{tot}}+\mathcal W_{\mathrm{ext}}
                     -\int_{B_1}dV_1\,N(P+Ts)~,
\end{equation}
where the explicit shift and gauge-potential contribution is
\begin{equation}\label{eq:global_tension_source_terms}
\begin{aligned}
    \mathcal W_{\mathrm{ext}}=\int_{B_1}dV_1\,n_a\Big[
        &-T^{ab}\nu_b+(\alpha_k-\alpha_\xi)J^a\\
        &+\left(2\eta_{kb}-\eta_{\xi b}
                         -2\kappa\Phi_0 A_b\right)\mathcal J^{ab}\Big]~.
\end{aligned}
\end{equation}
For example, the Gibbs relation rewrites the left-hand side of \eqref{eq:global_smarr} as $3G-M+\Phi_0Q_{(0)}+2\Phi_1Q_{(1)}$. Substituting the current integrals into this expression gives \eqref{eq:global_tension_smarr_identity}. The last integral retains the dependence on a general equation of state. For the ungauged black-string equation of state used in this paper, $P=-Ts$, as verified in Section~\ref{sec:non-ext_string}, and this integral vanishes. In a slicing with $\nu^a=0$ and vanishing gauge sources and reference compensators, $\mathcal W_{\mathrm{ext}}=0$, recovering the tensional Smarr relation of~\cite{Emparan:2011hg}. In the flux backgrounds considered here the explicit source terms must be retained. For an isolated balanced compact blackfold in Minkowski space with no external fields, the extrinsic equations imply $\mathcal T_{\mathrm{tot}}=0$. This does not require the local thermal-frame tension $\tau$ to vanish.

\section{The black string of five-dimensional minimal supergravity} \label{ch: THERMOGENERAL}
We now specialise this effective description to the black string used in our subsequent constructions. Section~\ref{sec:blackfold_summary} reviews the blackfold approach. Section~\ref{sec:non-ext_string} presents the non-extremal string geometry and thermodynamics in the thermal rest frame, identifying the quantities entering the constitutive relations of Section~\ref{sec:equilibrium_action}. Section~\ref{sec:susy_lim} takes the supersymmetric and null-wave limits of the geometry, stress tensor and currents. Finally, Section~\ref{sec:string_validity} determines the scale hierarchy required for a controlled blackfold approximation.

\subsection{Summary of the blackfold approach}\label{sec:blackfold_summary}
The blackfold approach is an effective field theory (EFT) for the long-wavelength dynamics of black holes whose horizons are locally approximated by black branes~\cite{Emparan:2009cs,Emparan:2009at}. Its starting point is a separation between the transverse size of the horizon and the longer scales over which its local state and embedding vary. A thin black ring, for example, resembles a straight black string when examined over distances much smaller than its bending radius~\cite{Emparan:2007wm}. The nonlinear geometry near the horizon is retained in this local approximate solution, while its slow deformations are described by a theory on the worldvolume.

To obtain this theory, one integrates out the transverse structure of the local brane geometry. At the classical level, this means solving for its short-distance gravitational and gauge fields and encoding their effects in an effective stress tensor, charge currents and response coefficients \cite{Emparan:2009at,Armas:2016mes}. The retained variables are the collective fields that specify the local thermodynamic state, velocity and position of the brane. In five-dimensional minimal supergravity, the local string geometry has a transverse horizon $\mathbb{S}^2$, and integrating over this sphere leaves a two-dimensional worldsheet $\Sigma_2$. Its intrinsic fields describe a charged fluid, while $X^\mu(\sigma)$ describes transverse displacements. The resulting EFT couples fluid dynamics to the dynamics of the embedding~\cite{Emparan:2011hg}.

Let $r_c$ denote the largest intrinsic length scale of the black string geometry, including its charge scales. The expansion requires
\begin{equation}\label{eq:blackfold_scale_hierarchy}
    r_c\ll\mathcal R\quad,\quad
    \mathcal R=\min\!\left(R_{\mathrm{bend}},L_{\mathrm{var}},L_{\mathrm{bg}}\right)~,
\end{equation}
where $R_{\mathrm{bend}}$ is the smallest bending radius, $L_{\mathrm{var}}$ the shortest scale of variation of the worldsheet fields, and $L_{\mathrm{bg}}$ the shortest local scale associated with the background curvature and gauge field \cite{Armas:2016mes}. This is an expansion in small gradients and curvatures, allowing finite local charge densities and velocities. In the case of gauged supergravity, which we briefly address at the end of this paper, the AdS radius must also be large compared with $r_c$, so the ungauged string supplies the leading local approximate solution \cite{Armas:2010hz}.

The EFT is organised through a deriavative expansion. At leading derivative order, each worldsheet point is assigned the stress tensor and currents of a uniform string in the corresponding local state. The monopole approximation represents these quantities as sources supported on $\Sigma_2$, as in \eqref{eq: multipole}. Hydrodynamic corrections describe transport within the fluid~\cite{Camps:2010br,DiDato:2015bta}. Higher multipoles capture the response of the finite transverse structure, including bending moments and electric polarisation~\cite{Armas:2011uf,Armas_2013,Armas_2013ele}. These corrections introduce transport and elastic response coefficients that must be determined from gravity. The leading constitutive relations are instead fixed once the local black brane thermodynamics and the relevant symmetries are specified.

Matching to the gravitational description makes this organisation concrete \cite{Camps:2012hw,Armas:2016mes}. In the near zone, one promotes the parameters and velocity of the uniform string to slowly varying fields, replaces the flat worldsheet metric by $\gamma_{ab}$, and allows the embedding to bend. Corrections to the metric and gauge field are then determined order by order, imposing horizon regularity. In the far zone, the effective sources generate weak perturbations of the prescribed supergravity background. Both descriptions apply at transverse distances $r_c\ll r\ll\mathcal R$, where their fields are matched. This fixes the effective coefficients and incorporates the gravitational and electromagnetic backreaction systematically.

At monopole order, the consistency conditions are the intrinsic and extrinsic equations \eqref{eq: blackfoldstress}--\eqref{eq: blackfoldcurrents}. The former govern the evolution of the local state and currents, while the latter balance the string stress against the external forces. In minimal supergravity, the Chern-Simons coupling ties the electric and magnetic currents into the 2-group of Section~\ref{sec: Ward} \cite{Armas:2024caa}. Electric charge obeys the modified conservation law, while the magnetic top-form coefficient is constant on each connected worldsheet. The transverse force involves both $F$ and $\mathcal F=-\star F$. Thus the local string parameters promoted to fields are subject to these equations, including the constraint on the string charge. In the next sections we detail the thermodynamics of the black string of \cite{Compere:2010fm} that we take as the starting point for constructing new black holes.

\subsection{The non-extremal black string}\label{sec:non-ext_string}
We now specialise the general string of Comp\`ere et al.~\cite{Compere:2010fm} to vanishing transverse rotation and express its geometry and thermodynamics in the thermal rest frame (in which $v_z=0$). The independent parameters are $m>0$, $\beta$ and $\delta$, which control non-extremality, magnetic charge and electric charge, respectively. As anticipated in Section~\ref{sec:equilibrium_action}, a 2-group fluid can carry an equilibrium energy flux in its thermal rest frame.\\
\\
\noindent\textbf{Geometry in the thermal rest frame.}
We write $c_\alpha=\cosh\alpha$, $s_\alpha=\sinh\alpha$ and introduce the dimensionless functions
\begin{equation}\label{eq:nonext_rest_parameters}
\begin{aligned}
 h_\delta&=\sqrt{1+3c_\delta^2}~,
 &\mathsf B&=c_\beta^2+s_\beta^2~,
 &\mathsf X&=\frac{2s_\beta c_\beta}{h_\delta}~,\\
 \mathsf A&=c_\beta^2+\frac{3s_\beta^2s_\delta^2}{h_\delta^2}~,
 &\mathsf Z&=\sqrt{\mathsf A^2-\mathsf X^2}~.
\end{aligned}
\end{equation}
The horizon generator of the solution with arbitrary $\gamma$ is $k=\partial_t+v_z\partial_z$, where
\begin{equation}\label{eq:nonext_rest_boost}
\begin{gathered}
 v_z=\frac{\mathsf A s_\gamma-\mathsf X c_\gamma}
              {\mathsf A c_\gamma-\mathsf X s_\gamma}
 \quad\Longrightarrow\quad
 \gamma=\chi\equiv\operatorname{arctanh}\frac{\mathsf X}{\mathsf A}~,\\
 c_\chi=\frac{\mathsf A}{\mathsf Z}
 \quad,\quad s_\chi=\frac{\mathsf X}{\mathsf Z}~.
\end{gathered}
\end{equation}
We implement this boost on the non-compact covering space and, when total charges are needed, identify $z\sim z+L$ in the resulting rest coordinates. Thus $L$ is the asymptotic spatial period in this frame. Even though we consider this possibility here as in \cite{Compere:2010fm} we are mostly interested in the non-compact version. Define the one-forms
\begin{equation}\label{eq:nonext_rest_coframe}
 \vartheta^0=c_\chi\,dt-s_\chi\,dz
 \quad,\quad
 \vartheta^1=c_\chi\,dz-s_\chi\,dt~.
\end{equation}
They relate the coordinates of the $\gamma=0$ solution to the rest coordinates and are not the thermal velocity and its spatial dual. The metric and gauge potential then take the form
\begin{equation}\label{eq:metricofnonextremal}
 ds^2=-\frac{H\Delta}{\xi}(\vartheta^0)^2
       +\frac{\xi}{H^2}
           \left(\vartheta^1+\frac{k_r}{\xi}\vartheta^0\right)^2
       +H\left(\frac{dr^2}{\Delta}+d\Omega_2^2\right)~,
\end{equation}
\begin{equation}\label{eq:nonext_rest_gauge}
 A=-\frac{\sqrt{3}}{H}\left(F_3\vartheta^0+F_2\vartheta^1\right)
       -Q_M\cos\theta\,d\phi~,
 \qquad
 Q_M=2\sqrt{3}\,m c_\delta^2\mathsf X~,
\end{equation}
where $d\Omega_2^2=d\theta^2+\sin^2\theta\,d\phi^2$ and
\begin{equation}\label{eq:nonext_metric_functions}
\begin{aligned}
 \Delta&=r(r-2m)~,
 &H&=F_1+\Delta~,\\
 \xi&=(F_4+\Delta)H-F_2^2~,
 &k_r&=F_5H-F_2F_3~.
\end{aligned}
\end{equation}
The five radial functions are
\begin{equation}\label{eq:nonext_radial_functions}
\begin{aligned}
 F_1={}&2mc_\delta^2\left[r\mathsf B
       +2ms_\beta^2\left(s_\beta^2
                 +\frac{s_\delta^2c_\beta^2}{h_\delta^2}\right)\right]~,\\
 F_2={}&2mc_\delta s_\delta\mathsf X
                       \left[r+m(c_\delta^2\mathsf B-1)\right]~,\\
 F_3={}&2mc_\delta s_\delta\left[-r\mathsf B
       +2ms_\beta^2\left(\frac{c_\beta^2(1+c_\delta^2)}{h_\delta^2}
                          -s_\beta^2\right)\right]~,\\
 F_4={}&2m\left[r\mathsf B(c_\delta^2+s_\delta^2)
       +2m\left((c_\beta^2s_\delta^2+s_\beta^2c_\delta^2)^2
                 +\frac{s_\delta^2s_\beta^2c_\beta^2}{h_\delta^2}\right)\right]~,\\
 F_5={}&2m\mathsf X\left[r+m(c_\delta^4\mathsf B-1)\right]~.
\end{aligned}
\end{equation}
In particular, both longitudinal components of $A$ have denominator $H=F_1+\Delta$. The magnetic potential is understood in two regular patches, obtained by replacing $-Q_M\cos\theta\,d\phi$ by $Q_M(1-\cos\theta)\,d\phi$ near the north pole and by $-Q_M(1+\cos\theta)\,d\phi$ near the south pole. With the standard orientation of the linking sphere, $\int_{S^2}F=4\pi Q_M$. The outer horizon is at $r_+=2m$, with generator $k=\partial_t$, and the solution approaches $-dt^2+dz^2+dr^2+r^2d\Omega_2^2$ at infinity.\\
\\
\noindent\textbf{Charges and horizon thermodynamics.}
Let $M,P_z,Q_E,Q_M$ and $\mathcal T$ denote the ADM mass, longitudinal momentum, electric charge, magnetic flux parameter and ADM tension in the thermodynamic conventions of~\cite{Compere:2010fm}. All quantities below are evaluated at $\gamma=\chi$. It is useful to define
\begin{equation}\label{eq:nonext_thermo_functions}
 \mathcal C=\frac{\Omega_2}{16\pi G_5}=\frac{1}{4G_5}
 \quad,\quad \mathsf K=1+3s_\delta^2
 \quad,\quad \mathsf Y=1+\frac{6c_\delta^2s_\beta^2}{h_\delta^2}~.
\end{equation}
The mass, momentum and electric charge per unit length, together with the tension, are
\begin{equation}\label{eq:nonext_rest_charges}
\begin{aligned}
 \frac{M}{L}&=2\mathcal C m
        \left[\mathsf B(1+\mathsf Kc_\chi^2)
                        -2\mathsf Xc_\chi s_\chi\right]~,\\
 \frac{P_z}{L}&=2\mathcal C m
        \left[\mathsf B\mathsf Kc_\chi s_\chi
                        -\mathsf X(c_\chi^2+s_\chi^2)\right]~,\\
 \frac{Q_E}{L}&=2\sqrt{3}\,\mathcal C m c_\delta s_\delta
                         (\mathsf Bc_\chi+\mathsf Xs_\chi)~,\\
 \mathcal T&=2\mathcal C m
        \left[\mathsf B(1-\mathsf Ks_\chi^2)
                        +2\mathsf Xc_\chi s_\chi\right]~.
\end{aligned}
\end{equation}
The magnetic parameter $Q_M$ is given in \eqref{eq:nonext_rest_gauge} and is not multiplied by $L$. There is no transverse angular momentum, $J_\phi=0$, and $\Omega_\phi=0$. Defining $\mathcal A_*=4m^2c_\delta^3c_\beta^2\mathsf Z$, the horizon area, entropy and temperature are
\begin{equation}\label{eq:nonext_rest_horizon}
 A_H=4\pi L\mathcal A_*\quad,\quad
 S_{\mathrm{BH}}=4\pi\mathcal C L\mathcal A_*\quad,\quad
 T_H=\frac{m}{2\pi\mathcal A_*}
       =\frac{1}{8\pi m c_\delta^3c_\beta^2\mathsf Z}~.
\end{equation}
The potentials conjugate to $Q_E$ and $Q_M$ are
\begin{equation}\label{eq:nonext_rest_potentials}
\begin{aligned}
 \Phi_E&=\frac{\sqrt{3}\,s_\delta\mathsf Y}{c_\delta\mathsf Z}~,\\
 \Phi_M&=\mathcal C L\Psi_M\quad,\quad
 \Psi_M=\frac{\sqrt{3}\,\mathsf X}{c_\delta^2\mathsf Z^2}
                    \left(1-\frac{\mathsf A s_\delta^2}{2c_\beta^2}\right)~.
\end{aligned}
\end{equation}
Here $\Psi_M$ is only a convenient expression for the magnetic potential per normalized charge and per unit length. The normalization appropriate to the currents of Section~\ref{sec:highergroup} will be specified below. The expression for $\Phi_M$ follows by substituting the other charges into the Smarr relation and also satisfies the first law for independent variations of $m,\beta,\delta$ and $L$. On the family with $v_z=0$, these relations read
\begin{equation}\label{eq:nonext_rest_first_law}
 dM=T_H\,dS_{\mathrm{BH}}+\Phi_E\,dQ_E
                    +\Phi_M\,dQ_M+\mathcal T\,dL~,
\end{equation}
\begin{equation}\label{eq:nonext_rest_smarr}
 2M=3T_HS_{\mathrm{BH}}+2\Phi_EQ_E+\Phi_MQ_M+\mathcal T L~.
\end{equation}
Although $P_z$ can vary within this family, its conjugate velocity is zero. Consequently there is no momentum work term in \eqref{eq:nonext_rest_first_law}.\\
\\
\noindent\textbf{Identification with the 2-group fluid.}
We now use the source and Hodge conventions of Section~\ref{sec:highergroup}. Choose the bulk orientation $dt\wedge dz\wedge dr\wedge\sin\theta\,d\theta\wedge d\phi$ and the worldsheet volume form $dt\wedge dz$. The thermal vectors on the asymptotic worldsheet are then
\begin{equation}\label{eq:nonext_thermal_vectors}
 \gamma_{ab}=\eta_{ab}\quad,\quad
 u^a=(1,0)\quad,\quad v^a=(0,-1)\quad,\quad
 \epsilon^{tz}=-1~.
\end{equation}
In particular $u^a$ is future directed and $v^a=\epsilon^{ab}u_b$, as in \eqref{eq:equilibrium_invariants}. With this orientation, the magnetic source equation \eqref{eq:Bianchi} gives $\rho_1=\mathcal C Q_M$. The sign of the electric density follows independently from \eqref{eq:EOM}. The asymptotic extraction of the currents and stress tensor from the geometry and gauge field is detailed in Appendix~\ref{app:stress_currents}. It gives
\begin{equation}\label{eq:nonext_rest_electric_current}
\begin{aligned}
 J^t&=-2\sqrt{3}\,\mathcal C m c_\delta s_\delta
                 (\mathsf Bc_\chi+\mathsf Xs_\chi)=-\frac{Q_E}{L}~,\\
 J^z&=-2\sqrt{3}\,\mathcal C m c_\delta s_\delta
                 (\mathsf Xc_\chi+\mathsf Bs_\chi)~.
\end{aligned}
\end{equation}
In the gauge of \eqref{eq:nonext_rest_gauge}, $A_t$ vanishes at infinity and $A_t(2m)=\Phi_E$. Hence the chemical potential in Section~\ref{sec:highergroup} is $\mu_0=A_t(\infty)-A_t(2m)=-\Phi_E$. Reversing both the electric charge and its conjugate potential leaves their thermodynamic work term invariant.\\
\\
The complete local dictionary between the 2-group fluid of Sec.~\ref{sec:highergroup} and the thermodynamics of the black brane is given in Table~\ref{tab:string-a0-charges}. The quantities in its last column always refer to the string with  $v_z=0$.
\begin{table}[htbp]
\centering
\renewcommand{\arraystretch}{1.35}
\begin{tabular}{@{}lcc@{}}
\toprule
Quantity & Section~\ref{sec:highergroup} & Black string thermodynamics \\
\midrule
Energy density & $\epsilon$ & $M/L$ \\
Entropy density & $s$ & $S_{\mathrm{BH}}/L$ \\
Temperature & $T$ & $T_H$ \\
Electric density & $\rho_0$ & $-Q_E/L$ \\
String charge & $\rho_1=Q_{(1)}$ & $\mathcal C Q_M$ \\
Electric chemical potential & $\mu_0$ & $-\Phi_E$ \\
String chemical potential & $\mu_1$ & $\Phi_M/(\mathcal C L)=\Psi_M$ \\
Thermodynamic pressure & $P$ & $-2\mathcal C m$ \\
Longitudinal stress & $\Pi=P-\mu_1\rho_1$ & $-\mathcal T$ \\
Energy-flux coefficient & $h=-\kappa\mu_0^2\rho_1$ & $-P_z/L$ \\
\bottomrule
\end{tabular}
\caption{Thermal-rest-frame dictionary in the conventions of Section~\ref{sec:highergroup}, with $\kappa=1/\sqrt{3}$ and the orientation \eqref{eq:nonext_thermal_vectors}. The string charge is a normalized flux, independent of the spatial period $L$.}
\label{tab:string-a0-charges}
\label{tab:string-a0-potentials}
\end{table}
The asymptotic metric and gauge field reproduce the constitutive relations
\begin{equation}\label{eq:stresstensornonextremal}
\begin{aligned}
 T^{ab}&=\epsilon u^au^b+\Pi v^av^b
                         -2\kappa\mu_0^2\rho_1u^{(a}v^{b)}~,\\
 J^a&=\rho_0u^a-2\kappa\mu_0\rho_1v^a~,
 &\mathcal J^{ab}&=\rho_1(u^av^b-u^bv^a)~.
\end{aligned}
\end{equation}
For example, the off-diagonal stress and the spatial electric current satisfy
\begin{equation}\label{eq:nonext_rest_flux_check}
 \frac{P_z}{L}=\kappa\Phi_E^2\mathcal C Q_M
 \quad,\quad J^z=-2\kappa\Phi_E\mathcal C Q_M
 \quad,\quad \mathcal T=2\mathcal C m+\Psi_M\mathcal C Q_M~.
\end{equation}
These equalities follow by inserting \eqref{eq:nonext_rest_boost} into \eqref{eq:nonext_rest_charges} and \eqref{eq:nonext_rest_electric_current}. Thus the nonzero momentum at $v_z=0$ is precisely the energy flux required by the 2-group constitutive relations. The thermal frame need not be a frame of vanishing momentum. The distinction between the coframe \eqref{eq:nonext_rest_coframe} and the thermal vectors \eqref{eq:nonext_thermal_vectors} is essential for this matching.\\
\\
To check the thermodynamic relations, first hold $L$ fixed in \eqref{eq:nonext_rest_first_law}. Using the dictionary gives
\begin{equation}\label{eq:nonext_local_thermodynamics}
\begin{aligned}
 d\epsilon&=T\,ds+\mu_0\,d\rho_0+\mu_1\,d\rho_1~,\\
 \epsilon+P&=Ts+\mu_0\rho_0+\mu_1\rho_1~,
 &dP&=s\,dT+\rho_0\,d\mu_0+\rho_1\,d\mu_1~.
\end{aligned}
\end{equation}
The Euler relation can be checked directly from the charges. Together with \eqref{eq:nonext_rest_smarr} and $\Pi=-\mathcal T$, it yields $P=-Ts=-2\mathcal C m$. This equality characterises the equation of state of this string. In particular, the pressure entering the equilibrium action is $P$, while the pressure in the extrinsic equation is $\Pi=P-\mu_1\rho_1$. The fixed string charge potential of \eqref{eq:equilibrium_fixed_charge_pressure} is $\widehat P=\Pi$. Equations~\eqref{eq:nonext_rest_charges}-\eqref{eq:nonext_rest_potentials} and the dictionary therefore supply a parametric equation of state for the fluid of Section~\ref{sec:equilibrium_action}.\\
\\
Finally, choosing a vanishing asymptotic longitudinal gauge sources and the corresponding zero reference compensators in Section~\ref{sec: conserved_charges}, the global dictionary is
\begin{equation}\label{eq:nonext_global_dictionary}
\begin{aligned}
 Q_{(0)}&=L\rho_0=-Q_E~,
 &Q_{(1)}&=\rho_1=\mathcal C Q_M~,\\
 \Phi_0&=\mu_0=-\Phi_E~,
 &\Phi_1&=L\mu_1=\frac{\Phi_M}{\mathcal C}~.
\end{aligned}
\end{equation}
It follows that $\Phi_0\,dQ_{(0)}=\Phi_E\,dQ_E$ and $\Phi_1\,dQ_{(1)}=\Phi_M\,dQ_M$. Thus \eqref{eq:nonext_rest_first_law} and \eqref{eq:nonext_rest_smarr} agree with the global first law and scaling relation of Section~\ref{sec: conserved_charges}, including the work $\mathcal T\,dL$ when the period varies. The corresponding Gibbs free energy is
\begin{equation}\label{eq:nonext_gibbs_free_energy}
 G=M-T_HS_{\mathrm{BH}}-\Phi_EQ_E-\Phi_MQ_M
       =-LP=2\mathcal C mL~.
\end{equation}
Variations of $\rho_1$ in these relations compare different string charge sectors. Within a given connected worldsheet its value remains constant, as required by the top-form conservation law.

\subsection{Supersymmetric and null-wave limits}\label{sec:susy_lim}
The non-extremal string of the previous section admits two regular extremal limits with vanishing transverse rotation~\cite{Compere:2010fm}. They differ in the behaviour of the thermal velocity. On the generic dyonic supersymmetric branch it remains timelike, and the thermal rest frame of Section~\ref{sec:non-ext_string} has a smooth limit. In the second limit the velocity approaches the speed of light in a fixed asymptotic frame. The latter is a charged version of the null-wave limit of~\cite{Emparan:2011hg}, with an ideal description in terms of null matter~\cite{Armas:2025nullmatter}. We take both limits directly in the metric, gauge potential and constitutive relations.

\subsubsection{Supersymmetric limit at finite thermal velocity}
The first limit is
\begin{equation}\label{eq:string_susy_scaling}
 m\longrightarrow0\quad,\quad \beta\longrightarrow\infty
 \quad,\quad me^{2\beta}=2\mathfrak p~,
 \qquad \mathfrak p>0\quad,\quad \delta>0~,
\end{equation}
with $\mathfrak p,\delta$ fixed. In the thermal rest frame $\gamma=\chi$, and the functions in \eqref{eq:nonext_rest_parameters} give
\begin{equation}\label{eq:string_susy_boost_limit}
 c_{\chi_{\mathrm B}}=\frac{3c_\delta^2-1}{3c_\delta s_\delta}
 \quad,\quad
 s_{\chi_{\mathrm B}}=\frac{h_\delta}{3c_\delta s_\delta}~.
\end{equation}
Thus this boost stays finite for fixed $\delta>0$. Define the two rest-frame charge parameters
\begin{equation}\label{eq:string_susy_parameters}
 Q_{\mathrm R}=2\mathfrak p c_\delta^2
 \quad,\quad q=\frac{4\mathfrak p c_\delta^2}{h_\delta}~.
\end{equation}
For example, the full radial function $H$, including its constant term, has the limit
\begin{equation}\label{eq:string_susy_radial_limit}
 \Delta(r)\longrightarrow r^2
 \quad,\quad H\longrightarrow r^2+Q_{\mathrm R}r+\frac{q^2}{4}~.
\end{equation}
Substituting the remaining radial functions and \eqref{eq:string_susy_boost_limit} into \eqref{eq:metricofnonextremal}--\eqref{eq:nonext_rest_gauge} gives
\begin{equation}\label{eq:string_susy_geometry}
\begin{aligned}
 ds^2={}&-f^2(dt+\omega\,dz)^2
             +f^{-1}(dz^2+dr^2+r^2d\Omega_2^2)~,\\
 A={}&\sqrt{3}(1-f)\,dt
       -\sqrt{3}\left(f\omega-\frac{q}{2r}\right)dz
       -\frac{\sqrt{3}q}{2}\cos\theta\,d\phi~,
\end{aligned}
\end{equation}
where
\begin{equation}\label{eq:string_susy_functions}
 f^{-1}=1+\frac{Q_{\mathrm R}}{r}+\frac{q^2}{4r^2}
 \quad,\quad
 \omega=\frac{3q}{2r}+\frac{3qQ_{\mathrm R}}{4r^2}
                         +\frac{q^3}{8r^3}~.
\end{equation}
Here $t,z$ are the thermal rest coordinates. The longitudinal potential vanishes at infinity, and the magnetic potential is understood patchwise. The horizon is at $r=0$, with generator $\partial_t$. Its spatial section has sphere radius $q/2$ and longitudinal metric coefficient $3(Q_{\mathrm R}^2-q^2)/q^2$. Regularity requires $Q_{\mathrm R}>q$, ensured by $\delta>0$.\\
\\
The same substitution in the charges and potentials of Section~\ref{sec:non-ext_string} yields
\begin{equation}\label{eq:string_susy_thermodynamics}
\begin{gathered}
 \epsilon=3\mathcal C Q_{\mathrm R}\quad,\quad
 \rho_0=-\sqrt{3}\,\mathcal C Q_{\mathrm R}\quad,\quad
 \rho_1=\frac{\sqrt{3}\,\mathcal C q}{2}~,\\
 T=P=\mu_1=0\quad,\quad \mu_0=-\sqrt{3}\quad,\quad
 s=\pi\sqrt{3}\,\mathcal C q\sqrt{Q_{\mathrm R}^2-q^2}~.
\end{gathered}
\end{equation}
In particular, the entropy remains finite while $Ts=2\mathcal C m$ vanishes. The first law reduces to $d\epsilon=-\sqrt{3}\,d\rho_0$, with the string charge still labelling distinct sectors. The limiting stress tensor and currents are
\begin{equation}\label{eq:string_susy_monopoles}
\begin{aligned}
 T^{ab}&=\mathcal C
      \begin{pmatrix}3Q_{\mathrm R}&3q/2\\3q/2&0\end{pmatrix}~,\\
 J^a&=-\sqrt{3}\,\mathcal C(Q_{\mathrm R},q)~,
 &\mathcal J^{tz}&=-\frac{\sqrt{3}\,\mathcal C q}{2}~.
\end{aligned}
\end{equation}
These expressions follow equally by taking \eqref{eq:string_susy_thermodynamics} in \eqref{eq:stresstensornonextremal}. The rest-frame tension $\mathcal T=-T^{zz}$ vanishes, while the energy flux and spatial electric current remain nonzero. They are the finite 2-group contributions proportional to $\mu_0^2\rho_1$ and $\mu_0\rho_1$.\\
\\
To relate this result to the three-parameter supersymmetric string, retain a finite relative rapidity $\eta=\gamma-\chi_{\mathrm B}$. The rest coordinates are related to canonical asymptotic coordinates by $t_{\mathrm R}=c_\eta t-s_\eta z$ and $z_{\mathrm R}=c_\eta z-s_\eta t$. The parameters of~\cite{Compere:2010fm,Kim:2010bf} are then
\begin{equation}\label{eq:string_susy_three_parameter_map}
\begin{gathered}
 Q=Q_{\mathrm R}c_\eta+q s_\eta
 \quad,\quad \Delta=-Q s_\eta
 \quad,\quad S=Q c_\eta=\sqrt{Q^2+\Delta^2}~,\\
 Q_{\mathrm R}=\frac{Q^2+q\Delta}{S}~.
\end{gathered}
\end{equation}
Together with \eqref{eq:string_susy_parameters}, this identifies all three parameters in terms of the non-extremal family. Applying the inverse rest-frame boost to \eqref{eq:string_susy_geometry} gives the full three-parameter geometry. The transformed stress and current are
\begin{equation}\label{eq:string_susy_boosted_monopoles}
\begin{aligned}
 T^{ab}&=3\mathcal C
   \begin{pmatrix}
     S&q/2-\Delta\\
     q/2-\Delta&\Delta(\Delta-q)/S
   \end{pmatrix}~,\\
 J^a&=-\sqrt{3}\,\mathcal C
         \left(Q,\frac{Q(q-\Delta)}{S}\right)~,
 &\mathcal J^{tz}&=-\frac{\sqrt{3}\,\mathcal C q}{2}~.
\end{aligned}
\end{equation}
The horizon velocity is $\tanh\eta=-\Delta/S$. In Appendix~\ref{app:susy_lim} we give the coordinate transformations, including the relation to the original supersymmetric coordinates and the effect of the boost on a compact identification.

\noindent\textbf{Global charges.}
The charges of Section~\ref{sec: conserved_charges} can now be specialised to this equation of state. For an embedded string, $A,B$ in the following expressions are the external sources, with the compensators and reference potentials fixed as in that section. The covariant form of \eqref{eq:string_susy_monopoles} is obtained with the local thermal vectors $u^a,v^a$. For either $X=\xi$ or $X=\chi_i$, its conserved-current density is
\begin{equation}\label{eq:string_susy_global_density}
\begin{aligned}
 \frac{n_a\widetilde P_X^a}{\mathcal C}
 ={}&3Q_{\mathrm R}(n\cdot u)(u\cdot X)
       -\frac{3q}{2}\left[(n\cdot u)(v\cdot X)
                          +(n\cdot v)(u\cdot X)\right]\\
 &-\sqrt{3}\,\alpha_X\left[Q_{\mathrm R}(n\cdot u)-q(n\cdot v)\right]
       -\frac{\sqrt{3}q}{2}e^a\eta_{Xa}~.
\end{aligned}
\end{equation}
Here $X_a$ is the pulled-back metric dual of the symmetry generator, and $e^a=-n_b\epsilon^{ba}$ retains the orientation of Section~\ref{sec: conserved_charges}. The mass and angular momenta are therefore
\begin{equation}\label{eq:string_susy_global_charges}
 M=\int_{B_1}dV_1\,n_a\widetilde P_\xi^a
 \quad,\quad
 J_i=-\int_{B_1}dV_1\,n_a\widetilde P_{\chi_i}^a~,
\end{equation}
with the integrand given explicitly by \eqref{eq:string_susy_global_density}. The electric charge, string charge and entropy become
\begin{equation}\label{eq:string_susy_global_electric_entropy}
\begin{aligned}
 Q_{(0)}&=\sqrt{3}\,\mathcal C\int_{B_1}dV_1
       \left[Q_{\mathrm R}(n\cdot u)-q(n\cdot v)\right]
                         -\mathcal Cq\oint_{B_1}A~,\\
 Q_{(1)}&=\frac{\sqrt{3}\,\mathcal Cq}{2}~,\\
 S_{\mathrm{BH}}&=-\pi\sqrt{3}\,\mathcal Cq
       \int_{B_1}dV_1\sqrt{Q_{\mathrm R}^2-q^2}\,(n\cdot u)~.
\end{aligned}
\end{equation}
In particular, $q$ is constant on a connected worldsheet, whereas $Q_{\mathrm R}$ and the thermal vectors can depend on position. The Page improvement survives the extremal limit. The corresponding potentials and mechanical tensional energy are
\begin{equation}\label{eq:string_susy_global_potentials}
 \Phi_0=-\sqrt{3}|k|-\alpha_k
 \quad,\quad
 \Phi_1=\int_{B_1}dV_1\,e^a
       \left(\sqrt{3}k_a-\eta_{ka}
                              +\frac{2\Phi_0}{\sqrt{3}}A_a\right)~,
\end{equation}
\begin{equation}\label{eq:string_susy_global_tension}
 \mathcal T_{\mathrm{tot}}=3\mathcal C\int_{B_1}dV_1\,N
       \left[q(e\cdot u)(e\cdot v)-Q_{\mathrm R}(e\cdot u)^2\right]~.
\end{equation}
Thus vanishing local $\mu_1$ and rest-frame tension do not imply vanishing global magnetic potential or tensional energy in a moving frame or in external fields.\\
\\
For the homogeneous string with period $L$ in the asymptotic frame of \eqref{eq:string_susy_boosted_monopoles}, and with vanishing external sources and reference compensators, these expressions reduce to
\begin{equation}\label{eq:string_susy_flat_global_charges}
\begin{gathered}
 M=3\mathcal C L S\quad,\quad
 P_z=3\mathcal C L\left(\frac q2-\Delta\right)\quad,\quad
 Q_{(0)}=-\sqrt{3}\,\mathcal C LQ
 \quad,\quad Q_{(1)}=\frac{\sqrt{3}\,\mathcal Cq}{2}~,\\
 S_{\mathrm{BH}}=\pi\sqrt{3}\,\mathcal C Lq
                        \sqrt{Q^2-q^2+2q\Delta}~,
 \qquad
 \mathcal T_{\mathrm{tot}}
       =\frac{3\mathcal C L\Delta(q-\Delta)}{S}~.
\end{gathered}
\end{equation}
Here $S=\sqrt{Q^2+\Delta^2}$ is the parameter in \eqref{eq:string_susy_three_parameter_map}, distinct from the entropy $S_{\mathrm{BH}}$. With $k=\partial_t+v_z\partial_z$, the potentials are
\begin{equation}\label{eq:string_susy_flat_global_potentials}
 v_z=-\frac{\Delta}{S}\quad,\quad
 \Phi_0=-\frac{\sqrt{3}Q}{S}\quad,\quad
 \Phi_1=\frac{\sqrt{3}L\Delta}{S}\quad,\quad T_0=0~.
\end{equation}
They satisfy $M=v_zP_z+\Phi_0Q_{(0)}+\Phi_1Q_{(1)}$ and, at fixed $L$, $dM=v_z\,dP_z+\Phi_0\,dQ_{(0)}+\Phi_1\,dQ_{(1)}$. In the thermal rest frame $\Delta=0$, so $M=-\sqrt{3}Q_{(0)}$, $\Phi_1=0$ and $\mathcal T_{\mathrm{tot}}=0$, while $P_z=3\mathcal C Lq/2$ remains nonzero.There is a purely magnetic endpoint but requires a divergent relative boost and is treated with the null-wave limits in the next Section~\ref{sec:string_null_limits}.

\subsubsection{Null-wave limits}\label{sec:string_null_limits}
For the second limit, we keep a fixed asymptotic Minkowski frame and boost the thermal vectors along the positive $z$ direction,
\begin{equation}\label{eq:string_null_vectors}
\begin{gathered}
 u^a=(c_\eta,s_\eta)\quad,\quad v^a=(-s_\eta,-c_\eta)
 \quad,\quad v_z=\tanh\eta\longrightarrow1~,\\
 \ell^a=(1,1)\quad,\quad \tilde\ell^a=\tfrac12(1,-1)
 \quad,\quad \ell^2=\tilde\ell^2=0\quad,\quad
 \ell\cdot\tilde\ell=-1~.
\end{gathered}
\end{equation}
Here $\eta=\gamma-\chi$ is the rapidity relative to the thermal rest frame. We use the null pair $\ell^a,\tilde\ell^a$ of~\cite{Armas:2025nullmatter}, with $\ell^t=1$ fixing its normalisation. The thermal vectors satisfy $u^a=e^\eta\ell^a/2+e^{-\eta}\tilde\ell^a$ and $v^a=-e^\eta\ell^a/2+e^{-\eta}\tilde\ell^a$. Substituting in the constitutive relations gives
\begin{equation}\label{eq:string_null_decomposition}
\begin{aligned}
 T^{ab}={}&\frac{\Pi-\epsilon}{2}\gamma^{ab}
   +\frac{e^{2\eta}}4
       \left[Ts+\mu_0(\rho_0+2\kappa\mu_0\rho_1)\right]\ell^a\ell^b\\
 &+e^{-2\eta}
       \left[Ts+\mu_0(\rho_0-2\kappa\mu_0\rho_1)\right]
                    \tilde\ell^a\tilde\ell^b~,\\
 J^a={}&\frac{e^\eta}{2}(\rho_0+2\kappa\mu_0\rho_1)\ell^a
       +e^{-\eta}(\rho_0-2\kappa\mu_0\rho_1)\tilde\ell^a~.
\end{aligned}
\end{equation}
The string current remains $\mathcal J^{ab}=\rho_1\epsilon^{ab}$ because a longitudinal boost preserves the worldsheet volume form. A finite null wave requires the coefficients of $\ell^a\ell^b$ and $\ell^a$ to stay finite as $\eta\to\infty$, while those of the opposite null direction vanish. In the electrically neutral case this is the scaling of $Ts\,u^au^b$ in~\cite{Emparan:2011hg}. For a charged wave the electric and 2-group terms in \eqref{eq:string_null_decomposition} also contribute. A realisation of this limit in the non-extremal solution, parametrised by $q>0$, $\lambda>0$ and finite $Q$, is
\begin{equation}\label{eq:string_null_parameter_scaling}
 m=\lambda e^{-2\eta}\quad,\quad
 e^{2\beta}=\frac{q}{m}\quad,\quad
 \delta=\frac{4Qm}{3q\sqrt{q\lambda}}
 \quad,\quad \eta\longrightarrow\infty~.
\end{equation}
The parameter $q$ fixes the magnetic flux, $Q_M\to\sqrt{3}q/2$, and $Q$ fixes the electric current. In this scaling $\chi-\beta\to0$, and the local thermodynamic quantities behave as
\begin{equation}\label{eq:string_null_thermal_scaling}
\begin{aligned}
 T&=\frac{\sqrt m}{\pi q^{3/2}}+o(\sqrt m)~,
 &s&=2\pi\mathcal C q^{3/2}\sqrt m+o(\sqrt m)~,\\
 \mu_0&=-\frac{\sqrt{3}Q}{q}\sqrt{\frac m\lambda}
                      +o(\sqrt m)~,
 &\rho_0&=\mathcal C q\mu_0+o(\sqrt m)~,\\
 \epsilon&\longrightarrow\frac{3\mathcal C q}{2}~,
 &\Pi&\longrightarrow-\frac{3\mathcal C q}{2}~,
 \qquad\rho_1\longrightarrow\frac{\sqrt{3}\,\mathcal C q}{2}~.
\end{aligned}
\end{equation}
Using $Ts=2\mathcal C m$, the surviving coefficients in \eqref{eq:string_null_decomposition} are
\begin{equation}\label{eq:string_null_wave_data}
\begin{aligned}
 \mathcal E&=\lim\frac{e^{2\eta}}4
          \left[Ts+\mu_0(\rho_0+2\kappa\mu_0\rho_1)\right]
          =\mathcal C\left(\frac\lambda2+\frac{3Q^2}{2q}\right)~,\\
 \lim\frac{e^\eta}{2}(\rho_0+2\kappa\mu_0\rho_1)
          &=-\sqrt{3}\,\mathcal C Q~.
\end{aligned}
\end{equation}
The coefficients of the opposite null direction vanish. The limiting stress tensor and currents therefore take the form
\begin{equation}\label{eq:string_null_monopoles}
 T^{ab}=-\frac{3\mathcal C q}{2}\gamma^{ab}+\mathcal E\ell^a\ell^b
 \quad,\quad J^a=-\sqrt{3}\,\mathcal C Q\ell^a
 \quad,\quad
 \mathcal J^{ab}=\frac{\sqrt{3}\,\mathcal C q}{2}\epsilon^{ab}~.
\end{equation}
Following~\cite{Armas:2025nullmatter}, $\mathcal E$ is the null energy density and $\mathcal E\ell^a$ is the null momentum current. The metric term describes a string of intrinsic tension $3\mathcal C q/2=\sqrt{3}\rho_1$. In this asymptotic frame,
\begin{equation}\label{eq:string_null_adm_charges}
 \frac ML=\frac{3\mathcal C q}{2}+\mathcal E
 \quad,\quad \frac{P_z}{L}=\mathcal E
 \quad,\quad \mathcal T=\frac{3\mathcal C q}{2}-\mathcal E~.
\end{equation}
The wave can thus cancel the longitudinal tension, while the intrinsic string tension remains finite. The coefficient of $\gamma^{ab}$ in \eqref{eq:string_null_monopoles} is the pressure appearing in the null-matter stress tensor. It differs from the parent thermodynamic pressure $P=-2\mathcal C m$, which vanishes. There is no finite thermal rest frame for the limiting wave \cite{Armas:2025nullmatter}. 

The full geometry follows by retaining the terms amplified by the boost before setting $m$ and $\delta$ to zero. Since $\Delta(r)\to r^2$ and $H\to(r+q/2)^2$, the result is
\begin{equation}\label{eq:string_null_geometry}
\begin{aligned}
 ds^2={}&V^2(dr^2+r^2d\Omega_2^2)
             +V^{-1}(-dt^2+dz^2)+W(dt-dz)^2~,\\
 A={}&\frac{\sqrt{3}Q}{rV^2}\left(1+\frac{q}{4r}\right)(dt-dz)
                   -\frac{\sqrt{3}q}{2}\cos\theta\,d\phi~,
\end{aligned}
\end{equation}
where
\begin{equation}\label{eq:string_null_metric_functions}
 V=1+\frac{q}{2r}
 \quad,\quad
 W=\frac{\lambda}{2rV}
    +\frac{3Q^2}{2qrV^4}\left(1+\frac{q}{2r}+\frac{q^2}{12r^2}\right)~.
\end{equation}
The horizon generator is $\ell=\partial_t+\partial_z$. The expansion $W=\mathcal E/(\mathcal C r)+\mathcal O(r^{-2})$, together with the gauge potential, independently reproduces \eqref{eq:string_null_monopoles} using Appendix~\ref{app:stress_currents}. At $r=0$, the sphere radius is $q/2$ and the longitudinal metric coefficient is $\lambda/q$. Thus the horizon has finite area for $\lambda>0$. This geometry is a type of charged pp-wave, generalising \cite{Armas:2026mul}. For a spatial period $L$ in these coordinates,
\begin{equation}\label{eq:string_null_entropy}
 \frac{S_{\mathrm{BH}}}{L}=\pi\mathcal C\sqrt{q^3\lambda}
 \quad,\quad s u^a\longrightarrow\frac{S_{\mathrm{BH}}}{L}\ell^a
 \quad,\quad T_H=\frac{T}{c_\eta}\longrightarrow0~.
\end{equation}
The vanishing proper entropy density is compensated by the divergent boost. Eliminating $\lambda$ gives the wave equation of state
\begin{equation}\label{eq:string_null_equation_of_state}
 \mathcal E=\frac{1}{2\pi^2\mathcal C q^3}
                      \left(\frac{S_{\mathrm{BH}}}{L}\right)^2
                    +\frac{3\mathcal C Q^2}{2q}~.
\end{equation}
At fixed $q$, its differential is
\begin{equation}\label{eq:string_null_first_law}
 d\mathcal E=\frac{S_{\mathrm{BH}}/L}{\pi^2\mathcal C q^3}
                  \,d\!\left(\frac{S_{\mathrm{BH}}}{L}\right)
             +\frac{3\mathcal C Q}{q}\,dQ~.
\end{equation}
These finite coefficients follow from the rescaled local temperature and electric chemical potential. The asymptotic Hawking temperature itself vanishes.\\
\\
To match the non-supersymmetric string of~\cite{Compere:2010fm}, its parameters $\Delta$ and $\Diamond$ are
\begin{equation}\label{eq:string_null_compere_map}
 \Delta=\frac q2+\frac{Q^2}{2q}+\frac\lambda6
 \quad,\quad
 \Diamond=-\frac34q^2(2\Delta q-q^2-Q^2)
                 =-\frac{q^3\lambda}{4}<0~.
\end{equation}
Here $\Delta$ is a constant charge parameter, distinct from the radial function $\Delta(r)$. The change $\widetilde z=z-t$, with $t$ unchanged, puts \eqref{eq:string_null_geometry} in the form of eqs.~(4.18)--(4.22) of that reference, whose radial function is $g=V^3+V^4W$. In particular, $\lambda>0$ selects the negative-quartic-invariant branch, and \eqref{eq:string_null_adm_charges} gives $M/L=3\mathcal C\Delta$, $P_z/L=3\mathcal C(\Delta-q/2)$ and $\mathcal T=3\mathcal C(q-\Delta)$.\\
\\
The scaling in the original solution parameters is
\begin{equation}\label{eq:string_null_original_scaling}
 me^{2\beta}\longrightarrow q
 \quad,\quad me^\gamma\longrightarrow\frac{2\sqrt{-\Diamond}}{q}
 \quad,\quad \frac\delta m\longrightarrow\frac{2Q}{3\sqrt{-\Diamond}}~.
\end{equation}
Consequently, the constants denoted by $q,Q,\Diamond$ in eq.~(4.2) of~\cite{Compere:2010fm} as printed are $q/2,Q/2,\Diamond/16$ in our flux normalisation. The physical limit is the speed-of-light scaling in \eqref{eq:string_null_decomposition}--\eqref{eq:string_null_wave_data}, with finite null energy and charge currents.

\noindent\textbf{Global charges.}
For an embedded null wave, the mass and angular momenta again take the form \eqref{eq:string_susy_global_charges}, now with
\begin{equation}\label{eq:string_null_global_density}
 n_a\widetilde P_X^a
 =-\frac{3\mathcal Cq}{2}(n\cdot X)
   +\left[\mathcal E(\ell\cdot X)-\sqrt{3}\,\mathcal C Q\alpha_X\right]
          (n\cdot\ell)
   -\frac{\sqrt{3}\,\mathcal Cq}{2}e^a\eta_{Xa}~.
\end{equation}
The null vector and its coefficient are inherited from the limiting thermal frame, with the normalisation used in \eqref{eq:string_null_vectors} in a local orthonormal frame. Substitution in the remaining charge integrals gives
\begin{equation}\label{eq:string_null_global_electric_entropy}
\begin{aligned}
 Q_{(0)}&=\sqrt{3}\,\mathcal C\int_{B_1}dV_1\,Q(n\cdot\ell)
                                      -\mathcal Cq\oint_{B_1}A~,\\
 Q_{(1)}&=\frac{\sqrt{3}\,\mathcal Cq}{2}~,\\
 S_{\mathrm{BH}}&=-\pi\mathcal C\int_{B_1}dV_1
                         \sqrt{q^3\lambda}\,(n\cdot\ell)~,
\end{aligned}
\end{equation}
and the mechanical tensional energy is
\begin{equation}\label{eq:string_null_global_tension}
 \mathcal T_{\mathrm{tot}}
   =\int_{B_1}dV_1\,N
       \left[\frac{3\mathcal Cq}{2}-\mathcal E(n\cdot\ell)^2\right]~.
\end{equation}
In deriving this result we used $(e\cdot\ell)^2=(n\cdot\ell)^2$. The finite entropy current, rather than the vanishing proper density $s$, enters the global entropy. Likewise, the electric Page charge remains nonzero although $\rho_0\to0$ in the parent thermal frame.\\
\\
The limiting equilibrium generator $k^a$ is null and proportional to $\ell^a$. Along the positive-$z$ branch, $|k|u^a\to k^a$, $|k|v^a\to-k^a$, $\mu_0\to0$ and $\mu_1\to\sqrt{3}$. Thus the global potentials are
\begin{equation}\label{eq:string_null_global_potentials}
 \Phi_0=-\alpha_k
 \quad,\quad
 \Phi_1=\int_{B_1}dV_1\,e^a
       \left(-\sqrt{3}k_a-\eta_{ka}
                              +\frac{2\Phi_0}{\sqrt{3}}A_a\right)~.
\end{equation}
These expressions are limits of the equilibrium potentials, with their closure conditions retained. They do not require a unit timelike thermal vector at extremality. In the homogeneous asymptotic frame without external sources, the charges \eqref{eq:string_null_adm_charges} and \eqref{eq:string_null_entropy} are supplemented by
\begin{equation}\label{eq:string_null_flat_global_charges}
\begin{gathered}
 Q_{(0)}=-\sqrt{3}\,\mathcal C LQ
 \quad,\quad Q_{(1)}=\frac{\sqrt{3}\,\mathcal Cq}{2}
 \quad,\quad
 \mathcal T_{\mathrm{tot}}=L\left(\frac{3\mathcal Cq}{2}-\mathcal E\right)~,\\
 v_z=1\quad,\quad \Phi_0=0\quad,\quad
 \Phi_1=\sqrt{3}L\quad,\quad T_0=0~.
\end{gathered}
\end{equation}
Consequently $M-P_z=\Phi_1Q_{(1)}$ and, at fixed $L$, $dM=dP_z+\Phi_1\,dQ_{(1)}$. The vanishing global electric potential is compatible with the $Q$ dependence of \eqref{eq:string_null_equation_of_state}, since the electric contribution changes $M$ and $P_z$ equally.\\
\\
\noindent\textbf{Supersymmetric magnetic null wave.}
The null-wave sector also contains the purely magnetic endpoint of the supersymmetric family and was not explicitly given in \cite{Compere:2010fm}. Take $m\to0$, $\beta\to\infty$ with $me^{2\beta}=2\mathfrak p$ fixed, $\delta=0$ and finite $\gamma$. Here $\chi=\beta$, so the relative rapidity satisfies $\eta=\gamma-\chi\to-\infty$ and the thermal velocity approaches $v_z=-1$. Thus this endpoint has no finite thermal rest frame. We define
\begin{equation}\label{eq:string_magnetic_bps_parameters}
 q=2\mathfrak p\quad,\quad
 \lambda_{\mathrm B}=q e^{-2\gamma}>0~.
\end{equation}
Here $\lambda_{\mathrm B}$ parametrises the momentum wave, while $q$ determines the magnetic flux. Taking the limit in the fixed asymptotic frame gives
\begin{equation}\label{eq:string_magnetic_bps_endpoint}
\begin{aligned}
 ds^2={}&V^2(dr^2+r^2d\Omega_2^2)+V^{-1}(-dt^2+dz^2)
                +\frac{\lambda_{\mathrm B}}{2rV}(dt+dz)^2~,\\
 V={}&1+\frac{q}{2r}\quad,\quad
 A=-\frac{\sqrt{3}q}{2}\cos\theta\,d\phi~.
\end{aligned}
\end{equation}
The corresponding currents are
\begin{equation}\label{eq:string_magnetic_bps_currents}
\begin{gathered}
 T^{ab}=-\frac{3\mathcal C q}{2}\gamma^{ab}
        +\frac{\mathcal C\lambda_{\mathrm B}}2
                   \ell^a\ell^b
 \quad,\quad \ell^a=(1,-1)~,\\
 J^a=0\quad,\quad
 \mathcal J^{ab}=\frac{\sqrt{3}\,\mathcal C q}{2}\epsilon^{ab}~.
\end{gathered}
\end{equation}
This supersymmetric magnetic wave has horizon generator $\partial_t-\partial_z$ and entropy density $\pi\mathcal C\sqrt{q^3\lambda_{\mathrm B}}$. It is not obtained by setting $\delta=0$ in the finite-rest-frame metric \eqref{eq:string_susy_geometry} while keeping those coordinates fixed. Its momentum has the opposite sign to the branch in \eqref{eq:string_null_monopoles}. Null motion by itself therefore does not distinguish the supersymmetric and non-supersymmetric families.
For a homogeneous string of period $L$ without external sources, its charges are
\begin{equation}\label{eq:string_magnetic_bps_global_charges}
\begin{gathered}
 M=\frac{\mathcal C L}{2}(3q+\lambda_{\mathrm B})
 \quad,\quad P_z=-\frac{\mathcal C L\lambda_{\mathrm B}}2
 \quad,\quad Q_{(0)}=0
 \quad,\quad Q_{(1)}=\frac{\sqrt{3}\,\mathcal Cq}{2}~,\\
 S_{\mathrm{BH}}=\pi\mathcal C L\sqrt{q^3\lambda_{\mathrm B}}
 \quad,\quad
 \mathcal T_{\mathrm{tot}}=\frac{\mathcal C L}{2}(3q-\lambda_{\mathrm B})~.
\end{gathered}
\end{equation}
The conjugate quantities are $v_z=-1$, $\Phi_0=0$, $\Phi_1=\sqrt{3}L$ and $T_0=0$. The charges in external fields follow from \eqref{eq:string_null_global_density}--\eqref{eq:string_null_global_tension} by taking $Q=0$, $\lambda=\lambda_{\mathrm B}$ and the oppositely directed $\ell^a$. In this endpoint $|k|\mu_0\to0$ and $|k|v^a\to k^a$, so the potentials are $\Phi_0=-\alpha_k$ and the magnetic period in \eqref{eq:string_susy_global_potentials}.

For compact stationary blackfolds with rotational generators, these extremal limits specialise the thermodynamic identities of Section~\ref{sec: Smarr} to
\begin{equation}\label{eq:string_extremal_global_thermodynamics}
\begin{aligned}
 G&=M-\sum_i\Omega_iJ_i-\Phi_0Q_{(0)}-\Phi_1Q_{(1)}=0~,\\
 dM&=\sum_i\Omega_i\,dJ_i+\Phi_0\,dQ_{(0)}+\Phi_1\,dQ_{(1)}~,\\
 \boldsymbol{\mathcal T}&=\Phi_1Q_{(1)}-\sum_i\Omega_iJ_i~.
\end{aligned}
\end{equation}
The first law here holds along families of equilibria at fixed external sources. The last equality follows from the scaling Smarr relation, so $\boldsymbol{\mathcal T}$ includes the source contributions of \eqref{eq:global_tension_source_terms} and need not equal the mechanical $\mathcal T_{\mathrm{tot}}$. Although $P\to0$ and $T_0\to0$, the entropy and all the charges above are obtained by taking the limit of the non-extremal family before evaluating these identities.

\subsection{Validity of the blackfold approach}\label{sec:string_validity}
We now formulate the validity conditions for a general smooth worldsheet embedding $X^\mu(\sigma)$, with the non-extremal string discussed above supplying the local seed geometry. The separation of scales required by the blackfold expansion concerns the whole gravitational field of this black string. For a charged string, the coordinate position $r_+=2m$ alone does not measure this scale. Following the energy-density and charge-density analysis of~\cite{Caldarelli_2011}, we require a region in which the string field is weak while the embedding and background still appear approximately constant. In five dimensions the fields of a straight string decay as $1/r$, so an energy or charge density divided by $\mathcal C$ defines a length. In the thermal rest frame a useful choice is
\begin{equation}\label{eq:string_validity_energy_radius}
 r_\epsilon\equiv\frac{\epsilon}{\mathcal C}
   =2m\left[\mathsf B(1+\mathsf Kc_\chi^2)
                        -2\mathsf Xc_\chi s_\chi\right]
   \sim m c_\delta^2\cosh(2\beta)~.
\end{equation}
Here and below $\sim$ suppresses numerical factors of order unity. Inserting the rest-frame boost \eqref{eq:nonext_rest_boost}, including the cancellation between its two terms, gives the bounds $3mc_\delta^2\mathsf B\leq r_\epsilon\leq6mc_\delta^2\mathsf B$. The charge lengths $|\rho_0|/\mathcal C$ and $|\rho_1|/\mathcal C$ are at most of this order. Equivalently, the asymptotic coefficients of the metric and gauge field in Appendix~\ref{app:stress_currents} become small at $r\gg r_\epsilon$.

The transverse horizon radius can be read off independently from the exact radial functions,
\begin{equation}\label{eq:string_validity_sphere_radius}
 R_{S^2}=\sqrt{H(2m)}
       =2m c_\delta c_\beta
            \sqrt{c_\beta^2+\frac{s_\delta^2s_\beta^2}{h_\delta^2}}
       \lesssim r_\epsilon~.
\end{equation}
This is the radius of the local transverse horizon sphere, independent of the shape of the embedding. At large electric boost $R_{S^2}$ can be parametrically smaller than $r_\epsilon$, so requiring only $R_{S^2}$ or $2m$ to be small would miss the extent of the charged field. The long scale $\mathcal R(\sigma)$ is determined locally as in \eqref{eq:blackfold_scale_hierarchy}. It includes the bending scales set by $K_{ab}{}^i$, the background curvature and the proper scales of variation of the string state and external fields~\cite{Emparan:2009at,Armas:2015minimal,Armas:2016mes}. An overlap region must exist at every worldsheet point,
\begin{equation}\label{eq:string_validity_hierarchy}
 r_\epsilon(\sigma)\ll r\ll\mathcal R(\sigma)
 \quad\Longrightarrow\quad
 \frac{m}{\mathcal R}\ll
             \frac{1}{c_\delta^2\cosh(2\beta)}~.
\end{equation}
All quantities on the right-hand side are evaluated at the same point. In terms of local geometric data, the corresponding requirements include
\begin{equation}\label{eq:string_validity_local_geometry}
 r_\epsilon|K_{\hat a\hat b}{}^i|\ll1
 \quad,\quad
 r_\epsilon^2|R^{\mathrm{bg}}_{\hat\mu\hat\nu\hat\rho\hat\sigma}|\ll1
 \quad,\quad
 \frac{r_\epsilon}{L_{\mathrm{var}}}\ll1~.
\end{equation}
Here $L_{\mathrm{var}}$ is the shortest proper variation scale of the local state and velocity fields. On a null branch this scale is defined using the finite limiting data $\mathcal E,Q,\ell^a$. Hatted indices denote orthonormal components, with each inequality imposed on all components in the same frame as the frame chosen to describe the local geometry. External fields must also perturb the near-string solution weakly,
\begin{equation}\label{eq:string_validity_fluxes}
 r_\epsilon|F_{\mathrm{bg}}|\ll1
 \quad,\quad r_\epsilon|\mathcal F_{\mathrm{bg}}|\ll1~,
\end{equation}
with their variation lengths included in $\mathcal R$. The norms here denote physical frame components. In minimal supergravity these two field strengths are related by a Hodge dual, see ~\ref{sec: sugra_ward_identities} and \ref{sec: sugra_implementation}. Separations from other horizon components, or between non-neighbouring portions of the same string, must also be large compared with the local scales of the black string.

Large longitudinal boosts require particular care. In a fixed orthonormal frame with time direction $n^a$, the relevant energy length is
\begin{equation}\label{eq:string_validity_frame_energy}
 r_\epsilon(n)\equiv\frac{T^{ab}n_an_b}{\mathcal C}~.
\end{equation}
This provides a sufficient weak-field scale for \eqref{eq:string_validity_hierarchy}--\eqref{eq:string_validity_fluxes}, with the metric coefficients and variation scales evaluated in that frame. For a bounded boost it changes \eqref{eq:string_validity_energy_radius} only by finite factors. In a speed-of-light limit it retains the wave contribution that would be lost by using the proper rest energy alone. The two extremal metrics make these requirements explicit.\\
\\
For the finite-velocity supersymmetric branch, \eqref{eq:string_susy_functions} gives
\begin{equation}\label{eq:string_validity_susy}
 r_\epsilon=3Q_{\mathrm R}\quad,\quad
 R_{S^2}=\frac q2\quad,\quad
 Q_{\mathrm R}>q
 \quad\Longrightarrow\quad Q_{\mathrm R}\ll\mathcal R~.
\end{equation}
Both $f^{-1}-1$ and $\omega$ are small in the overlap region $r\gg Q_{\mathrm R}$. In terms of the original limiting parameters, this condition is $2\mathfrak p c_\delta^2\ll\mathcal R$, even though $m$ has vanished. For a finite boost to the three-parameter frame, $r_\epsilon(n)=3S$, consistently with \eqref{eq:string_susy_boosted_monopoles}.\\
\\
For the null-wave branch, the asymptotic functions $V-1$ and $W$ instead give
\begin{equation}\label{eq:string_validity_null}
 r_\epsilon(n)=\frac{3q}{2}+\frac\lambda2+\frac{3Q^2}{2q}
 \quad\Longrightarrow\quad
 q\ll\mathcal R\quad,\quad
 \lambda\ll\mathcal R\quad,\quad
 \frac{Q^2}{q}\ll\mathcal R~.
\end{equation}
The electric length $|Q|$ is also small under these conditions. The horizon sphere still has radius $q/2$, but it does not detect an arbitrarily large momentum wave. For the supersymmetric magnetic endpoint the same reasoning gives $\max(q,\lambda_{\mathrm B})\ll\mathcal R$. All these inequalities are pointwise conditions on the local state and embedding. The shape and force balance must satisfy the blackfold equations, while this hierarchy tests whether the string supplies a controlled local approximation. Regular finite-area horizons additionally require $Q_{\mathrm R}>q$ on the dyonic supersymmetric branch and $\lambda>0$ or $\lambda_{\mathrm B}>0$ on the null branches. 

\section{Black rings and black Saturns}\label{sec:constructed_horizons}
We now construct asymptotically flat black rings and black Saturns using the black strings of Section~\ref{sec:non-ext_string} and their extremal limits. Section~\ref{sec:ring_flat} gives two non-extremal ring branches and four extremal branches, one of which is BPS. Sections~\ref{sec:susy_saturn} and \ref{sec:nonext_saturn} construct supersymmetric and non-extremal black Saturns, including families in thermal equilibrium and limits in which both horizons are extremal. In turn, in the related Appendix~\ref{Appendix_matching}, we compare these constructions with exact solutions and develops the BPS di-ring and bi-ring configurations. In Appendix~\ref{app:nonext_multirings} we construct di-rings and bi-rings in an exact non-extremal ring background, study their extremal limits and identify thermal di-ring families. Finally, section~\ref{sec:phase_diagrams} collects all these solution families in phase diagrams.

\subsection{Black rings in flat space}\label{sec:ring_flat}
We first construct dipole (black rings) with spatial horizon topology $\mathbb{S}^1\times\mathbb{S}^2$ in five-dimensional asymptotically Minkowski space, using the non-extremal string of Section~\ref{sec:non-ext_string}. We follow the stationary construction of~\cite{Armas:2017myl,Armas:2015qsv}, specifying the embedding and equilibrium fields before solving the extrinsic equation and evaluating the thermodynamic quantities. The non-extremal black rings constructed here provide new classes of solutions that have limited overlap with exact solutions. We will also perturbatively recover known BPS and non-extremal dipole rings in minimal supergravity but we will also uncover 3 new non-BPS extremal dipole rings that have not yet been found in the literature. We compare these different solutions in various phase diagrams.

\subsubsection{Embedding and equilibrium fields}
We consider asymptotically flat space in coordinates given by
\begin{equation}\label{eq:metric_flat}
 ds^2=-d\tau^2+dr^2+r^2\left(d\theta^2+\cos^2\theta\,d\psi^2
                                      +\sin^2\theta\,d\phi^2\right)~,
\end{equation}
with $0\leq\theta\leq\pi/2$ and both azimuthal angles of period $2\pi$. In worldsheet coordinates $\sigma^a=(t,\psi)$, we embed the string around a circle of radius $R$ such that
\begin{equation}\label{eq:embedding_flat}
 X^\tau=t\quad,\quad X^r=R\quad,\quad X^\theta=0
 \quad,\quad X^\psi=\psi~.
\end{equation}
The transverse horizon sphere belongs to the local black string geometry in the three-dimensional normal space of this embedding. Pulling back \eqref{eq:metric_flat} gives
\begin{equation}\label{eq:flat_ring_induced}
 \gamma_{ab}d\sigma^a d\sigma^b=-dt^2+R^2d\psi^2
             =-dt^2+dz^2\quad,\quad z=R\psi\quad,\quad L=2\pi R~.
\end{equation}
Here $L$ is the spatial period of the effective string. On a constant-$t$ slice, $n=\partial_t$, $dV_1=R\,d\psi$ and the lapse is $N=1$. We keep the orientation of Section~\ref{sec:non-ext_string}, so that $\epsilon^{t\psi}=-1/R$ and the oriented unit tangent is $e=-R^{-1}\partial_\psi=-\partial_z$.

Providing the embedding is not sufficient for prescribing the blackfold, we must also choose a general stationary Killing vector tangent to the worldsheet, with unit asymptotic time normalisation. The most general choice is
\begin{equation}\label{eq:flat_ring_killing}
 k=\partial_\tau+\Omega\partial_\psi
   =\partial_t+\Omega\partial_\psi
   =\partial_t+w\partial_z
 \quad,\quad w=\Omega R\quad,\quad |k|=\sqrt{1-w^2}~.
\end{equation}
For the non-extremal string $|w|<1$. The thermal velocity and its associated spacelike vector are therefore
\begin{equation}\label{eq:uada}
 u=\frac{1}{|k|}\left(\partial_t+\Omega\partial_\psi\right)
 \quad,\quad
 v=-\frac{1}{|k|}\left(\Omega R\,\partial_t+\frac1R\partial_\psi\right)~.
\end{equation}
They obey $u^2=-1$, $v^2=1$, $u\cdot v=0$ and $v^a=\epsilon^{ab}u_b$. In the orthonormal coordinates $(t,z)$, their components are $u^a=(1,w)/|k|$ and $v^a=(-w,-1)/|k|$. The physical rapidity relative to the thermal rest frame is $\eta$, with $w=\tanh\eta$.\footnote{In terms of the auxiliary metric boost of Appendix~\ref{app:nonext_monopoles}, $\eta=\gamma-\chi$ and $\Omega R=\tanh(\gamma-\chi)$, where $\chi$ is fixed by \eqref{eq:nonext_rest_boost}.}

We choose vanishing external potentials $A=B=0$ and zero reference compensators for the conserved generators. Equations~\eqref{eq:global_thermal_zero_modes}--\eqref{eq:global_magnetic_potential} reduce to
\begin{equation}\label{eq:flat_ring_thermal_potentials}
\begin{aligned}
 T&=\frac{T_0}{|k|}\quad,\quad
 \Phi_0=T_0\Lambda_\beta=|k|\mu_0~,\\
 \zeta_a&=T_0(\Lambda_{\beta a}-\nabla_a\varphi)
             =|k|\left(\mu_1v_a+\kappa\mu_0^2u_a\right)~,\\
 \Phi_1&=\int_{B_1}dV_1\,e^a\zeta_a
             =L\left(\mu_1-\kappa\mu_0^2w\right)~.
\end{aligned}
\end{equation}
Thus the local state at fixed global potentials is specified by
\begin{equation}\label{eq:flat_ring_local_state}
 T=\frac{T_0}{\sqrt{1-w^2}}\quad,\quad
 \mu_0=\frac{\Phi_0}{\sqrt{1-w^2}}\quad,\quad
 \mu_1=\frac{\Phi_1}{L}+\kappa\mu_0^2w~.
\end{equation}
The last term is required by the 2-group coupling even in a flat background with vanishing external fluxes. All these quantities are constant on the homogeneous ring.

\subsubsection{Blackfold equations and the two equilibrium branches}\label{sec:flat_ring_branches}
The intrinsic equations \eqref{eq: blackfoldstress}--\eqref{eq: blackfoldcurrents} reduce to conservation of $T^{ab}$, $J^a$ and the constant top-form coefficient $\rho_1$. They are solved by the homogeneous state above. With outward radial normal, the only nonzero extrinsic curvature component for 5d flat space is
\begin{equation}\label{eq:flat_ring_extrinsic_curvature}
 K_{\psi\psi}{}^r=-R\quad,\quad K_{zz}{}^r=-\frac1R~.
\end{equation}
There is no external force, so the remaining blackfold equation is $T^{zz}=0$. To evaluate it, we use $\Pi=P-\mu_1\rho_1$ and $h=-\kappa\mu_0^2\rho_1$, as defined in \eqref{eq:equilibrium_mechanical_pressure}. Substitution of \eqref{eq:uada} in the constitutive relations gives
\begin{equation}\label{eq:flat_ring_components}
\begin{aligned}
 T^{tt}&=\frac{\epsilon+\Pi w^2-2hw}{1-w^2}~,
 &T^{tz}&=\frac{(\epsilon+\Pi)w-h(1+w^2)}{1-w^2}~,\\
 T^{zz}&=\frac{\epsilon w^2-2hw+\Pi}{1-w^2}~,
 &\mathcal J^{tz}&=-\rho_1~,\\
 J^t&=\frac{\rho_0+2\kappa\mu_0\rho_1w}{\sqrt{1-w^2}}~,
 &J^z&=\frac{\rho_0w+2\kappa\mu_0\rho_1}{\sqrt{1-w^2}}~.
\end{aligned}
\end{equation}
Consequently the balance of forces requires
\begin{equation}\label{eq:extrinsic_flat}
 \epsilon w^2-2hw+\Pi=0~,
\end{equation}
and its two solutions are
\begin{equation}\label{eq:OmegaR}
 w_\pm=\Omega_\pm R
       =\frac{h\pm\sqrt{h^2-\epsilon\Pi}}{\epsilon}~.
\end{equation}
The local quantities in this expression are evaluated in the thermal rest frame of Section~\ref{sec:non-ext_string}. For the regular non-extremal equation of state, both roots are real and subluminal. Together with \eqref{eq:flat_ring_local_state}, this determines the equilibrium ring at prescribed global potentials. Equivalently, one may choose $(m,\beta,\delta,R)$, evaluate the local state from Section~\ref{sec:non-ext_string}, and obtain the two angular velocities and their global potentials. We take $\beta,\delta\geq0$ to fix the charge orientation. The two roots carry opposite longitudinal momentum, as will be explicit below. Also we will show that there are 4 possible extremal limits of these solutions.

The same equation follows directly from the equilibrium action \eqref{eq:equilibrium_action} of Section~\ref{sec:equilibrium_action}. On the circular embedding it becomes
\begin{equation}\label{eq:flat_ring_action}
 S_{\mathrm{eq}}[R]
   =\int dt\int_0^{2\pi}d\psi\,R\,P(T,\mu_0,\mu_1)
   =\int dt\,L P(T,\mu_0,\mu_1)~.
\end{equation}
We vary the radius at fixed worldsheet coordinates $(t,\psi)$, keeping $T_0,\Omega,\Phi_0,\Phi_1$ fixed. The local temperature and chemical potentials therefore vary according to \eqref{eq:flat_ring_local_state}. Since $A=B=0$, the embedding variation changes only the induced metric, with $\delta_R\gamma_{\psi\psi}=2R\,\delta R$. Using the stress tensor obtained by varying the action in Section~\ref{sec:equilibrium_action} gives
\begin{equation}\label{eq:flat_ring_variational_balance}
\begin{aligned}
 \delta_R S_{\mathrm{eq}}
   &=\frac12\int_{\Sigma_2}d^2\sigma\sqrt{-\gamma}\,
                  T^{ab}\delta_R\gamma_{ab}
     =2\pi\int dt\,T^{zz}\delta R\\
   &=2\pi\int dt\,
             \frac{\epsilon w^2-2hw+\Pi}{1-w^2}\,\delta R~.
\end{aligned}
\end{equation}
Here $R^2T^{\psi\psi}=T^{zz}$. Stationarity of the action under radial deformations thus gives $T^{zz}=0$, reproducing \eqref{eq:extrinsic_flat}. The total mechanical tensional energy is $\mathcal T_{\mathrm{tot}}=-LT^{zz}$ and vanishes on either balanced branch, consistently with the tension interpretation of~\cite{Armas:2015qsv}.  \\

\noindent\textbf{Global charges and thermodynamics.}
Using the currents of Section~\ref{sec: conserved_charges}, the charges and entropy for this configuration are
\begin{equation}\label{eq:charges_flat}
\begin{gathered}
 M=L T^{tt}\quad,\quad J=LR T^{tz}\quad,\quad
 Q_{(0)}=LJ^t\quad,\quad Q_{(1)}=\rho_1~,\\
 S_{\mathrm{BH}}=\frac{Ls}{\sqrt{1-w^2}}~.
\end{gathered}
\end{equation}
The string charge is a magnetic dipole flux and is not multiplied by $L$. On the two roots, \eqref{eq:flat_ring_components} simplifies to
\begin{equation}\label{eq:flat_ring_branch_charges}
\begin{aligned}
 M_\pm&=L(\epsilon-\Pi)~,
 &J_\pm&=\pm LR\sqrt{h^2-\epsilon\Pi}~,\\
 Q_{(0),\pm}&=\frac{L(\rho_0+2\kappa\mu_0\rho_1w_\pm)}
                          {\sqrt{1-w_\pm^2}}~,
 &Q_{(1),\pm}&=\rho_1~,\\
 S_{\mathrm{BH},\pm}&=\frac{Ls}{\sqrt{1-w_\pm^2}}~.
\end{aligned}
\end{equation}
Their conjugate potentials are
\begin{equation}\label{eq:flat_ring_branch_potentials}
\begin{gathered}
 \Omega_\pm=\frac{w_\pm}{R}\quad,\quad
 T_{0,\pm}=T\sqrt{1-w_\pm^2}\quad,\quad
 \Phi_{0,\pm}=\mu_0\sqrt{1-w_\pm^2}~,\\
 \Phi_{1,\pm}=L\left(\mu_1-\kappa\mu_0^2w_\pm\right)~.
\end{gathered}
\end{equation}
In particular, the two solutions at fixed local state have the same mass and magnetic charge, but generally different electric charge, entropy and potentials. They are not obtained from each other by reversing $J$ alone.

As a check, the neutral limit $\beta=\delta=0$ has $h=0$, $\epsilon=4\mathcal C m$ and $\Pi=-2\mathcal C m$. Equations~\eqref{eq:OmegaR} and \eqref{eq:flat_ring_branch_charges} then give $\Omega_\pm R=\pm1/\sqrt2$, $M=3\pi mR/G_5$ and $J_\pm=\pm\sqrt2\pi mR^2/G_5$, with $Q_{(0)}=Q_{(1)}=0$, reproducing the leading neutral black ring \cite{Emparan:2009vd}. On either branch, these thermodynamic quantities satisfy the general relations of Section~\ref{sec: Smarr}, including the Gibbs relation, first law and five-dimensional Smarr relation, with vanishing total tension. Below we test the BPS relation for the extremal limits of this configuration.

\subsubsection{Extremal branches and the BPS test}\label{sec:flat_ring_extremal}
For an isolated asymptotically flat solution of minimal supergravity, the BPS bound in our charge conventions is~\cite{Compere:2010fm}
\begin{equation}\label{eq:flat_ring_bps_bound}
 M\geq\sqrt{3}\,|Q_E|=\sqrt{3}\,|Q_{(0)}|~.
\end{equation}
Here $Q_E=-Q_{(0)}$. For the non-extremal roots, substitution of \eqref{eq:flat_ring_branch_charges} gives
\begin{equation}\label{eq:flat_ring_nonext_bps_gap}
 M_\pm-\sqrt{3}|Q_{(0),\pm}|
 =L\left[\epsilon-\Pi
       -\frac{\sqrt{3}\,|\rho_0+2\kappa\mu_0\rho_1w_\pm|}
                    {\sqrt{1-w_\pm^2}}\right]>0~,
\end{equation}
for finite $m>0$, $\beta$ and $\delta$ in the regular family. We now evaluate the limiting charges explicitly in various extremal regimes and note that all the cases below can be obtained as direct limits of \eqref{eq:OmegaR}.\footnote{Among these four extremal configurations, the non-rotating thin limit of the Suzuki--Tomizawa family~\cite{Suzuki:2024generalring} reaches only a constrained subset of the BPS rings. Its additional extremal sectors involve transverse rotation. We explain this restriction in Appendix~\ref{app:ST_extremal_audit}.} Finding a strictly positive gap in these relations at fixed local charge ratios excludes a BPS completion in the controlled thin-ring limit that we work with here.\\

\noindent\textbf{Finite-velocity limits of the supersymmetric string.}
Take the limit \eqref{eq:string_susy_scaling} with $Q_{\mathrm R}>q>0$. Section~\ref{sec:susy_lim} gives
\begin{equation}\label{eq:flat_ring_susy_local_state}
\begin{gathered}
 \epsilon=3\mathcal C Q_{\mathrm R}\quad,\quad \Pi=0
 \quad,\quad h=-\frac{3\mathcal C q}{2}~,\\
 \rho_0=-\sqrt{3}\,\mathcal C Q_{\mathrm R}
 \quad,\quad \rho_1=\frac{\sqrt{3}\,\mathcal Cq}{2}
 \quad,\quad \mu_0=-\sqrt{3}\quad,\quad T=\mu_1=0~.
\end{gathered}
\end{equation}
The balance equation factorises, giving
\begin{equation}\label{eq:flat_ring_susy_velocities}
 w(Q_{\mathrm R}w+q)=0
 \quad\Longrightarrow\quad
 w_+=0\quad,\quad w_-=-\frac{q}{Q_{\mathrm R}}~.
\end{equation}
In the three-parameter asymptotic frame of \eqref{eq:string_susy_three_parameter_map}, these are respectively $\Delta=0$, $Q=Q_{\mathrm R}$ and $\Delta=q$, $Q=\sqrt{Q_{\mathrm R}^2-q^2}$. Thus the two pressureless strings correspond to different balanced rings and are thus not two coordinate descriptions of one compact solution~\cite{Compere:2010fm}. On the positive momentum branch the charges and potentials are
\begin{equation}\label{eq:flat_ring_susy_plus}
\begin{gathered}
 M_+=3\mathcal C LQ_{\mathrm R}\quad,\quad
 J_+=\frac{3\mathcal C LRq}{2}~,\\
 Q_{(0),+}=-\sqrt{3}\,\mathcal C LQ_{\mathrm R}
 \quad,\quad Q_{(1),+}=\frac{\sqrt{3}\,\mathcal Cq}{2}~,\\
 S_{\mathrm{BH},+}=\pi\sqrt{3}\,\mathcal C Lq
                         \sqrt{Q_{\mathrm R}^2-q^2}~,\\
 \Omega_+=T_{0,+}=\Phi_{1,+}=0\quad,\quad \Phi_{0,+}=-\sqrt{3}~.
\end{gathered}
\end{equation}
The mass saturates the BPS bound, $M_+=-\sqrt{3}Q_{(0),+}$, and the first law reduces to $dM_+=-\sqrt{3}\,dQ_{(0),+}$. The horizon angular velocity vanishes even though the 2-group energy flux gives $J_+\ne0$. This is the branch realised by the thin supersymmetric black ring of~\cite{Elvang:2004rt}. Section~\ref{sec:flat_ring_comparison} and Appendix~\ref{app:exact_susy_ring} compare its charges, horizon area and two angular momenta with the blackfold result. On the negative momentum branch one instead finds
\begin{equation}\label{eq:flat_ring_susy_minus}
\begin{gathered}
 M_-=3\mathcal C LQ_{\mathrm R}\quad,\quad
 J_-=-\frac{3\mathcal C LRq}{2}~,\\
 Q_{(0),-}=-\sqrt{3}\,\mathcal C L\sqrt{Q_{\mathrm R}^2-q^2}
 \quad,\quad Q_{(1),-}=\frac{\sqrt{3}\,\mathcal Cq}{2}~,\\
 S_{\mathrm{BH},-}=\pi\sqrt{3}\,\mathcal C LqQ_{\mathrm R}~,
 \qquad T_{0,-}=0~,\\
 \Omega_-=-\frac{q}{RQ_{\mathrm R}}\quad,\quad
 \Phi_{0,-}=-\frac{\sqrt{3}\sqrt{Q_{\mathrm R}^2-q^2}}{Q_{\mathrm R}}
 \quad,\quad \Phi_{1,-}=\frac{\sqrt{3}Lq}{Q_{\mathrm R}}~.
\end{gathered}
\end{equation}
The BPS gap is strictly positive,
\begin{equation}\label{eq:flat_ring_susy_minus_gap}
 M_--\sqrt{3}|Q_{(0),-}|
       =3\mathcal C L\left(Q_{\mathrm R}
                           -\sqrt{Q_{\mathrm R}^2-q^2}\right)>0~.
\end{equation}
Thus the supersymmetry of the local straight string does not persist as supersymmetry of this asymptotically flat ring branch. Both finite-velocity branches satisfy the thermodynamic relations of Section~\ref{sec: Smarr} with $T_0=G=0$. In particular, on the negative-momentum branch $\Omega_-J_-=\Phi_{1,-}Q_{(1),-}$, and these nonzero work terms are essential for the first law and Smarr relation.

The two finite-velocity branches presented above constitute the asymptotically flat black rings whose infinite radius limit corresponds to the two extremal pressureless limits of the general black string, in the terminology of \cite{Compere:2010fm}. This reference had predicted the existence of such rings. It also observed that there is a bound relating the electric charges of both branches of strings, and here we can note that the same bound holds for their finite radius counterpart; this is
\begin{equation}\label{eq:QE_bound}
    Q_{E,+}\geq Q_{E,-}\quad ,\quad \text{at fixed }M,\,Q_{(1)}\,,|J|\,.
\end{equation}
This can be interpreted as the statement that only one of the finite-velocity branches can saturate the BPS bound\footnote{This holds for a given fixed $M$, finite $Q_\text{R}$ and $q<Q_{\text{R}}$, in which case the inequality \eqref{eq:QE_bound} is not saturated.} and consequently reproduce the behaviour of the supersymmetric black ring \cite{Elvang:2004rt}, as will be seen in \ref{sec:flat_ring_comparison}. Moreover, the two branches are disjoint, given that the absolute value of the momentum is bounded from below by that of the neutral ring, $|J_\pm|\geq \sqrt{2}\pi m R^2/G_5$ for $\delta=\beta=0$. At fixed charges, their absolute value of momentum is bounded from above by $|J|=|J_c|$, as displayed in Figure \ref{fig:BR_comparison}.
\\

\noindent\textbf{Null-wave limits.}
The other regular extremal regime of Section~\ref{sec:string_null_limits} has a null equilibrium generator. On the positive-momentum branch it is
\begin{equation}\label{eq:flat_ring_null_generator}
 k=\ell=\partial_t+\frac1R\partial_\psi
       =\partial_t+\partial_z\quad,\quad \Omega_+=\frac1R~.
\end{equation}
We use the finite limiting stress tensor and currents \eqref{eq:string_null_monopoles}, with the inherited worldsheet orientation. Their radial equation gives
\begin{equation}\label{eq:string_validity_null_ring}
 T^{zz}=-\frac{3\mathcal Cq}{2}+\mathcal E=0
 \quad\Longrightarrow\quad
 \mathcal E=\frac{3\mathcal Cq}{2}
 \quad,\quad \lambda=\frac{3(q^2-Q^2)}q>0~.
\end{equation}
Regularity requires $|Q|<q$. The global charges, entropy and potentials are
\begin{equation}\label{eq:flat_ring_null_charges}
\begin{gathered}
 M_+=3\mathcal C Lq\quad,\quad J_+=\frac{3\mathcal C LRq}{2}~,\\
 Q_{(0),+}=-\sqrt{3}\,\mathcal C LQ
 \quad,\quad Q_{(1),+}=\frac{\sqrt{3}\,\mathcal Cq}{2}~,\\
 S_{\mathrm{BH},+}=\pi\sqrt{3}\,\mathcal C Lq\sqrt{q^2-Q^2}~,\\
 T_{0,+}=\Phi_{0,+}=0\quad,\quad \Omega_+=\frac1R
 \quad,\quad \Phi_{1,+}=\sqrt{3}L~.
\end{gathered}
\end{equation}
The potentials are the limits of the finite-temperature ones, as derived in Section~\ref{sec:string_null_limits}. They give $M_+=\Omega_+J_++\Phi_{1,+}Q_{(1),+}$ and satisfy the first law and Smarr relation of Section~\ref{sec: Smarr}. The BPS gap is
\begin{equation}\label{eq:flat_ring_null_gap}
 M_+-\sqrt{3}|Q_{(0),+}|=3\mathcal C L(q-|Q|)>0~.
\end{equation}
Equality is reached only at the zero-area boundary $|Q|=q$. Hence the regular null-wave rings on this branch are not supersymmetric. In turn, the negative-momentum branch has the purely magnetic endpoint of \eqref{eq:string_magnetic_bps_endpoint}, with $\ell=\partial_t-\partial_z$. Balance fixes $\lambda_{\mathrm B}=3q$, and its thermodynamics is
\begin{equation}\label{eq:flat_ring_magnetic_null_charges}
\begin{gathered}
 M_-=3\mathcal C Lq\quad,\quad J_-=-\frac{3\mathcal C LRq}{2}~,\\
 Q_{(0),-}=0
 \quad,\quad Q_{(1),-}=\frac{\sqrt{3}\,\mathcal Cq}{2}~,\\
 S_{\mathrm{BH},-}=\pi\sqrt{3}\,\mathcal C Lq^2~,
 \qquad T_{0,-}=\Phi_{0,-}=0~,\\
 \Omega_-=-\frac1R\quad,\quad \Phi_{1,-}=\sqrt{3}L~.
\end{gathered}
\end{equation}
These expressions also follow by taking $Q_{\mathrm R}\to q$ in \eqref{eq:flat_ring_susy_minus}, keeping the global quantities finite. Although the local magnetic wave is supersymmetric, the balanced ring has positive mass and zero electric charge, so it does not saturate \eqref{eq:flat_ring_bps_bound}. The distinction between supersymmetry of a straight string and of the completed ring is therefore visible in both negative-momentum limits.

\subsubsection{Validity of the dipole ring configurations}\label{sec:flat_ring_validity}
For this homogeneous embedding the external curvature and fluxes vanish, and the bending scale in the asymptotic frame is $R$. The criterion of Section~\ref{sec:string_validity} reduces to
\begin{equation}\label{eq:flat_ring_validity}
 r_\epsilon(n)=\frac{T^{tt}}{\mathcal C}\ll R
 \quad,\quad n=\partial_t~.
\end{equation}
On either non-extremal equilibrium branch this length is $3g_1=6mc_\delta^2\cosh(2\beta)$. This condition measures the extent of the charged geometry as well as the transverse horizon size~\cite{Caldarelli_2011,Armas:2015minimal}. To compare thinness with the angular-momentum scale, we define
\begin{equation}\label{eq:string_validity_ring_charges}
 \ell_M=\sqrt{G_5M}\quad,\quad
 \ell_J=\frac{|J|}{M}=R\frac{|T^{tz}|}{T^{tt}}~.
\end{equation}
This leads to
\begin{equation}\label{eq:string_validity_spin_ratio}
 \left(\frac{\ell_M}{\ell_J}\right)^2
       =\frac{\pi}{2}\frac{r_\epsilon(n)}R
                            \left(\frac{T^{tt}}{T^{tz}}\right)^2~.
\end{equation}
For both finite-velocity extremal branches \eqref{eq:flat_ring_susy_plus} and \eqref{eq:flat_ring_susy_minus},
\begin{equation}\label{eq:string_validity_susy_ring}
 Q_{\mathrm R}\ll R\quad,\quad
 \ell_J=\frac{Rq}{2Q_{\mathrm R}}
 \quad,\quad
 \left(\frac{\ell_M}{\ell_J}\right)^2
       =\frac{6\pi Q_{\mathrm R}^3}{Rq^2}~.
\end{equation}
Thus a thin ring need not be ultraspinning when $Q_{\mathrm R}/q$ is large. On either balanced null branch, the finite asymptotic energy length is instead $r_\epsilon(n)=3q$, giving
\begin{equation}\label{eq:flat_ring_null_validity}
 q\ll R\quad,\quad \ell_J=\frac R2
 \quad,\quad \left(\frac{\ell_M}{\ell_J}\right)^2
                         =\frac{6\pi q}{R}\ll1~.
\end{equation}
Therefore the regular null-wave rings are ultraspinning solutions.

\subsubsection{Comparison with the exact supersymmetric ring}\label{sec:flat_ring_comparison}
Here we compare our approximate 4 extremal solutions with the exact SUSY ring solution of~\cite{Elvang:2004rt}, which provides a check of both the charges and the horizon area of the positive finite-velocity branch. Appendix~\ref{app:exact_susy_ring} gives its geometry and the parameter map $Q_{\mathrm{ring}}=2RQ_{\mathrm R}$, with the same magnetic parameter $q$ between the blackfold SUSY ring construction and the exact solution. Under this map the mass, electric and dipole charges agree with \eqref{eq:flat_ring_susy_plus}. With the angular orientations aligned, the blackfold angular momentum $J_+$ is compared with the exact longitudinal angular momentum $J_\psi$, associated with the same asymptotic rotation generator $\partial_\psi$. It reproduces the leading term of $J_\psi$ for $R\gg Q_{\mathrm R}$ at fixed $q/Q_{\mathrm R}$. The nonzero transverse angular momentum $J_\phi$ and the sub-leading correction to $J_\psi$ are not captured by the present monopole approximation.

To display the four extremal limiting branches together, we define the reduced area and angular momentum following~\cite{Emparan:2007wm,Armas:2017myl}
\begin{equation}\label{eq:flat_ring_reduced_definitions}
 a_H=c_a\frac{\mathcal A}{(G_5M)^{3/2}}
 \quad,\quad j=c_j\frac{G_5J}{(G_5M)^{3/2}}
 \quad,\quad
 c_a=\frac3{16}\sqrt{\frac3\pi}
 \quad,\quad c_j=\sqrt{\frac{27\pi}{32}}~,
\end{equation}
where $\mathcal A=4G_5S_{\mathrm{BH}}$. We plot $|j|$ so that the negative-momentum branch can be compared on the same axes. For the two finite-velocity branches, \eqref{eq:flat_ring_susy_plus} and \eqref{eq:flat_ring_susy_minus} give
\begin{equation}\label{aj_ring}
\begin{gathered}
 |j_\pm|=\frac{3q\sqrt R}{8Q_{\mathrm R}^{3/2}}~,\\
 a_{H,+}=\frac{\sqrt6\,q\sqrt{Q_{\mathrm R}^2-q^2}}
                         {4\sqrt R\,Q_{\mathrm R}^{3/2}}
 \quad,\quad
 a_{H,-}=\frac{\sqrt6\,q}{4\sqrt{RQ_{\mathrm R}}}~.
\end{gathered}
\end{equation}
The charged null-wave branch instead has
\begin{equation}\label{eq:flat_ring_null_reduced}
 j_{\mathrm{null}}=\frac38\sqrt{\frac Rq}
 \quad,\quad
 a_{H,\mathrm{null}}=\frac{\sqrt6}{4}\sqrt{\frac qR}
                              \sqrt{1-\frac{Q^2}{q^2}}~.
\end{equation}
\begin{figure}[t]
 \centering
 \includegraphics[width=.92\linewidth]{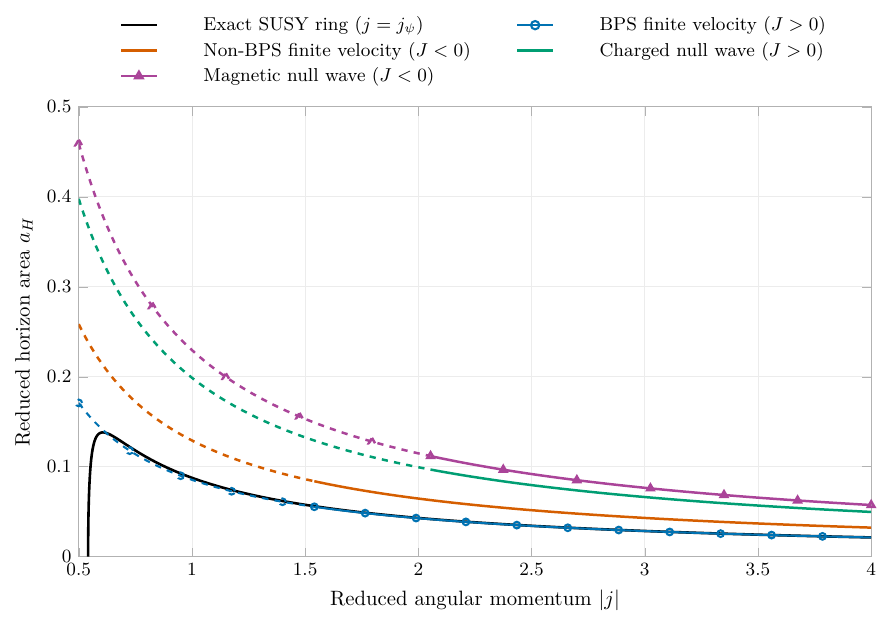}
 \caption{The four extremal blackfold branches and the exact SUSY ring, plotted as reduced area versus angular momentum magnitude. The exact ring and both finite-velocity branches have $q/Q_{\mathrm R}=3/4$. The positive null branch has $|Q|/q=1/2$, while the negative magnetic null branch has $Q_{(0)}=0$. Both null rings are non-BPS. The black curve uses the exact $J_\psi$ and approaches the BPS blackfold curve with increasing radius. Solid blackfold segments satisfy the illustrative thinness cutoff $r_\epsilon(n)/R\leq0.1$, where $r_\epsilon(n)$ estimates the transverse size of the string gravitational field and $R$ is the ring radius. Dashed segments are extrapolations beyond this cutoff.}
 \label{fig:extremal_local_ratios}
\end{figure}
The purely magnetic negative null wave is the $Q_{\mathrm R}\to q$ endpoint of the negative finite-velocity branch in \eqref{aj_ring}. Its reduced quantities are
\begin{equation}\label{eq:flat_ring_magnetic_null_reduced}
 j_{\mathrm{null},-}=-\frac38\sqrt{\frac Rq}
 \quad,\quad
 a_{H,\mathrm{null},-}=\frac{\sqrt6}{4}\sqrt{\frac qR}~.
\end{equation}
Its area and $|j|$ equal those of the positive null branch only at $Q=0$. Since the charged null curve plotted has $|Q|/q=1/2$, we display the magnetic null branch separately. 

The exact ring has two angular momenta. For its curve we use the full $J_\psi$ and compare it directly with the blackfold $J_+$, retaining the subleading contribution in the exact curve. Introducing $x=q/Q_{\mathrm R}$ and $\widehat R=R/Q_{\mathrm R}$ only to parameterise this comparison gives
\begin{equation}\label{eq:exact_ring_reduced}
\begin{aligned}
 j_{\psi,\mathrm{ex}}
 &=\frac{3x\sqrt{\widehat R}}{8}
                  \left(1+\frac1{\widehat R}-\frac{x^2}{6\widehat R^2}\right)~,\\
 a_{H,\mathrm{ex}}
 &=\frac{\sqrt6\,x}{4\sqrt{\widehat R}}
                  \sqrt{\left(1-\frac{x^2}{2\widehat R}\right)^2-x^2}~,
 \qquad \widehat R>\frac{x^2}{2(1-x)}~.
\end{aligned}
\end{equation}
At fixed $x$, the exact curve approaches the BPS blackfold curve as $\widehat R\to\infty$. At the stated local charge ratios, including zero electric charge for the magnetic null branch, the four blackfold curves obey
\begin{equation}\label{eq:flat_ring_reduced_products}
\begin{aligned}
 |j_+|a_{H,+}&=\frac{3\sqrt6}{32}x^2\sqrt{1-x^2}~,\\
 |j_-|a_{H,-}&=\frac{3\sqrt6}{32}x^2~,\\
 j_{\mathrm{null}}a_{H,\mathrm{null}}
             &=\frac{3\sqrt6}{32}\sqrt{1-Q^2/q^2}~,\\
 |j_{\mathrm{null},-}|a_{H,\mathrm{null},-}
             &=\frac{3\sqrt6}{32}~.
\end{aligned}
\end{equation}
Thus each displayed slice falls as $1/|j|$, with a coefficient determined by the limiting state. The two null branches have the same thinness threshold because both have $r_\epsilon(n)=3q$ and the same relation between $|j|$ and $R/q$.

Figure~\ref{fig:extremal_local_ratios} displays these curves at fixed local charge ratios. The exact supersymmetric curve approaches the BPS blackfold prediction at large radius, without adjusting a fit parameter. The two curves describe the same branch at different orders in the thin-ring expansion. Their separation includes both the correction to the area and the difference between $J_\psi$ and the leading blackfold angular momentum. The other extremal curves illustrate the different equations of state, but equal $|j|$ on this plot does not fix the mass, electric charge, dipole charge or both signed angular momenta. Their vertical ordering therefore does not determine a thermodynamically preferred phase.

\subsubsection{Extremal rings at fixed electric and dipole charges}\label{sec:flat_ring_fixed_charges}
Figure~\ref{fig:extremal_local_ratios} compares slices at different electric charges. We now fix the same electric charge $Q_{(0)}<0$ and dipole charge $Q_{(1)}>0$ on each family. This comparison tests the supersymmetric blackfold approximation at matched charges and shows how the extremal branches are arranged as the angular momentum varies. It also identifies common limiting values of the charges and area, whose interpretation depends on horizon regularity and the validity of the approximation. It is useful to define
\begin{equation}\label{eq:ring_fixed_charge_scales}
\begin{gathered}
 q=\frac{8G_5Q_{(1)}}{\sqrt3}\quad,\quad
 Q_\star=\frac{4G_5|Q_{(0)}|}{\sqrt3\pi}
 \quad,\quad M_{\mathrm{BPS}}=\sqrt3|Q_{(0)}|
                                  =\frac{3\pi Q_\star}{4G_5}~,\\
 R_c=\frac{Q_\star}{2q}\quad,\quad y=\frac{R}{R_c}
 \quad,\quad J_c=\frac{3\pi qR_c^2}{4G_5}
 \quad,\quad \mathcal A_c=\pi^2\sqrt3\,qQ_\star~.
\end{gathered}
\end{equation}
Here $Q_\star$ equals $Q_{\mathrm{ring}}$ for the exact supersymmetric ring. Keeping the global electric charge fixed requires distinguishing the definition of parameters on the three blackfold branches that admit nonzero electric charge,
\begin{equation}\label{eq:ring_fixed_charge_local_parameters}
\begin{aligned}
 \text{BPS finite velocity:}&\qquad Q_{\mathrm R}=\frac{Q_\star}{2R}~,\\
 \text{non-BPS finite velocity:}&\qquad
 Q_{\mathrm R}=\sqrt{q^2+\frac{Q_\star^2}{4R^2}}~,\\
 \text{charged null wave:}&\qquad Q=\frac{Q_\star}{2R}~.
\end{aligned}
\end{equation}
Substitution in \eqref{eq:flat_ring_susy_plus}, \eqref{eq:flat_ring_susy_minus} and \eqref{eq:flat_ring_null_charges} gives the following comparison. For the exact solution the angular momentum in the table is $J_\psi$, and $J_\phi=\pi q(3Q_\star-q^2)/(8G_5)$ is fixed along its curve.
\begin{equation}\label{eq:ring_fixed_charge_table}
\renewcommand{\arraystretch}{1.65}
\begin{array}{c|c|c|c}
 \text{family}&M/M_{\mathrm{BPS}}&J/J_c&\mathcal A/\mathcal A_c\\ \hline
 \text{exact SUSY}&1&y^2+J_\phi/J_c
             &\sqrt{(1-q^2/Q_\star)^2-y^2}\\
 \text{BPS finite velocity}&1&y^2&\sqrt{1-y^2}\\
 \text{non-BPS finite velocity}&\sqrt{1+y^2}&-y^2&\sqrt{1+y^2}\\
 \text{charged null wave}&y&y^2&\sqrt{y^2-1}
\end{array}
\end{equation}
The exact ring requires $0<y<1-q^2/Q_\star$ and $q^2<Q_\star$. The BPS blackfold has $0<y<1$, the negative momentum branch has $y>0$, and the charged null wave has $y>1$. The fourth branch, the negative magnetic null wave in \eqref{eq:flat_ring_magnetic_null_charges}, has identically $Q_{(0)}=0$ and is therefore excluded from this fixed nonzero electric-charge comparison. 

\begin{figure}[t]
 \centering
 \includegraphics[width=\linewidth]{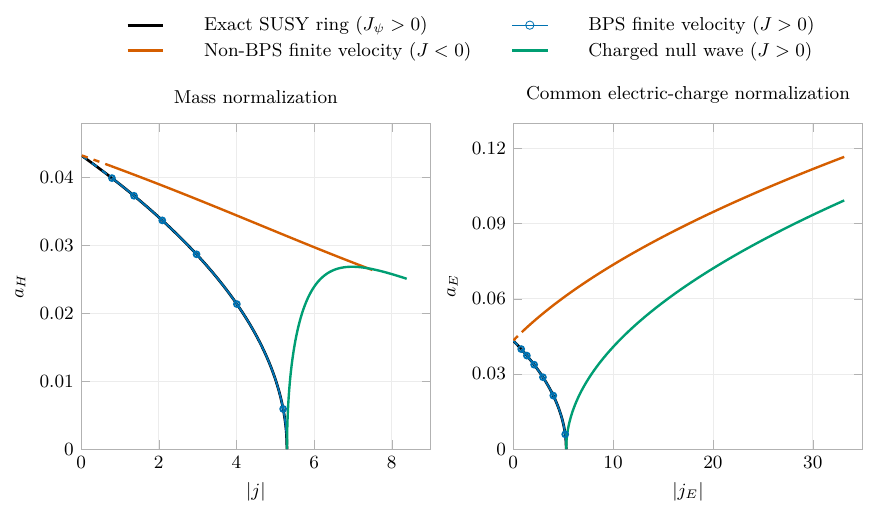}
 \caption{Extremal rings at common $Q_{(0)}<0$ and $Q_{(1)}>0$, with $q^2/Q_\star=1/400$. The magnetic null branch requires $Q_{(0)}=0$ and is excluded. The left panel uses each solution's mass to define $a_H$ and $j$. The right panel uses the common electric charge scale $M_{\mathrm{BPS}}$, so equal $|j_E|$ means equal $|J|$ and $a_E$ is proportional to area. The exact curve uses $J_\psi$. The BPS and charged null blackfold branches reach a common zero area boundary at $M=M_{\mathrm{BPS}}$, $J=J_c$. The orange branch has $J<0$, and its apparent meeting with the BPS curve as $R\to0$ lies outside the thin-ring approximation. Solid segments satisfy the illustrative cutoff $r_\epsilon(n)/R\leq0.1$, with $r_\epsilon(n)=T^{tt}/\mathcal C$. Dashed segments are extrapolations beyond it.}
 \label{fig:BR_comparison}
\end{figure}

An immediate consequence is that the two positive momentum blackfold limits occupy complementary intervals,
\begin{equation}\label{eq:ring_fixed_charge_spin_domains}
 \text{BPS:}\quad 0<J<J_c
 \qquad,\qquad
 \text{null wave:}\quad J>J_c~.
\end{equation}
The BPS branch approaches $y=1$ from below and the charged null branch from above. Their leading charges and areas tend to the same values, $(M,J,\mathcal A)=(M_{\mathrm{BPS}},J_c,0)$, at $R=R_c$. Figure~\ref{fig:BR_comparison} therefore displays a common zero-area boundary between these two positive momentum regimes. The BPS and charged null blackfold branches share the same zero-area
endpoint at $y=1$, where
$(M,J,\mathcal A)=(M_{\mathrm{BPS}},J_c,0)$.
This equality holds exactly within the leading blackfold description,
although the endpoint does not represent a regular finite-area horizon.
The exact supersymmetric ring instead reaches zero area at
$y=1-q^2/Q_\star$, so its endpoint differs at finite $q^2/Q_\star$.

As $y\to0$, the two finite-velocity branches formally approach $M=M_{\mathrm{BPS}}$, $J=0$ and $\mathcal A=\mathcal A_c$. However, this sends $R\to0$ at fixed charges, and the ratios in \eqref{eq:ring_fixed_charge_thinness} diverge. The blackfold approximation fails before reaching this apparent meeting point. For every $R>0$, the negative momentum branch has $J<0$ and $M>M_{\mathrm{BPS}}$, whereas the other two blackfold branches have $J>0$. Plotting $|J|$ can hide this distinction. Crossings in the left panel, which also uses a different mass to normalise each state, need not match either the mass or the angular momentum. Therefore they do not establish a connection between solutions or an entropy comparison at the same conserved charges.

The mass dependence also matters for the axes of a reduced area plot. At fixed $Q_{(0)}$ and $Q_{(1)}$, equal values of $|j|$ need not represent equal $|J|$ when the masses differ. Alongside \eqref{eq:flat_ring_reduced_definitions}, we therefore introduce a common normalisation,
\begin{equation}\label{eq:ring_electric_reduced_variables}
\begin{gathered}
 a_E=c_a\frac{\mathcal A}{(G_5M_{\mathrm{BPS}})^{3/2}}
 \quad,\quad
 j_E=c_j\frac{G_5J}{(G_5M_{\mathrm{BPS}})^{3/2}}~,\\
 a_H=a_E\left(\frac{M_{\mathrm{BPS}}}{M}\right)^{3/2}
 \quad,\quad
 j=j_E\left(\frac{M_{\mathrm{BPS}}}{M}\right)^{3/2}~.
\end{gathered}
\end{equation}
Equal $|j_E|$ now means equal angular momentum magnitude across the fixed charge families, and $a_E$ is directly proportional to their areas. Their masses remain those in \eqref{eq:ring_fixed_charge_table}. In particular, the decreasing reduced area of the negative-momentum branch in the left panel of Figure~\ref{fig:BR_comparison} accompanies an increasing physical area in the right panel.

For a representative thin-ring comparison we choose $q^2/Q_\star=1/400$. The approximation requires the transverse size of the string gravitational field to be much smaller than the ring radius $R$. We estimate that size by $r_\epsilon(n)=T^{tt}/\mathcal C$, as in Section~\ref{sec:flat_ring_validity}. Its ratio to $R$ is
\begin{equation}\label{eq:ring_fixed_charge_thinness}
 \frac{r_\epsilon(n)}R=\frac{6q^2}{Q_\star}
 \begin{cases}
 y^{-2} & \text{BPS finite velocity}~,\\
 \sqrt{1+y^2}\,y^{-2} & \text{non-BPS finite velocity}~,\\
 y^{-1} & \text{charged null wave}~.
 \end{cases}
\end{equation}
Solid blackfold segments satisfy the illustrative cutoff $r_\epsilon(n)/R\leq0.1$, while dashed segments extend the formulas to thicker rings. 

To determine whether the extremal curves describe competing phases in the microcanonical ensemble, one must also fix the mass and angular momentum. At fixed nonzero $Q_{(0)}$ and $Q_{(1)}$, the scales in \eqref{eq:ring_fixed_charge_scales} are common to all three branches. Fixing the mass ratio $\nu=M/M_{\mathrm{BPS}}>1$ in \eqref{eq:ring_fixed_charge_table} eliminates the radius and gives
\begin{equation}\label{eq:ring_extremal_microcanonical_test}
\begin{array}{c|c|c}
 \text{family}&J/J_c&\mathcal A/\mathcal A_c\\ \hline
 \text{non-BPS finite velocity}&-(\nu^2-1)&\nu\\
 \text{charged null wave}&\nu^2&\sqrt{\nu^2-1}
\end{array}
\end{equation}
The angular momenta differ in sign, and even their magnitudes differ by $J_c$. These states cannot compete at the same mass and angular momentum within the chosen charge sector. The BPS branch only exists at $\nu=1$. The other two branches approach singular or degenerate endpoints there, not a second regular ring of the same charges. Thus there is no regular pair among these three leading-order extremal branches satisfying the fixed dipole microcanonical constraints. 

\subsubsection{Non-extremal rings and comparison with exact solutions}\label{sec:flat_ring_nonext_plots}\label{sec:exact_nonext_comparison}
For non-extremal rings, fixing the electric and dipole charges leaves a two-parameter family on each momentum branch. Holding the mass fixed gives a one-parameter family, which we label by
\begin{equation}\label{eq:ring_mass_excess}
 \nu=\frac{M}{M_{\mathrm{BPS}}}>1~.
\end{equation}
\begin{figure}[t]
 \centering
 \includegraphics[width=\linewidth]{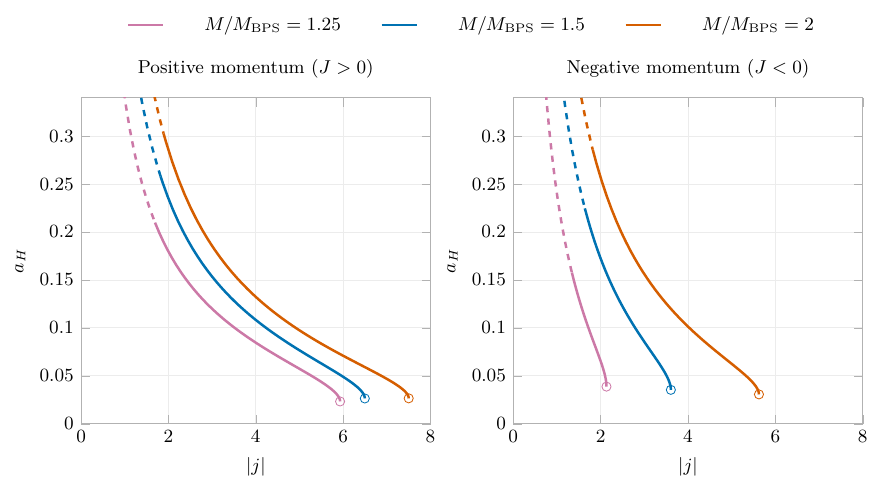}
 \caption{Non-extremal blackfold rings on the two equilibrium branches at fixed electric and dipole charges, with $q^2/Q_\star=1/400$. Magenta, blue and orange denote $M/M_{\mathrm{BPS}}=1.25,1.5,2$, respectively, with the same mass assigned to each colour in both panels. The left panel shows positive momentum and the right panel negative momentum, using $|j|$ on the horizontal axes. Open circles mark the zero-temperature endpoints \eqref{eq:ring_nonext_curve_endpoints}, namely the charged null wave for $J>0$ and the non-BPS finite-velocity ring for $J<0$. Solid segments satisfy the illustrative cutoff $r_\epsilon(n)/R\leq0.1$, and dashed segments are extrapolations beyond it. The temperature varies along each curve.}
 \label{fig:nonext_fixed_charges}
\end{figure}
\noindent Figure~\ref{fig:nonext_fixed_charges} displays the two equilibrium branches arising from the roots $w_\pm$ in \eqref{eq:OmegaR}. At the fixed electric and magnetic charge orientations used here, they carry $J>0$ and $J<0$, respectively, as follows from \eqref{eq:flat_ring_branch_charges}. The curves have the same $M,Q_{(0)},Q_{(1)}$ in the two panels for each colour, but different signs of their angular momenta. The temperature varies along each family. We use $\beta$ to parametrise the states while adjusting the remaining string parameters to keep these three charges fixed. Appendix~\ref{app:nonext_fixed_charge_curves} gives the construction and plotting details. At fixed $\nu>1$, the limit $\beta\to\infty$ gives $T_0\to0$ and the endpoints
\begin{equation}\label{eq:ring_nonext_curve_endpoints}
\begin{array}{c|c|c|c}
 &R/R_c&J/J_c&\mathcal A/\mathcal A_c\\ \hline
 +\ \text{branch}&\nu&\nu^2&\sqrt{\nu^2-1}\\
 -\ \text{branch}&\sqrt{\nu^2-1}&-(\nu^2-1)&\nu
\end{array}
\end{equation}
The positive branch approaches the charged null wave, with $Q/q=1/\nu$. The negative branch approaches the finite-velocity extremal solution, with $Q_{\mathrm R}/q=\nu/\sqrt{\nu^2-1}$. They reproduce the corresponding entries of \eqref{eq:ring_fixed_charge_table}. The BPS branch lies at $\nu=1$ and is not reached while keeping a strictly positive mass excess fixed.
\begin{figure}[t]
 \centering
 \includegraphics[width=\linewidth]{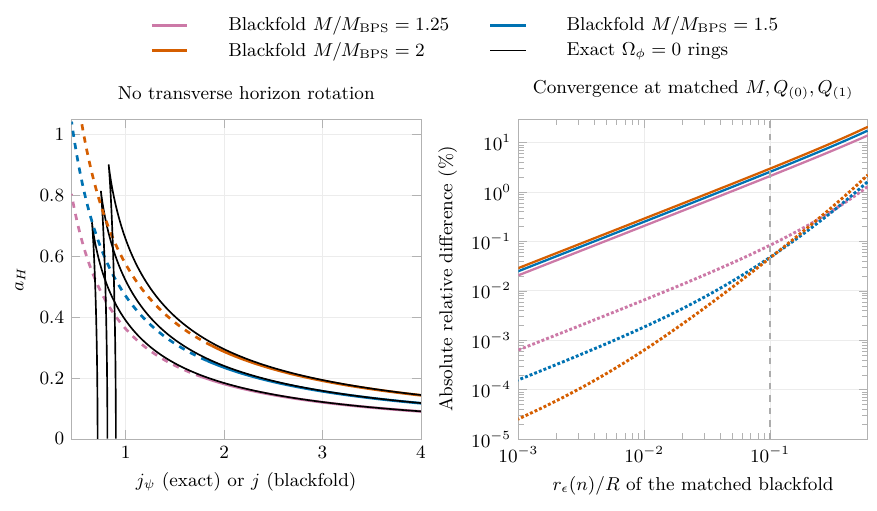}
 \caption{Exact non-extremal rings with $\Omega_\phi=0$ and the corresponding blackfold solutions obtained here. Each curve fixes $M,Q_{(0)}$ and varies the dipole, which is matched between each exact and blackfold pair. Magenta, blue and orange denote $M/M_{\mathrm{BPS}}=1.25,1.5,2$, respectively, and the black curves are the exact solutions. In the left panel, coloured solid segments satisfy the illustrative cutoff $r_\epsilon(n)/R\leq0.1$ and dashed segments are extrapolations beyond it. The right panel shows $100|\Delta_J|$ as solid curves and $100|\Delta_{\mathcal A}|$ as dotted curves, with the relative differences defined in \eqref{eq:ST_zero_transverse_errors} at matched $M,Q_{(0)},Q_{(1)}$. Its vertical dashed line marks the same thickness cutoff. The comparison imposes zero transverse horizon angular velocity, not zero total $J_\phi$.}
 \label{fig:exact_nonext_comparison}
\end{figure}
The exact Suzuki-Tomizawa rings~\cite{Suzuki:2024generalring} provide an independent check. We impose vanishing transverse horizon angular velocity, $\Omega_\phi=0$, so their thin limit contains the non-rotating string used here. The electromagnetic field can still carry transverse angular momentum, and $J_\phi$ need not vanish. The two blackfold branches give new approximate ring solutions beyond this exact family. For nonzero electric and dipole charges with the orientations held fixed, the negative momentum branch and generic states on the positive momentum branch are not covered by its non-rotating thin limit. The overlap is confined to a restricted subset of the positive momentum branch, because the Suzuki--Tomizawa family imposes an additional relation between the local charge parameters. Its regular extremal string limit reaches only a restricted BPS family, not either of the fixed-$\nu>1$ endpoints in \eqref{eq:ring_nonext_curve_endpoints}. Appendix~\ref{app:exact_nonext_limit} gives the parameter map and detailed comparison, and Appendix~\ref{app:ST_extremal_audit} explains the extremal restriction.

Figure~\ref{fig:exact_nonext_comparison} compares the area and longitudinal angular momentum for $\nu=1.25,1.5,2$. Each curve fixes $M,Q_{(0)}$ and varies the dipole, matching $Q_{(1)}$ between each exact ring and its blackfold counterpart. Unlike Figure~\ref{fig:nonext_fixed_charges}, the dipole therefore varies along a curve. The area and angular momentum are then predictions, with the blackfold spin compared to the exact $J_\psi$ for the same asymptotic rotation generator. The left panel shows agreement at large reduced angular momentum. The right panel compares the relative differences at matched charges and shows both tending to zero as $r_\epsilon(n)/R\to0$.

\subsection{Supersymmetric black Saturns}\label{sec:susy_saturn}
We now place the supersymmetric string of Section~\ref{sec:susy_lim} around a BMPV black hole~\cite{Breckenridge:1996is}. The resulting configuration has a central spherical horizon and a disconnected ring horizon. Both components must preserve the same supersymmetry. This requirement selects the thermal Killing vector of the ring and fixes its electric potential relative to that of the central black hole. Although the ring has nonzero angular momentum, its horizon angular velocity vanishes. The background fluxes are essential to its equilibrium, with the electric and magnetic forces cancelling the gravitational and rotational contributions to the extrinsic equation. We first derive this cancellation and the conserved charges, then compare with exact supersymmetric solutions using~\cite{Gauntlett:2002nw,Gauntlett:2004wh}.

\subsubsection{Background, embedding and equilibrium potentials}
We choose the angular orientations so that the two angular momenta of the BMPV solution have the same sign and introduce
\begin{equation}\label{eq:saturn_background_functions}
 f(r)=1-\frac{\mu}{r^2}\quad,\quad b(r)=\frac{\mathfrak j}{r^2}
 \quad,\quad \sigma=\sin^2\theta\,d\phi+\cos^2\theta\,d\psi~.
\end{equation}
Here $\mathfrak j$ has dimensions of length cubed and is distinct from the reduced angular momentum $j$. In the gauge $A\to0$ at infinity, the metric and gauge potential are~\cite{Breckenridge:1996is}
\begin{equation}\label{eq:saturn_BMPV}
\begin{aligned}
 ds^2={}&-(f\,d\tau+b\sigma)^2+\frac{dr^2}{f^2}
     +r^2\left(d\theta^2+\sin^2\theta\,d\phi^2+\cos^2\theta\,d\psi^2\right)~,\\
 A={}&\sqrt3\left[(1-f)d\tau-b\sigma\right]~.
\end{aligned}
\end{equation}
The coefficient of $A$ follows the normalisation of Section~\ref{sec:sugra_ward_identities}. The exterior has $r>\sqrt\mu$, with $\mu>0$ and $\mathfrak j^2<\mu^3$. The horizon generator is $\xi=\partial_\tau$, and its temperature and angular velocities vanish. Its charges and area are
\begin{equation}\label{eq:saturn_central_charges}
\begin{gathered}
 M_{\mathrm H}=\frac{3\pi\mu}{4G_5}\quad,\quad
 Q_{(0),\mathrm H}=-\frac{\sqrt3\pi\mu}{4G_5}\quad,\quad
 J_{\psi,\mathrm H}=J_{\phi,\mathrm H}=\frac{\pi\mathfrak j}{4G_5}~,\\
 \mathcal A_{\mathrm H}=2\pi^2\sqrt{\mu^3-\mathfrak j^2}~.
\end{gathered}
\end{equation}
The static charged black hole is obtained by setting $\mathfrak j=0$. The dual potential $B$ introduced in Section \ref{sec: Ward} is needed both for the force and for the global charges. We use two gauge-related expressions to keep this potential regular at the two rotational axes. The subscripts $N$ and $S$ denote overlapping coordinate regions containing $\theta=0$ and $\theta=\pi/2$, respectively. With orientation $d\tau\wedge dr\wedge d\theta\wedge d\phi\wedge d\psi>0$, we choose
\begin{equation}\label{eq:saturn_dual_patches}
\begin{aligned}
 B_N&=\sqrt3 b\,d\tau\wedge\sigma
                   +\sqrt3\mu\sin^2\theta\,d\phi\wedge d\psi~,\\
 B_S&=B_N-\sqrt3\mu\,d\phi\wedge d\psi~.
\end{aligned}
\end{equation}
Here $B_N$ is regular at the axis where the $\phi$ circle shrinks, while $B_S$ is regular where the $\psi$ circle shrinks. They describe the same field strength, with $dB_N=dB_S=-\star F+\kappa A\wedge F$ and $\kappa=1/\sqrt3$. The nonzero flux $\int_{S^3}dB=4\pi^2\sqrt3\mu$ through a sphere at fixed $r$ and $\tau$ prevents a globally regular two-form potential on that sphere. In a local angular chart in the overlap, $B_N-B_S=d\Lambda_{NS}$, where $\Lambda_{NS}=\sqrt3\mu\phi\,d\psi$. This transition relates two descriptions of the same fields. Although the ring below lies entirely in the region where $B_N$ is regular, the rotational charge is defined using the fields throughout the exterior. We explain how this fixes its representation near the ring in \eqref{eq:saturn_rotation_regular}--\eqref{eq:saturn_transverse_spin}.

We now choose worldsheet coordinates $(\tau,\psi)$ and the circular embedding
\begin{equation}\label{eq:saturn_embedding}
\begin{gathered}
 X^\tau=\tau\quad,\quad X^r=R\quad,\quad X^\theta=0
 \quad,\quad X^\psi=\psi~,\\
 X^\phi=\mathrm{constant}
 \quad,\quad \psi\sim\psi+2\pi~.
\end{gathered}
\end{equation}
Writing $f_R=f(R)$ and $b_R=b(R)$, the induced metric is
\begin{equation}\label{eq:saturn_induced}
 ds^2_{\Sigma_2}=-(f_Rd\tau+b_Rd\psi)^2+R^2d\psi^2
 \quad,\quad \sqrt{-\gamma}=f_RR~.
\end{equation}
A useful orthonormal coframe for this geometry that we will be using later is $e^{\hat0}=f_Rd\tau+b_Rd\psi$, $e^{\hat1}=R\,d\psi$. These are local frame forms. In particular, integrating $e^{\hat0}$ does not give a globally single-valued time coordinate on the circle. As in the case of black rings, we pick a general stationary thermal generator tangent to the ring,
\begin{equation}\label{eq:saturn_general_thermal}
 k=\partial_\tau+\Omega\partial_\psi\quad,\quad
 |k|^2=(f_R+b_R\Omega)^2-R^2\Omega^2
 \quad,\quad w=\frac{R\Omega}{f_R+b_R\Omega}~.
\end{equation}
For $f_R+b_R\Omega>0$ and $|w|<1$, the oriented unit vectors are
\begin{equation}\label{eq:saturn_uv}
\begin{aligned}
 u^a&=\frac{(1,\Omega)}{|k|}~,\\
 v^a&=\frac{\left(b_R(f_R+b_R\Omega)-R^2\Omega,
                              -(f_R+b_R\Omega)\right)}{f_RR|k|}~.
\end{aligned}
\end{equation}
The supersymmetric Killing vector of the BMPV background is $\xi=\partial_\tau$. It is an asymptotically unit-normalized, future-directed Killing vector. Aligning the local supersymmetric string with it selects
\begin{equation}\label{eq:saturn_susy_generator}
 \Omega=0\quad,\quad |k|=f_R\quad,\quad
 u^a=\left(\frac1{f_R},0\right)
 \quad,\quad v^a=\left(\frac{b_R}{f_RR},-\frac1R\right)~.
\end{equation}
This also follows from matching the electric potential of the BMPV horizon with that of the probe string. With $\mu_0=-\sqrt3$, Section~\ref{sec:susy_lim} gives
\begin{equation}\label{eq:saturn_electric_potential}
 \Phi_0=|k|\mu_0-\iota_kA
       =-\sqrt3\left[|k|+1-f_R-b_R\Omega\right]~.
\end{equation}
Equating it to the central value $-\sqrt3$ requires $|k|=f_R+b_R\Omega$, and hence $\Omega=0$. A longitudinal boost of the local string therefore does not in general preserve its supersymmetry in this background. The local state is \eqref{eq:string_susy_thermodynamics}, with $Q_{\mathrm R}>q>0$ and $\mathcal C=1/(4G_5)$. In the coframe above the stress tensor and currents are
\begin{equation}\label{eq:saturn_local_monopoles}
 T^{\hat a\hat b}=\mathcal C
 \begin{pmatrix}3Q_{\mathrm R}&3q/2\\3q/2&0\end{pmatrix}
 \quad,\quad J^{\hat a}=-\sqrt3\mathcal C(Q_{\mathrm R},q)
 \quad,\quad \mathcal J^{\hat0\hat1}=-\frac{\sqrt3\mathcal Cq}{2}~.
\end{equation}
Thus the rest-frame tension vanishes while the momentum density is nonzero. We use $B_N$, which is regular along the ring. Since $A$ and $B_N$ are independent of $\tau$, we choose time translations to act without an additional gauge transformation, setting $\lambda_\xi=0$ and $\Lambda_{\xi,N}=0$ in \eqref{eq:global_background_symmetry}. The zero-temperature limit of the thermal invariants in Section~\ref{sec:equilibrium_action} then gives
\begin{equation}\label{eq:saturn_global_potentials}
 T_0=0\quad,\quad \Phi_0=-\sqrt3\quad,\quad
 \zeta=-\sqrt3\,d\tau\quad,\quad \Phi_1=\oint_{B_1}\zeta=0~.
\end{equation}
The vanishing magnetic potential includes the background and 2-group terms in \eqref{eq:global_local_potentials}, rather than following from $\mu_1=0$ alone.

\subsubsection{Blackfold equations and the absence of a radial force}
The induced sources and state are constant, and the pullback of $F$ vanishes. Consequently the intrinsic equations of Section~\ref{sec: Ward} are satisfied. The nontrivial extrinsic equation is radial. Keeping its normal index in the coordinate basis $r$, while using the orthonormal tangent frame, the curvature components are
\begin{equation}\label{eq:saturn_extrinsic_components}
 K_{\hat0\hat0}{}^r=ff'=\frac{2\mu f}{R^3}
 \quad,\quad K_{\hat0\hat1}{}^r=\frac{f}{2R}(fb'-bf')
                                  =-\frac{f\mathfrak j}{R^4}
 \quad,\quad K_{\hat1\hat1}{}^r=-\frac{f^2}{R}~,
\end{equation}
where all functions and derivatives are evaluated at $R$. For brevity we denote these components by $K_{00},K_{01},K_{11}$. The relevant field components are
\begin{equation}\label{eq:saturn_force_components}
 F^r{}_{\hat0}=-\sqrt3 K_{00}\quad,\quad
 F^r{}_{\hat1}=-2\sqrt3 K_{01}\quad,\quad
 \mathcal F^r{}_{\hat0\hat1}=2\sqrt3 K_{01}~.
\end{equation}
The rotating BMPV background exerts both an electric and a magnetic force on the string. Keeping both terms and using the aligned supersymmetric state with $\Omega=0$, substitution into \eqref{eq:equilibrium_embedding_equation} yields
\begin{equation}\label{eq:saturn_no_force}
\begin{aligned}
 T^{\hat a\hat b}K_{\hat a\hat b}{}^r
     &=3\mathcal C Q_{\mathrm R}K_{00}+3\mathcal CqK_{01}~,\\
 F^r{}_{\hat a}J^{\hat a}
     &=3\mathcal C Q_{\mathrm R}K_{00}+6\mathcal CqK_{01}~,\\
 \frac12\mathcal F^r{}_{\hat a\hat b}\mathcal J^{\hat a\hat b}
     &=-3\mathcal CqK_{01}~.
\end{aligned}
\end{equation}
For $\mathfrak j\ne0$, the mixed curvature $K_{01}=-f_R\mathfrak j/R^4$ is nonzero throughout the exterior. Nevertheless, the sum of the electric and magnetic forces in \eqref{eq:saturn_no_force} is $3\mathcal C Q_{\mathrm R}K_{00}+(6-3)\mathcal CqK_{01}$, exactly equal to the extrinsic stress. The radial blackfold equation is therefore an identity for every allowed central angular momentum, including $K_{01}\ne0$. In the static limit $\mathfrak j=0$, the terms proportional to $K_{01}$ disappear and the same identity reduces to electric and gravitational force balance.

Thus force balance does not select a radius for the aligned supersymmetric ring in the rotating BMPV background. Within the concentric circular embedding considered here, the ring must lie outside the central horizon,
\begin{equation}\label{lower_bound}
 R>\sqrt\mu~.
\end{equation}
This inequality only ensures that the ring is outside the central horizon. The central horizon itself is excluded because $k$ becomes null there and the timelike worldsheet construction degenerates. Horizon regularity, exterior causality and the validity conditions below further restrict the admissible radii.

\subsubsection{Conserved charges and thermodynamics}
We evaluate the charges of Section~\ref{sec: conserved_charges} on the closed slice $\tau=\mathrm{constant}$. Its normal and volume element obey
\begin{equation}\label{eq:saturn_slice}
 n_a=\left(-\frac{f_RR}{\sqrt{R^2-b_R^2}},0\right)
 \quad,\quad dV_1=\sqrt{R^2-b_R^2}\,d\psi
 \quad,\quad -n_a u^a\,dV_1=R\,d\psi~.
\end{equation}
The background bound ensures $R^2>b_R^2$. The shift in the induced metric must be kept in these integrals. We again use the potential $B_N$ regular along the ring and choose no additional gauge transformation for time translations or rotations along $\psi$. Equations~\eqref{modified_J} and \eqref{Ptilde} then give
\begin{equation}\label{eq:saturn_charge_currents}
 \widetilde J^\tau=-\frac{\sqrt3\mathcal C Q_{\mathrm R}}{f_R}
 \quad,\quad \widetilde P_\xi^\tau=-\frac{3\mathcal C Q_{\mathrm R}}{f_R}
 \quad,\quad \widetilde P_{\chi_\psi}^\tau=\frac{3\mathcal CqR}{2f_R}~.
\end{equation}
The ring contributions are consequently
\begin{equation}\label{eq:saturn_ring_charges}
\begin{gathered}
 M_{\mathrm{ring}}=6\pi\mathcal C RQ_{\mathrm R}\quad,\quad
 Q_{(0),\mathrm{ring}}=-2\pi\sqrt3\mathcal C RQ_{\mathrm R}
 \quad,\quad Q_{(1)}=\frac{\sqrt3\mathcal Cq}{2}~,\\
 J_{\psi,\mathrm{ring}}=3\pi\mathcal CqR^2\quad,\quad
 \mathcal A_{\mathrm{ring}}=2\pi^2\sqrt3 Rq\sqrt{Q_{\mathrm R}^2-q^2}
 \quad,\quad S_{\mathrm{ring}}=\frac{\mathcal A_{\mathrm{ring}}}{4G_5}~.
\end{gathered}
\end{equation}
The electric and dual potential terms remove the apparent redshift dependent binding contribution to the mass. In this fixed gauge branch $M_{\mathrm{ring}}=-\sqrt3 Q_{(0),\mathrm{ring}}$.

There is also an interaction contribution to the transverse angular momentum $J_\phi$. The Killing vector $\chi_\psi=\partial_\psi$ generates rotations along the ring, while $\chi_\phi=\partial_\phi$ generates rotations in the transverse plane. At $\theta=0$ the latter vector vanishes, since the ring lies on the fixed axis of this rotation. Its mechanical orbital contribution to $J_\phi$ therefore vanishes. The electric field of the central hole and the magnetic field of the ring can nevertheless carry angular momentum, as in the exact supersymmetric Saturn solutions of~\cite{Bena:2005sliding}.

To evaluate this contribution, recall that $\eta_{\chi_\phi}$ in \eqref{eq:global_symmetry_potentials} is a one-form associated with the Killing vector $\chi_\phi$. In particular, $\eta_{\chi_\phi,\psi}$ denotes its component along the ring. The label $\chi_\phi$ identifies the rotation whose charge is $J_\phi$, whereas the index $\psi$ specifies the component contracted with the string current. We use the invariant Maxwell potential in \eqref{eq:saturn_BMPV} and set $\lambda_{\chi_\phi}=0$. On the embedding, both $(\chi_\phi)_b$ and $\alpha_{\chi_\phi}=\iota_{\chi_\phi}A$ vanish. The conserved current \eqref{Ptilde} therefore reduces to
\begin{equation*}
\begin{aligned}
 \widetilde P^a_{\chi_\phi}
 &=T^{ab}(\chi_\phi)_b+\alpha_{\chi_\phi}J^a
                         +\eta_{\chi_\phi,b}\mathcal J^{ab}
   =\eta_{\chi_\phi,b}\mathcal J^{ab}~,\\
 \widetilde P^\tau_{\chi_\phi}
 &=\eta_{\chi_\phi,\psi}\mathcal J^{\tau\psi}~.
\end{aligned}
\end{equation*}
The remaining task is to determine $\eta_{\chi_\phi,\psi}$, which need not vanish with $\chi_\phi$ on the ring.

Because $B$ requires the two gauge descriptions in \eqref{eq:saturn_dual_patches}, the rotation must act consistently in both. A symmetry of a form potential is specified by a spacetime transformation together with a gauge parameter~\cite{Compere:2007pform}. For $P=N,S$, the symmetry condition and the associated one-form are
\begin{equation}\label{eq:saturn_rotation_generator}
\begin{aligned}
 \mathcal L_{\chi_\phi}B_P+d\Lambda_{\chi_\phi,P}&=0~,\\
 \eta_{\chi_\phi,P}
 &=\iota_{\chi_\phi}B_P+\Lambda_{\chi_\phi,P}
               -\kappa(\iota_{\chi_\phi}A)A~.
\end{aligned}
\end{equation}
Here $\Lambda_{\chi_\phi,P}$ accompanies the rotation, while $\Lambda_{NS}$ relates the two descriptions of $B$. Since $\mathcal L_{\chi_\phi}B_P=0$, the local symmetry condition only requires $d\Lambda_{\chi_\phi,P}=0$. A closed parameter leaves $B_P$ unchanged but can still contribute to the charge through its integral around the string. Its value must therefore be fixed by the global rotation, rather than by the fields restricted to the worldsheet.

We require $\eta_{\chi_\phi}$ to be smooth over both rotational axes of the background. The reason is transparent for a static centre, $\mathfrak j=0$. Equation~\eqref{eq:global_symmetry_potential_identities} admits the one-form choice
\begin{equation*}
 \eta_{\chi_\phi}
   =\left(\sqrt3\mu\sin^2\theta+c_0\right)d\psi~.
\end{equation*}
The $\psi$ circle collapses at $\theta=\pi/2$, so regularity there fixes $c_0=-\sqrt3\mu$. Thus $\eta_{\chi_\phi}=-\sqrt3\mu\cos^2\theta\,d\psi$, whose pullback to the ring at $\theta=0$ is nonzero. For a rotating centre, the same calculation gives
\begin{equation}\label{eq:saturn_rotation_regular}
 \eta_{\chi_\phi}
 =-\sqrt3\left[bf\sin^2\theta\,d\tau
       +\mu\cos^2\theta\,d\psi+b^2\sin^2\theta\,\sigma\right]~.
\end{equation}
Here $\sigma=\sin^2\theta\,d\phi+\cos^2\theta\,d\psi$ is the one-form defined in \eqref{eq:saturn_background_functions}. The terms involving $b$ vanish on the ring, so its pullback is again $\eta_{\chi_\phi}=-\sqrt3\mu\,d\psi$. Adding a globally smooth closed one-form cannot change its integral around the ring, since this circle is contractible within the surrounding $S^3$.

Expressing the same \eqref{eq:saturn_rotation_regular} in the two descriptions of $B$ gives
\begin{equation}\label{eq:saturn_axial_lift}
 \Lambda_{\chi_\phi,S}=0\quad,\quad
 \Lambda_{\chi_\phi,N}=-\mathcal L_{\chi_\phi}\Lambda_{NS}
              =-\sqrt3\mu\,d\psi~.
\end{equation}
The difference of symmetry parameters cancels $\iota_{\chi_\phi}(B_N-B_S)=\sqrt3\mu\,d\psi$, giving the same $\eta_{\chi_\phi}$ in both descriptions. On the worldsheet, the contractions with $A$ and $B_N$ vanish and the surviving term is precisely $\Lambda_{\chi_\phi,N}$. This parameter represents the same global rotation in the gauge regular near the ring. It does not change the background fields. Finally, \eqref{eq:saturn_local_monopoles} and \eqref{eq:saturn_slice} give
\begin{equation*}
 \mathcal J^{\tau\psi}=-\frac{Q_{(1)}}{f_RR}
 \quad,\quad n_\tau\,dV_1=-f_RR\,d\psi~.
\end{equation*}
Substituting the reduced current into \eqref{eq:momentum}, and integrating over $\psi$, therefore yields
\begin{equation}\label{eq:saturn_transverse_spin}
\begin{aligned}
 J_{\phi,\mathrm{ring}}
 &=-Q_{(1)}\int_0^{2\pi}\eta_{\chi_\phi,\psi}\,d\psi
   =2\pi\sqrt3\mu Q_{(1)}=3\pi\mathcal Cq\mu~,\\
 J_{\psi,\mathrm{ring}}-J_{\phi,\mathrm{ring}}
 &=3\pi\mathcal Cq(R^2-\mu)~.
\end{aligned}
\end{equation}
We checked that these charges agree with the leading ADM result of the exact BPS black Saturn solution in Appendix~\ref{app:exact_susy_saturn}.  Adding \eqref{eq:saturn_central_charges} gives the leading total charges and the sum of the horizon areas,
\begin{equation}\label{eq:saturn_total_charges}
\begin{gathered}
 M_{\mathrm{tot}}=\frac{3\pi}{4G_5}(\mu+2RQ_{\mathrm R})
       =-\sqrt3 Q_{(0),\mathrm{tot}}~,\\
 J_{\psi,\mathrm{tot}}=\frac{\pi}{4G_5}(\mathfrak j+3qR^2)
 \quad,\quad J_{\phi,\mathrm{tot}}=\frac{\pi}{4G_5}(\mathfrak j+3q\mu)~,\\
 \mathcal A_{\mathrm{tot}}=\mathcal A_{\mathrm H}+\mathcal A_{\mathrm{ring}}~.
\end{gathered}
\end{equation}
Both horizons have $T_0=0$, $\Omega_i=0$ and $\Phi_0=-\sqrt3$, while the ring dipole potential vanishes. Along the supersymmetric family the first law reduces to $dM_{\mathrm{tot}}=-\sqrt3\,dQ_{(0),\mathrm{tot}}$. 

\subsubsection{Comparison with the exact supersymmetric Saturn}\label{sec:saturn_exact_comparison}
The exact supersymmetric black Saturn~\cite{Gauntlett:2004wh}, with a central BMPV horizon and a surrounding ring horizon, provides a direct test of the charges and area obtained above. Appendix~\ref{app:exact_susy_saturn} gives its geometry and the detailed comparison. The parameters are related by
\begin{equation}\label{eq:saturn_radius_dictionary}
 R_{\mathrm b}^2=R^2-\mu\quad,\quad
 Q_{\mathrm{ring}}=2RQ_{\mathrm R}~.
\end{equation}
Here $R_{\mathrm b}$ is the circle radius in the flat base of the exact metric and $R$ is the embedding radius in the BMPV background. Matching the asymptotic electric charge fixes the exact ring parameter $Q_{\mathrm{ring}}$ through the second relation.

At the same $\mu,\mathfrak j,q$, this identification matches the total mass, electric charge and ring dipole charge exactly. The blackfold reproduces both angular momenta and the ring area to leading order, including the transverse interaction term in \eqref{eq:saturn_transverse_spin}. The finite corrections are given in \eqref{eq:saturn_finite_corrections}. This checks both the BPS relation and the coupling to the background fields.

The validity conditions of Section~\ref{sec:string_validity} compare the string thickness with the local bending, curvature, flux and separation scales~\cite{Armas:2015minimal,Armas:2015qsv}. The ring must also perturb the central hole only weakly. These conditions restrict $R$ without requiring $R\gg\sqrt\mu$. Appendix~\ref{app:exact_susy_saturn} gives the explicit bounds and shows analytically that the horizon regularity inequalities ensure exterior causality for this exact family.

\begin{figure}[t]
 \centering
 \includegraphics[width=\linewidth]{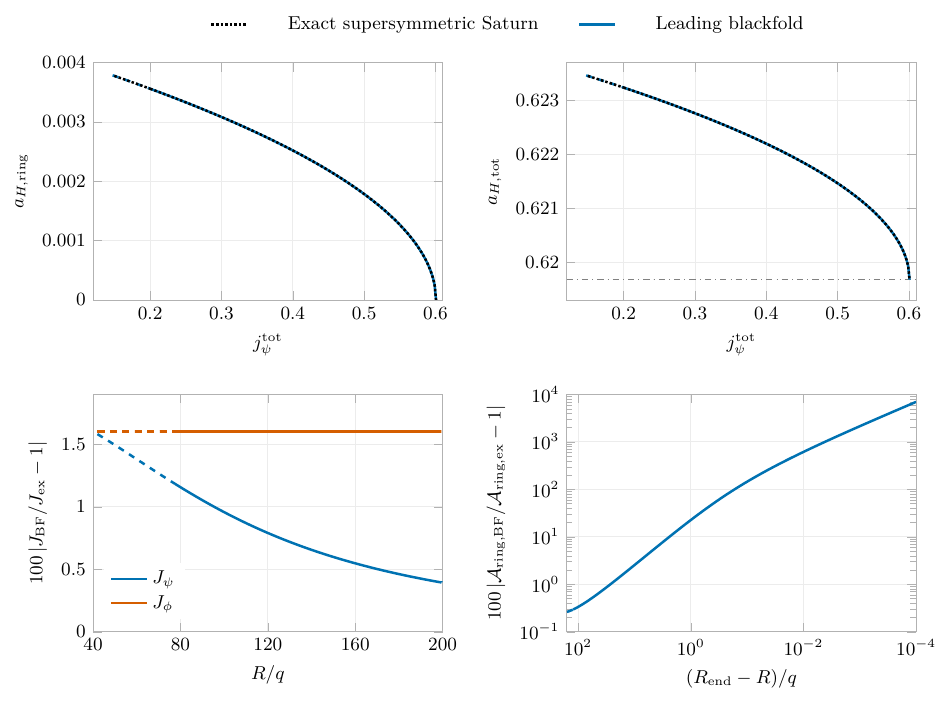}
 \caption{Exact and blackfold supersymmetric Saturns at fixed total mass, electric charge and ring dipole charge, with $\mu/q^2=1600$, $Q_{\mathrm{ring}}/q^2=400$ and $\mathfrak j/\mu^{3/2}=1/2$. \textit{Upper panels:} reduced ring area and total horizon area, using \eqref{eq:flat_ring_reduced_definitions}. Black dotted curves are exact, blue curves are the blackfold result, and the grey line is the central horizon contribution. \textit{Lower left:} absolute percentage errors in the total angular momenta, $100|J_{i,\mathrm{BF}}/J_{i,\mathrm{ex}}-1|$, at the same radius $R$, with $i=\psi$ in blue and $i=\phi$ in orange. \textit{Lower right:} the corresponding absolute percentage error in the ring area, plotted against $(R_{\mathrm{end}}-R)/q$. Here $R_{\mathrm{end}}=(Q_{\mathrm{ring}}-q^2)/(2q)=199.5q$ is the limiting embedding radius at which the exact ring area vanishes. Both axes of this panel are logarithmic, and the horizontal axis decreases to the right, so moving right approaches this endpoint. The growing error shows that the blackfold area loses relative accuracy there. Its own zero-area endpoint is $R/q=200$. Dashed coloured segments have $3Q_{\mathrm R}/R>0.1$, while solid segments meet this illustrative thickness cutoff.}
 \label{fig:susy_saturn_comparison}
\end{figure}

Figure~\ref{fig:susy_saturn_comparison} illustrates the agreement and its limits. The upper panels plot the ring and total areas against $J_\psi$. Their close agreement partly follows from a cancellation when eliminating the radius, derived in Appendix~\ref{app:exact_susy_saturn}. The lower panels compare states at the same radius and reveal the differences directly. There are two distinct limitations. At small radii, the string becomes too thick relative to the bending scale and its perturbation of the central geometry grows. Near the largest allowed radius, the exact ring area tends to zero, so a small absolute correction produces a large relative error even for a thin string.

\subsection{Non-extremal black Saturns}\label{sec:nonext_saturn}
We next surround a non-extremal charged black hole by the non-extremal string of Section~\ref{sec:non-ext_string}. We first keep the two central angular momenta equal, while allowing either direction of longitudinal motion of the string. This gives a family of stationary black Saturns at leading order in the blackfold approximation. In Section~\ref{sec:thermal_saturn_equilibrium} we instead choose a central horizon with vanishing transverse angular velocity and solve for the thermodynamic equilibrium conditions between the black hole and the ring. Towards the end, we explore the various extremal limits beyond the SUSY limit discussed in the previous subsection.

\subsubsection{Background and induced geometry}
For simplicity we take the equal angular momentum ungauged limit of the CCLP solution~\cite{Chong:2005hr}, written in the charge and angular orientations of Section~\ref{sec:susy_saturn} which we record in Appendix~\ref{Appendix_CCLP}. With $\sigma$ as in \eqref{eq:saturn_background_functions}, its fields are
\begin{equation}\label{eq:thermal_saturn_background}
\begin{aligned}
 ds^2={}&-dt^2+\frac{2a_{\mathrm H}q_{\mathrm H}}{r^2}\,dt\,\sigma
 +\left(\frac{2m_{\mathrm H}}{r^2}-\frac{q_{\mathrm H}^2}{r^4}\right)
                 (dt-a_{\mathrm H}\sigma)^2
 -\frac{2a_{\mathrm H}^2q_{\mathrm H}}{r^2}\sigma^2
 +\frac{dr^2}{\Delta_{\mathrm H}}+r^2d\Omega_3^2~,\\
 A={}&\frac{\sqrt3q_{\mathrm H}}{r^2}(dt-a_{\mathrm H}\sigma)~,
 \qquad
 \Delta_{\mathrm H}=1-\frac{2m_{\mathrm H}}{r^2}
     +\frac{q_{\mathrm H}^2+2a_{\mathrm H}^2(m_{\mathrm H}-q_{\mathrm H})}{r^4}~.
\end{aligned}
\end{equation}
Here $d\Omega_3^2=d\theta^2+\sin^2\theta\,d\phi^2+\cos^2\theta\,d\psi^2$. We take $m_{\mathrm H}>q_{\mathrm H}\geq0$ and $m_{\mathrm H}+q_{\mathrm H}>2a_{\mathrm H}^2$, and restrict to the causal exterior. The outer horizon satisfies
\begin{equation}\label{eq:thermal_saturn_horizon}
 r_{\mathrm H}^2=m_{\mathrm H}
   +\sqrt{(m_{\mathrm H}-q_{\mathrm H})
                    (m_{\mathrm H}+q_{\mathrm H}-2a_{\mathrm H}^2)}~.
\end{equation}
Appendix~\ref{Appendix_CCLP} gives its charges and potentials in the same conventions. We obtain a dual potential regular near $\theta=0$ that reads
\begin{equation}\label{eq:thermal_saturn_dual}
 B_N=\frac{\sqrt3a_{\mathrm H}q_{\mathrm H}}{r^2}\,dt\wedge\sigma
          +\sqrt3q_{\mathrm H}\sin^2\theta\,d\phi\wedge d\psi~,
 \qquad B_S=B_N-\sqrt3q_{\mathrm H}\,d\phi\wedge d\psi~.
\end{equation}
These expressions obey $dB=-\star F+\kappa A\wedge F$. We note, in particular, that the dual field must be included even though the background carries no magnetic charge at infinity.

We take the embedding is $X^t=t$, $X^r=R>r_{\mathrm H}$, $X^\theta=0$, $X^\psi=\psi$, with $\phi$ constant and $\psi$ of period $2\pi$, and  write its induced metric as
\begin{equation}\label{eq:stationary_circle_metric}
 \gamma_{ab}d\sigma^ad\sigma^b=-N^2dt^2+\mathcal R^2(d\psi-\omega dt)^2~,
 \qquad \sqrt{-\gamma}=N\mathcal R~.
\end{equation}
These functions are evaluated at $R$ and are explicitly given by
\begin{equation}\label{eq:thermal_saturn_lapse}
\begin{aligned}
 \mathcal R^2&=R^2+a_{\mathrm H}^2
       \left[\frac{2(m_{\mathrm H}-q_{\mathrm H})}{R^2}
                         -\frac{q_{\mathrm H}^2}{R^4}\right]~,\\
 \omega&=\frac{a_{\mathrm H}}{\mathcal R^2}
       \left[\frac{2m_{\mathrm H}-q_{\mathrm H}}{R^2}
                         -\frac{q_{\mathrm H}^2}{R^4}\right]~,
 &N^2&=\frac{R^2\Delta_{\mathrm H}(R)}{\mathcal R^2}~.
\end{aligned}
\end{equation}
Thus $\mathcal R$ is the proper radius of the constant-$t$ circle. A useful orthonormal frame is $E_0=N^{-1}(\partial_t+\omega\partial_\psi)$ and $E_1=\mathcal R^{-1}\partial_\psi$. With these choices, a general thermal Killing vector and associated spatial vector read
\begin{equation}\label{eq:stationary_circle_vectors}
\begin{gathered}
 k=\partial_t+\Omega\partial_\psi\quad,\quad
 w=\frac{\mathcal R(\Omega-\omega)}N\quad,\quad
 \Gamma=(1-w^2)^{-1/2}\quad,\quad |k|=\frac N\Gamma~,\\
 u=\Gamma(E_0+wE_1)\quad,\quad
 v=-\Gamma(wE_0+E_1)\quad,\quad
 \Omega=\omega+\frac N{\mathcal R}w~.
\end{gathered}
\end{equation}
These are local frame relations and preserve the original compact identification.

\subsubsection{Force equation and its solutions}
The induced metric, gauge potentials and local thermodynamic variables are constant along the angular direction, and the pullback of $F$ vanishes. The intrinsic equations are therefore satisfied. Regularity at $\theta=0$ also makes the force in the transverse plane vanish, leaving the radial extrinsic equation. We evaluate its stress, electric and magnetic contributions separately, as in \eqref{eq:saturn_no_force}. For brevity, we write $p=\Pi=P-\mu_1\rho_1$, $z=-h=\kappa\mu_0^2\rho_1$ and $j=2\kappa\mu_0\rho_1$. Here $j$ denotes the local spatial electric current, not a reduced angular momentum. In the thermal frame $(u,-v)$ the energy density, momentum density and longitudinal stress are $\epsilon,z,p$, while the electric current has components $(\rho_0,j)$. Using \eqref{eq:stationary_circle_vectors}, the constitutive relations of Section~\ref{sec:equilibrium_action} give their components in the frame $E_0,E_1$,
\begin{equation}\label{eq:circle_boosted_monopoles}
\begin{aligned}
 E&=\Gamma^2(\epsilon+2zw+pw^2)~,
 &P_{\parallel}&=\Gamma^2(p+2zw+\epsilon w^2)~,\\
 Z_{\parallel}&=\Gamma^2[z(1+w^2)+(\epsilon+p)w]~,
 &j_0&=\Gamma(\rho_0+jw)~,
 \qquad j_1=\Gamma(j+\rho_0w)~.
\end{aligned}
\end{equation}
Here $E=T^{\hat0\hat0}$, $Z_{\parallel}=T^{\hat0\hat1}$, $P_{\parallel}=T^{\hat1\hat1}$ and $(j_0,j_1)=(J^{\hat0},J^{\hat1})$, with the subscripts on $j_0,j_1$ serving only as component labels. The string current is unchanged by the boost, with $\mathcal J^{\hat0\hat1}=-\rho_1$ for the orientation in \eqref{eq:stationary_circle_vectors}. The equation of state is that of Table~\ref{tab:string-a0-charges}, with independent local parameters $m,\beta,\delta$.

As in the supersymmetric calculation, we first keep the upper radial index in the coordinate basis. We denote the resulting curvature and field components by $K_{AB}=K_{\hat A\hat B}{}^r/\Delta_{\mathrm H}$, $F_A=F_{r\hat A}$ and $H=\mathcal F_{r\hat0\hat1}$, where $A,B=0,1$. Explicitly,
\begin{equation}\label{eq:circle_radial_data}
\begin{aligned}
 K_{00}&=\frac{N'}N~,
 &K_{01}&=\frac{\mathcal R\omega'}{2N}~,
 &K_{11}&=-\frac{\mathcal R'}{\mathcal R}~,\\
 F_0&=\frac{A_t'+\omega A_\psi'}N~,
 &F_1&=\frac{A_\psi'}{\mathcal R}~,
 &H&=\frac{B_{t\psi}'+\kappa(A_tA_\psi'-A_\psi A_t')}
                                  {N\mathcal R}~.
\end{aligned}
\end{equation}
Here the primes mean radial differentiation followed by evaluation at $r=R$. In particular, $K_{AB}=-\tfrac12 E_A^a E_B^b\partial_R\gamma_{ab}$. The magnetic field is $\mathcal F=dB-\kappa A\wedge F=-\star F$, so its radial component includes both terms displayed in $H$. For \eqref{eq:thermal_saturn_background}, $A_\psi=-a_{\mathrm H}A_t$ makes $A_tA_\psi'-A_\psi A_t'=0$, and the field components reduce to
\begin{equation}\label{eq:thermal_saturn_force_fields}
 F_0=-\frac{2\sqrt3q_{\mathrm H}(1-a_{\mathrm H}\omega)}{NR^3}
 \quad,\quad F_1=\frac{2\sqrt3a_{\mathrm H}q_{\mathrm H}}{\mathcal R R^3}
 \quad,\quad H=-\frac{2\sqrt3a_{\mathrm H}q_{\mathrm H}}{N\mathcal R R^3}~.
\end{equation}
The three contributions to the radial blackfold equation are consequently
\begin{equation}\label{eq:thermal_saturn_force_terms}
\begin{aligned}
 \frac{T^{\hat a\hat b}K_{\hat a\hat b}{}^r}{\Delta_{\mathrm H}}
   ={}&\Gamma^2\Big[(\epsilon+2zw+pw^2)\frac{N'}N
             -(p+2zw+\epsilon w^2)\frac{\mathcal R'}{\mathcal R}\\
      &\hspace{2.2em}+\big[z(1+w^2)+(\epsilon+p)w\big]
                         \frac{\mathcal R\omega'}N\Big]~,\\[2pt]
 \frac{F^r{}_{\hat a}J^{\hat a}}{\Delta_{\mathrm H}}
   ={}&-\frac{2\sqrt3q_{\mathrm H}\Gamma}{R^3}
       \left[\frac{1-a_{\mathrm H}\omega}{N}(\rho_0+jw)
             -\frac{a_{\mathrm H}}{\mathcal R}(j+\rho_0w)\right]~,\\[2pt]
 \frac{\mathcal F^r{}_{\hat a\hat b}\mathcal J^{\hat a\hat b}}
             {2\Delta_{\mathrm H}}
   ={}&-\rho_1H
     =\frac{2\sqrt3a_{\mathrm H}q_{\mathrm H}\rho_1}
                        {N\mathcal R R^3}~.
\end{aligned}
\end{equation}
The first expression contains the radial gravitational and bending terms together with the contribution from the radial variation of the rotational shift $\omega$. The second is the force on the electric current. The third is the force on the string current, whose sign follows from $\mathcal J^{\hat0\hat1}=-\rho_1$. It is nonzero when $a_{\mathrm H}q_{\mathrm H}\rho_1\ne0$, even though the central black hole has no magnetic charge at infinity. For a static central hole, $\omega=K_{01}=F_1=H=0$, leaving gravitational, bending and electric forces.

Equating the extrinsic stress to the sum of the two gauge field forces, as required by \eqref{eq:equilibrium_embedding_equation}, gives
\begin{equation}\label{eq:circle_force}
 EK_{00}+2Z_{\parallel}K_{01}+P_{\parallel}K_{11}
               -j_0F_0-j_1F_1+\rho_1H=0~.
\end{equation}
The sign of the last term reflects moving the magnetic force $-\rho_1H$ to the left-hand side. Both the mass and the charge of the central object enter this equation. For fixed $R,m,\beta,\delta$, it is an algebraic problem so we can define
\begin{equation}\label{eq:circle_force_polynomial}
\begin{aligned}
 A(w)={}&\epsilon K_{00}+2zK_{01}+pK_{11}
       +2w[z(K_{00}+K_{11})+(\epsilon+p)K_{01}]\\
       &+w^2[pK_{00}+2zK_{01}+\epsilon K_{11}]~,\\
 B(w)={}&(\rho_0+jw)F_0+(j+\rho_0w)F_1~,\\
 D(w)={}&A(w)+\rho_1H(1-w^2)~.
\end{aligned}
\end{equation}
Thus, all equilibrium velocities are the real roots of
\begin{equation}\label{eq:circle_quartic}
 D(w)^2-(1-w^2)B(w)^2=0\quad,\quad
 |w|<1\quad,\quad D(w)=B(w)\sqrt{1-w^2}~.
\end{equation}
The last condition removes unnecessary roots introduced by squaring. It also includes the case $B=0$, where the original equation is simply $D=0$. At large $R$, \eqref{eq:circle_force} reduces to $p+2zw+\epsilon w^2=0$, recovering the two branches of Section~\ref{sec:flat_ring_branches}. Each simple flat-space root continues to a sufficiently distant circle in the black hole background. Closer to the horizon one must solve \eqref{eq:circle_quartic} and require timelike velocities and validity of the configurations. We note that solving this equation analytically does not bring any insight so we use numerics. For example, in units $G_5=1$, taking $(m_{\mathrm H},q_{\mathrm H},a_{\mathrm H})=(2,1,0.3)$ and $(m,\beta,\delta)=(10^{-3},0.7,0.5)$, the horizon is at $r_{\mathrm H}\simeq1.918$, and the two equilibrium velocities are
\begin{equation}\label{eq:thermal_saturn_examples}
\begin{array}{c|rr|rr}
 R&w_-&w_+&\Omega_-&\Omega_+\\ \hline
 4&-0.890453&0.527302&-0.189809&0.117877\\
 8&-0.816348&0.421762&-0.098595&0.051270\\
 16&-0.799994&0.398747&-0.049594&0.024740
\end{array}
\end{equation}
The rest frame energy scale is $\epsilon/\mathcal C\simeq0.01252$, which is much smaller than these radii and the local background scales so these solutions are within the expected regime of validity.

\subsubsection{Charges, potentials and thermal equilibrium}
We now compute the charges associated with these configurations. The worldsheet slice has $n_a=(-N,0)$ and $dV_1=\mathcal R\,d\psi$. Evaluating the improved currents of Section~\ref{sec: Smarr} gives
\begin{equation}\label{eq:circle_global_charges}
\begin{aligned}
 Q_{(0),\mathrm r}&=2\pi\mathcal R j_0+4\pi\kappa\rho_1A_\psi~,
 \qquad Q_{(1),\mathrm r}=\rho_1~,\\
 J_{\psi,\mathrm r}&=2\pi\mathcal R^2Z_{\parallel}
       +2\pi\mathcal R A_\psi j_0+2\pi\kappa\rho_1A_\psi^2~,\\
 M_{\mathrm r}&=2\pi N\mathcal R E+2\pi\omega\mathcal R^2Z_{\parallel}
       -2\pi\mathcal R A_tj_0\\
       &\quad-2\pi\kappa\rho_1A_tA_\psi
       +2\pi\rho_1B_{t\psi}~,\\
 S_{\mathrm r}&=2\pi\mathcal R s\Gamma~,
 \qquad\mathcal A_{\mathrm r}=4G_5 S_{\mathrm r}~.
\end{aligned}
\end{equation}
The transverse angular momentum, fixed as in \eqref{eq:saturn_axial_lift}, gives $J_{\phi,\mathrm r}=2\pi\sqrt3q_{\mathrm H}\rho_1$. This is an interaction contribution arising from the non-trivial background gauge fields. Adding the reference black hole charges of Appendix~\ref{Appendix_CCLP} gives the total charges of the black Saturn. The global temperature and electric potential are
\begin{equation}\label{eq:thermal_saturn_potentials}
 T_0=\frac N\Gamma T\quad,\quad
 \Phi_0=\frac N\Gamma\mu_0-A_t-\Omega A_\psi~,
\end{equation}
while the magnetic potential is
\begin{equation}\label{eq:circle_magnetic_potential}
 \Phi_1=2\pi N\mathcal R(\mu_1-\kappa\mu_0^2w)
       +2\pi B_{t\psi}
       -2\pi\kappa(A_t+\Omega A_\psi+2\Phi_0)A_\psi~.
\end{equation}
The orientation agrees with the positive $\Phi_1$ in \eqref{eq:flat_ring_susy_minus}. These charges obey the fixed-background thermodynamic identities of Section~\ref{sec: Smarr}. 

A stationary black Saturn need not be in thermal equilibrium \cite{Elvang:2007rd}. Thermodynamic equilibrium is a stronger requirement as it imposes a common $T_0$, electric potential and both horizon angular velocities. Equations~\eqref{eq:circle_quartic} and \eqref{eq:thermal_saturn_potentials} supply the longitudinal and electric conditions. An independent transverse rotational sector is absent from the present string equation of state, so equality of the remaining horizon angular velocity cannot be inferred from monopole balance. We now remove this particular rotational obstruction by considering CCLP backgrounds with a single non-vanishing angular velocity.

\subsubsection{Black Saturns in thermodynamic equilibrium}\label{sec:thermal_saturn_equilibrium}
The local string of Section~\ref{sec:non-ext_string} has zero transverse horizon angular velocity. To retain this string while matching both horizon rotations, we choose a central black hole with $\Omega_{\phi,\mathrm H}=0$. This is a regular subfamily of the ungauged CCLP solution~\cite{Chong:2005hr}, whose geometry and thermodynamics are given in Appendix~\ref{app:central_zero_transverse}. In the conventions of that appendix,
\begin{equation}\label{eq:thermal_saturn_zero_transverse}
 b_{\mathrm H}(r_+^2+a_{\mathrm H}^2)=a_{\mathrm H}q_{\mathrm H}
 \quad\Longleftrightarrow\quad \Omega_{\phi,\mathrm H}=0~.
\end{equation}
The central horizon therefore rotates only along $\psi$, although $J_{\phi,\mathrm H}$ is generally nonzero. When $q_{\mathrm H}=0$, this reduces to the singly rotating Myers--Perry background. We use Boyer-Lindquist coordinates for this subsection and embed the ring at $r=R>r_+$, $\theta=0$, along $\psi$. The relation to the radial coordinate used for the equal spin background is explained in Appendix~\ref{app:central_zero_transverse}. Its induced metric still has the form \eqref{eq:stationary_circle_metric}, now with the functions and gauge potentials in \eqref{eq:central_zero_transverse_axis}. The vectors $u^a,v^a$ are consequently those of \eqref{eq:stationary_circle_vectors}. Equilibrium requires
\begin{equation}\label{eq:thermal_saturn_matching}
 \Omega=\Omega_{\psi,\mathrm H}\quad,\quad
 |k|T=T_{\mathrm H}\quad,\quad
 |k|\mu_0-A_t-\Omega_{\psi,\mathrm H}A_\psi=\Phi_{0,\mathrm H}~,
 \qquad \Omega_{\phi,\mathrm r}=\Omega_{\phi,\mathrm H}=0~.
\end{equation}
There is no additional equality of magnetic potentials, since the spherical central horizon does not carry the ring's dipole charge. We keep that dipole fixed when constructing the Saturn family below. These conditions reduce the equilibrium problem to a small set of scalar equations. The common angular velocity fixes
\begin{equation}\label{eq:thermal_saturn_matching_velocity}
 w(R)=\frac{\mathcal R(R)}{N(R)}
                [\Omega_{\psi,\mathrm H}-\omega(R)]
 \quad,\quad |k(R)|=N(R)\sqrt{1-w(R)^2}~,
\end{equation}
in which we note that radii must satisfy $|w|<1$. At fixed central parameters and $\beta$, the electric condition determines $\delta$ as a function of $R$. The local temperature scales as $T=\Theta(\beta,\delta)/m$, where \eqref{eq:nonext_rest_horizon} gives $\Theta=mT$. Hence matching the two temperatures fixes
\begin{equation}\label{eq:thermal_saturn_matching_scale}
 \mu_0(\beta,\delta)=
 \frac{\Phi_{0,\mathrm H}+A_t+\Omega_{\psi,\mathrm H}A_\psi}{|k|}
 \quad,\quad
 m=\frac{|k|\Theta(\beta,\delta)}{T_{\mathrm H}}~.
\end{equation}
All monopole stresses and currents are linear in $m$. After dividing out this common factor, the remaining equation for $R$ is precisely the unsquared force equation \eqref{eq:circle_force}, evaluated with the new background. Both flux forces are non-trivial. In particular, \eqref{eq:central_zero_transverse_fluxes} shows that $\Omega_{\phi,\mathrm H}=0$ does not remove the magnetic force when the central hole is charged.

The two stationary velocity branches of the force equation do not automatically give two solutions in thermodynamic equilibrium with the same central hole. At a fixed radius, the equality $\Omega=\Omega_{\psi,\mathrm H}$ already fixes $w$ through \eqref{eq:thermal_saturn_matching_velocity}. Only a force balanced state with this velocity can satisfy all the matching conditions. The family below has $a_{\mathrm H}>0$ and follows the co-rotating branch. Where a second stationary velocity root exists at the same radius and local state, it has $\Omega<0$ in this sample and cannot match the central horizon. For each central state and $\beta$ that we tried, we found only one equilibrium radius. 

Once the equilibrium conditions are solved, the ring charges and entropy are obtained from \eqref{eq:circle_global_charges}, using the induced fields of \eqref{eq:central_zero_transverse_axis}. Adding the central black hole contributions gives the total charges and entropy at leading order. The background gauge fields contribute $J_{\phi,\mathrm r}=2\pi\sqrt3q_{\mathrm H}\rho_1$ to the transverse angular momentum, by the same mechanism as in \eqref{eq:saturn_transverse_spin}. Thus the configuration can carry $J_\phi\ne0$ even though both transverse horizon angular velocities vanish.

\begin{figure}[!hb]
 \centering
 \includegraphics[width=\textwidth]{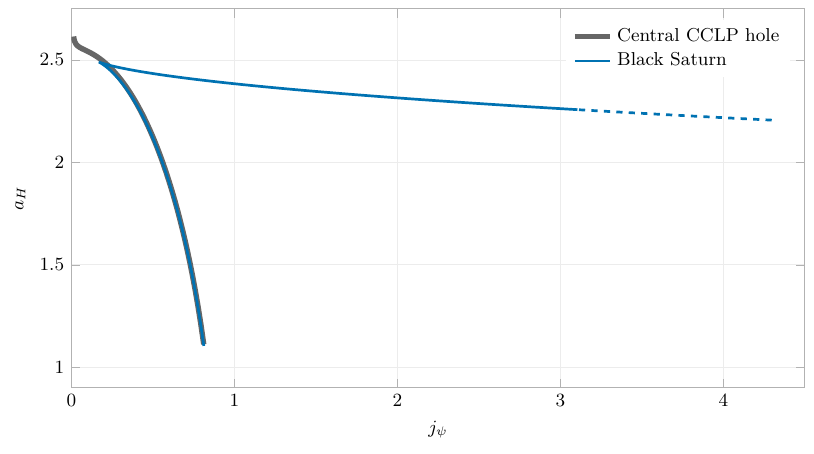}
 \caption{Reduced area versus longitudinal angular momentum for the central CCLP black hole alone (grey) and the full non-extremal black Saturn (blue). The reference black holes use the same central parameters as the Saturns, and each geometry is normalised by its own mass. The Saturns satisfy \eqref{eq:thermal_saturn_matching} and the fixed total charge ratios \eqref{eq:thermal_saturn_plot_ensemble}. The central black holes alone have a varying electric charge to mass ratio and no ring dipole. The range is $0.014\leq a_{\mathrm H}/\sqrt{\mathcal U_{\mathrm H}}\leq0.9$, with \eqref{eq:central_zero_transverse_parameters} defining $\mathcal U_{\mathrm H}$. Solid blue segments retain a thin, separated ring with $M_{\mathrm r}/M_{\mathrm H}\leq0.1$. The dashed segment relaxes this mass restriction. }
 \label{fig:thermal_saturn_equilibrium}
\end{figure}

Solving the force and matching equations gives configurations in which both horizons are non-extremal and have the same temperature, electric potential and angular velocities, to the order considered. Figure~\ref{fig:thermal_saturn_equilibrium} shows their total reduced area versus longitudinal angular momentum. We use \eqref{eq:flat_ring_reduced_definitions}, with the total mass, $\mathcal A=\mathcal A_{\mathrm H}+\mathcal A_{\mathrm r}$ and $J=J_{\psi,\mathrm H}+J_{\psi,\mathrm r}$, and fix
\begin{equation}\label{eq:thermal_saturn_plot_ensemble}
 \frac{\sqrt3|Q_{(0)}|}{M}=\frac12
 \quad,\quad
 \frac{G_5Q_{(1)}}{\sqrt{G_5M}}=10^{-4}~.
\end{equation}
We rescale each configuration to the same total mass, which also fixes its electric and dipole charges through \eqref{eq:thermal_saturn_plot_ensemble}. We generate the curve by varying the central rotation parameter and solving the equilibrium conditions while preserving these charges. For comparison, Figure~\ref{fig:thermal_saturn_equilibrium} also shows each central CCLP black hole on its own, before adding the ring. We use \eqref{eq:flat_ring_reduced_definitions} with $M_{\mathrm H},\mathcal A_{\mathrm H},J_{\psi,\mathrm H}$ for this reference curve, and the total quantities for the Saturn. The ring area is very small in this sample, while its angular momentum can be substantial because of its large radius. The overlay therefore shows how adding the ring extends the configuration to larger angular momentum. Although the total area increases, its reduced value need not increase because the total mass used in the denominator also changes.

The Saturn curve turns near $j_\psi\simeq0.168$. Its two segments describe different radii and central black hole parameters within this co-rotating family. At the same $j_\psi$ they also have different transverse angular momenta $J_\phi$, so their relative areas do not determine an entropy preference at fixed values of all conserved charges.

The approximation requires the string thickness to be small compared with the local bending, curvature and flux lengths, and with the proper radial separation $d_{\rm sep}$ from the central horizon,
defined in \eqref{eq:central_zero_transverse_distance} of Appendix~\ref{Appendix_CCLP},
so that $r_\epsilon/d_{\rm sep}\ll1$~\cite{Armas:2015minimal,Armas:2017myl}. The mass restriction in the caption additionally keeps the ring's contribution small relative to the central hole. These conditions select the displayed regime. In particular, the local strings in this sample are close to extremality, with $Ts/\epsilon\sim10^{-7}$-$10^{-6}$. Corrections that are small relative to their total energy can therefore matter for their thermal excitation.
 
Finally, we would like to comment on a putative entropy comparison and thermodynamic preference. For such comparison at fixed conserved charges instead requires a spherical CCLP black hole~\cite{Chong:2005hr} with the total charges of the black Saturn. This differs from the central background plotted in Figure~\ref{fig:thermal_saturn_equilibrium}. For each black Saturn, we must choose the central black hole to satisfy $(M,Q_{(0)},J_\psi,J_\phi)_{\mathrm{CCLP}}=(M,Q_{(0)},J_\psi,J_\phi)_{\mathrm{Sat}}$, as well as require a regular outer horizon. Our numerical analysis suggests that the CCLP black hole is thermodynamically preferred for lower values of $j_\psi$ while the black Saturns are preferred for higher values of $j_\psi$, at least within the regime of validity of the black Saturn configurations. This preliminary conclusion is sensitive to higher-order corrections to the blackfold approach, which we leave for future work.

\subsubsection{Extremal rings around a non-extremal black hole}\label{sec:thermal_saturn_ring_limits}
We return to the equal spin background \eqref{eq:thermal_saturn_background} and study some of its extremal limits. There are two independent extremal limits in this construction: that of the local string and that of the central black hole. We first keep the central parameters non-extremal and take the string limits of Section~\ref{sec:susy_lim}, at a fixed circle outside the horizon. This gives an extremal ring around a black hole with $T_{\mathrm H}>0$. Such a Saturn configuration is stationary, but its two components are neither both extremal nor in thermal equilibrium. \\

\noindent\textbf{Finite thermal velocity.}
Addressing first the limit \eqref{eq:string_susy_scaling} gives
\begin{equation}\label{eq:thermal_saturn_finite_limit}
 (\epsilon,p,z,\rho_0,j,\rho_1)
 =\mathcal C\left(3Q_{\mathrm R},0,\frac{3q}{2},
       -\sqrt3Q_{\mathrm R},-\sqrt3q,\frac{\sqrt3q}{2}\right)~,
 \qquad Q_{\mathrm R}>q>0~.
\end{equation}
The radial equation, with both flux forces retained, becomes
\begin{equation}\label{eq:thermal_saturn_finite_force}
\begin{aligned}
0={}&\Gamma^2\Big[Q_{\mathrm R}(K_{00}+2wK_{01}+w^2K_{11})
       +q\big(w(K_{00}+K_{11})+(1+w^2)K_{01}\big)\Big]\\
 &+\frac{\Gamma}{\sqrt3}
       \Big[(Q_{\mathrm R}+qw)F_0+(q+Q_{\mathrm R}w)F_1\Big]
       +\frac{q}{2\sqrt3}H~.
\end{aligned}
\end{equation}
Its admissible roots are obtained from \eqref{eq:circle_quartic}. For large $R$ they approach $w_+=0$ and $w_-=-q/Q_{\mathrm R}$. At a simple root the balance equation depends smoothly on the local non-extremality parameter, so each solution is reached from a nearby non-extremal ring. Near the central black hole, the number of timelike roots and their validity must be checked numerically. In this limit the ring temperature vanishes, while its entropy is finite, that is
\begin{equation}\label{eq:thermal_saturn_finite_entropy}
 T_{0,\mathrm r}=0\quad,\quad
 S_{\mathrm r}=2\pi^2\sqrt3\,\mathcal C\mathcal R q
                      \sqrt{Q_{\mathrm R}^2-q^2}\,\Gamma~.
\end{equation}
Its remaining charges follow from \eqref{eq:circle_global_charges}, and its potentials from \eqref{eq:thermal_saturn_potentials}--\eqref{eq:circle_magnetic_potential} with $\mu_0=-\sqrt3$ and $\mu_1=0$. In particular, neither $\Phi_0=-\sqrt3$ nor $\Phi_1=0$ follows from this local equation of state in a general background. For $(m_{\mathrm H},q_{\mathrm H},a_{\mathrm H})=(2,1,0.3)$, $(Q_{\mathrm R},q)=(0.005,0.002)$ and $R=8$, the two roots are $w_-=-0.468183$ and $w_+=0.070638$. \\

\noindent\textbf{Null waves.}
The other two extremal ring limits have
\begin{equation}\label{eq:thermal_saturn_null_generator}
 \ell=E_0+\varsigma E_1\quad,\quad k=N\ell
 \quad,\quad \Omega=\omega+\varsigma\frac N{\mathcal R}~.
\end{equation}
Here $\varsigma=+1$ describes the charged null branch, while $\varsigma=-1$ with $Q=0$ describes the magnetic endpoint of the local supersymmetric string. With $\tau=3\mathcal Cq/2$ and $\rho_1=\sqrt3\mathcal Cq/2$, the finite limiting currents are
\begin{equation}\label{eq:thermal_saturn_null_currents}
 T^{ab}=-\tau\gamma^{ab}+\mathcal E\ell^a\ell^b
 \quad,\quad J^a=-\sqrt3\mathcal C Q\ell^a
 \quad,\quad \mathcal J^{ab}=\rho_1\epsilon^{ab}~.
\end{equation}
The force equation is linear in the null energy and gives
\begin{equation}\label{eq:circle_null_balance}
\begin{aligned}
 \mathcal E&=
 \frac{-\tau(K_{00}-K_{11})
       -\sqrt3\mathcal C Q(F_0+\varsigma F_1)-\rho_1H}
             {K_{00}+2\varsigma K_{01}+K_{11}}~,\\
 \lambda&=\frac{2\mathcal E}{\mathcal C}-\frac{3Q^2}{q}>0~.
\end{aligned}
\end{equation}
In the case of the magnetic endpoint,  $Q=0$, we denote the resulting parameter by $\lambda_{\mathrm B}$. To obtain the thermodynamics, substitute
\begin{equation}\label{eq:circle_null_charge_substitution}
\begin{gathered}
 E=\tau+\mathcal E\quad,\quad Z_{\parallel}=\varsigma\mathcal E
 \quad,\quad P_{\parallel}=-\tau+\mathcal E~,\\
 j_0=-\sqrt3\mathcal C Q\quad,\quad
 j_1=-\varsigma\sqrt3\mathcal C Q\quad,\quad
 s\Gamma\longrightarrow\pi\mathcal C\sqrt{q^3\lambda}
\end{gathered}
\end{equation}
in \eqref{eq:circle_global_charges}, replacing $\lambda$ by $\lambda_{\mathrm B}$ on the magnetic branch. Thus $S_{\mathrm r}=2\pi^2\mathcal C\mathcal R\sqrt{q^3\lambda}$ and $T_{0,\mathrm r}=0$. The limits of the potentials are
\begin{equation}\label{eq:thermal_saturn_null_potentials}
\begin{aligned}
 \Phi_0&=-A_t-\Omega A_\psi~,\\
 \Phi_1&=2\pi\sqrt3N\mathcal R+2\pi B_{t\psi}
                   +2\pi\kappa(A_t+\Omega A_\psi)A_\psi~.
\end{aligned}
\end{equation}
Although $J^a=0$ on the magnetic branch, its global electric Page charge need not vanish. The background contribution in \eqref{eq:circle_global_charges} gives
\begin{equation}\label{eq:thermal_saturn_magnetic_page_charge}
 Q_{(0),\mathrm r}\big|_{Q=0}
       =-\frac{2\pi\sqrt3\,\mathcal Cq\,a_{\mathrm H}q_{\mathrm H}}{R^2}~.
\end{equation}
For the same non-extremal background and radius as above, $q=0.002$ and $Q=0.001$ give $\lambda=0.00509125$ and $\Omega=0.121263$ on the charged branch. The magnetic branch instead has $\lambda_{\mathrm B}=0.00689509$ and $\Omega=-0.120825$. At large $R$, these results approach $\lambda=3(q^2-Q^2)/q$ and $\lambda_{\mathrm B}=3q$. Consequently both null branches exist sufficiently far from the central hole for $|Q|<q$ on the charged branch.

All four branches therefore approach the corresponding extremal rings in flat space of Section~\ref{sec:flat_ring_extremal}. At fixed central black hole parameters, taking $R$ large gives $N\to1$, $\mathcal R/R\to1$ and $\omega\to0$, while the background flux forces become sub-leading in the radial equation. The two finite-velocity roots reduce to \eqref{eq:flat_ring_susy_velocities}, and the two null branches recover the balance conditions in \eqref{eq:string_validity_null_ring} and \eqref{eq:flat_ring_magnetic_null_charges}. The leading large radius ring mass, longitudinal angular momentum, electric charge and entropy likewise recover the flat space expressions. 

\subsubsection{Extremal black holes and extremal rings}\label{sec:thermal_saturn_both_extremal}
We now explore the possibility of having both horizons in an extremal regime. To make both horizons extremal, the preceding limits must be accompanied by a double root of $\Delta_{\mathrm H}$. Within the charge orientation used here, \eqref{eq:thermal_saturn_horizon} gives two regular possibilities,
\begin{equation}\label{eq:thermal_saturn_central_extremality}
\begin{aligned}
 \text{BMPV:}\qquad &m_{\mathrm H}=q_{\mathrm H}=\mu>0~,
                       &a_{\mathrm H}^2&<\mu~,\\
 \text{Rotational:}\qquad &m_{\mathrm H}>q_{\mathrm H}\geq0~,
                       &2a_{\mathrm H}^2&=m_{\mathrm H}+q_{\mathrm H}~.
\end{aligned}
\end{equation}
In both cases $r_{\mathrm H}^2=m_{\mathrm H}$ and $\Delta_{\mathrm H}=(1-m_{\mathrm H}/r^2)^2$. Their common boundary $m_{\mathrm H}=q_{\mathrm H}=a_{\mathrm H}^2$ has zero central horizon area and is excluded. At fixed $R>r_{\mathrm H}$, the exterior coefficients of the force equation have smooth limits on either regular branch. Extremality of the CCLP black hole in the center therefore does not by itself eliminate either pair of ring limits. \\

\noindent\textbf{A BMPV centre.}
We first study the case in which the extremal limit of the CCLP is the BMVP black hole. We set $\mathfrak j=a_{\mathrm H}\mu$, $f_R=1-\mu/R^2$ and $b_R=\mathfrak j/R^2$, as in Section~\ref{sec:susy_saturn}. The aligned finite-velocity branch is the supersymmetric Saturn already constructed there,
\begin{equation}\label{eq:thermal_saturn_BMPV_aligned}
 w_{\mathrm{aligned}}=-\frac{b_R}{R}
       =-\frac{\mathfrak j}{R^3}\quad,\quad \Omega=0~.
\end{equation}
Here $w$ is measured relative to the rotating frame $E_0,E_1$, so it can be nonzero even though the horizon angular velocity $\Omega$ vanishes. The force equation is satisfied at every radius allowed by regularity and blackfold validity, including for a rotating BMPV centre with $K_{01}\ne0$. The second finite-velocity branch instead requires solving \eqref{eq:thermal_saturn_finite_force} for its angular velocity at each radius. It approaches $w=-q/Q_{\mathrm R}$ far from the central hole.

To distinguish extremality from supersymmetry, consider the BPS gap $\mathcal G=M_{\mathrm{tot}}+\sqrt3Q_{(0),\mathrm{tot}}$ for the inherited negative total electric charge. The BMPV centre contributes zero to this gap. For the finite-velocity ring, let $u^{\hat0}=\cosh\chi$ and $u^{\hat1}=\sinh\chi$ in the coframe $e^{\hat0}=f_Rdt+b_Rd\psi$, $e^{\hat1}=R\,d\psi$ used in Section~\ref{sec:susy_saturn}. Thus $\chi=0$ corresponds to the aligned state with $\Omega=0$. Equation~\eqref{eq:circle_global_charges} gives
\begin{equation}\label{eq:thermal_saturn_BMPV_finite_gap}
\begin{aligned}
 \mathcal G={}&6\pi\mathcal C f_R(\cosh\chi-1)\Big[
 R(Q_{\mathrm R}\cosh\chi+q\sinh\chi)\\
 &\hspace{43mm}-b_R(q\cosh\chi+Q_{\mathrm R}\sinh\chi)\Big]~.
\end{aligned}
\end{equation}
The square bracket is positive for $Q_{\mathrm R}>q$ and $R>|b_R|$. Hence only the aligned branch saturates the BPS bound. The second branch has $\mathcal G>0$ despite both horizons having zero temperature.

The charged and magnetic null branches follow from \eqref{eq:circle_null_balance} on the same background. Their BPS gaps are
\begin{equation}\label{eq:thermal_saturn_BMPV_null_gaps}
\begin{aligned}
 \mathcal G_+&=\pi\mathcal C f_R\left[
 (R+b_R)\lambda+\frac3q
       \left(\sqrt{R+b_R}\,Q-\sqrt{R-b_R}\,q\right)^2\right]~,\\
 \mathcal G_-&=\pi\mathcal C f_R(R-b_R)(3q+\lambda_{\mathrm B})~.
\end{aligned}
\end{equation}
Both are strictly positive for regular rings, since $f_R>0$, $R>|b_R|$ and $\lambda,\lambda_{\mathrm B}>0$. Neither null branch gives a supersymmetric Saturn.

Unlike the aligned BPS branch, the other branches need not exist at every exterior radius. For example, for a static BMPV centre, $\mathfrak j=0$, balance for the magnetic branch requires
\begin{equation}\label{eq:thermal_saturn_static_null_radius}
 \lambda_{\mathrm B}=3q\,\frac{R^2+\mu}{R^2-3\mu}>0
 \quad\Longrightarrow\quad R>\sqrt{3\mu}~.
\end{equation}
The ring must therefore lie beyond $R=\sqrt{3\mu}$ and not just outside the horizon $R=\sqrt\mu$.\\

\noindent\textbf{A rotating non-supersymmetric extremal centre.}
On the second central branch in \eqref{eq:thermal_saturn_central_extremality}, the ring again obeys \eqref{eq:thermal_saturn_finite_force} or \eqref{eq:circle_null_balance}. Both finite-velocity branches and both null branches exist sufficiently far from the centre hole, where they approach the flat space rings. The central hole thermodynamics follows from Appendix~\ref{Appendix_CCLP}, and the total BPS gap is
\begin{equation}\label{eq:thermal_saturn_rotational_gap}
 \mathcal G=\frac{3\pi}{4G_5}(m_{\mathrm H}-q_{\mathrm H})
                         +M_{\mathrm r}+\sqrt3Q_{(0),\mathrm r}~.
\end{equation}
At fixed $m_{\mathrm H}-q_{\mathrm H}>0$, the positive central contribution cannot be cancelled by a parametrically small ring perturbation. These Saturns are therefore extremal but not supersymmetric in the regime considered here. A simultaneous limit in which the central BPS gap becomes comparable to the ring correction would require including backreaction of the ring, which we do not consider here.

For either type of centre, the total charges and entropy are the sum of the central and ring contributions given above. Having $T_{\mathrm H}=T_{0,\mathrm r}=0$ does not by itself establish thermodynamic equilibrium, since that also requires common angular velocities and electric potentials. Except for the aligned BPS family, whose exact completion is discussed in Appendix~\ref{app:exact_susy_saturn}, the constructions here give evidence for a much richer phase space of extremal Saturn solutions.

\subsection{Phase diagrams}\label{sec:phase_diagrams}
We now collect the asymptotically flat configurations in area versus angular momentum diagrams. For disconnected horizons the area is the sum of the component areas, and the mass and angular momentum include the interaction terms in the improved charges. We use the normalisations \eqref{eq:flat_ring_reduced_definitions},
\begin{equation}\label{eq:phase_overview_variables}
 a_H=\frac3{16}\sqrt{\frac3\pi}
          \frac{\mathcal A_{\rm tot}}{(G_5M_{\rm tot})^{3/2}}
 \quad,\quad
 j_\parallel=\sqrt{\frac{27\pi}{32}}
          \frac{G_5J_{\parallel,\rm tot}}{(G_5M_{\rm tot})^{3/2}}
 \quad,\quad
 e=\frac{\sqrt3|Q_{(0),\rm tot}|}{M_{\rm tot}}~.
\end{equation}
Throughout this section we use the ADM angular momentum $J_\psi$, retaining its sign, so $j_\parallel=j_\psi$. The same rotation generator is used for every family, including the bi-rings. These solution families have more parameters than the two plotted quantities. We therefore specify a one-parameter section through each family. All curves can be rescaled to the same mass, but they do not all have the same electric charge, second angular momentum or dipoles.

\subsubsection{Extremal black holes}\label{sec:phase_extremal}
\begin{figure}[!htbp]
 \centering
 \includegraphics[width=\textwidth]{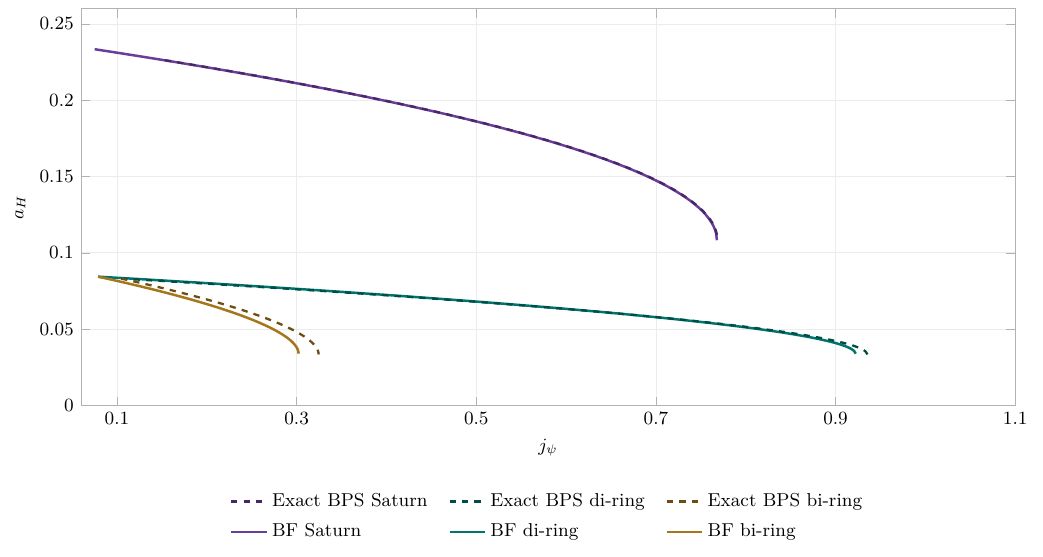}
 \caption{BPS Saturns, di-rings and bi-rings, with dashed exact curves and solid blackfold curves. The exact families have $G_5=1$, $\mathcal Q=1$ and $q_T=1/5$, using the notation of Appendix~\ref{app:susy_multirings}. The Saturn has $(\mu,\mathfrak j,q,\Lambda)=(1/4,1/16,1/5,71/40)$ and varying radius, with the map \eqref{eq:saturn_radius_dictionary}. Its ring area vanishes at $R=1.775$ in the exact solution and $R=1.875$ in the blackfold description. The di-ring and bi-ring have $\mu=\mathfrak j=0$, $q_1=q_2=1/10$, $\Lambda=12/5$ and $R_1=3R_2/5$, with $0<R_2<\Lambda$. Their leading blackfold mass corresponds to $\mathcal Q_{\mathrm{BF}}=0.99$.}
 \label{fig:phase_bps_comparison}
\end{figure}

The relevant picture for phase space of extremal solution is Figure~\ref{fig:extremal_ph_sp}, which appears in the Introduction. It compares the four isolated extremal blackfold rings of Section~\ref{sec:flat_ring_extremal} with exact rings, the BMPV black hole, the BPS black Saturns of Section~\ref{sec:susy_saturn} and Appendix~\ref{app:exact_susy_saturn}, and the BPS di-rings and bi-rings of Appendix~\ref{app:susy_multirings}. The finite-velocity branches are denoted by $F_\pm$, the charged null branch by $N_+$ and the opposite magnetic null branch by $N_-$. Solid curves show blackfold results throughout, while dashed, dotted and dash-dotted curves show exact solutions. The isolated blackfold curves have $Q_{\mathrm R}/q=5/2$ on $F_\pm$ and $Q/q=1/2$ on $N_+$. Their electric charge to mass ratios are\footnote{For reference, in Figure~\ref{fig:extremal_ph_sp}, the isolated
blackfold rings use $q=1$, with $75\leq R\leq1200$ for
$F_\pm$ and $30\leq R\leq1200$ for $N_\pm$.
The local charge ratios are $Q_{\mathrm R}/q=5/2$
on $F_\pm$, $Q/q=1/2$ on $N_+$, and $Q=0$ on $N_-$.
These samples satisfy $r_\epsilon(n)/R\leq0.1$.
The exact BPS ring is paired with the blackfold $F_+$
curve using $q=1$ and
$Q_{\mathrm{ring}}=2RQ_{\mathrm R}=5R$.
Their mass, electric charge and dipole charge agree at
each radius, although the mass and electric charge vary
along the curves.
The exact ST curves use $b=0.4$ and
$0.07\leq\eta\leq0.97$ in
\eqref{eq:ST_multiring_rotational_scalings}.
The BPS composites instead have fixed charge parameters,
specified in the caption of
Figure~\ref{fig:phase_bps_comparison}, where their curves
are enlarged. The null-ring and ST curves extend beyond
the displayed spin range.}
\begin{equation}\label{eq:phase_extremal_electric_ratios}
 e_{F_+}=1\quad,\quad e_{F_-}=\frac{\sqrt{21}}5
 \quad,\quad e_{N_+}=\frac12\quad,\quad e_{N_-}=0~.
\end{equation}
 The exact extremal Suzuki-Tomizawa rings retain transverse horizon rotation and likewise do not complete the non-rotating null or negative finite-velocity strings~\cite{Suzuki:2024generalring}. The BMPV black hole~\cite{Breckenridge:1996is} has equal angular momenta and obeys
\begin{equation}\label{eq:phase_exact_hole_curves}
 a_H=\sqrt{1-8j_\psi^2}
 \quad,\quad e=1
 \quad,\quad |j_\psi|<\frac1{2\sqrt2}~.
\end{equation}
\begin{figure}[!htbp]
 \centering
 \includegraphics[width=\textwidth]{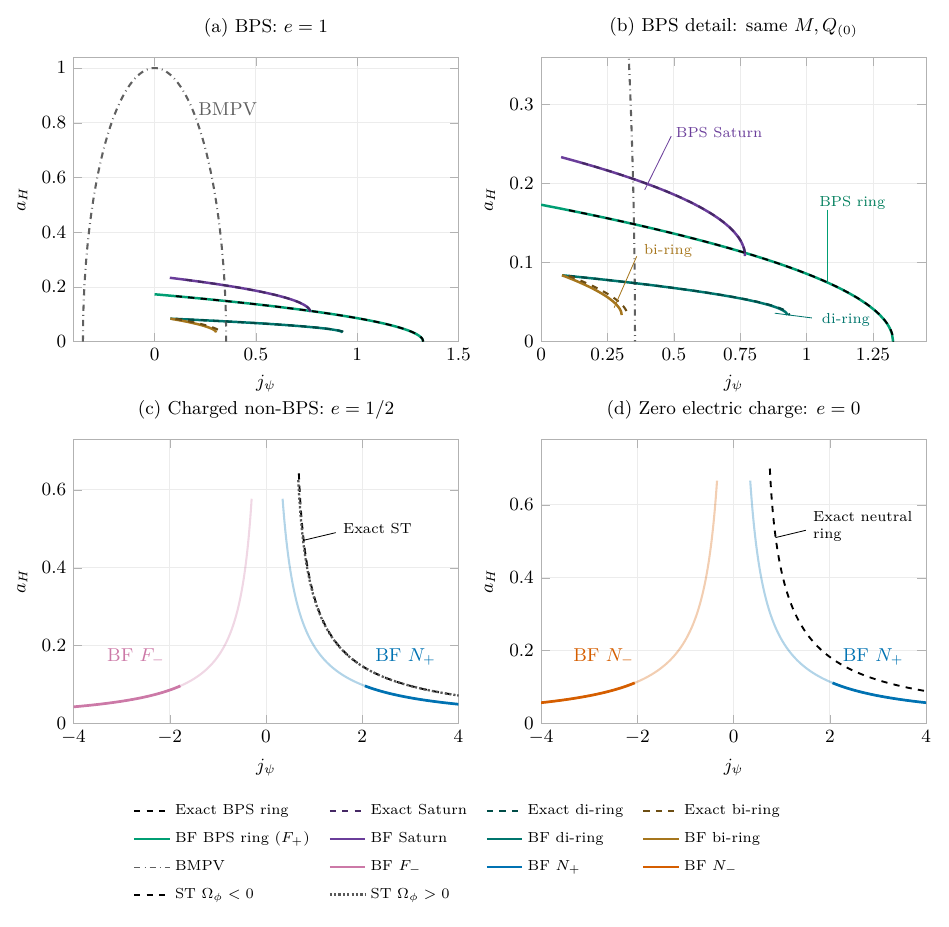}
 \caption{Extremal configurations at common mass $M=3\pi/4$
with $G_5=1$. The electric charge is fixed within each panel:
$e=1$ in (a,b), $e=1/2$ in (c), and $e=0$ in (d),
where $e=\sqrt3|Q_{(0)}|/M$. Panel (b) enlarges the BPS
ring and composite curves in (a). Solid curves show blackfold
results, while dashed, dotted and dash-dotted curves show
exact solutions. Parameter choices and validity restrictions
are described below.}
 \label{fig:phase_extremal_exact}
\end{figure}
Figure~\ref{fig:phase_extremal_exact} compares extremal
configurations at fixed mass and electric charge within
each panel.\footnote{In Figure~\ref{fig:extremal_ph_sp},
the isolated blackfold ring curves instead keep the local
string parameters fixed while varying the radius, so their
mass and electric charge generally vary along the curves.
The exact BPS ring is matched to its blackfold counterpart
at each radius, while the BPS composites have fixed charge
parameters. The different shapes of the BPS ring curves
in the two figures therefore reflect different choices
of parameters held fixed, rather than different solution
branches.}
All configurations have $G_5=1$ and $M=3\pi/4$, while
$e=\sqrt3|Q_{(0)}|/M$ equals $1$ in panels (a,b),
$1/2$ in (c), and $0$ in (d).
The exact BPS ring and its blackfold approximation use
$q=1/5$ and $Q_{\mathrm{ring}}=1$, fixing both the electric
and dipole charges. Their curves terminate at their
respective zero-area boundaries as the radius increases.
The Saturn has
$(\mu,\mathfrak j,q,Q_{\mathrm{ring}})
=(1/4,1/16,1/5,3/4)$.
The di-rings and bi-rings use $q_1=q_2=1/10$ and
$R_1=3R_2/5$. Their common parameter
$\Lambda=(Q_i-q_i^2)/(2q_i)$, defined in
Appendix~\ref{app:susy_multirings}, is $12/5$ for the exact
solutions and $97/40$ for the blackfold constructions,
ensuring equal total mass and electric charge.
Panel (c) uses $Q_{\mathrm R}/q=2/\sqrt3$ on $F_-$,
$Q/q=1/2$ on $N_+$, and $b=2-\sqrt3$ for the ST curves.
Panel (d) instead has $Q=0$ on both null branches and
$b=0$ for the ST solution.
Dark blackfold segments in (c,d) satisfy the illustrative
thinness cutoff $r_\epsilon/R\leq0.1$, while pale segments
are extrapolations. The BPS blackfold curves are also
extrapolations for the selected parameters.
Although mass and electric charge are matched within each
panel, the second angular momentum and dipole charges vary
across families. The figure therefore illustrates their
relative locations in the area--spin diagram without
establishing thermodynamic dominance.

Figure~\ref{fig:phase_bps_comparison} isolates the exact BPS Saturn, di-ring and bi-ring families and their blackfold counterparts, using the charge identifications of Appendix~\ref{app:exact_susy_comparison}. The Saturn descriptions have the same mass, electric charge and dipole. For the two-ring systems, the blackfold mass and angular momenta retain only the leading contribution of the added ring. Each reduced curve uses the mass predicted by its description. The close agreement is especially pronounced for the Saturn and di-ring, while the bi-ring shows a more visible difference in $j_\psi$. Similar phenomenology appears in extremal solutions of vacuum gravity with disconnected horizons \cite{Armas:2017myl}.

This agreement does not establish accuracy throughout the plotted range. In particular, the chosen Saturn has $r_\epsilon/R\gtrsim0.36$, so its string thickness is not much smaller than its radius. Its near-coincidence with the exact curve is partly due to the cancellation at fixed $J_\psi$ in \eqref{eq:saturn_area_fixed_spin}. The validity conditions of Section~\ref{sec:susy_saturn} and Appendix~\ref{app:susy_multirings} must still hold.

\subsubsection{Non-extremal black holes}\label{sec:phase_nonextremal}
\begin{figure}[!htbp]
 \centering
 \includegraphics[width=\textwidth]{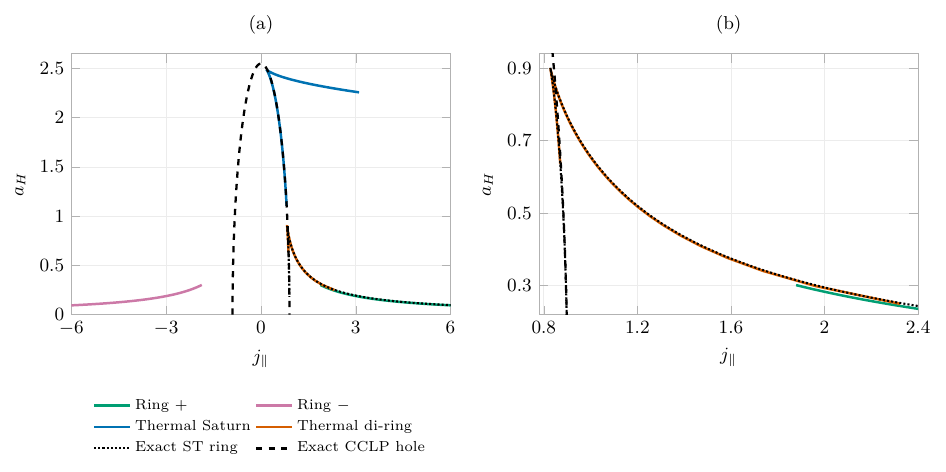}
 \caption{Non-extremal reduced area versus reduced angular curves at $e=1/2$. Solid curves denote blackfold constructions, the dotted curve the exact ST ring and the dashed curve the exact ungauged CCLP hole. Panel (b) enlarges the positive-spin ring region of panel (a). Isolated blackfold rings satisfy $r_\epsilon(n)/R\leq0.1$. The plots of the composite solutions obey the validity conditions of their constructions.}
 \label{fig:phase_nonextremal}
\end{figure}

Figure~\ref{fig:phase_nonextremal} compares the two non-extremal blackfold ring branches with the thermal Saturns of Section~\ref{sec:thermal_saturn_equilibrium} and the thermal di-rings of Appendix~\ref{app:ST_multiring_thermal}. In each composite, the two horizons have equal temperatures, electric potentials and angular velocities at leading order. For these blackfold families we fix
\begin{equation}\label{eq:phase_nonextremal_ensemble}
 e=\frac12\quad,\quad
 \frac{G_5Q_{(1),\mathrm r}}{\sqrt{G_5M_{\rm tot}}}=10^{-4}~.
\end{equation}
Here $Q_{(1),\mathrm r}$ is the dipole of the single ring or of the added ring in a composite. No thermal bi-ring is available within the present ansatz, since its angular-velocity matching condition has no regular non-extremal solution in the chosen ST family, as discussed in Appendix~\ref{app:ST_multiring_thermal}. The exact ST ring and ungauged CCLP black hole~\cite{Chong:2005hr} are shown at the same $e=1/2$, with $\Omega_\phi=0$. This condition does not require $J_\phi=0$. The CCLP curve follows from Appendix~\ref{app:central_zero_transverse} and becomes static when $a_H\simeq2.55$.
The large separation between the black Saturn and ring curves reflects how the area is distributed between their horizons. In the displayed black Saturns, almost all of the area comes from the central black hole. The di-rings instead receive almost all of their area from the central ring, and the small contribution of the outer ring leaves their curve close to the exact ST curve. Thus the different area scales are consistent with these particular constructions. At a common total mass, all curves have the same electric charge, but their transverse angular momenta and independent dipoles are not matched. In particular, the exact ST dipole and the inner di-ring dipole vary, while the CCLP hole has no ring dipole. The small thermal excitation of the added rings makes relative corrections to their temperatures and potentials potentially important, even when the geometric thinness conditions hold, as discussed in Section~\ref{sec:thermal_saturn_equilibrium} and Appendix~\ref{app:ST_multiring_thermal}.

\section{Extremal black rings in \texorpdfstring{AdS$_5$}{AdS5}}\label{sec:ads_rings}
As a further application of the framework discussed in this paper, we consider minimal U(1) gauged supergravity in five dimensions, which allows for configurations with AdS$_5$ asymptotics. The bosonic truncation of this theory reads
\begin{equation}\label{eq:Lg}
    S=\frac{1}{16\pi G_5}\int\left(R\star 1-\frac{1}{2}\star F\wedge F+\frac{\kappa}{3}F\wedge F\wedge A + \frac{12}{L^2} \star 1\right)\,,
\end{equation}
and differs from \eqref{eq:L}  by the addition of a
 cosmological constant term with $\Lambda = -6/L^2$. This allows us to consider the bending of black strings in global AdS$_5$, hence generating genuinely novel AdS black ring configurations. Here, we restrict our analysis to ring configurations, leaving a more comprehensive study of composite solutions, such as AdS$_5$ black Saturns and more intricate black objects, to future work. 
 
 At leading order, the same ungauged black string that we used throughout the paper supplies the local equation of state when its gravitational and charge scales are much smaller than $L$. We work under the assumption that the BPS bound of the resulting ring will be the same as that for the Gutowski-Reall black hole, namely
 \begin{equation} \label{BPSbound}
 M\geq |J_\psi|/L+\sqrt3|Q_{(0)}|
 \end{equation}
 where the conserved quantities are computed appropriately: the mass is obtained via holographic renormalization or, equivalently (at least in this framework) by subtracting the AdS vacuum energy subtracted at leading order where $J_\phi=0$~\cite{Cvetic:2005AdSBounds,Chong:2005hr}.
  
 This condition applies at any temperature, including to extremal configurations that do not preserve supersymmetry\footnote{Even though, as it is well known, the extremality bound and the BPS bound are different in AdS spacetimes.}. We therefore distinguish solutions of the leading balance equation from admissible candidates for complete AdS black rings. Thin rings in AdS were studied in~\cite{Caldarelli:2008pz}, including an obstruction to obtaining a supersymmetric ring by bending a supersymmetric string. Here we extend that analysis to three candidate branches of extremal rings that are generically non-supersymmetric. 

 Before concluding, we would like to point out that the bound \eqref{BPSbound} follows from the \(\mathfrak{su}(2,2|1)\) superalgebra, and was shown to be saturated by the Gutowski-Reall black hole solutions \cite{Gutowski:2004ez}. Nevertheless, in 5d gauged supergravity there exist other AdS$_5$ black string configurations \cite{Klemm:2000nj,Naka:2002jz,Benini:2012cz,Karndumri:2013iqa,Benini:2015bwz,Klemm:2016kxw}, whose near horizon geometry was found analytically. These configurations preserve supersymmetry via the topological twist, in particular, contrarily to the Gutowski-Reall solutions, they are characterized by a nonvanishing magnetic flux at the boundary \cite{Chamseddine:1999xk} and have no flat space limit. Their BPS bound must be derived using the superalgebra and boundary conditions appropriate to the string configuration, and in principle could be different than the one in \eqref{BPSbound}. This is what happens for instance in four-dimensional gauged supergravity, where two different BPS bounds exist, realizing two different superalgebras \cite{Hristov:2011ye,Hristov:2011qr}\footnote{Indeed there exist two branches of black hole solutions distinguished by the asymptotic value of the R-symmetry flux, which can be vanishing or non-vanishing. The microstates of these black holes are counted by the superconformal index and the topologically twisted index respectively (see \cite{Zaffaroni:2019dhb} for a review).}. We have verified that the charges of the string considered here are not compatible with the topological twist condition in minimal five-dimensional gauged supergravity \cite{Chamseddine:1999xk} hence we proceed with examining the naive bound \eqref{BPSbound}.

\subsection{Embedding and equilibrium equation}\label{AdSring}
We take the AdS background and embedding to be
\begin{equation}\label{eq:ads_background_embedding}
\begin{gathered}
 ds^2=-V(r)dt^2+\frac{dr^2}{V(r)}+r^2d\Omega_3^2
 \quad,\quad V(r)=1+\frac{r^2}{L^2}\quad,\quad A=B=0~,\\
 X^t=t\quad,\quad X^r=R\quad,\quad X^\theta=0
 \quad,\quad X^\psi=\psi\quad,\quad X^\phi=\mathrm{constant}~.
\end{gathered}
\end{equation}
Here $d\Omega_3^2=d\theta^2+\cos^2\theta\,d\psi^2+\sin^2\theta\,d\phi^2$, with $\psi$ of period $2\pi$. Set $x=R/L$ and $N=\sqrt{1+x^2}$. The induced metric and a static orthonormal frame on the ring are
\begin{equation}\label{eq:ads_induced_frame}
 \gamma_{ab}d\sigma^a d\sigma^b=-N^2dt^2+R^2d\psi^2
 \quad,\quad E_0=\frac{1}{N}\partial_t
 \quad,\quad E_1=\frac1R\partial_\psi~.
\end{equation}
The stationary thermal generator and local vectors are
\begin{equation}\label{eq:ads_thermal_vectors}
\begin{gathered}
 k=\partial_t+\Omega\partial_\psi\quad,\quad
 w=\frac{R\Omega}{N}\quad,\quad
 \Gamma=(1-w^2)^{-1/2}\quad,\quad |k|=\frac N\Gamma~,\\
 u=\Gamma(E_0+wE_1)\quad,\quad
 v=-\Gamma(wE_0+E_1)~.
\end{gathered}
\end{equation}
We choose $\epsilon^{t\psi}=-1/(NR)$, so $v^a=\epsilon^{ab}u_b$. For a timelike thermal velocity, $|w|<1$. With vanishing external fields, the equilibrium invariants of Section~\ref{sec:equilibrium_action} give the homogeneous local state
\begin{equation}\label{eq:ads_local_state}
 T=\frac{T_0\Gamma}{N}\quad,\quad
 \mu_0=\frac{\Phi_0\Gamma}{N}\quad,\quad
 \mu_1=\frac{\Phi_1}{2\pi RN}+\kappa\mu_0^2w~.
\end{equation}
Constant local data solve the intrinsic equations. The remaining equation follows by varying the radius in the action~\eqref{eq:equilibrium_action}, which on this embedding is
\begin{equation}\label{eq:ads_ring_action}
 S_{\mathrm{eq}}[R]=2\pi\int dt\,NR\,P(T,\mu_0,\mu_1)~,
\end{equation}
while keeping  $T_0,\Omega,\Phi_0,\Phi_1$ and $L$ fixed, with the local state varying according to~\eqref{eq:ads_local_state}. Since $A=B=0$, the embedding variation acts only through the induced metric. Writing $N'=dN/dR$, its nonzero variations are $\delta_R\gamma_{tt}=-2NN'\delta R$ and $\delta_R\gamma_{\psi\psi}=2R\delta R$. We define $E=T^{\hat0\hat0}=N^2T^{tt}$ and $P_{\parallel}=T^{\hat1\hat1}=R^2T^{\psi\psi}$ such that the stress tensor obtained from the action then gives
\begin{equation}\label{eq:ads_variational_balance}
\begin{aligned}
 \delta_R S_{\mathrm{eq}}
 &=\frac12\int dt\,d\psi\,NR\,T^{ab}\delta_R\gamma_{ab}\\
 &=2\pi\int dt\,N\left(P_{\parallel}-\frac{RN'}N E\right)\delta R~.
\end{aligned}
\end{equation}
The radial blackfold equation is therefore
\begin{equation}\label{eq:ads_balance}
 P_{\parallel}=\frac{RN'}N E=\frac{x^2}{1+x^2}E~.
\end{equation}
The right-hand side is the longitudinal pressure needed to balance the AdS attraction and vanishes in the flat-space limit.

To express this equation in local thermodynamic variables, we use $\Pi=P-\mu_1\rho_1$ and $h=-\kappa\mu_0^2\rho_1$ from Section~\ref{sec:equilibrium_action}. Substituting~\eqref{eq:ads_thermal_vectors} into~\eqref{eq:equilibrium_constitutive_relations} gives
\begin{equation}\label{eq:ads_frame_stresses}
 E=\Gamma^2(\epsilon+\Pi w^2-2hw)
 \quad,\quad
 P_{\parallel}=\Gamma^2(\epsilon w^2+\Pi-2hw)~.
\end{equation}
Thus the non-extremal balance condition becomes the quadratic equation
\begin{equation}\label{eq:ads_nonext_balance}
 (\epsilon-\alpha\Pi)w^2-2h(1-\alpha)w+\Pi-\alpha\epsilon=0
 \quad,\quad \alpha=\frac{x^2}{1+x^2}~.
\end{equation}
Only real roots with $|w|<1$ are retained. For the neutral string, $h=0$ and $\epsilon=-2\Pi$, so $w^2=(1+3x^2)/(2+3x^2)$, reproducing~\cite{Caldarelli:2008pz}. The balance equation for the string in flat space \eqref{eq:extrinsic_flat} is reproduced when $\alpha\to 0$. The momentum and electric charge densities in the same static frame are $Z_{\parallel}=T^{\hat0\hat1}=\Gamma^2[(\epsilon+\Pi)w-h(1+w^2)]$ and $j_0=J^{\hat0}=\Gamma(\rho_0+2\kappa\mu_0\rho_1w)$. The general formulae of Section~\ref{sec: Smarr} then give the thermodynamic charges
\begin{equation}\label{eq:ads_general_charges}
\begin{gathered}
 M=2\pi RN E\quad,\quad J_\psi=2\pi R^2 Z_{\parallel}
 \quad,\quad Q_{(0)}=2\pi Rj_0\quad,\quad Q_{(1)}=\rho_1~,\\
 S=2\pi R s\Gamma\quad,\quad T_0=\frac N\Gamma T
 \quad,\quad \Phi_0=\frac N\Gamma\mu_0
 \quad,\quad \Phi_1=2\pi RN(\mu_1-\kappa\mu_0^2w)~.
\end{gathered}
\end{equation}
At finite thickness, the fields of the charged ring can contribute a subleading transverse angular momentum $J_\phi$. The first law holds at fixed $L$, with an additional cosmological work term if $L$ is varied. Equation~\eqref{eq:ads_nonext_balance} applies to non-extremal rings, but we leave the study of finite-temperature solutions for future work and focus below on their extremal limits.

\subsection{Finite-velocity extremal branches}\label{sec:ads_finite}
To check whether we can find extremal branches with finite-velocity, we insert the supersymmetric local state \eqref{eq:flat_ring_susy_local_state} and write $t_q=q/Q_{\mathrm R}$, with $0<t_q<1$. The balance equation \eqref{eq:ads_nonext_balance} gives
\begin{equation}\label{eq:ads_finite_roots}
 (1+x^2)w^2+t_qw-x^2=0\quad\Longrightarrow\quad
 w_\pm=\frac{-t_q\pm\sqrt{t_q^2+4x^2(1+x^2)}}{2(1+x^2)}~.
\end{equation}
Both roots are timelike. They reduce to $0$ and $-q/Q_{\mathrm R}$ as $L\to\infty$ at fixed $R$. To express them in the three-parameter frame of~\eqref{eq:string_susy_three_parameter_map}, we identify the canonical orthonormal frame of the straight string with the local static frame at the ring and set $w=\tanh\eta$. Writing $S=\sqrt{Q^2+\Delta^2}$ for the metric parameter used in that map, the boosted stress tensor~\eqref{eq:string_susy_boosted_monopoles} gives
\begin{equation}\label{eq:ads_three_parameter_balance}
\begin{gathered}
 E=3\mathcal C S\quad,\quad
 P_{\parallel}=3\mathcal C\frac{\Delta(\Delta-q)}{S}
 \quad,\quad w=-\frac{\Delta}{S}~,\\
 \Delta(\Delta-q)=\frac{x^2}{1+x^2}(Q^2+\Delta^2)~.
\end{gathered}
\end{equation}
The second line is the AdS balance equation. It requires positive longitudinal pressure for $R>0$, so the pressureless values $\Delta=0$ and $\Delta=q$ do not balance a ring at finite $R/L$. On either balanced branch, $S=\Gamma^2(Q_{\mathrm R}+qw)=Q_{\mathrm R}(1+x^2)$, and therefore
\begin{equation}\label{eq:ads_three_parameter_branches}
 \Delta_\pm=-Q_{\mathrm R}(1+x^2)w_\pm
 =\frac{q\mp\sqrt{q^2+4Q_{\mathrm R}^2x^2(1+x^2)}}{2}~.
\end{equation}
Thus $\Delta_+<0$ and $\Delta_->q$ for $x>0$. In the flat-space limit $L\to\infty$ at fixed $R$, these become $\Delta_+\to0$ and $\Delta_-\to q$, respectively, recovering the two finite-velocity rings of~\eqref{eq:flat_ring_susy_velocities}.

The electrically and magnetically charged supersymmetric string bent in Section~6 of~\cite{Caldarelli:2008pz} corresponds to $\Delta=0$ in our notation, up to the longitudinal orientation convention. It is the same local extremal string used here, but its boost relative to the static AdS frame differs from both balanced roots at finite $R/L$. Neither of our two AdS branches therefore coincides with their configuration. The formal positive branch approaches it only in the flat-space limit. The vanishing longitudinal pressure of the $\Delta=0$ string is precisely the obstruction they identify to balancing it in AdS. Here the relative boost supplies the pressure required by~\eqref{eq:ads_three_parameter_balance}, but supersymmetry of the local string geometry does not imply supersymmetry of the resulting ring. The asymptotic charge test below determines which branches can remain candidates. The global charges of these branches are
\begin{equation}\label{eq:ads_finite_charges}
\begin{aligned}
 M_\pm&=6\pi\mathcal C RQ_{\mathrm R}N^3~,\\
 J_{\psi,\pm}&=\pm3\pi\mathcal C R^2
                   \sqrt{q^2+4Q_{\mathrm R}^2x^2(1+x^2)}~,\\
 Q_{(0),\pm}&=-2\pi\sqrt3\mathcal C R
                         \Gamma_\pm(Q_{\mathrm R}+qw_\pm)~,
 &Q_{(1)}&=\frac{\sqrt3\mathcal Cq}{2}~,\\
 \mathcal A_\pm&=2\pi^2\sqrt3 Rq\sqrt{Q_{\mathrm R}^2-q^2}\,\Gamma_\pm~,
 &\Omega_\pm&=\frac N Rw_\pm~.
\end{aligned}
\end{equation}
Both have $T_0=0$, $\Phi_0=-\sqrt3N/\Gamma_\pm$ and $\Phi_1=-2\pi\sqrt3RNw_\pm$. These satisfy $M=\Omega J_\psi+\Phi_0Q_{(0)}+\Phi_1Q_{(1)}$ and the extremal first law at fixed $L$.

\noindent\textbf{Energy bound.} There is an additional global obstruction. With the AdS vacuum energy subtracted, positivity of the supercharge anticommutator requires, at the order where $J_\phi=0$,
\begin{equation}\label{eq:ads_energy_bound}
 \mathcal B\equiv M-\frac{|J_\psi|}{L}-\sqrt3|Q_{(0)}|\geq0~.
\end{equation}
The coefficient of $Q_{(0)}$ is fixed by our source normalisation. This is the corresponding eigenvalue bound of~\cite{Cvetic:2005AdSBounds,Chong:2005hr,Gutowski:2004ez}, allowing either orientation of $J_\psi$. To evaluate it, we define $b_R=x/N$ and $v=|w|$ so that Equation~\eqref{eq:ads_finite_charges} gives
\begin{equation}\label{eq:ads_finite_gap}
 \frac{\mathcal B}{M}
 =1-\sqrt{(1-b_R^2)(1-v^2)}
                      -\frac{b_R}{2v}(b_R^2+v^2)~.
\end{equation}
On the positive branch $0<w_+<b_R$, and this expression is negative. Indeed it vanishes at $v=b_R$ and its derivative is positive for $0<v\leq b_R$. On the negative branch $|w_-|>b_R$, and it is positive. One can see the latter directly by subtracting $b_Rv$ from the last term and using $1-b_Rv\geq\sqrt{(1-b_R^2)(1-v^2)}$.

Consequently the positive branch solves the leading balance equation but violates the asymptotic energy bound. This gives a necessary consistency test for a possible completion. Suppose a family of regular AdS rings approached the leading charges~\eqref{eq:ads_finite_charges} as all local gravitational and charge lengths became small compared with $R$ and $L$, at fixed $R/L>0$ and $0<q/Q_{\mathrm R}<1$. If the corrections to these charges were relatively small, including $J_\phi/(LM)\to0$, the normalised energy gap would approach the strictly negative value in~\eqref{eq:ads_finite_gap}. This contradicts the exact asymptotic energy bound. Thus no regular family can have these leading charges with corrections that vanish in this limit.  

The negative branch, on the other hand, has $\mathcal B>0$ and passes this necessary test for being globally well-defined. However, a numerical analysis shows that \eqref{eq:ads_finite_gap} does not vanish for equilibrium velocities $w_-$ derived from the equilibrium condition \eqref{eq:ads_three_parameter_branches} and for parameters $Q_\text{R}$ and $q$ satisfying area positivity, $Q_\text{R}>q$. The next subsection considers two other null-wave branches.

\subsection{Null-wave branches}\label{sec:ads_null}
To take this limit we now pick $\ell=E_0+\varsigma E_1$, with $\ell^{\hat0}=1$. For the charged null wave $\varsigma=+1$, while the opposite magnetic endpoint has $\varsigma=-1$ and $Q=0$. Equation~\eqref{eq:ads_balance} fixes the null energy and the horizon parameter,
\begin{equation}\label{eq:ads_null_balance}
 \mathcal E=\frac{3\mathcal Cq}{2}(1+2x^2)\quad,\quad
 \lambda=3q(1+2x^2)-\frac{3Q^2}{q}>0~.
\end{equation}
The last inequality is the local nonzero-area condition. In particular, the wave has nonzero momentum. A coordinate transformation of its metric cannot remove this physical charge. Its global thermodynamic data are
\begin{equation}\label{eq:ads_null_charges}
\begin{gathered}
 M=6\pi\mathcal C RqN^3\quad,\quad
 J_\psi=3\varsigma\pi\mathcal CqR^2(1+2x^2)~,\\
 Q_{(0)}=-2\pi\sqrt3\mathcal C RQ\quad,\quad
 Q_{(1)}=\frac{\sqrt3\mathcal Cq}{2}\quad,\quad
 \mathcal A=2\pi^2R\sqrt{q^3\lambda}~,\\
 \Omega=\varsigma\frac N R\quad,\quad
 T_0=\Phi_0=0\quad,\quad \Phi_1=2\pi\sqrt3RN~.
\end{gathered}
\end{equation}
For the opposite magnetic wave, $\lambda=\lambda_{\mathrm B}=3q(1+2x^2)$. 

\noindent\textbf{Energy bound.} The BPS energy bound imposes the further global condition
\begin{equation}\label{eq:ads_null_bound}
 \frac{|Q|}{q}\leq N^3-\frac{x}{2}(1+2x^2)~.
\end{equation}
Thus positivity of the local horizon area alone is insufficient. The purely magnetic waves ($Q=0$) satisfy the bound strictly at finite $x$ but never saturate it. The charged branch is admissible provided that \eqref{eq:ads_null_bound} holds. A numerical analysis shows that the BPS bound \eqref{eq:ads_null_bound} can be saturated for a wide range of parameters such that area positivity is satisfied, as required by \eqref{eq:ads_null_balance}. For example, if $Q=20$, $q=1$, the BPS bound is saturated for $x=R/L\approx 19.98$, with $\lambda\approx 1198$.

Saturation of the BPS bound does not imply the existence of a smooth supersymmetric ring, and a separate analysis of the subleading geometry would be required to address this. Note that regular supersymmetric black rings are excluded in minimal gauged supergravity by the near-horizon results of~\cite{Kunduri:2006uh,Grover:2013hja}. We therefore do not interpret the horizons presented here as new supersymmetric solutions, but it would be interesting to understand higher-order corrections to the blackfold approach in this context. Thus, in total we have identified 3 new branches of extremal rings. We compare their leading horizon areas in the following subsection.

\subsection{Comparison between extremal rings}\label{sec:ads_ring_comparison}
We now compare the negative finite-velocity branch of Section~\ref{sec:ads_finite} with the charged and magnetic null-wave branches of Section~\ref{sec:ads_null}. We use the reduced area $a_H$ and angular momentum $j$ defined in~\eqref{eq:flat_ring_reduced_definitions}, with $J=J_\psi$, and plot $a_H$ against $|j|$. We set the mass to be equal in all three curves. It is also useful to define
\begin{equation}\label{eq:ads_comparison_parameters}
 \mathfrak m=\frac{G_5M}{L^2}
 \quad,\quad \nu=\frac{\sqrt3|Q_{(0)}|}{M}~.
\end{equation}
The two charged curves have the same electric charge, whereas the magnetic null wave necessarily has $\nu=0$. There is therefore no common nonzero electric charge sector containing all three branches while the dipole charge is allowed to vary along each curve.

\begin{figure}[!htbp]
 \centering
 \includegraphics[width=\linewidth]{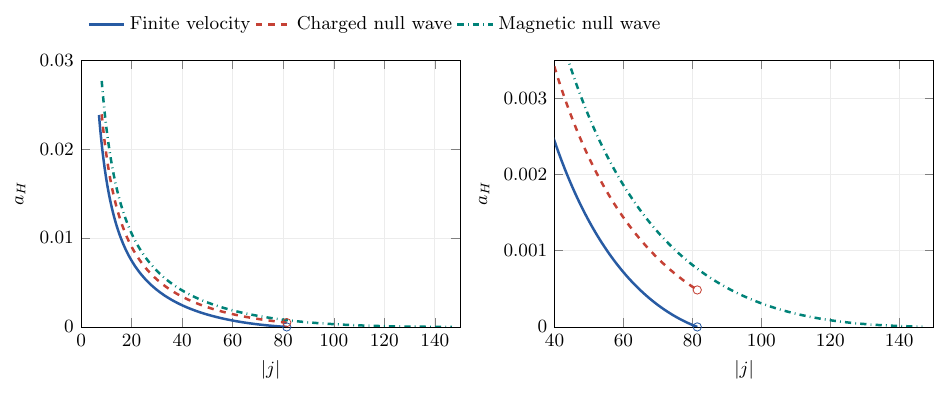}
 \caption{Comparison of the three extremal AdS ring candidates at equal $M$ with $G_5M/L^2=10^{-4}$. Blue solid and red dashed curves denote the negative finite-velocity and charged null-wave branches at the same electric charge, $\sqrt3|Q_{(0)}|/M=1/2$. The teal dash-dotted curve denotes the magnetic null wave with $Q_{(0)}=0$. For the positive magnetic parameter used in the formulas, the blue and teal branches have $J_\psi<0$ and the red branch has $J_\psi>0$. Both panels plot its magnitude. The right panel magnifies the small area region of the left panel. We display $R/L\geq0.1$ and retain only positive area states satisfying the leading energy bound. The magnetic curve is stopped at $R/L=3$ which is a plotting cutoff. Open circles mark formal limits of the charged curves at $|j|=c_j/(2\sqrt{\mathfrak m})\simeq81.41$. The finite-velocity area vanishes as $q\to0$ at $R/L\to1$, while the charged null wave reaches the leading energy bound at $R/L\simeq0.93336$ with nonzero leading area.}
 \label{fig:ads_ring_comparison}
\end{figure}
At fixed $M$ and $\nu$, the finite-velocity branch can be parametrised by $x=R/L$. Equations~\eqref{eq:ads_finite_roots} and~\eqref{eq:ads_finite_charges} give
\begin{equation}\label{eq:ads_comparison_finite}
 w_-=-\sqrt{1-\nu^2N^2}
 \quad,\quad \frac{q}{Q_{\mathrm R}}=\frac{\nu^2N^4-1}{w_-}
 \quad,\quad \frac{Q_{\mathrm R}}L=\frac{2\mathfrak m}{3\pi xN^3}~,
\end{equation}
where $N=\sqrt{1+x^2}$ and we retain $0<q/Q_{\mathrm R}<1$. For the null waves, the corresponding parameters follow from~\eqref{eq:ads_null_charges},
\begin{equation}\label{eq:ads_comparison_null}
 \frac qL=\frac{2\mathfrak m}{3\pi xN^3}
 \quad,\quad \frac Qq=\nu N^3
 \quad,\quad \frac{|J_\psi|}{ML}=\frac{x(1+2x^2)}{2N^3}~.
\end{equation}
Here we choose $Q\geq0$, so $Q_{(0)}\leq0$. The charged wave has $\varsigma=+1$, and the magnetic wave has $\varsigma=-1$ and $Q=0$. We impose positive local horizon area and the energy bound~\eqref{eq:ads_energy_bound}. The other positive finite-velocity solution to the blackfold equations that we found fails this bound and therefore is not included.

Figure~\ref{fig:ads_ring_comparison} shows that the charged null wave has a larger leading area than the finite-velocity branch throughout their displayed overlap at the same $|j|$. To align the signs of $J_\psi$, we reversed both rotation plane angles on the finite-velocity solution. This preserves the spacetime orientation and electric charge but reverses the dipole sign. The area ordering can therefore be interpreted as a leading entropy ordering at common $M,Q_{(0)}$ and $J_\psi$ when both dipole signs are allowed. It does not compare states at fixed dipole charge. The magnetic curve lies above the charged null curve at common $|j|$ but its different electric charge does not allow us to make statements about which extremal ring is thermodynamically preferred. We will be looking at these comparisons further in a future publication.


\section{Outlook and conclusions}
In this work, we have extended the black hole phase diagram of five-dimensional minimal gauged and ungauged supergravity by combining the blackfold approach with the framework of generalized symmetries. Starting from supergravity coupled to probe sources, we derived the local 2-group Ward identities and formulated the equilibrium theory of a two-dimensional 2-group fluid. The conserved currents in \eqref{modified_J} and \eqref{Ptilde} provide a systematic prescription for computing the global classical charges of the configurations studied here. Matching the fluid thermodynamics to that of the non-extremal black string of \cite{Compere:2010fm} then supplied the constitutive data for a symmetry-based blackfold effective theory. By solving its embedding equations in fixed supergravity backgrounds, we constructed non-extremal and extremal asymptotically flat black rings and composite configurations, together with extremal black ring candidates in AdS$_5$, at leading order in the blackfold expansion. \\ \\
\noindent
Our main results can be summarized as follows:

\begin{itemize}

\item \textbf{Asymptotically flat black rings.}
In Section~\ref{sec:ring_flat}, we constructed black rings with four extremal limiting branches. Two have finite equilibrium velocity and opposite longitudinal momentum, while the other two arise from infinite-boost limits in which the equilibrium generator becomes null. These null-wave branches comprise an electrically charged family and a purely magnetic family, the latter also arising as an endpoint of the negative-momentum finite-velocity branch. The positive-momentum finite-velocity branch saturates the BPS bound, and its leading thermodynamics agrees with the thin-ring limit of the exact supersymmetric black ring of \cite{Elvang:2004rt}, as illustrated in Figure~\ref{fig:extremal_local_ratios}.

\item \textbf{Composite configurations.}
We constructed stationary black Saturns, di-rings and bi-rings, including a seven-parameter family of non-extremal black Saturns with four independent asymptotic conserved charges and a magnetic dipole charge. We also identified non-extremal Saturn and di-ring families in mutual thermodynamic equilibrium. The dual potential $B$, together with the gauge transformations accompanying the spacetime symmetries, plays an essential role in the conserved charge construction, particularly in recovering the interaction contribution to the transverse angular momentum $J_\phi$. For the supersymmetric Saturn, the electric and magnetic forces balance the full extrinsic stress, making the radial equilibrium condition an identity. Figures~\ref{fig:susy_saturn_comparison} and \ref{fig:phase_bps_comparison} illustrate the comparison between our leading order predictions and the corresponding exact BPS families. Figure~\ref{fig:phase_extremal_exact} extends the area versus angular momentum diagrams to include non-supersymmetric extremal ring horizons.\footnote{The non-supersymmetric extremal branches discussed here are classical configurations. Studies of near-extremal black holes show that quantum effects can qualitatively modify the thermodynamics at sufficiently low temperatures \cite{Turiaci:2023wrh}. Assessing such effects for the configurations considered here requires a separate analysis.} In the non-extremal sector, Figure~\ref{fig:phase_nonextremal} adds Saturn and di-ring branches to the diagram containing isolated black rings and the ungauged limit of the CCLP black hole.

\item \textbf{Extremal black rings in AdS$_5$.}
The leading balance equation admits two finite-velocity extremal branches. The positive-velocity branch violates the asymptotic energy bound \eqref{eq:ads_energy_bound}, whereas the negative-velocity branch satisfies it strictly. Together with the charged and purely magnetic null-wave branches, this leaves three candidate branches for a regular extremal completion, subject to the validity conditions of the blackfold expansion. The magnetic null wave also satisfies the energy bound strictly. The charged null wave is admissible in the parameter range specified by \eqref{eq:ads_null_bound} and can saturate the bound at leading order. This leading equality does not establish supersymmetry: regular supersymmetric black rings are excluded in minimal gauged supergravity \cite{Kunduri:2006uh,Grover:2013hja}. Understanding the fate of this limiting locus therefore requires an analysis beyond leading order in the blackfold expansion. Figure~\ref{fig:ads_ring_comparison} further shows that the charged null-wave branch has a larger leading entropy than the admissible finite-velocity branch at common mass, electric charge and longitudinal angular momentum. 

\end{itemize}

These results provide perturbative constructions of new black hole families and demonstrate how their charges and thermodynamics can be organized through the local higher-group symmetry structure. Comparisons with known exact solutions test this description in the regimes where the blackfold approximation applies. An important next step is to determine higher-order corrections and assess the global regularity of the new branches, particularly in AdS$_5$. The asymptotically flat composite constructions could also be extended to backgrounds containing a central black lens \cite{PhysRevLett.113.211101,Tomizawa_2016} or a capped black hole \cite{Suzuki:2024abu}.

A further direction concerns the relation between supersymmetric configurations and protected state counting.\footnote{Related questions have been explored for other composite systems, including supersymmetric Grey Galaxies and black holes dressed by graviton or brane excitations \cite{Choi:2024xnv,Choi:2025lck,Amariti:2026lfc}.} Extending our framework to matter-coupled supergravity would introduce additional charges and scalar fields, providing a broader setting in which to search for supersymmetric black rings and composite configurations. Equivariant localization offers a complementary approach: it can predict on-shell actions and thermodynamic properties from topological and gauge data without requiring an explicit solution \cite{Colombo:2025yqy}. Comparing these predictions with perturbative constructions could constrain candidate families and guide the search for regular solutions.

Recent constructions of Euclidean black hole, black string, black ring and multi-center saddles provide a concrete starting point for investigating their contributions to gravitational supersymmetric indices \cite{Cassani:2024kjn,Cassani:2025iix,Boruch:2025qdq,Boruch:2025sie,Bandyopadhyay:2025jbc}. It would be valuable to identify controlled blackfold limits of these saddles and determine whether an appropriate Euclidean extension of our effective theory reproduces their on-shell actions. Where a holographic description is available, such a comparison could clarify how additional gravitational saddles are represented in protected observables of the dual field theory.

Finally, our AdS construction uses the ungauged black string as its local seed, an approximation justified when all intrinsic gravitational and charge length scales are much smaller than both the bending scale and the AdS radius $L$. Gauged supergravity also admits magnetically supported supersymmetric strings for which an R-symmetry flux implements a topological twist.\footnote{Examples with AdS$_3\times\Sigma_g$ near-horizon geometries were studied in \cite{Klemm:2000nj,Naka:2002jz,Benini:2012cz,Karndumri:2013iqa,Benini:2015bwz,Klemm:2016kxw}. Non-extremal black strings and their supersymmetric limits were recently constructed in \cite{Bobev:2026jqd}.} Determining whether these strings admit a controlled blackfold description requires identifying a suitable hierarchy of scales compatible with the twist and its associated boundary conditions. Such an extension could provide access to sectors of the gauged-supergravity phase diagram beyond those described by bending the (ungauged) black strings considered here.

\section*{Acknowledgements}

We would like to thank A. Gnecchi for collaboration at early stages, and D. Cassani, V. Dimitrov, R. Emparan, A. Gnecchi and A. Virmani for useful discussions and clarifications. The work of CT is supported by the MISU FNRS grant 40024018 ”Pushing horizons in Black hole Physics” and the Ministerio de
Ciencia, Innovacion y Universidades / Agencia Estatal de Investigacion (AEI), via the Ramon y Cajal
program, RYC2024-048886-I.  The work of PVM is supported by the Netherlands Organisation for Scientific Research (NWO) under the VICI grant VI.C.202.104. JA is partly funded by the Dutch Institute for Emergent Phenomena (DIEP) cluster
at the University of Amsterdam via the DIEP programme Foundations and Applications of
Emergence (FAEME) and the national NWA-ORC 2023 consortium EAAS. This work was performed in part at the Galileo Galilei Institute for Theoretical Physics (Florence), which is supported by INFN.


\appendix

\section{Asymptotic stresses and charge currents}\label{app:stress_currents}
In this appendix we derive the effective stress tensor and charge currents by matching the leading far fields of a uniform string to the monopole sources of Section~2. We first give the extraction in an asymptotically Cartesian frame, then apply it to the non-extremal solution of Section~\ref{sec:non-ext_string} and to its supersymmetric counterpart. Throughout this appendix the we consider geometries with vanishing transverse rotation. We use the normalization $\mathcal C=\Omega_2/(16\pi G_5)$ and the source conventions of \eqref{eq:EOM} and \eqref{eq:Bianchi}. The bulk orientation is $dt\wedge dz\wedge dV_\perp$, with $dV_\perp=R^2\sin\theta\,dR\wedge d\theta\wedge d\phi$, and the worldsheet orientation is $\epsilon_{tz}=1$.

\subsection{Extraction from the far region}\label{app:monopole_extraction}
Take $R=|\mathbf x_\perp|$ to be an isotropic radial coordinate in the three transverse directions and write the asymptotic metric as $g_{\mu\nu}=\eta_{\mu\nu}+h_{\mu\nu}$, where
\begin{equation}\label{eq:app_asymptotic_metric}
 h_{\mu\nu}dx^\mu dx^\nu
 =\frac{1}{R}\left(c_{tt}\,dt^2+2c_{tz}\,dt\,dz
             +c_{zz}\,dz^2+c_\perp\delta_{ij}\,dx^i dx^j\right)
       +\mathcal O(R^{-2})~.
\end{equation}
For the solutions considered here, $c_\perp=c_{tt}-c_{zz}$. The trace is consequently $h^\mu{}_{\mu}=2c_\perp/R$, and the trace-reversed perturbation $\bar h_{\mu\nu}=h_{\mu\nu}-\frac12\eta_{\mu\nu}h^\lambda{}_{\lambda}$ has $\bar h_{ij}=0$ at this order, that is, it obeys the linearised harmonic gauge condition. The longitudinal Einstein equations then reduce to
\begin{equation}\label{eq:stressformula}
 \nabla_\perp^2\bar h_{ab}
       =-16\pi G_5T_{ab}\delta^{(3)}(\mathbf x_\perp)
 \quad,\quad
 \nabla_\perp^2\frac{1}{R}
       =-\Omega_2\delta^{(3)}(\mathbf x_\perp)~.
\end{equation}
The delta functions represent effective sources whose fields reproduce the leading asymptotic geometry. Nonlinear gravitational and Maxwell contributions are already encoded in the coefficients of that geometry. It follows that
\begin{equation}\label{eq:app_adm_stress}
 T_{tt}=\mathcal C(2c_{tt}-c_{zz})
 \quad,\quad T_{tz}=\mathcal Cc_{tz}
 \quad,\quad T_{zz}=\mathcal C(2c_{zz}-c_{tt})~.
\end{equation}
If $z$ has spatial period $L$ in this asymptotic frame, the mass, momentum and mechanical tension are
\begin{equation}\label{eq:app_adm_charges}
 M=LT^{tt}=LT_{tt}
 \quad,\quad P_z=LT^{tz}=-LT_{tz}
 \quad,\quad \mathcal T=-T^{zz}=-T_{zz}~.
\end{equation}
The tension $\mathcal T$ has units of energy per length, while $L\mathcal T$ is its integrated tensional energy. These formulas implement the asymptotic ADM extraction used for the general string~\cite{Compere:2010fm}.\\
\\
For the electric current, we choose vanishing longitudinal gauge potentials at infinity and write
\begin{equation}\label{eq:app_asymptotic_gauge}
 A=\frac{a_t\,dt+a_z\,dz}{R}-Q_M\cos\theta\,d\phi
          +\mathcal O(R^{-2})~.
\end{equation}
The magnetic term is understood patchwise as in Section~\ref{sec:non-ext_string}. With the signs of \eqref{eq:EOM}, the linearised electric equation is
\begin{equation}\label{eq:app_electric_extraction}
\begin{gathered}
 \nabla_\perp^2A_a=-16\pi G_5J_a\delta^{(3)}(\mathbf x_\perp)
 \quad\Longrightarrow\quad J_a=\mathcal C a_a~,\\
 J^t=-\mathcal C a_t
 \quad,\quad J^z=\mathcal C a_z~.
\end{gathered}
\end{equation}
The Chern-Simons term is quadratic in the field strengths and does not enter this linear matching. Its contribution to the electric Page surface integral also vanishes at infinity in the chosen gauge, since $A_z=\mathcal O(R^{-1})$. Hence the electric charge of Section~\ref{sec: conserved_charges} is $Q_{(0)}=LJ^t=-Q_E$, where $Q_E=\mathcal C L a_t$ is the thermodynamic electric charge used in Section~\ref{sec:non-ext_string}.\\
\\
The magnetic charge is fixed directly by the flux. Distributionally,
\begin{equation}\label{eq:app_magnetic_extraction}
\begin{aligned}
 F_M&=Q_M\sin\theta\,d\theta\wedge d\phi~,
 &dF_M&=\Omega_2 Q_M\delta^{(3)}(\mathbf x_\perp)\,dV_\perp~,\\
 \star(dt\wedge dz)&=-dV_\perp~,
 &\mathcal J_{ab}&=\mathcal C Q_M\epsilon_{ab}~.
\end{aligned}
\end{equation}
The last equality follows from \eqref{eq:Bianchi}, including its factor of $16\pi G_5$. In particular,
\begin{equation}\label{eq:app_magnetic_charge}
 \rho_1=Q_{(1)}=\mathcal C Q_M
       =\frac{1}{16\pi G_5}\int_{S^2}F
 \quad,\quad \mathcal J^{tz}=-\rho_1~.
\end{equation}
The string charge is not integrated along $z$. A longitudinal Lorentz boost changes the components of $J^a$ and $T^{ab}$, while leaving this top-form coefficient unchanged.

\subsection{Non-extremal string}\label{app:nonext_monopoles}
The radial functions of \eqref{eq:nonext_radial_functions} are linear in $r$. For the asymptotic expansion it is useful to separate their coefficients as $F_i=f_i+rg_i$, with
\begin{equation}\label{eq:gi1}
\begin{aligned}
 g_1&=2mc_\delta^2\mathsf B~,
 &g_2&=2mc_\delta s_\delta\mathsf X~,
 &g_3&=-2mc_\delta s_\delta\mathsf B~,\\
 g_4&=2m(c_\delta^2+s_\delta^2)\mathsf B~,
 &g_5&=2m\mathsf X~.
\end{aligned}
\end{equation}
Here $\mathsf B$ and $\mathsf X$ are defined in \eqref{eq:nonext_rest_parameters}. The constants $f_i$ first contribute at order $R^{-2}$ in the asymptotically Cartesian metric and the longitudinal gauge potential, so the monopole charges depend only on the $g_i$. The full radial functions remain necessary for the horizon geometry.\\
\\
We retain an arbitrary longitudinal boost $\gamma$ during the extraction. We introduce the auxiliary coframe
\begin{equation}\label{eq: uv}
 \widetilde u_a=(c_\gamma,-s_\gamma)
 \quad,\quad \widetilde v_a=(-s_\gamma,c_\gamma)
 \quad,\quad
 \eta_{ab}=-\widetilde u_a\widetilde u_b
                    +\widetilde v_a\widetilde v_b~,
\end{equation}
which encode the coefficients of the coordinate one-forms after the boost, with $\widetilde u_a d\sigma^a=c_\gamma dt-s_\gamma dz$. At $\gamma=\chi$ they reduce to the thermal rest frame \eqref{eq:nonext_rest_coframe}. The isotropic radius is defined by
\begin{equation}\label{eq:app_isotropic_radius}
 r=R\left(1+\frac{m}{2R}\right)^2~.
\end{equation}
Expanding the metric \eqref{eq:metricofnonextremal} with $\chi$ replaced by $\gamma$ gives
\begin{equation}\label{eq:app_nonext_metric_perturbation}
\begin{aligned}
 h_{\mu\nu}dx^\mu dx^\nu
 ={}&\frac{1}{R}\left[g_4\widetilde u_a\widetilde u_b
          +(g_4-g_1)\widetilde v_a\widetilde v_b
          +2g_5\widetilde u_{(a}\widetilde v_{b)}\right]d\sigma^a d\sigma^b\\
 &+\frac{g_1}{R}\delta_{ij}\,dx^i dx^j+\mathcal O(R^{-2})~.
\end{aligned}
\end{equation}
In particular, $h^\mu{}_{\mu}=2g_1/R$. The longitudinal gauge potential has the expansion
\begin{equation}\label{eq:app_nonext_gauge_expansion}
 A_a=-\frac{\sqrt{3}}{R}
                  (g_3\widetilde u_a+g_2\widetilde v_a)
                   +\mathcal O(R^{-2})~.
\end{equation}
Applying \eqref{eq:stressformula} and \eqref{eq:app_electric_extraction} therefore yields
\begin{equation}\label{eq:nonext_auxiliary_monopoles}
\begin{aligned}
 T_{ab}&=\mathcal C\left[g_4\widetilde u_a\widetilde u_b
        +(g_4-g_1)\widetilde v_a\widetilde v_b
        +2g_5\widetilde u_{(a}\widetilde v_{b)}-g_1\eta_{ab}\right]~,\\
 J_a&=-\sqrt{3}\,\mathcal C
                (g_3\widetilde u_a+g_2\widetilde v_a)~,
 &\mathcal J_{ab}&=\mathcal C Q_M\epsilon_{ab}~.
\end{aligned}
\end{equation}
To display the charges in the chosen asymptotic frame, the required metric coefficients are
\begin{equation}\label{eq:app_nonext_metric_coefficients}
\begin{aligned}
 c_{tt}&=g_4c_\gamma^2+(g_4-g_1)s_\gamma^2
                       -2g_5c_\gamma s_\gamma~,\\
 c_{tz}&=g_5(c_\gamma^2+s_\gamma^2)
                       -(2g_4-g_1)c_\gamma s_\gamma~,\\
 c_{zz}&=g_4s_\gamma^2+(g_4-g_1)c_\gamma^2
                       -2g_5c_\gamma s_\gamma~.
\end{aligned}
\end{equation}
They satisfy $c_{tt}-c_{zz}=g_1$. The electric and magnetic charges are
\begin{equation}\label{eq:app_nonext_electric_magnetic}
\begin{aligned}
 \frac{Q_E}{L}&=-J^t
       =\sqrt{3}\,\mathcal C(g_2s_\gamma-g_3c_\gamma)~,\\
 J^z&=-\sqrt{3}\,\mathcal C(g_2c_\gamma-g_3s_\gamma)~,
 &Q_{(1)}&=\mathcal C Q_M=\sqrt{3}\,\mathcal C c_\delta^2g_5~.
\end{aligned}
\end{equation}
\noindent\textbf{Boost to the thermal rest frame.}
For a solution with arbitrary $\gamma$, the horizon velocity in \eqref{eq:nonext_rest_boost} is $v_z=\tanh(\gamma-\chi)$. Define the relative rapidity $\eta=\gamma-\chi$, keeping $m,\beta,\delta$ fixed. The coordinates adapted to the thermal rest frame are
\begin{equation}\label{eq:app_nonext_thermal_boost}
 t_{\mathrm R}=c_\eta t-s_\eta z
 \quad,\quad z_{\mathrm R}=c_\eta z-s_\eta t
 \quad,\quad \tanh\chi=\frac{\mathsf X}{\mathsf A}~.
\end{equation}
Equivalently, the thermal boost can be read directly from the asymptotic coefficients,
\begin{equation}\label{eq:app_thermal_boost_from_monopoles}
 \tanh\chi=\frac{(5g_1-g_4)g_5}
                  {(2g_1-g_4)(g_1+g_4+4m)}~.
\end{equation}
Indeed, the horizon generator and the unit thermal velocity transform as
\begin{equation}\label{eq:app_rest_generator}
 k=\partial_t+v_z\partial_z
      =\frac{1}{c_\eta}\partial_{t_{\mathrm R}}
 \quad,\quad u=c_\eta k=\partial_{t_{\mathrm R}}~.
\end{equation}
The temperature measured with the rest-frame generator is consequently $T_{\mathrm R}=c_\eta T_H(\gamma)$. Substituting the inverse of \eqref{eq:app_nonext_thermal_boost} into the auxiliary coframe gives
\begin{equation}\label{eq:app_nonext_rest_coframe_check}
\begin{aligned}
 c_\gamma dt-s_\gamma dz
       &=c_\chi dt_{\mathrm R}-s_\chi dz_{\mathrm R}~,\\
 c_\gamma dz-s_\gamma dt
       &=c_\chi dz_{\mathrm R}-s_\chi dt_{\mathrm R}~.
\end{aligned}
\end{equation}
Thus the full metric and gauge field become precisely \eqref{eq:metricofnonextremal} and \eqref{eq:nonext_rest_gauge}, with $t,z$ replaced by $t_{\mathrm R},z_{\mathrm R}$. The radial functions are unchanged. In particular, starting from $\gamma=0$ requires $\eta=-\chi$.\\
\\
The same change of coordinates acts on the contravariant monopole tensors through
\begin{equation}\label{eq:app_thermal_tensor_boost}
\begin{gathered}
 (\Lambda_\eta)^a{}_b=
    \begin{pmatrix}c_\eta&-s_\eta\\-s_\eta&c_\eta\end{pmatrix}~,\\
 T_{\mathrm R}^{ab}=(\Lambda_\eta)^a{}_c
                         (\Lambda_\eta)^b{}_d T^{cd}
 \quad,\quad J_{\mathrm R}^a=(\Lambda_\eta)^a{}_bJ^b
 \quad,\quad \rho_{1,\mathrm R}=\rho_1~.
\end{gathered}
\end{equation}
The top-form coefficient is unchanged because $\det\Lambda_\eta=1$. Applying these transformations to \eqref{eq:nonext_auxiliary_monopoles} is equivalent to replacing $\gamma$ by $\chi$, and reproduces the mass, momentum and tension in \eqref{eq:nonext_rest_charges}. In the following matrix, $M,P_z,L$ and $\mathcal T$ denote the quantities evaluated in this thermal rest frame,
\begin{equation}\label{eq:app_rest_stress_matrix}
 T^{ab}=\begin{pmatrix}
             M/L&P_z/L\\ P_z/L&-\mathcal T
           \end{pmatrix}
       =\begin{pmatrix}
             \epsilon&\kappa\mu_0^2\rho_1\\
             \kappa\mu_0^2\rho_1&P-\mu_1\rho_1
         \end{pmatrix}~.
\end{equation}
Together with \eqref{eq:app_nonext_electric_magnetic}, this gives the constitutive relations \eqref{eq:stresstensornonextremal} in the orientation $u^a=(1,0)$ and $v^a=(0,-1)$. In particular, vanishing horizon velocity does not require vanishing momentum density.

The boost is performed on the non-compact covering space, as in Section~\ref{sec:non-ext_string}. If the original solution is identified by $(t,z)\sim(t,z+L)$, its image is instead
\begin{equation}\label{eq:app_boosted_identification}
 (t_{\mathrm R},z_{\mathrm R})
   \sim(t_{\mathrm R}-Ls_\eta,z_{\mathrm R}+Lc_\eta)~.
\end{equation}
A compact string with a purely spatial period in the thermal rest frame is obtained by imposing $z_{\mathrm R}\sim z_{\mathrm R}+L_{\mathrm R}$ separately. 

\subsection{Supersymmetric string}\label{app:susy_lim}
We next apply the same extraction to the three-parameter supersymmetric string~\cite{Kim:2010bf,Compere:2010fm}. To distinguish its original coordinates from the canonical asymptotic ones, denote the former by $\hat t,\hat z$. For $Q>0$, introduce
\begin{equation}\label{eq:app_susy_parameters}
 S=\sqrt{Q^2+\Delta^2}
 \quad,\quad b=\frac{\Delta}{S}
 \quad,\quad \alpha=\frac{Q^2+q\Delta}{S}~.
\end{equation}
In these coordinates the full metric is
\begin{equation}\label{eq:app_susy_original_metric}
 ds^2=-f^2(d\hat t+\omega_z\,d\hat z)^2
       +f^{-1}(d\hat z^2+dr^2+r^2d\Omega_2^2)~,
\end{equation}
where
\begin{equation}\label{eq:app_susy_original_functions}
 f^{-1}=1+\frac{\alpha}{r}+\frac{q^2}{4r^2}
 \quad,\quad
 \omega_z=b+\frac{3q}{2r}+\frac{3q\alpha}{4r^2}
                         +\frac{q^3}{8r^3}~.
\end{equation}
The functions entering the metric behave as $f=1-\alpha/r+\mathcal O(r^{-2})$ and $\omega_z=b+3q/(2r)+\mathcal O(r^{-2})$. Thus the asymptotic longitudinal metric is $-(d\hat t+b\,d\hat z)^2+d\hat z^2$. It is flat, but $\hat t,\hat z$ are not orthonormal coordinates. The change
\begin{equation}\label{boost}
 \hat t=\frac{Q}{S}\,t
 \quad,\quad
 \hat z=\frac{S}{Q}\,z+\frac{\Delta}{Q}\,t~~,
\end{equation}
brings it to $-dt^2+dz^2$. Its Jacobian has determinant one, so it preserves the orientation and does not introduce a volume factor into the magnetic charge. The radial coordinate $r$ is already isotropic for this solution.\\
\\
After \eqref{boost}, the coefficients of the $1/r$ metric perturbation are
\begin{equation}\label{eq:app_susy_metric_coefficients}
\begin{aligned}
 c_{tt}&=\frac{\alpha(2+b^2)-3qb}{1-b^2}~,
 &c_{zz}&=\frac{\alpha(1+2b^2)-3qb}{1-b^2}~,\\
 c_{tz}&=\frac{3\alpha b-\frac32q(1+b^2)}{1-b^2}~,
 &c_\perp&=\alpha~.
\end{aligned}
\end{equation}
They again satisfy $c_\perp=c_{tt}-c_{zz}$. Equations~\eqref{eq:app_adm_stress} give
\begin{align}
 T_{tt}&=3\mathcal C S~,
       \label{eq:stresssusy11}\\
 T_{tz}&=-3\mathcal C\left(\frac q2-\Delta\right)~,
       \label{eq:stresssusy12}\\
 T_{zz}&=3\mathcal C\frac{\Delta(\Delta-q)}{S}~.
       \label{eq:stresssusy22}
\end{align}
The ADM quantities in this frame are consequently
\begin{equation}\label{eq:app_susy_adm_charges}
 \frac{M}{L}=3\mathcal C S
 \quad,\quad \frac{P_z}{L}=3\mathcal C\left(\frac q2-\Delta\right)
 \quad,\quad \mathcal T=3\mathcal C\frac{\Delta(q-\Delta)}{S}~.
\end{equation}
The tension vanishes for $\Delta=0$ and $\Delta=q$, subject to the regularity condition \eqref{eq:app_susy_rest_regularity}. The canonical asymptotic frame used here has horizon velocity $-b$, since the supersymmetric horizon generator is proportional to $\partial_t-b\partial_z$. It coincides with the thermal rest frame only when $\Delta=0$.
For completeness, the gauge potential in the original coordinates can be written as
\begin{equation}\label{gauge_field}
\begin{aligned}
 A={}&\sqrt{3}(1-f)\,d\hat t-\frac{\sqrt{3}q}{2}(1+\cos\theta)\,d\phi\\
 &-\sqrt{3}\left[f\left(\omega_z-b-\frac{b\alpha}{r}
                              -\frac{bq^2}{4r^2}\right)
                             -\frac{q}{2r}\right]d\hat z~.
\end{aligned}
\end{equation}
Its longitudinal part is $\sqrt{3}[\alpha\,d\hat t-(q-b\alpha)\,d\hat z]/r+\mathcal O(r^{-2})$. Applying \eqref{boost} gives
\begin{equation}\label{eq:app_susy_gauge_expansion}
 A=\frac{\sqrt{3}Q}{r}\,dt
      -\frac{\sqrt{3}Q(q-\Delta)}{Sr}\,dz
      -\frac{\sqrt{3}q}{2}(1+\cos\theta)\,d\phi
      +\mathcal O(r^{-2})~.
\end{equation}
The electric current in the conventions of Section~2 is therefore
\begin{equation}\label{J1_Delta}
 J^t=-\sqrt{3}\,\mathcal C Q
 \quad,\quad J^z=-\sqrt{3}\,\mathcal C\frac{Q(q-\Delta)}{S}~.
\end{equation}
The angular potential has $F_{\theta\phi}=\sqrt{3}q\sin\theta/2$, giving
\begin{equation}\label{eq:susytopcurrent}
 Q_M=\frac{\sqrt{3}q}{2}
 \quad,\quad
 \mathcal J_{ab}=\frac{\sqrt{3}\,\mathcal C q}{2}\epsilon_{ab}
 \quad,\quad Q_{(1)}=\frac{\sqrt{3}\,\mathcal C q}{2}~.
\end{equation}
For a compactification of period $L$ in the canonical coordinates, $Q_{(0)}=-\sqrt{3}\,\mathcal C LQ$ and $Q_E=\sqrt{3}\,\mathcal C LQ$. At fixed $Q,q,L$, varying $\Delta$ changes the momentum and tension while leaving both charge integrals unchanged. The spatial electric current still depends on $\Delta$, as required by its transformation between asymptotic frames.

\medskip
\noindent\textbf{Thermal rest frame of the three-parameter string.}
At zero temperature, the thermal rest frame is defined by the smooth limit of the finite-temperature velocity, aligned with the normalized supersymmetric horizon generator. Since its velocity in the canonical coordinates is $-b$, the required relative rapidity and Lorentz boost are
\begin{equation}\label{eq:app_susy_thermal_boost}
\begin{gathered}
 \eta_{\mathrm B}=-\operatorname{arctanh}\frac{\Delta}{S}
 \quad,\quad c_{\eta_{\mathrm B}}=\frac{S}{Q}
 \quad,\quad s_{\eta_{\mathrm B}}=-\frac{\Delta}{Q}~,\\
 t_{\mathrm R}=\frac{St+\Delta z}{Q}
 \quad,\quad z_{\mathrm R}=\frac{\Delta t+Sz}{Q}~.
\end{gathered}
\end{equation}
It sends $\partial_t-b\partial_z$ to $(Q/S)\partial_{t_{\mathrm R}}$. Combining it with \eqref{boost} gives a particularly simple map from the original coordinates of \eqref{eq:app_susy_original_metric},
\begin{equation}\label{eq:app_susy_original_to_rest}
 t_{\mathrm R}=\hat t+b\hat z
 \quad,\quad z_{\mathrm R}=\hat z~.
\end{equation}
The first transformation \eqref{boost} makes the asymptotic coordinates orthonormal, while \eqref{eq:app_susy_thermal_boost} aligns their time direction with the horizon generator.\\
\\
To see the effect on the full solution, define the rest-frame electric parameter
\begin{equation}\label{eq:app_susy_rest_parameter}
 Q_{\mathrm R}=\frac{Q^2+q\Delta}{S}=\alpha~.
\end{equation}
In the new coordinates the metric is
\begin{equation}\label{eq:app_susy_rest_metric}
 ds^2=-f_{\mathrm R}^2
        (dt_{\mathrm R}+\omega_{\mathrm R}dz_{\mathrm R})^2
       +f_{\mathrm R}^{-1}
          (dz_{\mathrm R}^2+dr^2+r^2d\Omega_2^2)~,
\end{equation}
where
\begin{equation}\label{eq:app_susy_rest_functions}
\begin{aligned}
 f_{\mathrm R}^{-1}&=1+\frac{Q_{\mathrm R}}{r}
                           +\frac{q^2}{4r^2}~,\\
 \omega_{\mathrm R}&=\omega_z-b
       =\frac{3q}{2r}+\frac{3qQ_{\mathrm R}}{4r^2}
                             +\frac{q^3}{8r^3}~.
\end{aligned}
\end{equation}
Using $f_{\mathrm R}^{-1}=1+\alpha/r+q^2/(4r^2)$ in \eqref{gauge_field}, the terms proportional to $b$ combine into the same shift of $\omega_z$. The gauge potential becomes
\begin{equation}\label{eq:app_susy_rest_gauge}
\begin{aligned}
 A={}&\sqrt{3}(1-f_{\mathrm R})\,dt_{\mathrm R}
       -\sqrt{3}\left(f_{\mathrm R}\omega_{\mathrm R}
                           -\frac{q}{2r}\right)dz_{\mathrm R}\\
 &-\frac{\sqrt{3}q}{2}(1+\cos\theta)\,d\phi~.
\end{aligned}
\end{equation}
Equations~\eqref{eq:app_susy_rest_metric}-\eqref{eq:app_susy_rest_gauge} are exactly the two-parameter supersymmetric string with parameters $Q_{\mathrm R},q$. This is an equivalence of the full local fields. The regularity condition is preserved because
\begin{equation}\label{eq:app_susy_rest_regularity}
 Q_{\mathrm R}^2-q^2
      =\frac{Q^2}{S^2}(Q^2-q^2+2q\Delta)>0~.
\end{equation}
On the regular branch with $Q,q>0$, this gives $Q_{\mathrm R}>q$.\\
\\
Applying \eqref{eq:app_thermal_tensor_boost} with $\eta=\eta_{\mathrm B}$ to the previously extracted stresses and currents gives
\begin{equation}\label{eq:app_susy_rest_monopoles}
\begin{aligned}
 T_{\mathrm R}^{ab}&=\mathcal C
        \begin{pmatrix}3Q_{\mathrm R}&3q/2\\3q/2&0\end{pmatrix}~,\\
 J_{\mathrm R}^a&=-\sqrt{3}\,\mathcal C(Q_{\mathrm R},q)~,
 &\mathcal J_{\mathrm R}^{t_{\mathrm R}z_{\mathrm R}}
       &=-\frac{\sqrt{3}\,\mathcal C q}{2}~.
\end{aligned}
\end{equation}
The electric parameter must therefore change from $Q$ to $Q_{\mathrm R}$ when passing to the rest-frame density. With $u_{\mathrm R}^a=(1,0)$ and $v_{\mathrm R}^a=(0,-1)$, the local dictionary is
\begin{equation}\label{eq:app_susy_rest_dictionary}
\begin{gathered}
 \epsilon=3\mathcal C Q_{\mathrm R}
 \quad,\quad \rho_0=-\sqrt{3}\,\mathcal C Q_{\mathrm R}
 \quad,\quad \rho_1=\frac{\sqrt{3}\,\mathcal C q}{2}~,\\
 T=0\quad,\quad P=0\quad,\quad
 \mu_0=-\sqrt{3}\quad,\quad \mu_1=0~.
\end{gathered}
\end{equation}
Thus the rest-frame tension vanishes. The momentum density is $3\mathcal C q/2$, and the spatial electric current is $-\sqrt{3}\,\mathcal C q$. These nonzero components reproduce the 2-group terms in \eqref{eq:stresstensornonextremal}. For the non-compact string the boost gives the two-parameter rest-frame description. If the original string is instead compactified by $(t,z)\sim(t,z+L)$, the same identification in rest-frame coordinates is
\begin{equation}\label{eq:app_susy_compact_identification}
 (t_{\mathrm R},z_{\mathrm R})
 \sim\left(t_{\mathrm R}+\frac{L\Delta}{Q},
           z_{\mathrm R}+\frac{LS}{Q}\right)~.
\end{equation}
The time shift follows directly from \eqref{eq:app_susy_thermal_boost}. Keeping this identification describes the same compact spacetime in new coordinates. Imposing a purely spatial period at fixed $t_{\mathrm R}$ would instead choose a different compactification when $\Delta\ne0$.\\
\\
The supersymmetric limit of the non-extremal boost provides a direct check. Write $me^{2\beta}=2\mathfrak p$ and take $m\to0$, $\beta\to\infty$ with $\mathfrak p$ and $\delta>0$ fixed. Then
\begin{equation}\label{eq:app_thermal_boost_susy_limit}
 \tanh\chi_{\mathrm B}=\frac{h_\delta}{3c_\delta^2-1}
 \quad,\quad Q_{\mathrm R}=2\mathfrak p c_\delta^2
 \quad,\quad q=\frac{4\mathfrak p c_\delta^2}{h_\delta}~.
\end{equation}
The identification of $Q,\Delta$ in eqs.~(4.10)--(4.12) of~\cite{Compere:2010fm} gives $\tanh(\gamma-\chi_{\mathrm B})=-\Delta/S$. Hence the relative rapidity in \eqref{eq:app_nonext_thermal_boost} tends to $\eta_{\mathrm B}$, and taking the regular supersymmetric limit commutes with passing to the thermal rest frame.

\section{Comparison with exact solutions}\label{Appendix_matching}

We compare the blackfold configurations with exact solutions. The supersymmetric cases include single rings, black Saturns, di-rings and bi-rings. We then consider the non-extremal Suzuki-Tomizawa rings. For each family we give the parameter maps and compare the charges and areas. We also discuss which extremal limits are accessible within the non-extremal exact family.

\subsection{Supersymmetric solutions}\label{app:exact_susy_comparison}

We first compare the isolated ring of Section~\ref{sec:flat_ring_extremal} with the supersymmetric solution of~\cite{Elvang:2004rt}. We then compare the black Saturn of Section~\ref{sec:susy_saturn} with the exact concentric solutions of~\cite{Gauntlett:2004wh}. Finally, we construct supersymmetric di-rings and bi-rings in these backgrounds and compare their charges and areas with the exact family.

\subsubsection{Blackfold limit of the SUSY ring}\label{app:exact_susy_ring}

We define $Q_{\mathrm{ring}}$, $q$ and $R$ as the parameters of the exact supersymmetric black ring in \cite{Elvang:2004rt}. The parameter $Q_{\mathrm{ring}}$ has dimensions of length squared, whereas $q$ and $R$ have dimensions of length. The latter is the radius of the circle in the flat base of the solution. Its ADM charges are~\cite{Elvang:2004rt}
\begin{equation}\label{charges_blackring}
\begin{gathered}
 M_{\mathrm{ex}}=\frac{3\pi Q_{\mathrm{ring}}}{4G_5}~,\\
 J_\phi=\frac{\pi q}{8G_5}(3Q_{\mathrm{ring}}-q^2)
 \quad,\quad
 J_\psi=\frac{\pi q}{8G_5}(6R^2+3Q_{\mathrm{ring}}-q^2)~.
\end{gathered}
\end{equation}
We quote the nonnegative angular momenta in the orientation of that reference and compare their magnitudes with the blackfold angular momenta. In the electric and string charge normalisations used throughout this work, we have
\begin{equation}\label{eq:exact_ring_charge_conventions}
 Q_{(0),\mathrm{ex}}=-\frac{\sqrt{3}\pi Q_{\mathrm{ring}}}{4G_5}
 \quad,\quad Q_{(1),\mathrm{ex}}=\frac{\sqrt{3}q}{8G_5}~.
\end{equation}
Thus $M_{\mathrm{ex}}=-\sqrt{3}Q_{(0),\mathrm{ex}}$, in agreement with \eqref{eq:flat_ring_bps_bound}. This uses the gauge field normalisation of Section~\ref{sec:sugra_ward_identities} when translating the charges of~\cite{Elvang:2004rt}. The horizon area and the proper radius of its circle are
\begin{equation}\label{area}
\begin{aligned}
 \mathcal A_{\mathrm{ex}}
 &=2\pi^2q^2L_H
 =\pi^2q\sqrt{3\big[(Q_{\mathrm{ring}}-q^2)^2-4q^2R^2\big]}~,\\
 L_H&=\sqrt{3\left[\frac{(Q_{\mathrm{ring}}-q^2)^2}{4q^2}-R^2\right]}~.
\end{aligned}
\end{equation}
Regular rings obey $q>0$ and $Q_{\mathrm{ring}}>q^2+2qR$. Their surface gravity and both horizon angular velocities vanish. Notice that $L_H$ is a proper horizon length, while $R$ specifies the embedding in the base geometry.

To extract the straight string, set $\widehat z=R\psi$ and $r=-R/y$ in the ring coordinates of~\cite{Elvang:2004rt}, and take
\begin{equation}\label{rescaling}
 Q_{\mathrm{ring}}=2R\bar Q\quad,\quad q=\bar q
 \quad,\quad R\longrightarrow\infty~,
 \qquad \bar Q,\bar q,r,\widehat z\ \text{fixed}~.
\end{equation}
The metric functions tend to
\begin{equation}\label{Rw}
\begin{aligned}
 f^{-1}&\longrightarrow1+\frac{\bar Q}{r}+\frac{\bar q^2}{4r^2}~,\\
 \omega_\psi d\psi
 &\longrightarrow-\left(\frac{3\bar q}{2r}
              +\frac{3\bar q\bar Q}{4r^2}
              +\frac{\bar q^3}{8r^3}\right)d\widehat z~,
 \qquad \omega_\phi\longrightarrow0~.
\end{aligned}
\end{equation}
After reversing the longitudinal coordinate, this is the metric in \eqref{eq:string_susy_geometry}. The parameter identification is simply
\begin{equation}\label{matching_ring}
 \bar Q=Q_{\mathrm R}\quad,\quad \bar q=q
 \quad,\quad R_{\mathrm{embedding}}=R~.
\end{equation}
Here the last equality fixes the leading-order embedding convention. It does not identify $R$ with $L_H$. Substituting this map in the exact charges and comparing with \eqref{eq:flat_ring_susy_plus} gives
\begin{equation}\label{charges_mu0}
\begin{aligned}
 M_{\mathrm{ex}}&=M_+=\frac{3\pi RQ_{\mathrm R}}{2G_5}~,\\
 J_+&=\frac{3\pi R^2q}{4G_5}~,\\
 J_\psi&=J_+\left(1+\frac{Q_{\mathrm R}}R-\frac{q^2}{6R^2}\right)~,\\
 J_\phi&=\frac{3\pi RqQ_{\mathrm R}}{4G_5}
                          -\frac{\pi q^3}{8G_5}~.
\end{aligned}
\end{equation}
Here $J_+$ is the leading blackfold charge for rotation around the ring. With the angular orientations aligned, it must be compared with the exact charge $J_\psi$ for the same asymptotic generator $\partial_\psi$. At fixed $0<q/Q_{\mathrm R}<1$, \eqref{charges_mu0} gives $J_\psi=J_+[1+\mathcal O(Q_{\mathrm R}/R)]$, while $J_\phi/J_\psi=\mathcal O(Q_{\mathrm R}/R)$. Thus the longitudinal angular momenta agree to the order retained. Both exact angular momenta grow with $R$, but the transverse one is parametrically smaller. Its nonzero value and the correction to $J_\psi$ require information beyond the present monopole approximation. In particular, the vanishing of $\omega_\phi$ in the local limiting metric does not imply a vanishing global $J_\phi$ at finite radius.

The identity $J_\psi-J_\phi=J_+$ also follows algebraically from \eqref{charges_mu0} under the chosen parameter map. It is not an identification of the blackfold angular momentum with the charge of $\partial_\psi-\partial_\phi$, nor does it establish matching beyond leading order. The curves in Section~\ref{sec:flat_ring_comparison} consequently compare $J_+$ with the full exact $J_\psi$.

The area comparison requires the longitudinal horizon metric as well as the transverse sphere. From \eqref{eq:string_susy_geometry},
\begin{equation}\label{A_bkfd}
\begin{aligned}
 \mathcal A_+
 &=\int_0^{2\pi R} dz\,4\pi\left(\frac q2\right)^2
                       \sqrt{\frac{3(Q_{\mathrm R}^2-q^2)}{q^2}}\\
 &=2\pi^2\sqrt{3}\,Rq\sqrt{Q_{\mathrm R}^2-q^2}
 =4G_5S_{\mathrm{BH},+}~.
\end{aligned}
\end{equation}
Consequently, $L_H/R\to\sqrt{3(Q_{\mathrm R}^2-q^2)}/q$.  At finite $R$ the exact area is
\begin{equation}\label{eq:exact_ring_area_matching}
\begin{aligned}
 \mathcal A_{\mathrm{ex}}
 &=2\pi^2\sqrt{3}\,Rq
       \sqrt{\left(Q_{\mathrm R}-\frac{q^2}{2R}\right)^2-q^2}~,\\
 \frac{\mathcal A_{\mathrm{ex}}}{\mathcal A_+}
 &=1-\frac{Q_{\mathrm R}q^2}{2R(Q_{\mathrm R}^2-q^2)}
                  +\mathcal O\left(\frac{Q_{\mathrm R}^2}{R^2}\right)~.
\end{aligned}
\end{equation}
The expansion holds for fixed $0<q/Q_{\mathrm R}<1$ and $Q_{\mathrm R}/R\ll1$. The expansion is  not uniform near the zero area boundary. Indeed the exact regularity condition is $Q_{\mathrm R}>q+q^2/(2R)$, which reduces to the string bound $Q_{\mathrm R}>q$ only in the thin limit. The mass, electric and dipole charges, leading angular momentum, entropy and vanishing horizon angular velocity therefore all agree with the positive finite-velocity branch of the blackfold configurations constructed in the main text. The other extremal branches in Section~\ref{sec:flat_ring_extremal} have a strictly positive BPS gap and do not approximate this supersymmetric ring.

\subsubsection{Exact supersymmetric black Saturn and its thin-ring limit}\label{app:exact_susy_saturn}
Here we specialise the supersymmetric Gibbons-Hawking construction of~\cite{Gauntlett:2002nw,Gauntlett:2004wh} to one central spherical horizon and one ring horizon. This also fixes the global meaning of the angular momenta in Section~\ref{sec:susy_saturn}. The metric of this solution reads
\begin{equation}\label{eq:saturn_GH_metric}
\begin{aligned}
 ds^2&=-Z^{-2}(d\tau+\omega)^2+Z\,ds_4^2~,\\
 ds_4^2&=H\left(d\varrho^2+\varrho^2d\vartheta^2
                      +\varrho^2\sin^2\vartheta\,d\varphi^2\right)
       +H^{-1}(d\chi+\cos\vartheta\,d\varphi)^2~,\\
 Z&=L+\frac{K^2}{H}\quad,\quad
 \omega=\omega_\chi(d\chi+\cos\vartheta\,d\varphi)
                                     +\widehat\omega_\varphi d\varphi~,\\
 \omega_\chi&=\mathcal M+\frac{3KL}{2H}+\frac{K^3}{H^2}~.
\end{aligned}
\end{equation}
Here $H=1/\varrho$, $\chi$ has period $4\pi$ and the flat four-dimensional radius is $\rho=2\sqrt\varrho$. The functions $K,L,\mathcal M$ are harmonic on the auxiliary flat three-dimensional space. The remaining equation is
\begin{equation}\label{eq:saturn_GH_curl}
 \nabla\times\widehat\omega
     =H\nabla\mathcal M-\mathcal M\nabla H
                         +\frac32(K\nabla L-L\nabla K)~.
\end{equation}
In our gauge convention, the corresponding Maxwell potential can be written locally as
\begin{equation}\label{eq:saturn_GH_gauge}
 A=\sqrt3\left[d\tau-Z^{-1}(d\tau+\omega)
          +\frac KH(d\chi+\cos\vartheta\,d\varphi)+\alpha\right]
 \quad,\quad \nabla\times\alpha=-\nabla K~.
\end{equation}
The one-form $\alpha$ is defined patchwise on the auxiliary space. These expressions use the same normalisation as \eqref{eq:saturn_BMPV}.

We place the ring pole at $\mathbf x=-d\hat z$, where $d=R_{\mathrm b}^2/4>0$, and define $\Sigma=|\mathbf x+d\hat z|$. A convenient parametrisation with no pole of $K$ at the central horizon is
\begin{equation}\label{eq:saturn_harmonic_functions}
\begin{aligned}
 K&=\frac{q}{2\Sigma}~,\\
 L&=1+\frac{\mu}{4\varrho}
                         +\frac{Q_{\mathrm{ring}}-q^2}{4\Sigma}~,\\
 \mathcal M&=-\frac{3q}{4}+\frac{\mathfrak j}{8\varrho}
                      +\frac{3q(R_{\mathrm b}^2+\mu)}{16\Sigma}~,\\
 \widehat\omega_\varphi
   &=-\frac{3q\varrho d\sin^2\vartheta}
                       {2\Sigma(\Sigma+\varrho+d)}~.
\end{aligned}
\end{equation}
Relative to the isolated ring, the additions are $\delta L=\mu H/4$ and $\delta\mathcal M=3\mu K/8+\mathfrak jH/8$. They cancel in the right-hand side of \eqref{eq:saturn_GH_curl}, so $\widehat\omega$ is unchanged. In particular, one must include the term proportional to $q\mu/\Sigma$ in $\mathcal M$ when adding the central black hole.

For completeness, if $m_1$ denotes the residue of $\mathcal M$ at the ring pole, absence of a string singularity in \eqref{eq:saturn_GH_curl} requires
\begin{equation}\label{eq:saturn_integrability}
 \frac{m_1}{d}-\frac{3q}{4}\left(1+\frac{\mu}{4d}\right)=0
 \quad\Longrightarrow\quad
 m_1=\frac{3q(R_{\mathrm b}^2+\mu)}{16}~.
\end{equation}
The central residue condition is then satisfied as well. This is a relation between the ring momentum residue and its position. Exterior causality follows from the horizon bounds by the argument below.

We now choose plane angles by $\chi=\psi+\phi$, $\varphi=\phi-\psi$ and $\vartheta=\pi-2\theta$. The ring lies at $\theta=0$ along $\psi$, as in \eqref{eq:saturn_embedding}. Turning off $q$ and $Q_{\mathrm{ring}}$ gives the BMPV metric with $r^2=\rho^2+\mu$. The asymptotic expansion of \eqref{eq:saturn_GH_metric} gives
\begin{equation}\label{eq:saturn_exact_charges}
\begin{aligned}
 M_{\mathrm{ex}}&=\frac{3\pi}{4G_5}(\mu+Q_{\mathrm{ring}})~,
 &Q_{(0),\mathrm{ex}}&=-\frac{\sqrt3\pi}{4G_5}(\mu+Q_{\mathrm{ring}})~,\\
 J_{\phi,\mathrm{ex}}&=\frac{\pi}{8G_5}
       \left[2\mathfrak j+6q\mu+q(3Q_{\mathrm{ring}}-q^2)\right]~,
 &Q_{(1)}&=\frac{\sqrt3q}{8G_5}~,\\
 J_{\psi,\mathrm{ex}}&=J_{\phi,\mathrm{ex}}
                         +\frac{3\pi qR_{\mathrm b}^2}{4G_5}~.
\end{aligned}
\end{equation}
The spherical horizon at $\varrho=0$ retains the area $2\pi^2\sqrt{\mu^3-\mathfrak j^2}$. Near the ring pole the horizon metric is the product of a two-sphere of radius $q/2$ and a circle of circumference $2\pi L_{\mathrm H}$, where
\begin{equation}\label{eq:saturn_exact_horizons}
\begin{aligned}
 L_{\mathrm H}^2&=3\left[\frac{(Q_{\mathrm{ring}}-q^2)^2}{4q^2}
                                      -R_{\mathrm b}^2-\mu\right]~,\\
 \mathcal A_{\mathrm{ring,ex}}&=2\pi^2q^2L_{\mathrm H}
   =\pi^2q\sqrt{3\left[(Q_{\mathrm{ring}}-q^2)^2
                                   -4q^2(R_{\mathrm b}^2+\mu)\right]}~.
\end{aligned}
\end{equation}\\
\noindent\textbf{Horizon regularity and exterior causality.}
Both horizons are generated by $\partial_\tau$ and have zero temperature and angular velocities. For $q>0$ and $Q_{\mathrm{ring}}-q^2>0$, positivity of their areas requires
\begin{equation}\label{eq:saturn_horizon_bounds}
 \mathfrak j^2<\mu^3\quad,\quad
 Q_{\mathrm{ring}}>q^2+2q\sqrt{R_{\mathrm b}^2+\mu}~.
\end{equation}
The ring must also lie outside the central horizon, so $R_{\mathrm b}^2>0$, or equivalently $R>\sqrt\mu$. At fixed $Q_{\mathrm{ring}}$ and $q$, the second inequality in \eqref{eq:saturn_horizon_bounds} gives an upper bound $R<(Q_{\mathrm{ring}}-q^2)/(2q)$. Thus the radius remains free under force balance but is restricted by horizon regularity.

In \cite{Gauntlett:2004wh} the absence of closed timelike curves was investigated through a numerical scan of selected concentric ring configurations. For the one-hole, one-ring family \eqref{eq:saturn_harmonic_functions}, these horizon bounds also ensure that the full exterior is stably causal. This can be shown directly, without a numerical scan. With $q>0$, $\mu>0$ and $R_{\mathrm b}>0$, the function $Z$ is manifestly positive. It therefore suffices to establish
\begin{equation}\label{eq:saturn_causality}
\begin{aligned}
 \mathcal I&\equiv HZ^3-H^2\omega_\chi^2
                  -\frac{\widehat\omega_\varphi^2}
                         {\varrho^2\sin^2\vartheta}~,\\
 &=\frac34K^2L^2-2K^3\mathcal M+HL^3
          -3HKL\mathcal M-H^2\mathcal M^2
                  -\frac{\widehat\omega_\varphi^2}
                         {\varrho^2\sin^2\vartheta}>0~.
\end{aligned}
\end{equation}
Indeed $g^{\tau\tau}=-\mathcal I/(HZ)$, so positivity of $\mathcal I$ makes $\tau$ a global time function. The integrability condition~\eqref{eq:saturn_integrability} ensures that the time fibration is free of Dirac-Misner strings, without requiring any periodic identification of $\tau$. We take $\tau$ to be globally defined and noncompact.

To prove positivity, introduce the auxiliary parameters
\begin{equation}\label{eq:saturn_causality_parameters}
 h=\frac{\mu}{4}\quad,\quad c=\frac{\mathfrak j}{8}
 \quad,\quad \lambda=\frac{Q_{\mathrm{ring}}-q^2}{2q}~.
\end{equation}
The bounds \eqref{eq:saturn_horizon_bounds} become $|c|<h^{3/2}$ and $\lambda>2\sqrt{d+h}$. Since $\varrho$ and $\Sigma$ are distances from two centres separated by $d$, the triangle inequalities give
\begin{equation}\label{eq:saturn_causality_triangle}
 x=\varrho+\Sigma-d\geq0\quad,\quad
 y=\varrho+d-\Sigma\geq0\quad,\quad
 z=\Sigma+d-\varrho\geq0~.
\end{equation}
In particular, $2\varrho=x+y$, $2\Sigma=x+z$ and $2d=y+z$. Substitution of the harmonic functions into \eqref{eq:saturn_causality} gives a polynomial in $q/2$ at fixed $\lambda$,
\begin{equation}\label{eq:saturn_causality_polynomial}
 \mathcal I=\sum_{n=0}^{4}\left(\frac q2\right)^n\mathcal I_n~.
\end{equation}
The triangle inequalities imply the following bounds on its coefficients,
\begin{equation}\label{eq:saturn_causality_coefficients}
\begin{aligned}
 \mathcal I_0&=\frac{(\varrho+h)^3-c^2}{\varrho^4}>0~,\\
 \mathcal I_1&\geq\frac{3(\varrho+h)}{\varrho^3\Sigma}
                  \left[\lambda(\varrho+h)-2|c|\right]>0~,\\
 \mathcal I_3&\geq\frac{\lambda[\lambda^2-3(d+h)]-2|c|}
                            {\varrho\Sigma^3}>0~,\\
 \mathcal I_4&=\frac{3[\lambda^2-4(d+h)+4\Sigma]}{4\Sigma^4}>0~.
\end{aligned}
\end{equation}
For the remaining coefficient, the same parameter bounds give
\begin{equation}\label{eq:saturn_causality_quadratic}
 4\varrho^2\Sigma^2\mathcal I_2
 \geq 3\left[x^2+3xy+2y^2+4z(x+y)
                      +12hx+6h(y+z)\right]\geq0~.
\end{equation}
To obtain this bound, the numerator is first bounded below by setting $c=h^{3/2}$ and $\lambda=2\sqrt{d+h}$. Its $\lambda$ dependence is increasing throughout the allowed range. One then uses $\sqrt{h(d+h)}\leq h+d/2$ and $xyz/(x+y+z)\leq xy$. Thus every term in \eqref{eq:saturn_causality_polynomial} is nonnegative and $\mathcal I_0$ is strictly positive. This proves $g^{\tau\tau}<0$ throughout the exterior, including the smooth limits on the axes away from the two horizons.

The inequalities \eqref{eq:saturn_horizon_bounds}, together with the stated positive charge branch and $R_{\mathrm b}>0$, are therefore sufficient analytic causality bounds for this Saturn family. The argument uses the specific harmonic functions \eqref{eq:saturn_harmonic_functions}. It does not require a bound on the total angular momenta of the form obeyed by an isolated BMPV black hole.\\

\noindent\textbf{Matching to the blackfold.}
To compare with the blackfold construction, we subtract the fixed central reference charges in \eqref{eq:saturn_central_charges}. The coordinate relation $r^2=\rho^2+\mu$ identifies $R_{\mathrm b}^2=R^2-\mu$, while matching the asymptotic electric charge fixes $Q_{\mathrm{ring}}=2RQ_{\mathrm R}$, as in \eqref{eq:saturn_radius_dictionary}. The exact mass, electric charge and dipole charge then reproduce \eqref{eq:saturn_ring_charges}. Each angular momentum is compared with the blackfold charge for the same asymptotic axial generator. Subtracting \eqref{eq:saturn_total_charges} from \eqref{eq:saturn_exact_charges}, and rewriting \eqref{eq:saturn_exact_horizons} in the blackfold parameters, gives
\begin{equation}\label{eq:saturn_finite_corrections}
\begin{aligned}
 J_{i,\mathrm{ex}}-J_{i,\mathrm{BF}}
       &=\frac{\pi q}{8G_5}(3Q_{\mathrm{ring}}-q^2)
                         \quad (i=\psi,\phi)~,\\
 \mathcal A_{\mathrm{ring,ex}}
       &=2\pi^2\sqrt3 Rq
           \sqrt{\left(Q_{\mathrm R}-\frac{q^2}{2R}\right)^2-q^2}~.
\end{aligned}
\end{equation}
The angular momentum correction is the same in both planes. Consequently the difference $J_\psi-J_\phi$ agrees exactly under this parameter map, although each angular momentum agrees only to leading blackfold order. The exact $J_\phi$ in~\eqref{eq:saturn_exact_charges} contains the mixed term $6q\mu$, which contributes $3\pi q\mu/(4G_5)$ to the angular momentum. The blackfold reproduces this ring-hole interaction term in~\eqref{eq:saturn_transverse_spin} through its coupling to the background gauge fields, using the regular one-form $\eta_{\chi_\phi}$ in~\eqref{eq:saturn_rotation_regular}. The ratio of this interaction term to the leading longitudinal ring momentum $J_{\psi,\mathrm{ring}}=3\pi qR^2/(4G_5)$ is $\mu/R^2$, so it survives the thin limit at fixed nonzero $\mu/R^2$. The corrections reduce to those of the isolated supersymmetric ring when $\mu=\mathfrak j=0$.

The blackfold entropy integral uses $-n\cdot u\,dV_1=R\,d\psi$. Comparing its area in \eqref{eq:saturn_ring_charges} with the exact result gives
\begin{equation}\label{eq:saturn_area_matching_prescription}
 \frac{\mathcal A_{\mathrm{ring,ex}}}{\mathcal A_{\mathrm{ring,BF}}}
   =\sqrt{\frac{\left(Q_{\mathrm R}-q^2/(2R)\right)^2-q^2}
                       {Q_{\mathrm R}^2-q^2}}
   =1-\frac{Q_{\mathrm R}q^2}{2R(Q_{\mathrm R}^2-q^2)}
                           +\mathcal O\left(\frac{Q_{\mathrm R}^2}{R^2}\right)~.
\end{equation}
The expansion holds at fixed $0<q/Q_{\mathrm R}<1$ and $Q_{\mathrm R}/R\ll1$, away from the zero area boundary. Matching instead the local pole coefficient would replace $Q_{\mathrm R}$ by $(Q_{\mathrm{ring}}-q^2)/(2R)$, which differs at subleading order from the asymptotic-charge prescription used here. The same happens for the isolated ring as discussed in Appendix~\ref{app:exact_susy_ring}.\\

\noindent\textbf{Validity of the approximation.}
For the aligned string, the thickness estimate of Section~\ref{sec:string_validity} is $r_\epsilon=3Q_{\mathrm R}$. Dividing \eqref{eq:saturn_extrinsic_components} by $f_R$ converts its radial index to a unit normal and gives the bending conditions
\begin{equation}\label{eq:saturn_validity}
 3Q_{\mathrm R}\max\left\{\frac{2\mu}{R^3},
                    \frac{|\mathfrak j|}{R^4},\frac{f_R}{R}\right\}\ll1~.
\end{equation}
The background curvature, flux and separation bounds in \eqref{eq:string_validity_local_geometry}--\eqref{eq:string_validity_fluxes} must also hold. For bounded $\mathfrak j/\mu^{3/2}$, $Q_{\mathrm R}/R\ll1$ controls the local expansion. An additional sufficient restriction for treating the central hole as a fixed background is $Q_{\mathrm{ring}}/R_{\mathrm b}^2\ll1$, which makes the ring at the centre weak. We note that these conditions do not require $R\gg\sqrt\mu$ but they fail sufficiently close to the central horizon at fixed ring charges. \\

\noindent\textbf{Interpretation of Figure~\ref{fig:susy_saturn_comparison}.}
The figure fixes $\mu/q^2=1600$, $Q_{\mathrm{ring}}/q^2=400$ and $\mathfrak j/\mu^{3/2}=1/2$. The exact and blackfold states therefore have the same total mass, electric charge and dipole charge, with the same central horizon area. In the upper panels, eliminating the radii separately in favour of the same $J_{\psi,\mathrm{tot}}$ gives
\begin{equation}\label{eq:saturn_area_fixed_spin}
 \left.\left(\mathcal A_{\mathrm{ring,ex}}^2
                  -\mathcal A_{\mathrm{ring,BF}}^2\right)
                       \right|_{J_{\psi,\mathrm{tot}}}
       =\pi^4q^6~.
\end{equation}
This cancellation explains the close agreement of the upper curves. Adding the common central horizon area leaves the difference between the total areas equal to the difference between the ring areas.  The lower panels instead compare states at the same radius. Their angular momentum differences follow directly from \eqref{eq:saturn_finite_corrections}, while the ring areas obey
\begin{equation}\label{eq:saturn_area_error}
 \frac{\mathcal A_{\mathrm{ring,BF}}}{\mathcal A_{\mathrm{ring,ex}}}
 =\left[1+\frac{2Q_{\mathrm{ring}}q^2-q^4}
 {(Q_{\mathrm{ring}}-q^2)^2-4q^2R^2}\right]^{1/2}~.
\end{equation}
At $R/q=199$, the blackfold ring area is about $42\%$ too large, although $3Q_{\mathrm R}/R\simeq0.015$ and the longitudinal angular momentum differs by only $0.40\%$. The exact area vanishes at $R/q=199.5$, whereas the blackfold area remains positive until $R/q=200$. Thus the relative area error grows without bound as the exact endpoint is approached. 

For the plotted parameters, the thickness cutoff $3Q_{\mathrm R}/R\leq0.1$ requires $R/q\geq\sqrt{6000}$. Smaller radii are shown by dashed segments. The condition $Q_{\mathrm{ring}}/(R^2-\mu)\ll1$ also fails as the ring approaches the central horizon, where its perturbation of the background is no longer weak. 

For the exact family in Figure~\ref{fig:susy_saturn_comparison}, $\mathfrak j/\mu^{3/2}=1/2$ satisfies the central causality bound. The bound on the ring parameters becomes $40<R/q<199.5$, where the lower limit places the ring outside the central horizon and the upper limit ensures positive ring area. The analytic argument above therefore establishes exterior causality throughout this interval, including the entire plotted range before the zero-area endpoint.

\subsubsection{Supersymmetric di-rings and bi-rings}\label{app:exact_susy_multirings}\label{app:susy_multirings}
We now replace the central black hole background by an exact supersymmetric ring and add a thin supersymmetric string. Wrapping the string in the same plane gives a di-ring, while wrapping it in the orthogonal plane gives a bi-ring. The exact solutions of~\cite{Gauntlett:2004wh} allow any number of rings and an optional central black hole, so the construction also applies to these more general backgrounds. The comparison is particularly direct: adding the blackfold ring corresponds to adding one small pole to the harmonic functions of the exact solution.\\

\noindent\textbf{The supersymmetric background.}
We use the metric and angular coordinates of \eqref{eq:saturn_GH_metric}, writing $t=\tau$. Each ring is represented by a pole at $\mathbf x_i=-s_i d_i\hat z$ in the auxiliary three-dimensional space, with $d_i>0$ and $s_i=\pm1$. In the physical four-dimensional base, $s_i=+1$ describes a circle along $\psi$, while $s_i=-1$ describes a circle along $\phi$. Define $\Sigma_i=|\mathbf x+s_i d_i\hat z|$, $R_i^2=4d_i$ and $q_T=\sum_iq_i$. For $N$ rings a convenient family is
\begin{equation}\label{eq:multi_susy_harmonics}
\begin{aligned}
 H&=\frac1\varrho~,
 &K&=\sum_{i=1}^N\frac{q_i}{2\Sigma_i}~,\\
 L&=1+\frac\mu{4\varrho}+\Lambda K~,
 &\mathcal M&=-\frac{3q_T}{4}+\frac{\mathfrak j}{8\varrho}
                  +\frac34\sum_{i=1}^N
                       \frac{q_i(d_i+\mu/4)}{\Sigma_i}~,\\
 \widehat\omega_\varphi&=\frac34\sum_{i=1}^Nq_i
        \left(1-\frac{\varrho+d_i}{\Sigma_i}\right)
                               (\cos\vartheta+s_i)~.
\end{aligned}
\end{equation}
The parameters $\mu$ and $\mathfrak j$ describe the optional central black hole and are set to zero when it is absent. Each $q_i>0$ is a magnetic parameter. The electric parameters are restricted by $Q_i-q_i^2=2\Lambda q_i$, with the same positive $\Lambda$ for all rings. This $\Lambda$ is distinct from the thermal gauge potentials of Section~\ref{sec:equilibrium_action}.

The common ratio makes the pairwise terms in \eqref{eq:saturn_GH_curl} cancel. The coefficients in $\mathcal M$ then ensure regularity of the time fibration on the rotation axes, removing the Dirac-Misner strings~\cite{Gauntlett:2004wh}. This specifies the integrability conditions used here. The electric parameters cannot be varied independently while keeping the remaining harmonic functions unchanged. For the charge comparison we choose the Maxwell potential to vanish at infinity,
\begin{equation}\label{eq:multi_susy_gauge}
\begin{aligned}
 A&=\sqrt3\bigg[dt-Z^{-1}(dt+\omega)
       +\frac KH(d\chi+\cos\vartheta\,d\varphi)\\
   &\hspace{2.6cm}+\alpha-\frac{q_T}{2}d\chi\bigg]~,\\
 \alpha&=-\frac12\sum_iq_i
             \frac{\varrho\cos\vartheta+s_id_i}{\Sigma_i}\,d\varphi~.
\end{aligned}
\end{equation}
The angular one-forms are defined in regular patches, as in Appendix~\ref{app:exact_susy_saturn}. In particular, the constant term proportional to $q_Td\chi$ fixes the reference gauge for the Page charge.\\

\noindent\textbf{Embedding and force balance.}
We now place the new string at $\mathbf x_*=-s_*d_*\hat z$, away from all existing poles, and define $R_*^2=4d_*$. A star on a background field denotes its value at this point. For a background ring along $\psi$, choosing $s_*=+1$ and $d_*>\max_i d_i$ adds an outer coplanar ring. Choosing $s_*=-1$ instead adds an orthogonal ring. The string wraps the Gibbons-Hawking fibre. In terms of the physical angle $\varphi_*$ of period $2\pi$, equal to $\psi$ or $\phi$ in the two cases, its induced metric is
\begin{equation}\label{eq:multi_susy_induced}
 ds^2_\Sigma=-Z_*^{-2}(dt+2\omega_{\chi,*}d\varphi_*)^2
                  +\mathcal R_*^2d\varphi_*^2
 \quad,\quad \mathcal R_*=2\sqrt{\frac{Z_*}{H_*}}~.
\end{equation}
The supersymmetric horizon generator is $k=\partial_t$. Thus $u=Z_*\partial_t$ and $|k|=Z_*^{-1}$, while $v$ is the unit tangent orthogonal to $u$, oriented along increasing $\varphi_*$. The temperature vanishes and the equilibrium potentials are $\Phi_0=-\sqrt3$ and $\Phi_1=0$. The quantity $\mathcal R_*d\varphi_*$ measures spatial distance orthogonal to $u$. Because of the cross term in the metric, $\mathcal R_*$ differs from the proper radius on a constant-$t$ slice.

To match the added string to the exact family, its local charges must include the effect of the background fields. Adding a small pole to $K$ changes $Z=L+K^2/H$ through both $\delta L$ and the mixed term $2K\delta K/H$. Converting to proper local distances gives
\begin{equation}\label{eq:multi_susy_local_dictionary}
 Q_{\mathrm R,*}=\frac{q_*}{2}\sqrt{\frac{H_*}{Z_*}}
                   \left(\Lambda+\frac{2K_*}{H_*}\right)~,
 \qquad q_{\rm local}=q_*~.
\end{equation}
Here $q_*$ is the magnetic parameter of the new ring. This relation fixes its local electric parameter $Q_{\mathrm R,*}$ for the chosen position.

In the orthonormal tangent frame $(u,v)$, the supersymmetric background satisfies
\begin{equation}\label{eq:multi_susy_force_identities}
 F_{i0}=-\sqrt3 K_{00i}\quad,\quad
 2F_{i1}+\mathcal F_{i01}=-2\sqrt3 K_{01i}~.
\end{equation}
for every normal direction $i$. Substituting the string stress tensor and currents from \eqref{eq:saturn_local_monopoles}, with $T^{11}=0$, gives
\begin{equation}\label{eq:multi_susy_no_force}
 T^{AB}K_{ABi}=3\mathcal C Q_{\mathrm R,*}K_{00i}
                         +3\mathcal Cq_*K_{01i}
 =F_{iA}J^A+\frac12\mathcal F_{iAB}\mathcal J^{AB}~.
\end{equation}
The gravitational, electric and dual magnetic forces therefore cancel for either embedding, including when $K_{01i}\ne0$. The force equation imposes no further condition on the radius. A nonzero horizon area does impose a restriction, since
\begin{equation}\label{eq:multi_susy_local_bound}
 Q_{\mathrm R,*}^2-q_*^2
   =\frac{q_*^2H_*}{4Z_*}
                   (\Lambda^2-4d_*-\mu)>0~.
\end{equation}
Thus the local string bound $Q_{\mathrm R,*}>q_*$ is precisely $R_*^2+\mu<\Lambda^2$.\\

\noindent\textbf{Charges and comparison with the exact solution.}
The exact background has asymptotic charges
\begin{equation}\label{eq:multi_susy_exact_charges}
\begin{aligned}
 \mathcal Q&=\mu+2\Lambda q_T+q_T^2~,
 \qquad M=\frac{3\pi\mathcal Q}{4G_5}=-\sqrt3Q_{(0)}~,\\
 J_\phi&=\frac\pi{8G_5}
 \left[2\mathfrak j+6\mu q_T+2q_T^3+6\Lambda q_T^2
                       +3\sum_iq_iR_i^2(1-s_i)\right]~,\\
 J_\psi&=J_\phi+\frac{3\pi}{4G_5}\sum_i s_iq_iR_i^2~.
\end{aligned}
\end{equation}
To identify the contribution of the new ring, we hold $\mu$, $\mathfrak j$, $\Lambda$ and all original ring parameters fixed. The blackfold charge integrals give
\begin{equation}\label{eq:multi_susy_probe_charges}
\begin{aligned}
 \delta M&=\frac{3\pi}{2G_5}(\Lambda+q_T)q_*
                      =-\sqrt3\delta Q_{(0)}~,\\
 \delta J_\phi&=\frac{\pi q_*}{8G_5}
       [6\mu+6q_T^2+12\Lambda q_T+3R_*^2(1-s_*)]~,\\
 \delta J_\psi&=\delta J_\phi+\frac{3\pi q_*R_*^2s_*}{4G_5}~,
 &Q_{(1),*}&=\frac{\sqrt3q_*}{8G_5}~.
\end{aligned}
\end{equation}
The gauge contributions to these currents are treated as in Appendix~\ref{app:exact_susy_saturn}. They include angular momentum carried by the surrounding fields, so the added ring generally changes both angular momenta. We compare each with the exact charge associated with the same asymptotic rotation generator.

The exact comparison now follows by replacing $q_T$ with $q_T+q_*$ in \eqref{eq:multi_susy_exact_charges} and including the new ring in the sums. Expanding to first order in $q_*$ reproduces \eqref{eq:multi_susy_probe_charges} while higher powers of $q_*$ lie beyond this leading blackfold comparison. For the area, each ring in the exact family has
\begin{equation}\label{eq:multi_susy_horizon_areas}
 \mathcal A_i=2\pi^2q_i^2\sqrt{3(\Lambda^2-R_i^2-\mu)}~,
\end{equation}
while the blackfold entropy integral for the added string yields
\begin{equation}\label{eq:multi_susy_probe_area}
 \mathcal A_*=2\pi^2\sqrt3\mathcal R_*q_*
                         \sqrt{Q_{\mathrm R,*}^2-q_*^2}
       =2\pi^2q_*^2\sqrt{3(\Lambda^2-R_*^2-\mu)}~.
\end{equation}
where the second equality uses \eqref{eq:multi_susy_local_bound}. It agrees with the exact area of the new ring under the stated charge identification. This equality is a property of this supersymmetric family and does not imply that all finite-thickness corrections vanish. The exact charges satisfy $M=-\sqrt3Q_{(0)}$, and the additional blackfold rings preserve this relation. These configurations are supersymmetric, not only because of the exact harmonic construction, but also because they satisfy the BPS bound. These comparisons underlie the di-ring and bi-ring curves in Figure~\ref{fig:phase_bps_comparison}. The validity regime of these configurations is discussed below.\\

\noindent\textbf{Regularity and validity.}
The free radius in the force equation does not mean that every placement gives a regular solution. All ring poles must be distinct and obey $R_i^2+\mu<\Lambda^2$. If the central black hole is present, it must satisfy $\mathfrak j^2<\mu^3$ and has area $2\pi^2\sqrt{\mu^3-\mathfrak j^2}$. Exterior causality must also satisfy \eqref{eq:saturn_causality}, evaluated with the full harmonic functions. The horizon inequalities and the BPS relation alone do not establish this condition for the general multi-ring family. Finally, the blackfold description requires the new string to be thin compared with the local bending, curvature and flux lengths, and with its proper separation from the existing horizons, as in Section~\ref{sec:string_validity}. Its perturbation of the original horizons must remain small. These conditions determine which part of the exact family is described by adding a thin string to a fixed background.

\subsection{Non-extremal solutions}\label{app:exact_nonext_limit}
We detail the comparison mentioned in Section~\ref{sec:exact_nonext_comparison}. The three-parameter Suzuki-Tomizawa solution~\cite{Suzuki:2024chargedring} provides a non-trivial check on the blackfold solutions, while the four-parameter family~\cite{Suzuki:2024generalring} permits vanishing transverse horizon angular velocity at finite radius. We give the local geometry and exact thermodynamics, compare states at matched charges and classify the extremal limits.

\subsubsection{Charge conventions and the local string limit}\label{app:ST_local_limit}
We denote the electric and magnetic surface integrals of~\cite{Suzuki:2024chargedring,Suzuki:2024generalring} by $Q_{\rm ST}$ and $q_{\rm ST}$. With the orientations used here, their relation to our charges is\footnote{In particular, their electric charge has twice our normalisation.}
\begin{equation}\label{eq:ST_charge_conventions}
 Q_{(0)}=\frac12Q_{\rm ST}\quad,\quad
 Q_{(1)}=-\frac{q_{\rm ST}}{4G_5}\quad,\quad
 q=-\frac{2q_{\rm ST}}{\sqrt3}~.
\end{equation}
We use $b\in(0,1)$ for their Harrison parameter, to distinguish it from the string parameter $\beta$. Both exact families obey
\begin{equation}\label{eq:ST_mass_ratio}
 \frac{M}{M_{\mathrm{BPS}}}=\frac{1+b^2}{2b}~.
\end{equation}
First consider the three-parameter solution, with scale $\ell_{\rm ST}$ and shape parameter $n$. Its local limit is obtained by taking
\begin{equation}\label{eq:ST_thin_scaling}
\begin{gathered}
 R=\sqrt2\ell_{\rm ST}\to\infty\quad,\quad
 n\to0\quad,\quad k=Rn\ \text{fixed}~,\\
 y_{\rm ST}=-\frac Rr\quad,\quad x_{\rm ST}=\cos\theta
 \quad,\quad z=R\psi~.
\end{gathered}
\end{equation}
Here $R$ is the embedding radius to leading order. It is not the proper radius of the horizon circle. The metric and gauge field reduce to the black string in \cite{Compere:2010fm}. The parameter map is
\begin{equation}\label{eq:ST_string_dictionary}
\begin{gathered}
 2m=k\quad,\quad \gamma=\gamma_+~,\\
 c_\delta^2\mathsf B=\frac{1+b^2}{1-b^2}
 \quad,\quad
 c_\delta^2\mathsf X=
       \frac{\sqrt2\,b^2}{\sqrt{(1-b^2)(1-b^6)}}~.
\end{gathered}
\end{equation}
The two equations in the second line fix $\beta$ and $\delta$ in terms of the single parameter $b$. They constitute a restriction on the local electric and magnetic charges. In the coordinates \eqref{eq:ST_thin_scaling}, we define $c=(1-b^2)^{-1/2}$, $s=bc$, $f=1-k/r$ and
\begin{equation}\label{eq:ST_local_functions}
\begin{aligned}
 U&=1-b^6+b^6\frac{k^2}{r^2}~,\\
 V&=(1-b^6)\left(1-\frac{2k}{r}\right)-b^6\frac{k^2}{r^2}~,\\
 N&=c^2U-s^2V~,\\
 W&=\frac{\sqrt2 k\sqrt{1-b^6}}r
       \left(\frac{c^3}{V}-\frac{s^3b^3f}{U}\right)~.
\end{aligned}
\end{equation}
Expanding the three-parameter metric of~\cite{Suzuki:2024chargedring} gives
\begin{equation}\label{eq:ST_local_metric}
 ds^2=-\frac{UV}{N^2}(dt+Wdz)^2
       +\frac{(1-b^6)fN}{UV}\,dz^2
       +\frac{N}{1-b^6}
          \left(\frac{dr^2}{f}+r^2d\Omega_2^2\right)~,
\end{equation}
with gauge potential
\begin{equation}\label{eq:ST_local_gauge}
\begin{aligned}
 A={}&\frac{\sqrt3cs(U-V)}N\,dt
 -\frac{\sqrt6cs k\sqrt{1-b^6}}{rN}(c-sb^3f)\,dz
 -Q_M(1+\cos\theta)\,d\phi~,\\
 Q_M={}&\frac{\sqrt6 k b^2}{\sqrt{(1-b^2)(1-b^6)}}~.
\end{aligned}
\end{equation}
The magnetic potential is written in a south-pole patch and has flux $4\pi Q_M$. Apparent poles at $V=0$ cancel between the terms of the longitudinal metric. The horizon is at $r=k$. In particular, all $dt\,d\phi$ and $dz\,d\phi$ components vanish in this limit, although the finite-radius ring has a nonzero $J_\phi$.

Under \eqref{eq:ST_string_dictionary}, the radial function of the black string solution \cite{Compere:2010fm} satisfies
\begin{equation}\label{eq:ST_radial_match}
 H=\frac{r^2N}{1-b^6}\quad,\quad \Delta=r(r-k)~.
\end{equation}
The longitudinal components of \eqref{eq:ST_local_metric} and \eqref{eq:ST_local_gauge} agree with \eqref{eq:metricofnonextremal} and \eqref{eq:nonext_rest_gauge} when their coframe uses $\gamma_+$ in place of $\chi$. This identifies the full non-rotating local geometry, not just its integrated charges. We extract the stress tensor and currents from the large-$r$ fields using Appendix~\ref{app:monopole_extraction}, and obtain the thermodynamic quantities from the horizon.

In the asymptotic frame of the ring, the limiting stress tensor and currents are
\begin{equation}\label{eq:ST_local_monopoles}
\begin{aligned}
 T^{tt}&=3\mathcal Ck\frac{1+b^2}{1-b^2}~,
 &T^{tz}&=\sqrt2\mathcal Ck\frac{\sqrt{1-b^6}}{(1-b^2)^{3/2}}~,
 &T^{zz}&=0~,\\
 J^t&=-2\sqrt3\mathcal Ck\frac{b}{1-b^2}~,
 &J^z&=-\sqrt6\mathcal Ck
       \frac{b(1+b^2)}{(1-b^2)\sqrt{1+b^2+b^4}}~,
 &\mathcal J^{tz}&=-\rho_1~,\\
 \rho_1&=\frac{\sqrt6\mathcal Ck b^2}
                    {\sqrt{(1-b^2)(1-b^6)}}~.
\end{aligned}
\end{equation}
Thus the extrinsic equation is satisfied directly by the limit of the exact metric. The horizon velocity and asymptotic temperature are
\begin{equation}\label{eq:ST_local_thermal}
 w=\frac{1-b^2}{\sqrt{2(1+b^2+b^4)}}
 \quad,\quad
 T_0=\frac{(1-b^2)^{3/2}\sqrt{1-b^6}}{4\sqrt2\pi k}~.
\end{equation}
Boosting by $\eta=\operatorname{arctanh}w$ gives the thermal rest frame of Section~\ref{sec:non-ext_string}. Integrating \eqref{eq:ST_local_monopoles} gives $M=LT^{tt}$, $J=LR T^{tz}$ and $Q_{(0)}=LJ^t$, while the leading area is
\begin{equation}\label{eq:ST_thin_area}
 \mathcal A_0=\frac{8\sqrt2\pi^2Rk^2}
                       {(1-b^2)^{3/2}\sqrt{1-b^6}}~.
\end{equation}
These quantities agree with the positive momentum solution of Section~\ref{sec:flat_ring_nonext_plots} after applying \eqref{eq:ST_string_dictionary}.

The four-parameter exact solution has an additional shape parameter $\lambda$ \cite{Suzuki:2024generalring}. Its thin limit takes $\lambda,n\to0$ with $\zeta=\lambda/n>0$ fixed. The appropriate local scale and the rotation parameter of the transverse Kerr seed are
\begin{equation}\label{eq:ST_spinning_thin}
 k=\frac{Rn(1+\zeta^2)}{2\zeta}
 \quad,\quad m=\frac k2
 \quad,\quad \frac{a_\perp}{m}=\frac{\zeta^2-1}{\zeta^2+1}~.
\end{equation}
The leading mass, longitudinal momentum, electric current and magnetic charge still obey \eqref{eq:ST_string_dictionary} and \eqref{eq:ST_local_monopoles}. The additional freedom is transverse rotation, not an independent local charge ratio. The limiting horizon data are
\begin{equation}\label{eq:ST_spinning_horizon}
\begin{gathered}
 \frac{\mathcal A}{\mathcal A_0}
 =\frac{(1+\zeta)^2+b^6(1-\zeta)^2}{2(1+\zeta^2)}~,\\
 T_0=\frac{\sqrt2\zeta(1-b^2)^{3/2}\sqrt{1-b^6}}
 {2\pi k\big[(1+\zeta)^2+b^6(1-\zeta)^2\big]}~,\\
 \omega_\phi=\frac{(\zeta^2-1)(1-b^2)^{3/2}\sqrt{1-b^6}}
 {\sqrt2 k\big[(1+\zeta)^2+b^6(1-\zeta)^2\big]}~.
\end{gathered}
\end{equation}
Our non-rotating local string therefore corresponds to $\zeta=1$. For generic $\zeta$, a blackfold theory including transverse spin is needed to reproduce the horizon area. A finite global $J_\phi$ does not by itself establish a finite local spin, since the three-parameter solution also has a subleading angular momentum carried by the fields.

\subsubsection{Exact thermodynamics and vanishing transverse rotation}\label{app:ST_exact_thermodynamics}
For reproducibility, we also specify the exact expressions used for the finite-radius comparison. In the four-parameter representation of~\cite{Suzuki:2024generalring}, write
\begin{equation}\label{eq:ST_exact_auxiliary}
\begin{gathered}
 h_1=\lambda^2+(1-\lambda^2)n^2\quad,\quad
 h_2=\lambda+(1-\lambda)n^2\quad,\quad
 D=(1-\lambda)\lambda(1-n^2)(1-b^6)~,\\
 h_{\pm4}=\lambda\pm(1-\lambda)n\quad,\quad
 h_{+6}=n(1-\lambda)(1-b^3)+\lambda(1+b^3)~.
\end{gathered}
\end{equation}
Then it follows that
\begin{equation}\label{eq:ST_exact_charge_formulae}
\begin{aligned}
 M&=\frac{3\pi\ell_{\rm ST}^2(c^2+s^2)h_1(D+h_1)}{2G_5Dh_2}~,\\
 Q_{\rm ST}&=-\frac{4Mb}{\sqrt3(1+b^2)}~,
 &q_{\rm ST}&=-\frac{\sqrt3\ell_{\rm ST}s^2h_1}{c\sqrt{Dh_2}}~.
\end{aligned}
\end{equation}
The horizon area is $\mathcal A=8\pi^2\ell_{\rm ST}^2n/\kappa_H$, with
\begin{equation}\label{eq:ST_exact_surface_gravity}
\begin{gathered}
 \kappa_H=\frac{2n(1-n^2)
       \sqrt{(1-b^6)(1-\lambda)^3\lambda^3h_2^3}}
                    {\ell_{\rm ST}h_1\mathcal D_H}~,\\
 \mathcal D_H=c^3(1+n)h_2h_{+4}^2
       +s^3(1-n)h_{-4}
          \big[(1-\lambda)\lambda(1-n^2)h_{+6}-h_2h_{+4}\big]~.
\end{gathered}
\end{equation}
We evaluate the angular momenta using Eqs.~(55)--(57) of~\cite{Suzuki:2024generalring}, restoring $G_5^{-1}$. In using the other horizon formulas of that paper, the combination required is
\begin{equation}\label{eq:ST_h5_check}
 h_{+5}=n(1+b^3)+h_2(1-b^3)~.
\end{equation}
The opposite sign printed in its definition (47) is inconsistent with the positive surface gravity and with the neutral non-rotating ring. Equation~\eqref{eq:ST_exact_surface_gravity} instead uses the equivalent denominator appearing in its magnetic potential formula, Eq.~(127) of version~3. With \eqref{eq:ST_h5_check}, these expressions obey the Smarr relation and the first law under independent variations of all four parameters.

The restriction of vanishing transverse horizon rotation is obtained from the exact angular velocities,
\begin{equation}\label{eq:ST_exact_angular_velocities}
 (\Omega_\psi,\Omega_\phi)
 =\frac{\kappa_H}{4(1-\lambda)\lambda n\sqrt{1-n^2}(1-b^6)h_2}
                   (h_1h_{+5}h_{+6},h_{+7})~,
\end{equation}
where
\begin{equation}\label{eq:ST_exact_h7}
\begin{aligned}
 h_3={}&(2-\lambda)\lambda+(1-\lambda)^2n^2~,\\
 h_{+7}={}&\lambda n(1+b^3)^2h_1
       -(1-\lambda)n(1-b^3)^2h_1h_2
       +(1-b^6)h_3h_{-4}h_{+4}~.
\end{aligned}
\end{equation}
In the non-extremal interior $0<n,\lambda,b<1$, the common prefactor in \eqref{eq:ST_exact_angular_velocities} is nonzero. Hence $\Omega_\phi=0$ is exactly $h_{+7}=0$. Substituting $\lambda=n+\alpha n^2+\mathcal O(n^3)$ gives
\begin{equation}\label{eq:ST_h7_thin_expansion}
 h_{+7}=4n^4\big[(1-b^6)(\alpha+1)+2b^3\big]
                         +\mathcal O(n^5)~,
\end{equation}
so $h_{+7}=0$ gives
\begin{equation}\label{eq:ST_zero_transverse_shape}
 \lambda=n-\left(1+\frac{2b^3}{1-b^6}\right)n^2
                      +\mathcal O(n^3)~.
\end{equation}
Thus $\lambda/n\to1$, and \eqref{eq:ST_spinning_thin} gives $a_\perp/m\to0$. The limiting geometry, charges and thermodynamics are precisely \eqref{eq:ST_string_dictionary}--\eqref{eq:ST_thin_area}. This expansion is taken at fixed $b$ and is not uniform as $b\to1$.

The three-parameter subfamily is instead the locus $\lambda=n/(1+n)$. There $h_{-4}=0$, $h_2=n$ and $h_{+7}=b^3h_1h_{+5}h_{+6}$, giving
\begin{equation}\label{eq:ST_three_angular_velocities}
 \Omega_\phi=b^3\Omega_\psi~.
\end{equation}
For a charged non-extremal ring it therefore has nonzero transverse horizon rotation at finite radius, even though its thin limit has $a_\perp=0$. It supplies the explicit local metric check above. The comparison in Figure~\ref{fig:exact_nonext_comparison} instead enforces $\Omega_\phi=0$ exactly. \footnote{There is a related normalisation issue in the three-parameter presentation. The direct expansion of its metric, using its asymptotically flat coordinates, gives $M=\frac{3\pi n\ell_{\rm ST}^2[1+n-(1-n)b^6]}{G_5(1-n^2)(1-b^6)} \frac{1+b^2}{1-b^2}$.
Relative to the expressions printed in Eqs.~(78)--(80) of~\cite{Suzuki:2024chargedring}, this divides the mass by $1+n$ and both angular momenta by $(1+n)^{3/2}$. The area (85) is unchanged. These are the expressions consistent with the metric, its reduced-area formula (93), and the first law. They also coincide with \eqref{eq:ST_exact_charge_formulae} and the four-parameter angular momenta at $\lambda=n/(1+n)$, with the same $\ell_{\rm ST}$. As an independent neutral check, they give $j^2=(1+n)^3/(8n)$ and $a_H=2\sqrt{n(1-n)}$ at $b=0$.}

\subsubsection{Comparison at matched charges}\label{app:ST_charge_matching}
We compare the $\Omega_\phi=0$ exact rings with the corresponding non-rotating blackfold. The total transverse angular momentum $J_\phi$ can be nonzero because the electromagnetic field carries angular momentum. The restriction on local charge ratios in \eqref{eq:ST_string_dictionary} selects a subset of the positive-momentum blackfold branch.

Writing $d=q^2/Q_\star$ and fixing a mass ratio $\nu>1$ as in \eqref{eq:ring_mass_excess}, the exact charge formulas \eqref{eq:ST_exact_charge_formulae} imply
\begin{equation}\label{eq:ST_exact_charge_constraint}
 \frac{q^2}{Q_\star}=\frac{b^3h_1}{D+h_1}~.
\end{equation}
Fixing $M,Q_{(0)},Q_{(1)}$ then gives
\begin{equation}\label{eq:ST_fixed_charge_curve}
\begin{gathered}
 b=\nu-\sqrt{\nu^2-1}\quad,\quad
 K=\frac{d(1-b^6)}{b^3-d}~,\\
 \frac{n^2}{1-n^2}=K\lambda-(K+1)\lambda^2
 \quad,\quad 0<\lambda<\frac K{K+1}~.
\end{gathered}
\end{equation}
We solve \eqref{eq:ST_fixed_charge_curve} together with $h_{+7}=0$, then fix $\ell_{\rm ST}$ by the prescribed $q$. This requires $d<b^3$. The corresponding blackfold state is selected by \eqref{eq:ST_string_dictionary}, with the same three charges. Its area and longitudinal angular momentum are predictions to be compared with those of the exact ring, rather than further matching conditions. No transverse-spin parameter is varied independently in this comparison.

The three-parameter subfamily would instead select
\begin{equation}\label{eq:ST_three_parameter_point}
 \lambda=\frac n{1+n}\quad,\quad
 n=\frac{d(1-b^6)}{2b^3-d(1+b^6)}~.
\end{equation}

To obtain continuous curves, we hold $M,Q_{(0)}$ fixed and vary $d$, matching the dipole separately between each exact ring and its blackfold counterpart. Thus $Q_{(1)}$ varies along a curve but agrees within every compared pair. This fixes $b$ and hence the local string parameters $\beta,\delta$. Using \eqref{eq:ST_local_monopoles} and \eqref{eq:ST_thin_area}, the selected blackfold has
\begin{equation}\label{eq:ST_zero_transverse_reduced}
 j_{\rm BF}^2=\frac{b^3}{4d(1+b^2)^3}
 \quad,\quad
 a_{H,\rm BF}^2=\frac{2d}{b^3(1+b^2)^3}~.
\end{equation}
In particular, $a_{H,\rm BF}j_{\rm BF}=1/[\sqrt2(1+b^2)^3]$. The thickness parameter of this matched blackfold is
\begin{equation}\label{eq:ST_zero_transverse_thinness}
 \frac{k}{R}=\frac{d(1-b^6)}{2b^3}
 \quad,\quad
 \frac{r_\epsilon(n)}R
 =\frac{3d(1+b^2)(1-b^6)}{2b^3(1-b^2)}~.
\end{equation}
The limit $d\to0$ at fixed $b$ therefore approaches a thin ring with large reduced angular momentum. These curves keep the local charge ratio fixed. The curves in Figure~\ref{fig:nonext_fixed_charges} instead hold the dipole fixed while varying the local string parameter $\beta$.

The right panel of Figure~\ref{fig:exact_nonext_comparison} compares states paired by their common $M,Q_{(0)},Q_{(1)}$, using
\begin{equation}\label{eq:ST_zero_transverse_errors}
 \Delta_J=\frac{J_\psi-J_{\rm BF}}{J_{\rm BF}}
 \quad,\quad
 \Delta_{\mathcal A}=\frac{\mathcal A_{\rm ex}-\mathcal A_{\rm BF}}
                              {\mathcal A_{\rm BF}}~.
\end{equation}
Both differences vanish as \eqref{eq:ST_zero_transverse_thinness} tends to zero. The exact temperature and longitudinal horizon angular velocity also approach the blackfold predictions. These checks compare the same local non-rotating equation of state with an exact solution that has no transverse horizon rotation at any radius. 

\subsubsection{Which extremal sectors occur in the exact family?}\label{app:ST_extremal_audit}
The distinction between longitudinal and transverse horizon rotation is essential. The angle $\psi$ runs around the ring, while $\phi$ parametrises transverse rotation. The family compared in Appendix~\ref{app:ST_charge_matching} obeys $\Omega_\phi=0$. There is no non-extremal $\Omega_\psi=0$ interior in the chosen charge and spin orientation, since every factor in $h_1h_{+5}h_{+6}$ and the prefactor of \eqref{eq:ST_exact_angular_velocities} is positive. Vanishing $\Omega_\psi$ can occur at a boundary, notably the BPS limit. We analyse the $\Omega_\phi=0$ slice below.

For fixed $0<b<1$, the equation $h_{+7}=0$ has the root used in Figure~\ref{fig:exact_nonext_comparison}. Its thin expansion selects $a_\perp/m\to0$, but also fixes a relation between the two local charge parameters. In particular, at fixed nonzero magnetic flux its regular zero-temperature string limit is
\begin{equation}\label{eq:ST_audit_BPS_local}
\begin{gathered}
 b\to1\quad,\quad k\to0\quad,\quad \beta\to\infty~,\\
 \cosh^2\delta\to\frac53
 \quad,\quad \frac{Q_{\mathrm R}}q\to\sqrt{\frac32}
 \quad,\quad w\to0~.
\end{gathered}
\end{equation}
The finite-radius completion is the restricted BPS family
\begin{equation}\label{eq:ST_audit_BPS_global}
 Q_{\mathrm{ring}}=q^2+\sqrt6qR_{\mathrm b}~,
 \qquad M=-\sqrt3Q_{(0)}~.
\end{equation}
Thus even the supersymmetric limit does not cover arbitrary $Q_{\mathrm R}/q$. At the fixed charges of Figure~\ref{fig:BR_comparison}, it selects the single point $y=\sqrt{2/3}(1-q^2/Q_\star)$ on the exact supersymmetric curve. Its leading blackfold counterpart is $y=\sqrt{2/3}$. The generic negative-momentum finite-velocity and charged null-wave states instead have $M>M_{\mathrm{BPS}}$ and cannot coincide with this endpoint. More directly, \eqref{eq:ST_string_dictionary} fixes $\delta$ to a nonzero constant as $\beta\to\infty$, excluding the charged null-wave scaling.

One can also exclude a regular non-BPS extremal endpoint of the $\Omega_\phi=0$ slice without invoking a thin expansion. The two non-BPS extremal ring boundaries of the full four-parameter solution are parametrised in~\cite{Suzuki:2024generalring} by $0<\eta<1$ and fixed $|b|<1$. Their transverse angular velocities are proportional, with nonzero finite-area denominators, to
\begin{equation}\label{eq:ST_audit_transverse_extremal}
 \Omega_{\phi,\mathrm{ext}}\ \propto\
 \pm(1-b^2)\left[2-(1+b^6)\eta^2\right]~.
\end{equation}
The bracket is strictly positive. These regular extremal rings therefore lie outside $\Omega_\phi=0$. In their controlled thin limit, $\zeta\to0$ or $\zeta\to\infty$ in \eqref{eq:ST_spinning_thin} gives the extremal transverse Kerr rotation $|a_\perp|=m$, rather than any of the $a_\perp=0$ limits of Section~\ref{sec:susy_lim}.

There is nevertheless a zero-temperature degeneration on $\Omega_\phi=0$ at fixed $b$ as $n\to1$. Direct substitution into $h_{+7}=0$ gives
\begin{equation}\label{eq:ST_audit_degenerate_endpoint}
 \lambda\to\frac{1-b^3}{2}\quad,\quad
 \frac{q^2}{Q_\star}\to b^3\quad,\quad
 a_H\to0\quad,\quad \kappa_H\sqrt{G_5M}\to0~,
\end{equation}
which is not a finite-area extremal string or ring.  The resulting classification, at the fixed charge orientations used in the main text, is
\begin{center}
\begin{tabular}{@{}ll@{}}
\toprule
Local extremal sector & $\Omega_\phi=0$ Suzuki-Tomizawa family\\
\midrule
Finite-velocity BPS branch & Present only at $Q_{\mathrm R}/q=\sqrt{3/2}$\\
Other finite-velocity branch & Absent as a regular limit\\
Charged positive null wave & Absent as a regular limit\\
Opposite magnetic null wave & Absent at nonzero magnetic flux\\
Transversely rotating extremal string & Present in the full family, outside this slice\\
\bottomrule
\end{tabular}
\end{center}
The conclusion regarding the magnetic branch also follows from \eqref{eq:ST_exact_charge_constraint}. At fixed finite mass, $b\to0$ gives both vanishing electric charge and vanishing magnetic flux. It cannot leave the purely magnetic null endpoint with $q\ne0$. Reversing an angular coordinate changes charge orientations and does not manufacture an independent missing electric to magnetic ratio. 

\subsubsection{Blackfold curves at fixed mass and charges}\label{app:nonext_fixed_charge_curves}
For reference, we construct the fixed-$M,Q_{(0)},Q_{(1)}$ curves of the general non-extremal blackfold family in Section~\ref{sec:flat_ring_nonext_plots}. Unlike the exact comparison in Appendix~\ref{app:ST_charge_matching}, this construction varies the local charge ratios on each momentum branch. The dipole charge fixes $m$, and the mass then fixes $R$. Using \eqref{eq:flat_ring_branch_charges} and the functions of Section~\ref{sec:non-ext_string}, the remaining electric charge condition is
\begin{equation}\label{eq:ring_nonext_fixed_charge_solver}
\begin{gathered}
 m=\frac{q}{4c_\delta^2\mathsf X}
 \quad,\quad
 R=\frac{\nu Q_\star\mathsf X}{q\mathsf B}~,\\
 \nu=\frac{c_\delta\mathsf B}
 {s_\delta(\mathsf Bc_{\gamma_\pm}+\mathsf Xs_{\gamma_\pm})}~.
\end{gathered}
\end{equation}
We vary $\beta>0$ and solve the last equation for $\delta>0$ on each branch, with $\gamma_\pm=\chi+\operatorname{arctanh}w_\pm$, where $w_\pm$ is given by \eqref{eq:OmegaR} and $\chi$ by \eqref{eq:nonext_rest_boost}. Equations~\eqref{eq:ring_nonext_fixed_charge_solver} then give $m$ and $R$ without further fitting. Every point has the prescribed $M$, $Q_{(0)}$ and $Q_{(1)}$.
The area and asymptotic temperature can be evaluated directly and read
\begin{equation}\label{eq:ring_nonext_plot_area}
\begin{aligned}
 \mathcal A_\pm
 &=32\pi^2Rm^2c_\delta^3c_\beta^2
                   (\mathsf A c_{\gamma_\pm}-\mathsf Xs_{\gamma_\pm})~,\\
 T_{0,\pm}
 &=\frac{1}{8\pi m c_\delta^3c_\beta^2
                   (\mathsf A c_{\gamma_\pm}-\mathsf Xs_{\gamma_\pm})}~.
\end{aligned}
\end{equation}
These follow from the horizon expressions in Section~\ref{sec:non-ext_string} and the boost $\eta=\gamma-\chi$. The angular momentum is evaluated with \eqref{eq:flat_ring_branch_charges}. As a check along each fixed-charge and fixed-mass curve, the first law reduces to
\begin{equation}\label{eq:ring_fixed_mass_curve_first_law}
 T_0\frac{dS_{\mathrm{BH}}}{d\beta}
       +\Omega\frac{dJ}{d\beta}=0~.
\end{equation}
The zero-temperature endpoints are given in \eqref{eq:ring_nonext_curve_endpoints}. They approach the charged null wave on the positive momentum branch and the finite-velocity extremal solution on the negative momentum branch. Figure~\ref{fig:nonext_fixed_charges} retains the electric and dipole charges of Figure~\ref{fig:BR_comparison}. We choose $\nu=1.25,1.5,2$, with the same mass assigned to curves of the same colour in the two panels. Their spin signs differ, so an area comparison at equal $|j|$ still does not fix the full signed angular momentum. The thinness criterion on either non-extremal curve is particularly simple,
\begin{equation}\label{eq:ring_nonext_fixed_charge_thinness}
 \frac{r_\epsilon(n)}R
       =\frac{6\nu q^2}{Q_\star y^2}~.
\end{equation}
The figure uses the same solid and dashed convention as the extremal comparison. In particular, the small-radius ends eventually leave the blackfold regime.

\section{CCLP black hole conventions and thermodynamics}\label{Appendix_CCLP}
We use two subfamilies of the ungauged CCLP black hole~\cite{Chong:2005hr} throughout the paper. The first has equal angular momenta and supplies the background for the stationary Saturns and their extremal limits. The second has vanishing transverse horizon angular velocity and is used for mutual equilibrium in Section~\ref{sec:thermal_saturn_equilibrium}. Both use the normalisation of Section~\ref{sec: sugra_ward_identities}.

\subsection{Equal angular momenta}\label{app:central_equal_spin}
Equation~\eqref{eq:thermal_saturn_background} uses the radial coordinate $r^2=r_{\rm BL}^2+a_{\mathrm H}^2$, with the charge and orientation chosen to approach \eqref{eq:saturn_BMPV}. The two angular momenta have the same sign. The useful determinant identity is
\begin{equation}\label{eq:central_determinant}
 \det g_{(t,\psi)}\big|_{\theta=0}=-r^2\Delta_{\mathrm H}~.
\end{equation}
Using $g_{\psi\psi}=\mathcal R^2$, this identity gives
$N^2=-\det g_{(t,\psi)}/g_{\psi\psi}
=R^2\Delta_{\mathrm H}(R)/\mathcal R^2$,
as used in \eqref{eq:thermal_saturn_lapse}. The charges are given by
\begin{equation}\label{eq:central_nonext_charges}
 M_{\mathrm H}=\frac{3\pi m_{\mathrm H}}{4G_5}
 \quad,\quad Q_{(0),\mathrm H}=-\frac{\sqrt3\pi q_{\mathrm H}}{4G_5}
 \quad,\quad J_{\psi,\mathrm H}=J_{\phi,\mathrm H}
       =\frac{\pi a_{\mathrm H}(2m_{\mathrm H}-q_{\mathrm H})}{4G_5}~.
\end{equation}
Introduce $\mathcal R_{\mathrm H}$ and let $\omega_{\mathrm H}$ denote \eqref{eq:thermal_saturn_lapse} evaluated at \eqref{eq:thermal_saturn_horizon}. Using this, the horizon generator and thermodynamics are
\begin{equation}\label{eq:central_nonext_thermodynamics}
\begin{gathered}
 k_{\mathrm H}=\partial_t+\Omega_{\mathrm H}(\partial_\psi+\partial_\phi)
 \quad,\quad \Omega_{\mathrm H}=\omega_{\mathrm H}~,\\
 \mathcal A_{\mathrm H}=2\pi^2r_{\mathrm H}^2\mathcal R_{\mathrm H}
 \quad,\quad S_{\mathrm H}=\frac{\mathcal A_{\mathrm H}}{4G_5}
 \quad,\quad T_{\mathrm H}=
              \frac{r_{\mathrm H}\Delta_{\mathrm H}'(r_{\mathrm H})}
                        {4\pi\mathcal R_{\mathrm H}}~,\\
 \Phi_{0,\mathrm H}=-\frac{\sqrt3q_{\mathrm H}}{r_{\mathrm H}^2}
                                      (1-a_{\mathrm H}\Omega_{\mathrm H})~,
\end{gathered}
\end{equation}
whcih satisfy the first law
\begin{equation}\label{eq:central_nonext_first_law}
 dM_{\mathrm H}=T_{\mathrm H}dS_{\mathrm H}
     +\Omega_{\mathrm H}(dJ_{\psi,\mathrm H}+dJ_{\phi,\mathrm H})
     +\Phi_{0,\mathrm H}dQ_{(0),\mathrm H}~.
\end{equation}
Causality requires $\Delta_{\mathrm H}>0$ and $\mathcal R^2>0$ outside the outer horizon. The supersymmetric limit of this solution is  $m_{\mathrm H}=q_{\mathrm H}=\mu$ and $\mathfrak j=a_{\mathrm H}\mu$. In this limit $T_{\mathrm H}=\Omega_{\mathrm H}=0$, $\Phi_{0,\mathrm H}=-\sqrt3$ and $\mathcal A_{\mathrm H}=2\pi^2\sqrt{\mu^3-\mathfrak j^2}$. In particular $M_{\mathrm H}=-\sqrt3Q_{(0),\mathrm H}$. There is another extremal rotating regime that satisfies $m_{\mathrm H}+q_{\mathrm H}=2a_{\mathrm H}^2$ and is not the BMPV supersymmetric limit.

\subsection{Vanishing transverse horizon angular velocity}\label{app:central_zero_transverse}
In this case introduce $a_{\mathrm H}$ and $b_{\mathrm H}$, associated with angular directions $\psi$ and $\phi$, respectively. In this subsection $r$ is the Boyer-Lindquist radius. It is useful to define
\begin{equation}\label{eq:central_general_functions}
\begin{gathered}
 \Sigma=r^2+a_{\mathrm H}^2\sin^2\theta+b_{\mathrm H}^2\cos^2\theta
 \quad,\quad f=\frac{2m_{\mathrm H}}\Sigma-\frac{q_{\mathrm H}^2}{\Sigma^2}~,\\
 \varpi=b_{\mathrm H}\sin^2\theta\,d\phi
                       +a_{\mathrm H}\cos^2\theta\,d\psi
 \quad,\quad \widetilde\varpi=a_{\mathrm H}\sin^2\theta\,d\phi
                       +b_{\mathrm H}\cos^2\theta\,d\psi~,\\
 \mathcal D(r)=(r^2+a_{\mathrm H}^2)(r^2+b_{\mathrm H}^2)
       +q_{\mathrm H}^2-2a_{\mathrm H}b_{\mathrm H}q_{\mathrm H}
       -2m_{\mathrm H}r^2~.
\end{gathered}
\end{equation}
In terms of these, the metric and electric potential in our conventions are
\begin{equation}\label{eq:central_general_metric}
\begin{aligned}
 ds^2={}&-dt^2+\frac{2q_{\mathrm H}}\Sigma
                       (dt-\varpi)\widetilde\varpi
          +f(dt-\varpi)^2+\frac{\Sigma r^2}{\mathcal D(r)}dr^2
          +\Sigma d\theta^2\\
       &+(r^2+b_{\mathrm H}^2)\sin^2\theta\,d\phi^2
         +(r^2+a_{\mathrm H}^2)\cos^2\theta\,d\psi^2~,\\
 A={}&\frac{\sqrt3q_{\mathrm H}}\Sigma(dt-\varpi)~.
\end{aligned}
\end{equation}
For $a_{\mathrm H}=b_{\mathrm H}$, the redefinition of radial coordinate $r^2=r_{\rm BL}^2+a_{\mathrm H}^2$ recovers \eqref{eq:thermal_saturn_background}. A dual potential regular near the ring's axis is
\begin{equation}\label{eq:central_general_dual}
\begin{aligned}
 B_N={}&\frac{\sqrt3q_{\mathrm H}}\Sigma\,dt\wedge\widetilde\varpi
       +\frac{\sqrt3q_{\mathrm H}(r^2+a_{\mathrm H}^2)\sin^2\theta}
                    {\Sigma}\,d\phi\wedge d\psi~,\\
 B_S={}&B_N-\sqrt3q_{\mathrm H}\,d\phi\wedge d\psi~.
\end{aligned}
\end{equation}
It obeys $dB=-\star F+\kappa A\wedge F$. In particular, the transition between the two gauge descriptions is the same as in \eqref{eq:thermal_saturn_dual}, which fixes the transverse interaction contribution to the ring angular momentum. Define $r_+$ as the largest positive root of $\mathcal D(r)=0$. The central transverse angular velocity and angular momentum can thus be expressed as
\begin{equation}\label{eq:central_general_transverse}
\begin{aligned}
 \Omega_{\phi,\mathrm H}
  &=\frac{b_{\mathrm H}(r_+^2+a_{\mathrm H}^2)-a_{\mathrm H}q_{\mathrm H}}
          {(r_+^2+a_{\mathrm H}^2)(r_+^2+b_{\mathrm H}^2)
                               -a_{\mathrm H}b_{\mathrm H}q_{\mathrm H}}~,\\
 J_{\phi,\mathrm H}
  &=\frac{\pi}{4G_5}(2m_{\mathrm H}b_{\mathrm H}-a_{\mathrm H}q_{\mathrm H})~.
\end{aligned}
\end{equation}
They vanish on different parameter surfaces when the black hole is charged. Solving $\Omega_{\phi,\mathrm H}=0$ together with the horizon equation gives the explicit parametrisation
\begin{equation}\label{eq:central_zero_transverse_parameters}
 \mathcal U_{\mathrm H}=m_{\mathrm H}
                   +\sqrt{m_{\mathrm H}^2-q_{\mathrm H}^2}
 \quad,\quad
 b_{\mathrm H}=\frac{a_{\mathrm H}q_{\mathrm H}}{\mathcal U_{\mathrm H}}
 \quad,\quad r_+^2=\mathcal U_{\mathrm H}-a_{\mathrm H}^2~.
\end{equation}
We take $m_{\mathrm H}>q_{\mathrm H}\geq0$ and $a_{\mathrm H}^2<\mathcal U_{\mathrm H}$. The inner root has $r_-^2=q_{\mathrm H}^2r_+^2/\mathcal U_{\mathrm H}^2<r_+^2$. The mass and electric charge retain the first two expressions in \eqref{eq:central_nonext_charges}, while
\begin{equation}\label{eq:central_zero_transverse_charges}
 J_{\psi,\mathrm H}=\frac{\pi a_{\mathrm H}\mathcal U_{\mathrm H}}{4G_5}
 \quad,\quad
 J_{\phi,\mathrm H}=\frac{\pi a_{\mathrm H}q_{\mathrm H}^3}
                              {4G_5\mathcal U_{\mathrm H}^2}~.
\end{equation}
The horizon generator is $k_{\mathrm H}=\partial_t+\Omega_{\psi,\mathrm H}\partial_\psi$, with
\begin{equation}\label{eq:central_zero_transverse_thermo}
\begin{gathered}
 \Omega_{\psi,\mathrm H}=\frac{a_{\mathrm H}}{\mathcal U_{\mathrm H}}
                   \left(1-\frac{q_{\mathrm H}^2}{\mathcal U_{\mathrm H}^2}\right)
 \quad,\quad
 T_{\mathrm H}=\frac{r_+}{2\pi\mathcal U_{\mathrm H}}
                   \left(1-\frac{q_{\mathrm H}^2}{\mathcal U_{\mathrm H}^2}\right)~,\\
 \Phi_{0,\mathrm H}=-\frac{\sqrt3q_{\mathrm H}}{\mathcal U_{\mathrm H}}
 \quad,\quad \mathcal A_{\mathrm H}=2\pi^2\mathcal U_{\mathrm H}r_+
 \quad,\quad S_{\mathrm H}=\frac{\mathcal A_{\mathrm H}}{4G_5}~.
\end{gathered}
\end{equation}
Along this subfamily the first law is $dM_{\mathrm H}=T_{\mathrm H}dS_{\mathrm H}+\Omega_{\psi,\mathrm H}dJ_{\psi,\mathrm H}+\Phi_{0,\mathrm H}dQ_{(0),\mathrm H}$. The coefficient of $dJ_{\phi,\mathrm H}$ vanishes, even though $J_{\phi,\mathrm H}$ itself need not vanish.

For the embedding $r=R$, $\theta=0$, we define $\Sigma_R=R^2+b_{\mathrm H}^2$ and $f_R=2m_{\mathrm H}/\Sigma_R-q_{\mathrm H}^2/\Sigma_R^2$. Pulling back the metric and potentials gives
\begin{equation}\label{eq:central_zero_transverse_axis}
\begin{gathered}
 \mathcal R^2=R^2+a_{\mathrm H}^2
       -\frac{2q_{\mathrm H}a_{\mathrm H}b_{\mathrm H}}{\Sigma_R}
       +a_{\mathrm H}^2f_R~,
 \qquad
 \omega=\frac{a_{\mathrm H}f_R-q_{\mathrm H}b_{\mathrm H}/\Sigma_R}
                  {\mathcal R^2}~,\\
 N^2=\frac{\mathcal D(R)}{\Sigma_R\mathcal R^2}
 \quad,\quad A_t=\frac{\sqrt3q_{\mathrm H}}{\Sigma_R}
 \quad,\quad A_\psi=-a_{\mathrm H}A_t
 \quad,\quad B_{t\psi}=b_{\mathrm H}A_t~.
\end{gathered}
\end{equation}
These functions enter \eqref{eq:circle_radial_data} without any further coordinate change. The resulting flux components are
\begin{equation}\label{eq:central_zero_transverse_fluxes}
 F_0=-\frac{2\sqrt3q_{\mathrm H}R(1-a_{\mathrm H}\omega)}{N\Sigma_R^2}
 \quad,\quad
 F_1=\frac{2\sqrt3a_{\mathrm H}q_{\mathrm H}R}{\mathcal R\Sigma_R^2}
 \quad,\quad
 H=-\frac{2\sqrt3b_{\mathrm H}q_{\mathrm H}R}{N\mathcal R\Sigma_R^2}~.
\end{equation}
The force equation is still \eqref{eq:circle_force}. To convert its covariant-normal coefficients to a unit radial normal, we multiply $K_{AB},F_A,H$ by $\sqrt{\mathcal D(R)/(\Sigma_RR^2)}$. Finally, the proper radial separation used in the validity check is
\begin{equation}\label{eq:central_zero_transverse_distance}
 d_{\rm sep}=\int_{r_+}^R dr\,
       \sqrt{\frac{(r^2+b_{\mathrm H}^2)r^2}{\mathcal D(r)}}~.
\end{equation}

\section{Di-rings and bi-rings on an exact ring background}\label{app:nonext_multirings}
We now apply the construction of Section~\ref{sec:nonext_saturn} with a black ring in place of the central black hole. We call the exact Suzuki-Tomizawa ring the \emph{seed}, and describe the added thin ring by wrapping the string. Placing the two rings in the same plane gives a di-ring, while placing them in orthogonal planes gives a bi-ring. The seed supplies the gravitational and gauge fields in which the added ring must balance. Its response to the added ring is neglected at this order. The two components have independent local parameters. In particular, the added string need not satisfy the restricted dictionary \eqref{eq:ST_string_dictionary} obeyed by the thin limit of the exact seed. We first determine its embedding and balance equation, then impose thermodynamic equilibrium. Finally, we take the extremal limits of the two horizons separately. This distinguishes stationary configurations from those with common horizon potentials and from those admitting an exact supersymmetric completion.

\subsection{Background and the two embeddings}
Our background is the four-parameter solution of Suzuki and Tomizawa~\cite{Suzuki:2024generalring}. We use the parameters $(\ell_{\rm ST},n,\lambda,b)$ and charge conventions of Appendix~\ref{app:exact_nonext_limit}. The metric and Maxwell potential have the form
\begin{equation}\label{eq:ST_background_block}
\begin{aligned}
 ds^2={}&-\frac{\mathscr H(y,x)}{D^2\mathscr H(x,y)}
                 (dt+\mathscr W_\psi d\psi+\mathscr W_\phi d\phi)^2\\
 &+\frac D{\mathscr H(y,x)}
       [\mathscr F(y,x)d\psi^2-2\mathscr J(x,y)d\psi d\phi
                                      -\mathscr F(x,y)d\phi^2]
       +g_{xx}dx^2+g_{yy}dy^2~,\\
 A={}&A_tdt+A_\psi d\psi+A_\phi d\phi~.
\end{aligned}
\end{equation}
The functions are given in Eqs.~(37)--(47) of \cite{Suzuki:2024generalring}. We retain the full background, without taking its thin-ring limit. The exterior has $-1\leq x\leq1$ and $-1/n<y<-1$, with the seed horizon at $y=-1/n$ and spatial infinity at $x=y=-1$. We embed the string on a circle along an axis where one azimuthal circle shrinks smoothly to zero. It wraps the other, non-vanishing circle,
\begin{equation}\label{eq:ST_two_embeddings}
\begin{array}{c|ccc|c}
 &x&y&\hbox{wrapped angle}&\hbox{worldsheet coordinates}\\ \hline
 \hbox{di-ring}&-1&y_*&\psi&(t,\psi)\\
 \hbox{bi-ring}&x_*&-1&\phi&(t,\phi)
\end{array}
\end{equation}
where $-1/n<y_*<-1$ and $-1<x_*<1$. In the first embedding, the $\phi$ circle collapses and the added ring lies outside the seed in the same plane. In the second, the $\psi$ circle collapses and the added ring lies in the orthogonal plane. We use  $\varphi$ for the wrapped angle and $s$ for the coordinate that locates the circle, namely $s=y_*$ or $s=x_*$, respectively. Pulling back the metric to the worldsheet $\Sigma$ gives
\begin{equation}\label{eq:ST_probe_induced}
\begin{gathered}
 \gamma_{ab}d\sigma^ad\sigma^b
       =-N^2dt^2+\mathcal R^2(d\varphi-\omega dt)^2~,
 \qquad \gamma_{ab}=g_{ab}\big|_\Sigma~,\\
 \mathcal R^2=\gamma_{\varphi\varphi}\quad,\quad
 \omega=-\frac{\gamma_{t\varphi}}{\gamma_{\varphi\varphi}}
 \quad,\quad N^2=-\frac{\det\gamma}{\gamma_{\varphi\varphi}}~.
\end{gathered}
\end{equation}
Here $\mathcal R$ is the proper radius on a constant-$t$ slice, $N$ is the lapse and $\omega$ is the angular shift due to frame dragging. A stationary ring has $k=\partial_t+\Omega\partial_\varphi$ and local velocity $w=\mathcal R(\Omega-\omega)/N$. Its vectors $u^a,v^a$ are given by \eqref{eq:stationary_circle_vectors}, with $\psi$ replaced by $\varphi$. The same replacement in \eqref{eq:thermal_saturn_potentials} and \eqref{eq:circle_magnetic_potential} gives its equilibrium potentials. The non-extremal string requires $|w|<1$.

\subsection{Solving the blackfold equations in the exact fields}\label{app:ST_multiring_force}
All pulled back fields and local state variables are constant along the wrapped circle, so the intrinsic equations hold. Smoothness of the axis removes the force normal to its plane. The remaining equation balances forces along $s$ and has the same form as \eqref{eq:circle_force},
\begin{equation}\label{eq:ST_probe_balance}
 EK_{00}+2Z_{\parallel}K_{01}+P_{\parallel}K_{11}
             =j_0F_0+j_1F_1-\rho_1H~.
\end{equation}
The left-hand side is the extrinsic force due to the energy, momentum and longitudinal stress of the string. The right-hand side contains the electric force and the force on its magnetic string charge. The components $E,Z_{\parallel},P_{\parallel},j_0,j_1$ follow from the boosted constitutive relations \eqref{eq:circle_boosted_monopoles}. The geometric and field coefficients are those of \eqref{eq:circle_radial_data}, with primes denoting $\partial_s$ and $H=\mathcal F_{s\hat0\hat1}$.

Both $F=dA$ and $\mathcal F=-\star F$ must be evaluated in the full five-dimensional metric before restricting to the axis. The angular coordinates degenerate there, so setting metric components to zero before taking the dual would discard a finite magnetic force. We use the orientation $-dt\wedge dx\wedge dy\wedge d\phi\wedge d\psi$, consistent with Section~\ref{sec:susy_saturn}. For fixed $(m,\beta,\delta)$ and a chosen circle in the seed background, equation~\eqref{eq:ST_probe_balance} determines the allowed velocities. One may solve the quartic \eqref{eq:circle_quartic}, retaining only timelike roots. Distant solutions can be understood without a numerical calculation. In the asymptotically flat region, the background fields become weak and balance reduces to $p+2zw+\epsilon w^2=0$, the flat-space equation of Section~\ref{sec:flat_ring_branches}. Each simple root persists at sufficiently large finite separation from the seed. This establishes distant di-ring and bi-ring branches, while their continuation towards the seed requires solving the full equation. We will be doing this numerically for representative examples.

The dual potential can be constructed explicitly from the same background fields. In the asymptotically vanishing Maxwell gauge of \eqref{eq:ST_background_block}, its time components are
\begin{equation}\label{eq:ST_probe_dual_time}
 B_{ti}=\sqrt3\,g_{ti}
       +\left(\kappa A_t-\frac{1+b^2}{b}\right)A_i
 \quad,\quad i\in\{\phi,\psi\}~.
\end{equation}
Here $b$ is the seed's charge parameter, with $0<b<1$. This expression follows from the electric Harrison transformation in Eqs.~(6) and (10) of~\cite{Suzuki:2024generalring}, in our orientation. It vanishes at infinity and obeys
\begin{equation}\label{eq:ST_probe_dual_quadrature}
 \partial_s B_{t\varphi}
 =\mathcal F_{st\varphi}
       -\kappa(A_t\partial_sA_\varphi-A_\varphi\partial_sA_t)~.
\end{equation}
Thus the pullback needed for either ring is simply
$B|_\Sigma=B_{t\varphi}|_\Sigma\,dt\wedge d\varphi$, without a further integration. The neutral limit of \eqref{eq:ST_probe_dual_time} is $B_{ti}=0$, obtained by taking $b\to0$ in the complete expression.

For completeness, the purely angular component fixes how this potential is continued between the two axes. A representative regular at the outer $\phi$ axis $x=-1$ is
\begin{equation}\label{eq:ST_dual_complete}
\begin{aligned}
 B^{(\phi)}={}&B_{t\phi}\,dt\wedge d\phi
              +B_{t\psi}\,dt\wedge d\psi
              +\mathscr B(x,y)\,d\phi\wedge d\psi~,\\
 \mathscr B(x,y)={}&\int_{-1}^{x}d\bar x\,
 \left[\sqrt{-g}\,g^{yy}g^{tI}\partial_yA_I
       +\kappa(A_\psi\partial_{\bar x}A_\phi
                         -A_\phi\partial_{\bar x}A_\psi)\right]_{(\bar x,y)}~.
\end{aligned}
\end{equation}
Here $I\in\{t,\phi,\psi\}$ is summed. All quantities in the integral are the explicit ST functions of \eqref{eq:ST_background_block}, evaluated before taking an axis limit. The Maxwell equation ensures that the other derivative is
$\partial_y\mathscr B=-\sqrt{-g}\,g^{xx}g^{tI}\partial_xA_I
+\kappa(A_\psi\partial_yA_\phi-A_\phi\partial_yA_\psi)$.
Consequently $dB^{(\phi)}=-\star F+\kappa A\wedge F$ throughout this patch. On the other outer axis, $y=-1$, $\mathscr B$ is a constant fixed by the seed's electric Page charge. The potential regular where the $\psi$ circle shrinks is therefore
\begin{equation}\label{eq:ST_dual_patch_transition}
\begin{gathered}
 B^{(\psi)}=B^{(\phi)}-c_B\,d\phi\wedge d\psi~,\\
 c_B=\frac1{4\pi^2}\int_{S^3_\infty}dB
       =-\frac{4G_5}{\pi}Q_{(0),\mathrm{seed}}~.
\end{gathered}
\end{equation}
We use $B^{(\phi)}$ for the di-ring and $B^{(\psi)}$ for the bi-ring. The patch change leaves $B_{t\varphi}$ unchanged.

The charge integrals can now be written explicitly. With $\Gamma=(1-w^2)^{-1/2}$ and the boosted densities in \eqref{eq:circle_boosted_monopoles}, the added ring carries
\begin{equation}\label{eq:ST_probe_charges}
\begin{aligned}
 Q_{(0),\mathrm r}&=2\pi\mathcal Rj_0+4\pi\kappa\rho_1A_\varphi~,\qquad
 Q_{(1),\mathrm r}=\rho_1~,\\
 J_{\varphi,\mathrm r}&=2\pi\left(\mathcal R^2Z_{\parallel}
        +\mathcal RA_\varphi j_0+\kappa\rho_1A_\varphi^2\right)~,\\
 M_{\mathrm r}&=2\pi\left[N\mathcal RE+\omega\mathcal R^2Z_{\parallel}
        -\mathcal RA_tj_0
        +\rho_1(B_{t\varphi}-\kappa A_tA_\varphi)\right]~,\\
 S_{\mathrm r}&=2\pi\mathcal R s\Gamma~,\qquad
 \mathcal A_{\mathrm r}=4G_5S_{\mathrm r}~.
\end{aligned}
\end{equation}
These expressions apply to each admissible root of \eqref{eq:ST_probe_balance}, with all background fields evaluated on that root's embedding.

There is also an angular momentum along the collapsing circle. Its mechanical contribution vanishes, but the gauge fields contribute through the current \eqref{Ptilde}. As in Section~\ref{sec:susy_saturn}, requiring the one-form associated with the asymptotic rotation generator to be regular at both outer axes fixes its pullback to $\eta_{\chi_\perp}=-c_Bd\varphi$. Thus
\begin{equation}\label{eq:ST_probe_transverse_charge}
 J_{\perp,\mathrm r}=2\pi c_B\rho_1
             =-8G_5Q_{(0),\mathrm{seed}}Q_{(1),\mathrm r}~,
 \qquad
 (\varphi,\perp)=
 \begin{cases}(\psi,\phi)&\text{di-ring}~,\\
              (\phi,\psi)&\text{bi-ring}~.
 \end{cases}
\end{equation}
This uses the same magnetic orientation as \eqref{eq:ST_probe_charges}. The total mass, electric charge and each angular momentum are the corresponding seed charge plus the added-ring contribution, while $\mathcal A_{\rm tot}=\mathcal A_{\rm seed}+\mathcal A_{\mathrm r}$ at this order. The two dipoles remain separate flux charges, each measured on a sphere linking its own horizon.

\subsection{Non-extremal di-rings in thermal equilibrium}\label{app:ST_multiring_thermal}
In this section we will be solving the blackfold equations in the previous subsection but selecting configurations that are in thermal equilibrium (equal temperatures, angular velocities and electric potentials) with the central ring. For a di-ring, which wraps $\psi$, we must therefore choose a seed with
\begin{equation}\label{eq:thermal_diring_seed}
 \Omega_{\phi,\mathrm H}=0\quad\Longleftrightarrow\quad
 h_{+7}(n,\lambda,b)=0~.
\end{equation}
The function $h_{+7}$ is the numerator determining $\Omega_{\phi,\mathrm H}$ in \eqref{eq:ST_exact_angular_velocities}. This condition fixes the seed shape parameter $\lambda$ at each $(n,b)$ and is imposed on the exact solution. It does not set its transverse ADM angular momentum or its magnetic field to zero~\cite{Suzuki:2024generalring}.

The outer ring lies at $x=-1$, $y=y_*$ and must corotate with the seed. Thus its generator and redshift are
\begin{equation}\label{eq:thermal_diring_generator}
 k=\partial_t+\Omega_{\psi,\mathrm H}\partial_\psi\quad,\quad
 w=\frac{\mathcal R}{N}(\Omega_{\psi,\mathrm H}-\omega)
 \quad,\quad |k|=N\sqrt{1-w^2}~.
\end{equation}
At a timelike circle, the remaining matching conditions are
\begin{equation}\label{eq:thermal_diring_matching}
\begin{gathered}
 T_{0,\mathrm r}=|k|T=T_{\mathrm H}>0\quad,\quad
 \Phi_{0,\mathrm r}=|k|\mu_0-A_t-\Omega_{\psi,\mathrm H}A_\psi
                         =\Phi_{0,\mathrm H}~,\\
 \Omega_{\psi,\mathrm r}=\Omega_{\psi,\mathrm H}
 \quad,\quad \Omega_{\phi,\mathrm r}=\Omega_{\phi,\mathrm H}=0~.
\end{gathered}
\end{equation}
We hold each dipole flux fixed in the equilibrium variations, so equality of the two magnetic potentials is not required. Allowing magnetic charge transfer between horizons would define a different thermodynamic problem.

These equations can be solved in a definite order. We choose the seed and the string parameter $\beta$, then consider a trial position $y_*$. The electric-potential equation determines $\delta$, and the temperature equation determines $m$,
\begin{equation}\label{eq:thermal_diring_local_parameters}
 \mu_0(\beta,\delta)
    =\frac{\Phi_{0,\mathrm H}+A_t+\Omega_{\psi,\mathrm H}A_\psi}{|k|}
 \quad,\quad
 m=\frac{|k|\Theta(\beta,\delta)}{T_{\mathrm H}}~,
 \qquad \Theta=mT~.
\end{equation}
The equation of state makes $\Theta$ independent of $m$. Substitution into \eqref{eq:ST_probe_balance} leaves one equation for $y_*$. The ring position is therefore determined by balance once the seed and $\beta$ are specified, although more than one admissible root may be possible. Both gauge forces remain in this equation, including when $\Omega_{\phi,\mathrm H}=0$. A representative solution, in units $G_5=\ell_{\rm ST}=1$, is
\begin{equation}\label{eq:thermal_diring_example}
\begin{gathered}
 (n,b,\lambda)\simeq(0.1,0.266409,0.0905692)~,
 \qquad (y_*,\beta)\simeq(-3.786283,6.682234)~,\\
 (\delta,m)\simeq(7.12962\times10^{-4},7.98588\times10^{-10})~.
\end{gathered}
\end{equation}
The added blackfold ring lies outside the seed horizon at $y=-10$, both areas are positive and the common temperature is $T_0\simeq0.320597$. Its mass is about $0.46\%$ of the seed mass, and its thickness is about $0.17\%$ of the proper separation between the circle and the seed horizon.

\begin{figure}[t]
 \centering
 \includegraphics[width=\textwidth]{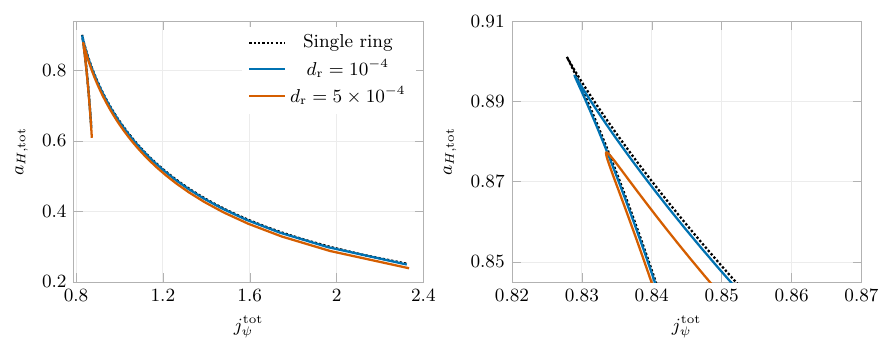}
 \caption{Reduced total area versus longitudinal angular momentum for non-extremal di-rings satisfying \eqref{eq:thermal_diring_matching}. The seed has $\Omega_{\phi,\mathrm H}=0$. Blue and orange curves have $d_{\mathrm r}=10^{-4}$ and $5\times10^{-4}$, respectively, at $e=1/2$. The dotted curve is a single exact ring at the same electric charge to mass ratio. The right panel enlarges the turning region. The sampled interval is $0.02\leq n\leq0.85$ and does not determine the full existence range. The inner dipole and transverse angular momentum vary along the curves, and the single ring has no outer dipole.}
 \label{fig:thermal_dirings}
\end{figure}

Figure~\ref{fig:thermal_dirings} shows continuous families obtained in this way. We use the total mass, area and longitudinal angular momentum in \eqref{eq:flat_ring_reduced_definitions}, and keep
\begin{equation}\label{eq:thermal_diring_plot_ensemble}
 e\equiv\frac{\sqrt3|Q_{(0),\rm tot}|}{M_{\rm tot}}=\frac12
 \quad,\quad
 d_{\mathrm r}\equiv
 \frac{G_5Q_{(1),\mathrm r}}{\sqrt{G_5M_{\rm tot}}}
     \in\{10^{-4},5\times10^{-4}\}~.
\end{equation}
We vary $n$ and adjust $b$ and $\beta$ to preserve these ratios while solving the matching and balance equations. Rescaling to a common total mass then fixes the total electric charge and outer dipole along each curve. The inner dipole and transverse angular momentum vary. The dashed curve is the curve for the single ring. The curves nearly coincide because the seed dominates the total area in this construction.

Across the plotted families, the added mass is less than $4.4\%$ of the seed mass, and the evaluated thickness, bending, flux and separation ratios are below $0.058$. The full validity conditions are stated explicitly in Appendix~\ref{app:ST_multiring_equilibrium}. For a bi-ring, the added string wraps $\phi$, so the transverse matching condition is instead $\Omega_{\psi,\mathrm H}=0$. Equation~\eqref{eq:ST_exact_angular_velocities} gives $\Omega_{\psi,\mathrm H}>0$ throughout the regular non-extremal interior of this seed family in the chosen orientations. The present string therefore gives thermal di-rings but no thermal bi-rings. This obstruction concerns mutual equilibrium, not the stationary bi-rings constructed above, and may be avoided by including transverse string rotation which we do not consider here.

\subsection{Regular extremal limits of the exact seed ring}\label{app:ST_multiring_seed_limits}
The seed and the added ring have independent temperatures, so either horizon can become extremal first. We begin with the exact seed. As in Appendix~\ref{app:ST_extremal_audit}, we retain only limits with finite mass, positive area and $S^1\times S^2$ horizon topology. There are three such extremal sectors in the chosen charge and angular orientations: a BPS ring and two rotational, non-BPS rings. The BPS sector requires taking the charge and shape parameters to their boundaries together~\cite{Suzuki:2024generalring},
\begin{equation}\label{eq:ST_multiring_BPS_scaling}
 b=1-\frac{\varepsilon}{3}\quad,\quad
 \lambda=\chi_1\varepsilon\quad,\quad
 n=\sqrt{\chi_1(\chi-\chi_1)}\,\varepsilon
 \quad,\quad \varepsilon\to0~,
 \qquad 0<\chi_1<\chi~.
\end{equation}
At fixed $\ell_{\rm ST}$ and $\chi$, the limiting geometry is independent of $\chi_1$. It is the restricted supersymmetric ring \eqref{eq:ST_audit_BPS_global}, whose radius and charge parameters are
\begin{equation}\label{eq:ST_multiring_BPS_parameters}
 R_{\mathrm b}=\sqrt2\ell_{\rm ST}\quad,\quad
 q_{\mathrm{seed}}=\sqrt3\ell_{\rm ST}\chi\quad,\quad
 Q_{\mathrm{ring}}=3\ell_{\rm ST}^2\chi(2+\chi)~.
\end{equation}
Its temperature and both angular velocities vanish. Taking $b\to1$ with the shape parameters fixed would not produce this finite-radius ring.

The two rotational sectors instead keep $0<b<1$. For $0<\eta<1$, they are reached by
\begin{equation}\label{eq:ST_multiring_rotational_scalings}
\begin{aligned}
 \Omega_{\phi,\mathrm H}<0:\qquad
 &n\to0\quad,\quad
 \lambda=\frac{1-\eta^2}{\eta^2(1-b^6)}n^2~,\\
 \Omega_{\phi,\mathrm H}>0:\qquad
 &n\to0\quad,\quad
 \lambda=\frac{(1-b^6)\eta^2}{1-b^6\eta^2}~.
\end{aligned}
\end{equation}
They have zero temperature and positive area, but do not saturate the BPS bound,
\begin{equation}\label{eq:ST_multiring_seed_gap}
 \frac{\sqrt3|Q_{(0),\mathrm{seed}}|}{M_{\mathrm{seed}}}
       =\frac{2b}{1+b^2}<1~.
\end{equation}
Both horizon angular velocities remain nonzero. In particular,
\begin{equation}\label{eq:ST_multiring_extremal_rotation_ratio}
 \frac{\Omega_{\phi,\mathrm H}}{\Omega_{\psi,\mathrm H}}
   =\pm\frac{2-(1+b^6)\eta^2}{(1-b^6)\eta^2}
 \quad,\quad \Omega_{\psi,\mathrm H}>0~.
\end{equation}
The numerator and denominator are positive, so the sign distinguishes the two rotational sectors. They are related by a reflection combined with a change of charge orientation~\cite{Suzuki:2024generalring}. We retain both with the charge orientation and asymptotic angular generators fixed. This classification also explains the effect of setting one angular velocity to zero. Imposing $h_{+7}=0$, and hence $\Omega_{\phi,\mathrm H}=0$, on the non-extremal seed leaves only the BPS ring as a regular finite-area extremal endpoint. The condition selects a path within \eqref{eq:ST_multiring_BPS_scaling}, without changing its limiting geometry. The other condition, $\Omega_{\psi,\mathrm H}=0$, cannot hold in the regular non-extremal interior, but is satisfied at the same BPS endpoint and does not lead to regular extremal branch.

\subsection{The four extremal limits of the added ring}\label{app:ST_multiring_probe_limits}
We next take the extremal limits of the local string while keeping its circle separated from the seed horizon. The seed may be non-extremal or belong to any of the three extremal sectors above. In either case its exterior fields remain smooth at the circle, and the balance equation is still \eqref{eq:ST_probe_balance}. The finite-velocity state \eqref{eq:thermal_saturn_finite_limit} gives the force equation \eqref{eq:thermal_saturn_finite_force}, evaluated in the exact ring background. Its distant solutions follow from the asymptotic radial coefficients. Using an outward unit normal,
\begin{equation}\label{eq:ST_multiring_asymptotic_balance}
 K_{00},K_{01},F_0,F_1,H=o(\mathcal R^{-1})
 \quad,\quad K_{11}=-\mathcal R^{-1}+o(\mathcal R^{-1})~.
\end{equation}
Thus balance reduces to $w(Q_{\mathrm R}w+q)=0$, giving the two branches
\begin{equation}\label{eq:ST_multiring_finite_roots}
 w_+\longrightarrow0\quad,\quad
 w_-\longrightarrow-\frac q{Q_{\mathrm R}}
 \quad,\quad Q_{\mathrm R}>q>0~.
\end{equation}
The roots are simple and timelike, so both persist at sufficiently large finite separation in either embedding. 

The remaining two branches have null generators $\ell=E_0+\varsigma E_1$, where $E_0=N^{-1}(\partial_t+\omega\partial_\varphi)$ and $E_1=\mathcal R^{-1}\partial_\varphi$. The charged branch has $\varsigma=+1$, while the opposite magnetic branch has $\varsigma=-1$ and $Q=0$. Their finite stress tensor and currents are \eqref{eq:thermal_saturn_null_currents}, and \eqref{eq:circle_null_balance} determines the null energy. Far from the seed this gives
\begin{equation}\label{eq:ST_multiring_null_roots}
 \lambda\longrightarrow\frac{3(q^2-Q^2)}q>0
 \quad,\quad
 \lambda_{\mathrm B}\longrightarrow3q>0~,
 \qquad |Q|<q~.
\end{equation}
Here $\lambda$ denotes the string's null-wave parameter, distinct from the seed shape parameter used above. Positivity gives nonzero entropy on both branches. Although the magnetic branch has no local electric current, it can have a nonzero Page charge because of the background gauge potential.

These are precisely the four extremal ring limits of Section~\ref{sec:flat_ring_extremal}, continued into the seed background. All have $T_{0,\mathrm r}=0$, with entropy and charges obtained from \eqref{eq:thermal_saturn_finite_entropy} and \eqref{eq:circle_null_charge_substitution}. Their common local charge scale can be reduced while preserving balance, making the added ring thin and weak at a fixed distant circle. 

\subsection{Counting the doubly extremal regimes}\label{app:ST_multiring_extremal_count}
For each of the three extremal seed sectors, either embedding admits the four distant ring limits just described. The count of extremal regimes is therefore $3\times4=12$ for di-rings and another twelve for bi-rings, as shown in Table~\ref{tab:ST_multiring_extremal_count}. Each entry denotes a continuous family of leading stationary configurations.

\begin{table}[htbp]
 \centering
 \begin{tabular}{lcc}
  \toprule
  Extremal seed & Di-ring sectors & Bi-ring sectors \\
  \midrule
  Restricted BPS ring & $4$ & $4$ \\
  Rotational, $\Omega_{\phi,\mathrm H}<0$ & $4$ & $4$ \\
  Rotational, $\Omega_{\phi,\mathrm H}>0$ & $4$ & $4$ \\
  \midrule
  Total & $12$ & $12$ \\
  \bottomrule
 \end{tabular}
 \caption{Combinations with both horizons extremal at leading blackfold order. Each entry contains two finite-velocity branches, a charged null branch and the opposite magnetic null branch. Requiring either seed angular velocity to vanish retains only the BPS row. Mutual thermodynamic equilibrium and an exact completion require the additional conditions discussed below.}
 \label{tab:ST_multiring_extremal_count}
\end{table}

Restricting the parent seed to $\Omega_{\phi,\mathrm H}=0$ retains only its BPS endpoint, giving four doubly extremal sectors per embedding, or eight in total. Requiring $\Omega_{\psi,\mathrm H}=0$ at the extremal endpoint selects the same row.  If only the added ring is extremal, there are four distant branches for a fixed non-extremal seed. If only the seed is extremal, its three sectors can each support the two distant non-extremal velocity branches, giving six combinations per embedding. Restricting the seed to the BPS sector reduces the latter count to two. 

\subsection{Supersymmetry, thermal equilibrium and validity}\label{app:ST_multiring_equilibrium}
The preceding count uses zero temperature and force balance. Supersymmetry is more restrictive. At fixed $0<b<1$, a rotational extremal seed has the finite positive BPS gap \eqref{eq:ST_multiring_seed_gap}, which cannot be removed by a parametrically weak added ring. With a BPS seed, the ring aligned with the supersymmetric Killing vector instead has
\begin{equation}\label{eq:ST_multiring_aligned_BPS}
 k=\partial_t\quad,\quad
 w=-\frac{\omega\mathcal R}{N}\quad,\quad
 \Omega=0\quad,\quad \Phi_0=-\sqrt3\quad,\quad\Phi_1=0~.
\end{equation}
This is the continuation of the $w_+\to0$ branch. The gravitational, electric and magnetic forces cancel by \eqref{eq:multi_susy_no_force} for either embedding. The other finite-velocity branch and the two null branches are not aligned with this Killing vector and do not inherit the seed's supersymmetry. The aligned branch can be matched to the exact supersymmetric construction of Appendix~\ref{app:susy_multirings}. Doing so requires choosing the added ring's local charges according to \eqref{eq:multi_susy_local_dictionary} and imposing the integrability, horizon and causality conditions given there. The charge and area comparison there verifies the exact total BPS relation $M=-\sqrt3Q_{(0)}$ in \eqref{eq:multi_susy_exact_charges}. Thus there is one aligned sector per embedding with a known supersymmetric completion.

The radius bound makes this distinction explicit. In the exact family used in Appendix~\ref{app:susy_multirings}, all rings share the charge ratio $\Lambda=(Q_i-q_i^2)/(2q_i)$. For the restricted BPS seed \eqref{eq:ST_multiring_BPS_parameters},
\begin{equation}\label{eq:ST_multiring_BPS_radius_bound}
 \Lambda=\frac{Q_{\mathrm{ring}}-q_{\mathrm{seed}}^2}
                   {2q_{\mathrm{seed}}}
        =\sqrt{\frac32}\,R_{\mathrm b}~,
 \qquad R_*^2<\Lambda^2~.
\end{equation}
Here $R_*$ is the radius parameter of the added ring in the harmonic description of Appendix~\ref{app:susy_multirings}. The inequality ensures positive horizon area. An outer coplanar ring in this completion must therefore lie in $R_{\mathrm b}<R_*<\sqrt{3/2}R_{\mathrm b}$. An orthogonal ring obeys the same area bound, but need not have $R_*>R_{\mathrm b}$. The additional causality conditions must still be checked numerically.

Thermal equilibrium also selects a smaller set than the stationary count. Even when both temperatures vanish, the electric potentials and both angular velocities must agree. With no independent transverse rotation on the added string, this requires $\Omega_{\phi,\mathrm H}=0$ for a di-ring or $\Omega_{\psi,\mathrm H}=0$ for a bi-ring. Both conditions select the BPS seed. Matching the remaining angular velocity then selects the aligned branch \eqref{eq:ST_multiring_aligned_BPS}, whose completed family has the common electric potential $-\sqrt3$. 

The validity requirements can be stated directly in terms of the exact background. The circle must lie in its regular exterior, with $N^2>0$, $\mathcal R^2>0$ and $|w|<1$ for a timelike branch. In the orthonormal frame $n=E_0$, $E_1=\mathcal R^{-1}\partial_\varphi$, the energy length and the extrinsic curvature along the unit normal $g_{ss}^{-1/2}\partial_s$ are
\begin{equation}\label{eq:ST_probe_validity_local_data}
\begin{gathered}
 r_\epsilon(n)=\frac E{\mathcal C}
       =\frac{\Gamma^2}{\mathcal C}(\epsilon+2zw+pw^2)~,\\
 K_{\hat a\hat b}^{\hat s}
  =\frac1{\sqrt{g_{ss}}}
     \begin{pmatrix}
       N'/N&\mathcal R\omega'/(2N)\\
       \mathcal R\omega'/(2N)&-\mathcal R'/\mathcal R
     \end{pmatrix}~,
 \qquad {}'=\partial_s~.
\end{gathered}
\end{equation}
Here $g_{ss}=g_{yy}|_{x=-1}$ for the di-ring and $g_{ss}=g_{xx}|_{y=-1}$ for the bi-ring. The required dimensionless quantities are
\begin{equation}\label{eq:ST_probe_validity_bounds}
\begin{gathered}
 \frac{r_\epsilon(n)}{\mathcal R}\ll1\quad,\quad
 r_\epsilon(n)\max_{\hat a,\hat b}|K_{\hat a\hat b}^{\hat s}|\ll1
 \quad,\quad r_\epsilon(n)^2\max|R^{\mathrm{bg}}_{\hat\mu\hat\nu\hat\rho\hat\sigma}|\ll1~,\\
 r_\epsilon(n)\max|F^{\mathrm{bg}}_{\hat\mu\hat\nu}|\ll1
 \quad,\quad r_\epsilon(n)\max|\mathcal F^{\mathrm{bg}}_{\hat\mu\hat\nu\hat\rho}|\ll1
 \quad,\quad \frac{r_\epsilon(n)}{L_{\mathrm{var}}}\ll1
 \quad,\quad \frac{r_\epsilon(n)}{d_{\mathrm{sep}}}\ll1~.
\end{gathered}
\end{equation}
The maxima include all physical components in the same orthonormal frame, not just the components entering the force equation. The length $L_{\mathrm{var}}$ is the shortest proper variation scale of the local state and external fields, as in Section~\ref{sec:string_validity}. The separation is the shortest distance to the seed horizon on the constant-$t$ spatial slice,
\begin{equation}\label{eq:ST_probe_separation}
 d_{\mathrm{sep}}=\inf_{\gamma:\Sigma\to\mathcal H_{\mathrm{seed}}}
       \int_\gamma\sqrt{g_{xx}\,dx^2+g_{yy}\,dy^2}~.
\end{equation}
The paths run through the exterior orbit space and end at $y=-1/n$. An extremal horizon can be at infinite proper distance on this slice. This does not remove the local curvature, flux or perturbative-backreaction requirements.

The energy length remains finite in the extremal limits and must include the wave energy. Explicitly,
\begin{equation}\label{eq:ST_probe_extremal_thickness}
 r_\epsilon(n)=
 \begin{cases}
 3\Gamma^2(Q_{\mathrm R}+qw)&\text{finite-velocity branches}~,\\[2pt]
 \dfrac12\left(3q+\lambda+\dfrac{3Q^2}{q}\right)
                              &\text{charged null branch}~,\\[4pt]
 \dfrac12(3q+\lambda_{\mathrm B})&\text{magnetic null branch}~.
 \end{cases}
\end{equation}
Regular local horizons further require $Q_{\mathrm R}>q>0$ on the finite-velocity branches, and $q>0$ with $\lambda>0$ or $\lambda_{\mathrm B}>0$ on the null branches. These restrictions must be imposed together with \eqref{eq:ST_probe_validity_bounds} after solving balance.

Finally, treating the seed as fixed requires the perturbation produced by the added ring to remain small near the seed horizon. We use the conservative restriction $|M_{\mathrm r}|/M_{\mathrm{seed}}\ll1$ in addition to the local conditions above. A small mass ratio alone does not guarantee a small response near an extremal horizon. In particular, control of relative corrections to small temperatures or to departures from extremal potentials requires a separate near-extremal analysis, beyond these leading stationary equations.

\bibliographystyle{JHEP}

\bibliography{references}

\end{document}